\documentclass[a4paper,fleqn,usenatbib]{mnras}

\usepackage{graphicx}	% Including figure files
\usepackage[normalem]{ulem}
\usepackage{booktabs}
\usepackage{newtxtext,newtxmath}
\usepackage[T1]{fontenc}
\usepackage{ae,aecompl}

\usepackage{amsmath}	% Advanced maths commands
\usepackage{amssymb}	% Extra maths symbols

\graphicspath{{figures/}}

\usepackage{float}
\usepackage{isotope}
\usepackage{capt-of}
\usepackage{longtable}
\usepackage{lscape}
\usepackage[usenames,dvipsnames]{color}

\usepackage{subfigure}

\usepackage{adjustbox}
\usepackage{hyperref}

\newcommand{\mppnp}{\textsc{mppnp}}
\newcommand{\MESA}{\textsc{MESA}}

\newcommand{\mzams}{\ensuremath{M_{\rm ZAMS}}}

\newcommand{\teff}{\ensuremath{T_{\rm eff}}}

\newcommand{\msun}{\ensuremath{\, {\rm M}_\odot}}
\newcommand{\lsun}{\ensuremath{\, {\rm L}_\odot}}

\newcommand{\spr}{\mbox{$s$-process}}

\newcommand{\ipr}{\mbox{$i$-process}}
\newcommand{\iprn}{\mbox{$i$ process}}

\newcommand{\czw}{\ensuremath{^{12}\mem{C}}}

\newcommand{\fig}[1]{Fig.\,\ref{#1}}

\newcommand{\tab}[1]{Table\,\ref{#1}}

\newcommand{\sect}[1]{Sect.\,\ref{#1}}
\newcommand{\app}[1]{Appendix \ref{#1}}
\newcommand{\cdr}{\ensuremath{^{13}\mem{C}}}

\newcommand{\mem}[1]{\ensuremath{\mathrm{ #1}}}

\newcommand{\etal}{et~al.\,}

\title[NuGrid stellar data set. II.]{NuGrid Stellar Data Set. II. 
Stellar Yields from H to Bi for Stellar Models with $\mzams = 1$ to $25\msun$ and $Z=0.0001$ to $0.02$}

\author[C. Ritter et al.]{
C. Ritter$^{1,5,7,8}$, \thanks{E-mail: critter@uvic.ca}
F. Herwig$^{1,7,8}$, \thanks{E-mail: fherwig@uvic.ca}
S. Jones$^{2,4,8}$,
M. Pignatari$^{3,8}$, 
C. Fryer$^{4,8}$ and 
R. Hirschi$^{5,6,8}$
\\
$^{1}$Department of Physics \& Astronomy, University of Victoria, Victoria, BC, V8P5C2 Canada. \\
$^{2}$Heidelberg Institute for Theoretical Studies, Schloss-Wolfsbrunnenweg 35, D-69118 Heidelberg, Germany. \\
$^{3}$E. A. Milne Centre for Astrophysics, Department of Physics \& Mathematics, University of Hull, HU6 7RX, United Kingdom. \\
$^{4}$Computational Physics and Methods (CCS-2), LANL, Los Alamos, NM, 87545, USA. \\
$^{5}$Keele University, Keele, Staffordshire ST5 5BG, United Kingdom. \\
$^{6}$Institute for the Physics and Mathematics of the Universe (WPI), University of Tokyo, 5-1-5 Kashiwanoha, Kashiwa 277-8583, Japan. \\
$^{7}$JINA-CEE, Michigan State University, East Lansing, MI, 48823, USA \\
$^{8}$NuGrid collaboration, \url{http://www.nugridstars.org} 
}

\date{Accepted XXX. Received YYY; in original form ZZZ}

\pubyear{2017}

\begin{document}
\label{firstpage}
\pagerange{\pageref{firstpage}--\pageref{lastpage}}
\maketitle

\begin{abstract}
%250 words as required for ApJS, MNRAS
We provide here a significant extension of the NuGrid Set 1 models
in mass coverage and toward lower metallicity, adopting the same physics
assumptions. The combined data set now includes the initial masses
$\mzams/\msun$ = 1, 1.65, 2, 3, 4, 5, 6, 7, 12, 15, 20, 25 for $Z =
0.02, 0.01, 0.006, 0.001, 0.0001$ with $\alpha$-enhanced composition
for the lowest three metallicities. These models are computed with the
\MESA\ stellar evolution code and are evolved up to the AGB, the white dwarf stage, or
until core collapse. The nucleosynthesis was calculated for all
isotopes in post-processing with the NuGrid \mppnp\ code.  Explosive
nucleosynthesis is based on semi-analytic 1D shock
models. Metallicity-dependent mass loss, convective boundary mixing in
low- and intermediate mass models and H and He core burning massive star
models is included.  Convective O-C shell mergers in some stellar
models lead to the strong production of odd-Z elements P, Cl, K and
Sc.  In AGB models with hot dredge-up the convective boundary mixing
efficiency is reduced to accommodate for its energetic feedback. In both
low-mass and massive star models at the lowest metallicity H-ingestion
events are observed and lead to i-process nucleosynthesis and
substantial \isotope[15]{N} production.  Complete yield data tables, derived data
products and online analytic data access are provided.
\end{abstract}


%%%%%%%%%%%%%%%%%%%%%%%%%%%%%%%%%%%%%%%%%%%%%%%%%%

%%%%%%%%%%%%%%%%% BODY OF PAPER %%%%%%%%%%%%%%%%%%
\section{Introduction}
\label{intro}

\noindent 

Stellar yields data are a fundamental input for galactic chemical
evolution models \citep[e.g.][]{romano:10,nomoto:13,molla:15},
hydrodynamic models \citep[e.g.][]{scannapieco:05} and chemodynamic models
\citep[e.g.][]{few:12,cote:13,schaye:15}.
\cite{gibson:02} and \cite{romano:10} showed that results of chemical
evolution models are strongly affected by uncertainties related to the
choice of the yield set: for example, yield sets lead to %\cite 
$0.6\,\mathrm{dex}$ differences in [C/O] ratio and $0.8\,\mathrm{dex}$ for
[C/Fe] in their galaxy models.  These yield studies couple separate yield
sets for massive and low-mass stars.  These two separate sets often
use different stellar evolution codes and different nuclear networks.
In this paper, we present yields based on stellar models of a
range of initial masses and metallicities calculated with the \MESA\ \citep{paxton:11}
stellar evolution code and post-processed with the NuGrid post-processing
network \citep[][P16]{pignatari:13a}.

This work builds upon the %recent 
study by P16, and includes important improvements over this study.
In this work, the same stellar code \MESA\ \citep{paxton:11} is used for the full stellar set, while
the yields set from P16 are calculated
with different stellar evolution codes: \MESA\ for
AGB star models and the Geneva stellar evolution code \citep[GENEC]{eggenberger:08} for massive star
models. In this work, we have extended the set of models by adding more low-mass, intermediate-mass and massive stars:
we provide models also for $\mzams=1\msun$, $6\msun$, $7\msun$ and $12\msun$ stars, including now low-mass supernova progenitors
and super-AGB models, not included in the P16 set. In particular, a finer grid for intermediate-mass stars is important for galactic chemical evolution applications of the yield set, since these stars are important producers of
\isotope[13]{C} and \isotope[14]{N}, in particular at low metallicity
\citep[e.g.][]{siess:10,ventura:11,karakas:12,ventura:13,gil-pons:13,doherty:14b}. Models in the narrow transition mass range from AGB stars to massive stars that may including electron-capture SN \citep{1996ApJ...459..701G,2013ApJ...772..150J,2014ApJ...797...83J}, as well as yields for Type Ia SN are beyond the scope of this work.
Finally, in addition to new masses the yield set is extended by adding models
with three lower metallicities for all initial masses.  Below $Z=0.01$
an $\alpha$-enhanced initial abundance is adopted which leads to
[Fe/H] = -1.24, -2.03 and -3.03 for $Z=0.006, 0.001$ and $Z=0.0001$.
%[Fe/H] = -1.18, -1.96 and -2.97 for $Z=0.006, 0.001$ and $Z=0.0001$.

The yields of massive AGB stars and super-AGB (S-AGB) stars
depend on the nucleosynthesis during hot-bottom burning \citep[HBB,][]{sackmann:92,lattanzio:96,doherty:10,garcia-hernandez:13,ventura:15}.
There are two options to resolve HBB in stellar models: either to
couple the mixing and burning operators or choose time steps smaller
than the convective turnover timescale $\tau_{\rm conv}$ of the envelope
(e.g. $\tau_{\rm conv} \sim \mathrm{hrs}$ for the $\mzams=4\msun$,
$Z=0.0001$ model).  The difficulty in modeling the HBB process is
that the large networks required for the heavy element nucleosynthesis
in HBB require considerable computing time.  But post-processing codes
which decouple mixing and burning operators need to resolve the extremely short mixing time scale when HBB
convective-reactive conditions are relevant. In this work we present
a nested-network post-processing approach in which mixing and burning operators are coupled.
With this approach we accurately calculate stellar yields also for isotopes affected by HBB conditions.
%nucleosynthesis of CNO species and s-process elements also in these HBB conditions.

%In addition to new masses the yield set is extended by adding models
%with three lower metallicities for all initial masses.  Below $Z=0.01$
%an $\alpha$-enhanced initial abundance is adopted which leads to
%[Fe/H] = -1.18, 1.96 and -2.97 for $Z=0.006, 0.001$ and $Z=0.0001$.
Ingestion events are common at low and zero-metallicity in AGB models
of low mass \citep[e.g.][]{fujimoto:00,cristallo:09b}, in He-core
flash in low-metallicity low-mass models \citep[e.g.][]{campbell:10}, and in
S-AGB models in a wide range of metallicities
\citep[e.g.][]{gil-pons:10,jones:16}.  The energy release as well as
nuclear burning on the convective turn-over time scale due to H
ingestion might violate the treatment of convection via mixing-length
theory (MLT) \citep{herwig:01a} and/or the assumption of hydrostatic
equilibrium \citep[e.g. in S-AGB models][]{jones:16}.  The
three-dimensional (3D) hydrodynamic simulations of H ingestion of the
post-AGB star Sakurai's object show that global and non-radial
instabilities can be triggered in such convective-reactive phases can
not be simulated in 1D stellar evolution
\citep{herwig:14}. \citet{herwig:11} and \citet{herwig:01a} also
reported that observational abundances and light curve of Sakurai's
object can not be explained with 1D models based on the MLT. Thus, the
predictive power of 1D stellar evolution models to describe H
ingestion events might be limited. The models nevertheless provide
information about the frequency of such events as well as their potential impact 
on the production of elements.

Yield tables are typically provided in the literature but in order to
trace back the underlying reasons for certain abundance features in
yield tables it is important to have access to the full stellar
models.  In this paper we provide full web access of the stellar
evolution and post-processing data including yield tables and an
interactive interface to analyze and retrieve data.  

The paper is
organized as follows: in Sect.\ \ref{methods} we describe the methods
used to perform the stellar evolution simulations, the semi-analytic models
of the core-collapse supernova (CCSN) shock and
post-processing.  In Sect.\ \ref{stellarevol} we introduce the general
properties of stellar models and features related to low
metallicity.  In Sect.\ \ref{ppnucleo} we analyze the final yields at
low metallicity. The latter are grouped by nucleosynthesis process.  We discuss
our assumptions in Sect.\ \ref{discussion} and compare the results
with available literature.  In Sect.\ \ref{sec:concl} we summarize the
results.

%\newpage

%methods
\section{Methods}
\label{methods}

\noindent The yields presented in this paper have been produced using 1D stellar
evolution calculations and a semi-analytic prescription for CCSN shock
propagation together with a post-processing nuclear reaction network. The
details of three steps are described in this section.

\subsection{Stellar evolution} \label{sec:stellarevol}
The stellar evolution calculations were performed using the \MESA\ stellar
evolution code \citep{paxton:11}, rev.~3709. The AGB models in NuGrid Set~1
\citep{pignatari:13a} were not recomputed, and those models used rev.~3372 of MESA. The AGB models in
this work adopt the same opacities as P16, in which case the two revisions
produce similar results. For example, the time-evolution of H-free core masses
agree to within $0.2\, \%$. A comparison of AGB models of newer \MESA~revisions
with the P16 models is presented in \citet{battino:16}. MESA rev.~3709 was also
used for the massive star models. This is in contrast with P16, who used GENEC \citep{eggenberger:08}.
A detailed comparison of GENEC and MESA (and KEPLER) massive star models at solar
metallicity was performed by \citet{jones:15}, who found that the CO core
masses are within $10\, \%$ to $15\, \%$ of one another and the elemental abundances produced
in the He core by the weak s-process agree within 30~\%.
The physics assumptions up to the end of core He burning in the massive star models are as in \citet{jones:15}.

\subsubsection{Initial composition and nuclear reaction network} \label{sec:iniabu}

\noindent We use solar-scaled initial abundance at $Z=0.02$ and
$Z=0.01$ as in P16, based on \cite{grevesse:93} and with the isotopic
ratios from \cite{lodders:03}. At $Z=0.006$ and below we enhance the
$\alpha$ isotopes \isotope[12]{C}, \isotope[16]{O}, \isotope[20]{Ne},
\isotope[24]{Mg}, \isotope[28]{Si}, \isotope[32]{S}, \isotope[36]{Ar},
\isotope[40]{Ca} and \isotope[48]{Ti}. The $\alpha$ enhancements were
derived from fits of halo and disk stars from \cite{reddy:06} and
references therein.  For each enhanced isotope ${\alpha}$ we apply
Eq.\ \ref{eq:alpha_ini} where $A_{\alpha}$ and $B_{\alpha}$ were
derived from the fits for metallicities $-1 \leq [\mathrm{Fe/H}] \leq
0$ \citep[][]{reddy:06}. For $[\mathrm{Fe/H}]< -1$ we
assume a constant $[X_{\alpha}/\mathrm{Fe}]$ of $[X_{\alpha}/\mathrm{Fe}] = -A_{\alpha} + B_{\alpha}$.
%instead of Eq.~\ref{eq:alpha_ini}.

\begin{equation}
[X_{\alpha}/\mathrm{Fe}] = A_{\alpha} [\mathrm{Fe/H}] + B_{\alpha} 
\label{eq:alpha_ini}
\end{equation}

\noindent For isotopes of Ne, S and Ar values from \cite{kobayashi:06} were
adopted.  The resulting $[X_{\alpha}/\mathrm{Fe}]$ and mass fractions for
$Z=0.0001$ are shown in \tab{tab:iniabuset1p5}.  The fit result gives $\mathrm{[O/Fe]} =
0.89$ which is close to the top of the [O/Fe] distribution but within the maximum
given in \citet[][]{reddy:06}.  For the initial abundance of Li in AGB models
with $\mzams>3\msun$ we choose as a lower limit the Li plateau
\citep{sbordone:10}. 
%In other stellar models the Li abundance was accidentally scaled as above.
In other stellar models the initial Li abundance was unintentionally scaled
down with metallicity as other light elements and unrealistic values were adopted.
An overview of the model assumptions is presented in the following sections, and a comparison with P16 is given in \tab{tab:overview_assump}.

In the low-mass stellar models up to $\mzams=3\msun$ we use the same
network in \MESA\ as in P16 (\texttt{agb.net}). For the massive AGB
and super-AGB models ($4\msun\leq\mzams\leq7\msun$) we use the network
\texttt{agbtomassive.net} which includes an extended network for C, O
and Ne burning and relevant electron-capture reactions. No significant rate updates have been adopted compared to P16.
%C-burning rate physics are the same??
The nuclear reaction network for stellar models with masses
$M_\mathrm{zams}\geq12\msun$ is the same as in \citet[][their Table\,2]{jones:15} from the pre-main sequence until the depletion of oxygen in
the core, at which point the network is reduced to \texttt{approx21.net} to
follow Si burning and deleptonisation in the Fe core.

\subsubsection{Mass loss}
\label{method_massloss}
Semi-empirical prescriptions for mass loss \citep[e.g.][]{vassiliadis:93,vanloon:05b} are still commonly used in stellar evolution. 
In order to stay consistent with P16 we apply the mass loss prescription by \cite{reimers:75}
for the red giant branch phase and the prescription of \cite{bloecker:95} for the AGB phase.
Both prescription are functions of the mass, luminosity and radius of the stellar model.
The efficiency parameter $\eta_{\rm Bloecker}$ is increased to mimic the effect of the C-rich dust-driven phase
as described in P16. A more realistic hydrodynamic approach to mass loss models \citep[e.g.][]{mattsson:10} in combination with observational calibrations taking into account better data now available \citep[e.g.][]{rosenfield:14} should ultimately be deployed for yield calculations. 

\begin{figure} %[!htbp]
\centering

\resizebox{8.0cm}{!}{\includegraphics{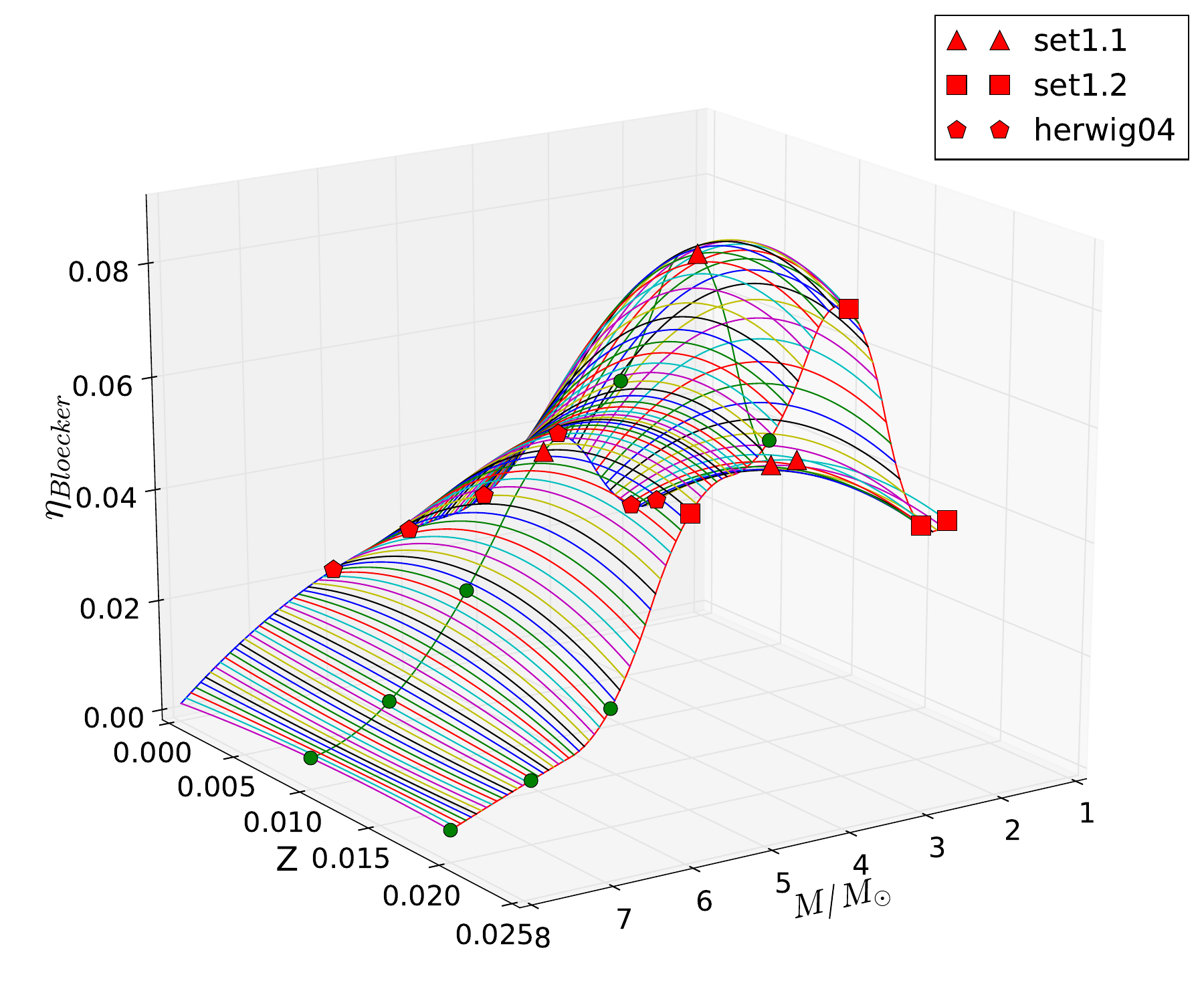}}
\resizebox{8.0cm}{!}{\includegraphics{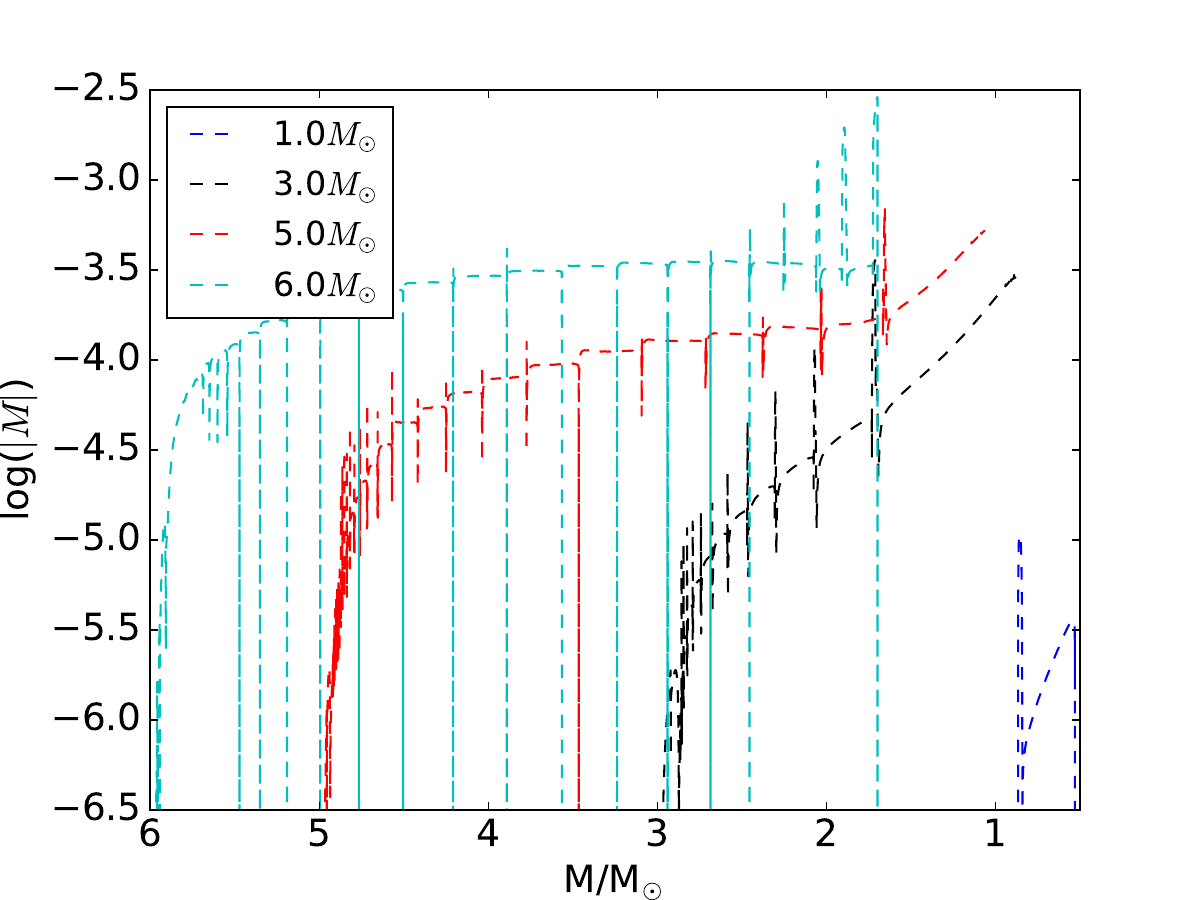}}
\
\caption{3D spline fit of  $\eta_{\rm Bloecker}$ dependent of mass and metallicity based on \citet{herwig:04a} and P16 (top). The green circles represent additional ad-hoc values.
Mass loss in $\msun/{\rm yr}$ for stars of $Z=0.0001$ based on the mass-metallicity fits of $\eta_{Bloecker}$ (bottom).
}
\label{fig:final_massloss_fit}
\end{figure}

Our approach  here aims to bridge the mass loss choice of P16 with that of \citet[H04]{herwig:04a} who adopted a metallicity dependent mass loss based on \cite{vanloon:00}. 
Since the H04 and these $Z=0.0001$ \MESA\ models are slightly different we derive values of $\eta_{\rm Bloecker}$ to be used in the \MESA\ models to obtain the same mass loss as in H04. 
We then fit $\eta_{\rm Bloecker}$ in the mass-metallicity plane to be constrained by the mass loss adopted in P16 for $Z=0.02$ and $0.01$ and by H04 for $Z=0.0001$. 
The resulting spline fit of $\eta_{\rm Bloecker}$ in the mass-metallicity plane is shown in \fig{fig:final_massloss_fit}.
We have added ad-hoc values for stellar models of $\mzams/\msun$ = 4, 6, 7, 8
for solar and half-solar metallicity to extrapolate the general trend of decreasing $\eta_{\rm Bloecker}$ at higher initial mass. 
The fit corresponds to the general notion that  $\eta_{\rm Bloecker}$, and with it the mass loss, 
 decreases for low-mass AGB stars with decreasing metallicity \citep[]{willson:00}. This contrasts with the observational findings of shorter
 AGB lifetimes with lower metallicity in low-mass AGB stars  \citep{rosenfield:14}. 
%This trend is in agreement with observational evidence of shorter AGB lifetimes with lower metallicity in low-mass AGB stars 

The mass loss prescription adopted in the massive star models depends on the effective temperature $\teff$ and
the surface hydrogen mass fraction X(H) as in \cite{glebbeek:09}. 
For $\teff\geq1.1\times10^4\, \mathrm{K}$ and $X({\rm H})\geq0.4$ we adopt the mass loss rate of \cite{vink:01}. At lower
temperatures the \cite{vink:01} rate transits into the \cite{dejager:88} rate and the latter is adopted below $\teff=10^4\, \mathrm{K}$.
If $X({\rm H})<0.4$ we adopt either \cite{nugis:00} when  $\teff<10^4\, \mathrm{K}$, otherwise \cite{dejager:88}.
The \cite{nugis:00} and \cite{vink:01} rates depend explicitly on
metallicity. See \cite{glebbeek:09} for further details.
A correction factor of 0.8 is adopted for mass loss rates of massive star models as deduced for MS OB stars in \cite{maeder:01}.
%This mass loss is similar to the one adopted in P16 \citep{jones:15}.}

\subsubsection{Hot-bottom burning} \label{subsecHBB}

\noindent HBB is the activation of the
CNO cycle at the bottom of the convective envelope in massive AGB and S-AGB stars \citep[][]{scalo:75,sackmann:92}.
%For stars of lower metallicity higher temperatures in the envelopes
%lead to the activation of HBB at lower initial mass which makes HBB more common at lower Z. 
Higher temperatures in the AGB envelopes at lower metallicity lead to the activation of HBB
at lower initial mass compared to AGB models of higher metallicity. This increases the number of stars
which experience HBB with decreasing metallicity.

During HBB the mixing timescale of the convective envelope $\tau_{\rm conv}$ and
nuclear timescales of CNO p-capture reactions $\tau_{p}$ become similar as shown for the
$\mzams=4\msun$, $Z=0.0001$ model in \fig{fig:hbb_timescales}. 
$\tau_{\rm conv}$ is calculated as $\tau_{\rm conv}=l_{\rm MLT}^2/D$ where $D$ is the diffusion coefficient and $l_{\rm MLT}$ is the
mixing length according to MLT.
The coupling of mixing and burning operators in stellar evolution codes allow to resolve HBB
correctly. Typically, post-processing codes decouple mixing and burning in %monsoon? \citep[][]{cristallo:15} in
order to solve differential equations for large reaction networks including
heavy elements. To model HBB in the decoupled approach it is necessary to
resolve the mixing timescale at the bottom of the convective envelope. This is
just hours, for example in this $\mzams=4\msun$, $Z=0.0001$ model
(\fig{fig:hbb_timescales}), which is short compared to the interpulse phases of
tens of thousands of years. \cite{cristallo:15} calculate heavy elements with a
large network in their stellar evolution code and approximate CNO production
due to HBB with a burn-mix-burn step. Our solution is to solve the coupled
reaction and diffusion equations for a subset of important isotopes (see \sect{sec:nugrid_codes}).

\subsubsection{Convective boundary mixing treatment} \label{subsecCBM}

\noindent We apply convective boundary mixing (CBM) at all convective
boundaries of the AGB models. CBM is modeled with an exponential-diffusive convective boundary mixing
model \citep[][]{freytag:96,herwig:00}.  A CBM efficiency of $f=0.014$ is used
at all convective boundaries of AGB models except for the bottom of the pulse-driven
convective zone (PDCZ)  and during the third dredge-up (TDUP) of the
thermal-pulse (TP)-AGB stage.  Motivated by 2D and 3D simulations of
\cite{herwig:07a} a lower CBM efficiency of $f_{\rm PDCZ}=0.008$ is applied at the
PDCZ bottom boundary.  An increased mixing efficiency of $f_{\rm CE}=0.126$ is
applied at the bottom of the convective envelope during the TDUP which is
calibrated for low-mass stellar models to produce the \cdr\ pocket
\citep{herwig:03}.  This approach is the same as in P16.

CBM is only accounted for in the massive star models from the pre-main sequence up to the end of core He burning. 
It is implemented as the exponential diffusion model of \citet[][]{freytag:96} with $f=0.022$ at all convective boundaries except for the bottom of convective shells in which nuclear fuel is burning, where $f=0.005$ was used.
From the extinction of core He burning, which we have defined as the time when the central mass fraction of helium falls below $10^{-5}$, %$X_\mathrm{c}(^4\mathrm{He}) < 10^{-5}$,
$f$ is set to zero, equivalent to assuming no CBM.

\begin{figure}
\centering
\includegraphics[width=\columnwidth]{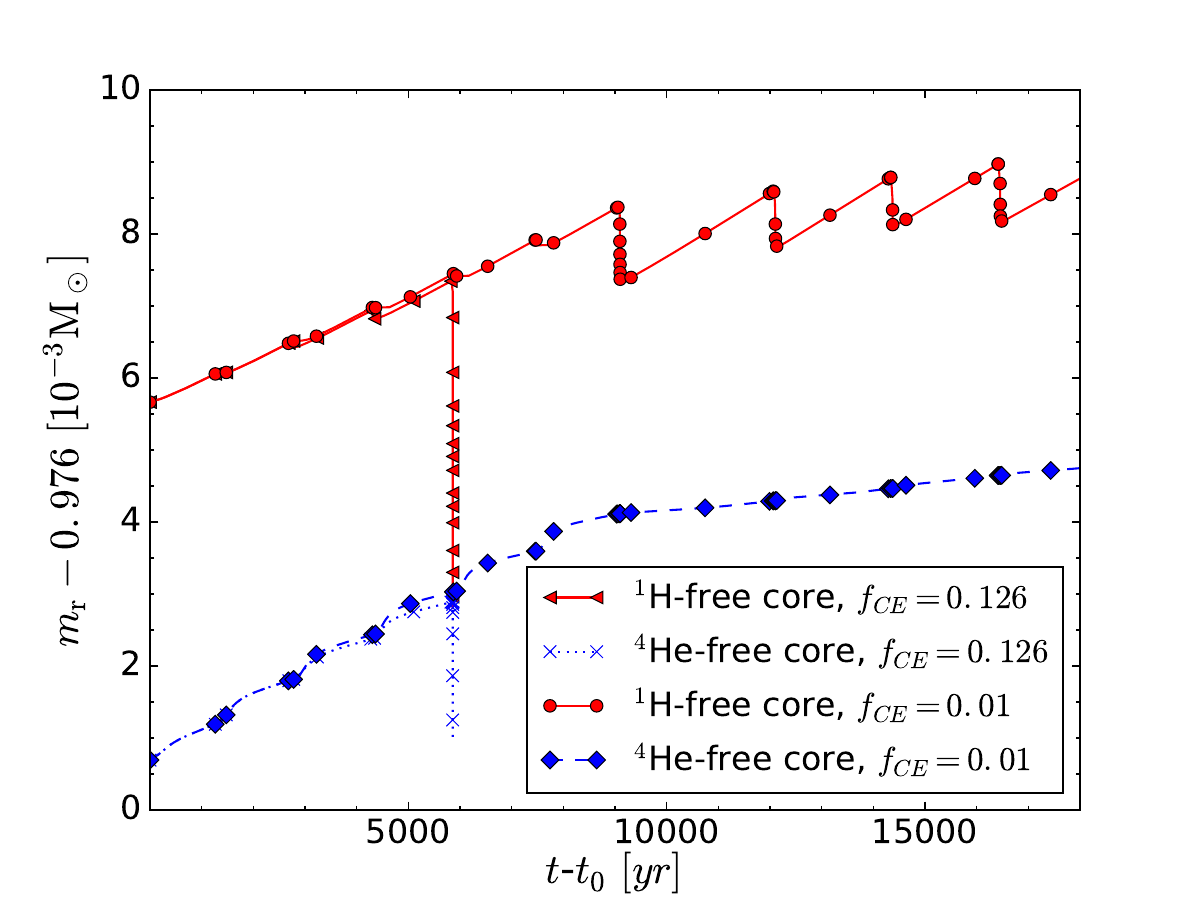}
\includegraphics[width=\columnwidth]{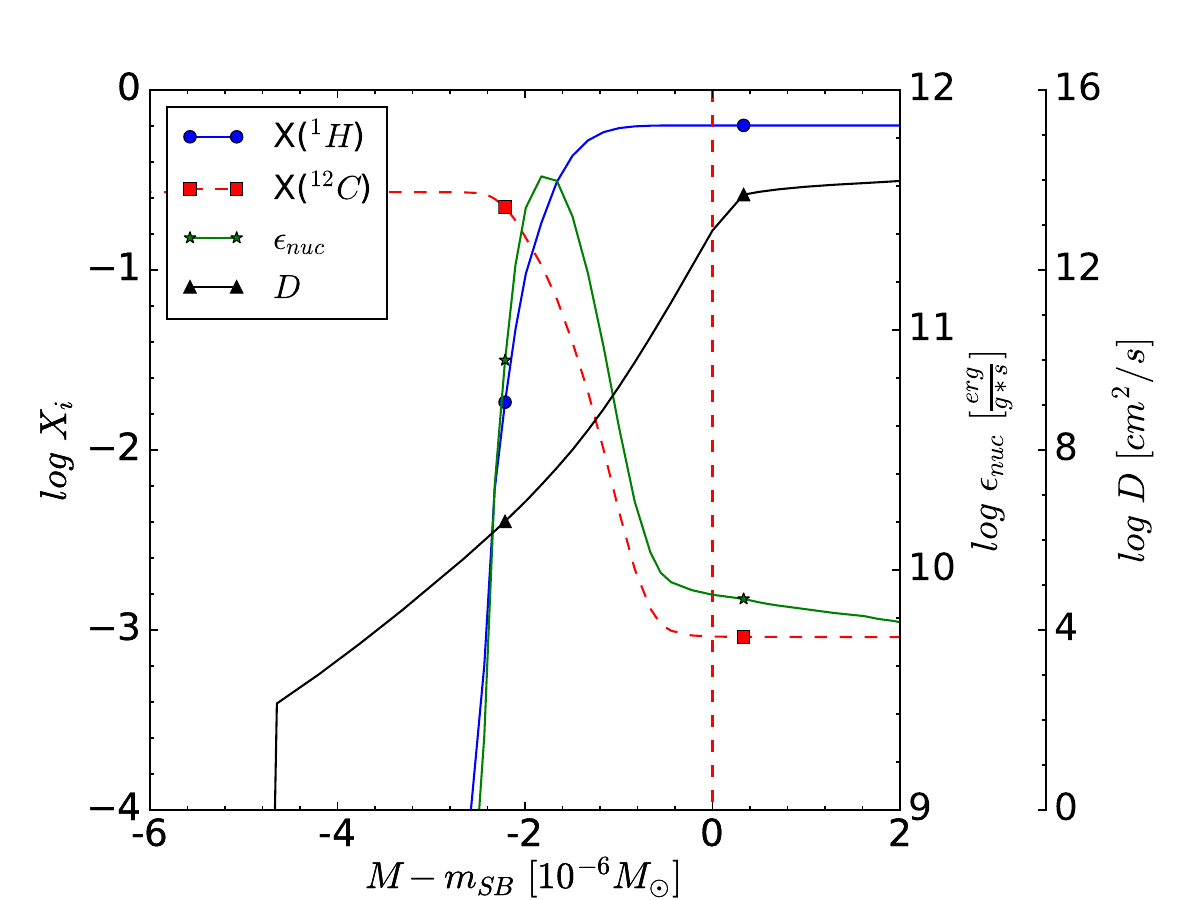}
\caption{
Evolution of H-free and He-free cores for $f_{\rm CE}=0.126$ and $f_{\rm CE}=0.01$ for $\mzams=5\msun$, $Z=0.0001$ models (top).
$t_0$ marks the beginning of the TP-AGB phase. Abundance profile and energy release due to H mixing
through the bottom of the convective envelope during HDUP at $\approx(t-t_0)=7800 \mathrm{yr}$ for the case of $f_{\rm CE}=0.01$ (bottom).
The vertical dashed line marks the position of the mass coordinate of the Schwarzschild boundary $m_{\mathrm{\rm SB}}$.} %after 6thT
\label{fig:HDUP}
\end{figure}

Corrosive H-burning during TDUP in low-metallicity massive AGB stars leads to an increase
of the TDUP efficiency and is referred to as hot dredge-up
\citep[HDUP,][]{herwig:04a}.  The application of CBM at the bottom of the
convective envelope results in strong burning of the mixed protons below the
envelope and extreme TDUP efficiencies in these massive AGB models at low metallicity.
In a $\mzams=5\msun$, $Z=0.0001$ test model with CBM parameter $f_{\rm CE}=0.126$
used for the \cdr-pocket formation in low-mass AGB stars the TDUP penetrates
into the C/O core after the sixth TP as shown in the Kippenhahn diagram in
\fig{fig:HDUP}. This finding is in agreement with \cite{herwig:03c} who found
that the HDUP can penetrate into the C/O core and terminate the AGB phase
\citep[see also][]{goriely:04}. The abundance profile during the  TDUP  at the
bottom of the convective envelope shows the peak of nuclear burning in the CBM
region which steepens the radiative gradient and hence leads to a deeper
penetration of the envelope into the He intershell (\fig{fig:HDUP}).
\cite{karakas:10} models do not experience HDUP because the authors do not model CBM
in the stellar evolution simulation. Instead, they introduce an ad-hoc partial
mixing zone for the formation of the \cdr-pocket in the post-processing simulations.

%circumvent the issue of HDUP by applying an ad-hoc partial
%mixing zone for the formation of the \cdr-pocket only in the post-processing
%simulations and not during the stellar evolution. 

One way to reduce the vigour of H burning during the HDUP is the reduction of $f_{\rm CE}$. 
The efficiency of CBM at the lower boundary
of the convective envelope in massive and S-AGB is not known. Investigations 
of the impact of CBM efficiency on structure and nucleosynthesis such as for S-AGB models by \cite{jones:16}  are required.
A physical interpretation of the assumption of a reduced CBM is based on the buoyancy of the mixed and burning material 
which hinders boundary mixing.
The situation is similar to the bottom of convective burning shells in the late stage of massive stars where
the energy release leads to a lower CBM and a stiffer boundary \citep[e.g.][]{cristini:16,jones:17}.
Following \citet{herwig:04a}, we limit CBM by reducing $f_{\rm CE}$ here to $0.01$ for $\mzams\geq4\msun$  models if the dredge-up after a thermal pulse is hot (\tab{tab:f_table}).
With this approach we prevent the termination of the AGB phase due to too extreme H burning during the TDUP. 
The limiting of $f_{\rm CE}$ in massive and S-AGB models is new in this work, compared to P16. Other choices of CBM efficiencies are as in P16 (\tab{tab:overview_assump}).

\subsection{Semi-analytic CCSN explosions} \label{sec:method:exp}

\noindent We use a semi-analytic approach for core-collapse supernova explosions as described in P16.  
The method drives a shock off the proto-neutron star based on a mass cut derived
from \citet[F12]{fryer:12}.  
%We then initiate a shock at the top of the mass 
%cut and propogate it through the star using Sedov blast wave assumptions.
The mass cuts are mass- and metallicity dependent and
are provided for delayed and a rapid explosion prescription. The mass coordinates
based on these models are shown in \tab{tab:coo_fallback}. For some massive star models
such as the $\mzams=15\msun$, $Z=0.006$ model the mass cut is deeper located than
than the outer edge of the Fe core as visible from the Fe-core masses in \tab{tab:fe_cores}.

One of the big uncertainties in the yields is the position of the
mass cut.  The data from F12 were based on fits to the
stellar structures produced by comparing the models from a range of
stellar evolution codes \citep{woosley:02,limongi:06,young:09}.  
These mass-cut prescriptions were then validated against the compact remnant 
mass distribution \citep{belczynski:12}. For these stellar evolution models, the mass cut is fairly similar for
models with $\mzams<25\msun$. However, in particular for the $\mzams=12\msun$ model,
the core from the \MESA\ model is much larger than that produced by the
Kepler code.  This corresponds to much higher densities in the inner
$2\msun$ and, based on the F12 results,
%with the F12 analysis, 
we expect the 
\MESA\ $\mzams=12\msun$ models to collapse down to a black hole rather than explode to
produce a low-mass neutron star. In this case, the $\mzams=12\msun$ stars would not provide SN yields and would contribute to the chemical evolution of the Galaxy only by stellar winds. In part, these results for the $12\msun$ stellar progenitors are caused by the use %this is caused by the use
of a small nuclear network in the \MESA\ code during Si burning. 
%\mpcom{CAN YOU EXPLAIN WHY WITH ONE SENTENCE? OR REFER TO WHERE THIS ISSUE HAVE BEEN DISCUSSED.}

At earlier times, the \MESA\ models with $\mzams=12\msun$ and GENEC models
with $\mzams=15\msun$ of P16 are very similar. Therefore, for the \MESA\ models with
$\mzams=12\msun$ we also use the mass cut prescription of F12 under the assumption
of $\mzams=15\msun$ as adopted for these GENEC models. % with initial mass of
This allows to provide a SN yield set of
these \MESA\ models at all metallicities. For more massive \MESA\ models, we use the mass cut prescription of F12 as in P16.
The same semi-analytic CCSN prescription as in P16 is applied, except the modification of the $\mzams=12\msun$ models.
%For the \MESA\ models with
%$\mzams=12\msun$ we use the mass cut prescription of F12 under the assumption
%of $\mzams=15\msun$ as adopted for these GENEC models. % with initial mass of
%This allows to provide a first estimate of the yields of
%these \MESA\ models and this assumption is applied to all metallicities.  The
%differences between the more massive \MESA\ models and GENEC models of P16 is
%less extreme and, for these models we use the mass cut prescription of F12 as in P16.

\subsection{Nucleosynthesis code and processed data}
\label{sec:nugrid_codes}

\noindent The temperature, density and diffusion coefficient (from the mixing
length theory of convection along with the convective boundary mixing model)
$T(m),~\rho(m)$ and $D(m)$ in the MESA stellar evolution models are saved every
time step and post-processed with the multi-zone NuGrid code \mppnp~using and
the same reaction network as in P16. To summarize: every stellar evolution time
step, the 1097-isotope nuclear reaction network is solved using a first-order
Newton-Raphson backward Euler integration, which is followed by an implicit
diffusion solve. The network adapts the problem size every time step (and every
computational grid cell) depending upon the reaction flux of each isotope at
current state. The AGB models of P16 were not post-processed again, but are part of the updated analysis presented here.

In \sect{subsecHBB} we described issues that arise with such an
operator-splitting method during hot bottom burning in models of
massive AGB and super-AGB stars.  To predict realistic abundances in
these conditions we have implemented a nested-network method to solve
the coupled mixing and burning equations for a small network which
includes species which are affected by HBB.  We solve the small
network for zones of the convective envelope and a large decoupled
network for the whole stellar model. After each time step the
abundances from the coupled solution replace the abundances from the
large network.  The coupled solution is merged into the large network
by normalizing the total abundance of isotopes of the small network to
be equal to the total abundance of the corresponding isotopes of the
large network. Here, the coupled solution includes mixing and burning, and as 
in all of the post-processing the
  structure is provided by MESA.  Just as a reminder, MESA solves
 structure, mixing and burning operators together.  The small network
models the Cameron-Fowler transport mechanism and \isotope[7]{Li}
production \citep{cameron:71}, CNO, NeNa and MgAl cycles and includes
isotopes up to \isotope[35]{Cl} similar to \cite{siess:10}. Heavier
isotopes, which are only included in the large network, do not take
part in HBB nucleosynthesis according to the present state-of-the-art
\citep[e.g. review by][]{herwig:05}. As such we don't expect the
heavier isotopes to be affected by our choice of decoupling of burning
and mixing.

\begin{figure}
\centering
\includegraphics[width=\columnwidth]{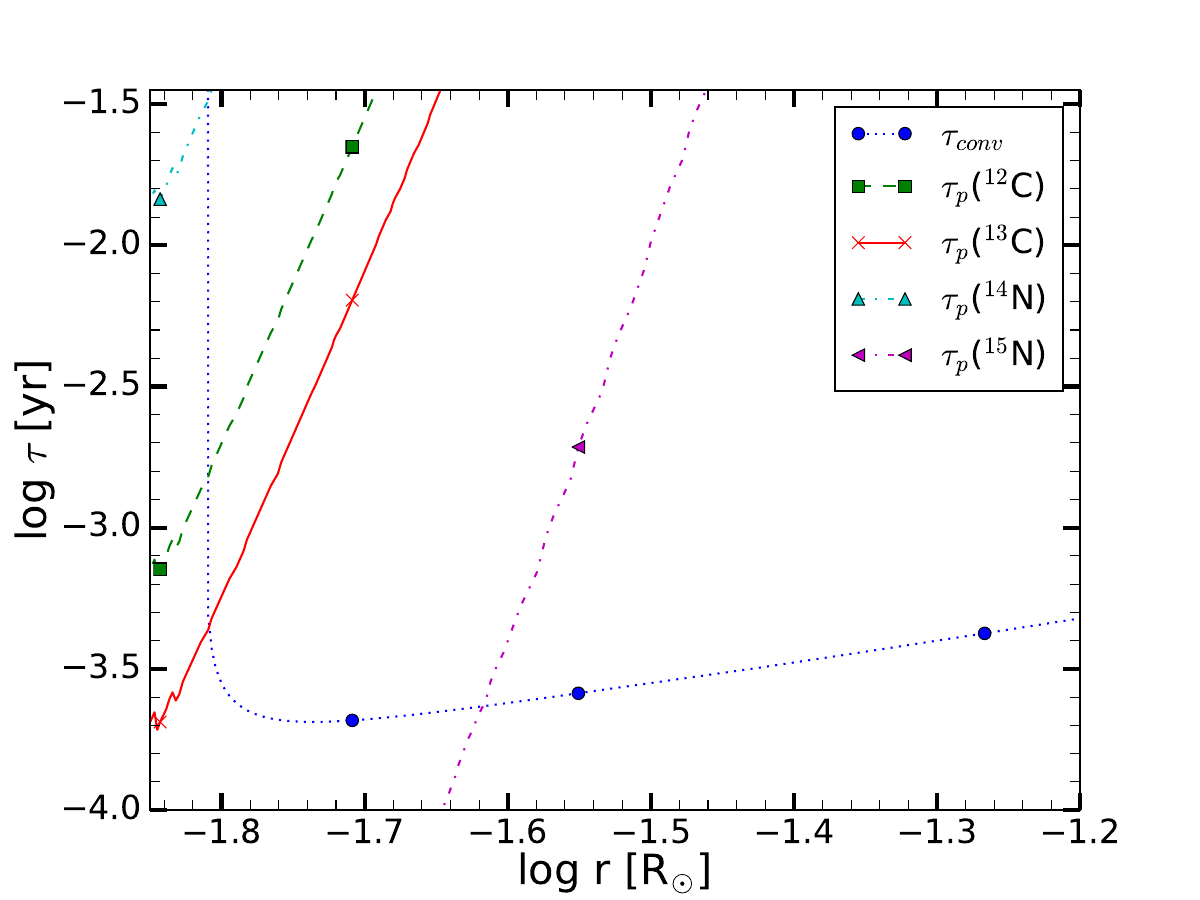}
\includegraphics[width=\columnwidth]{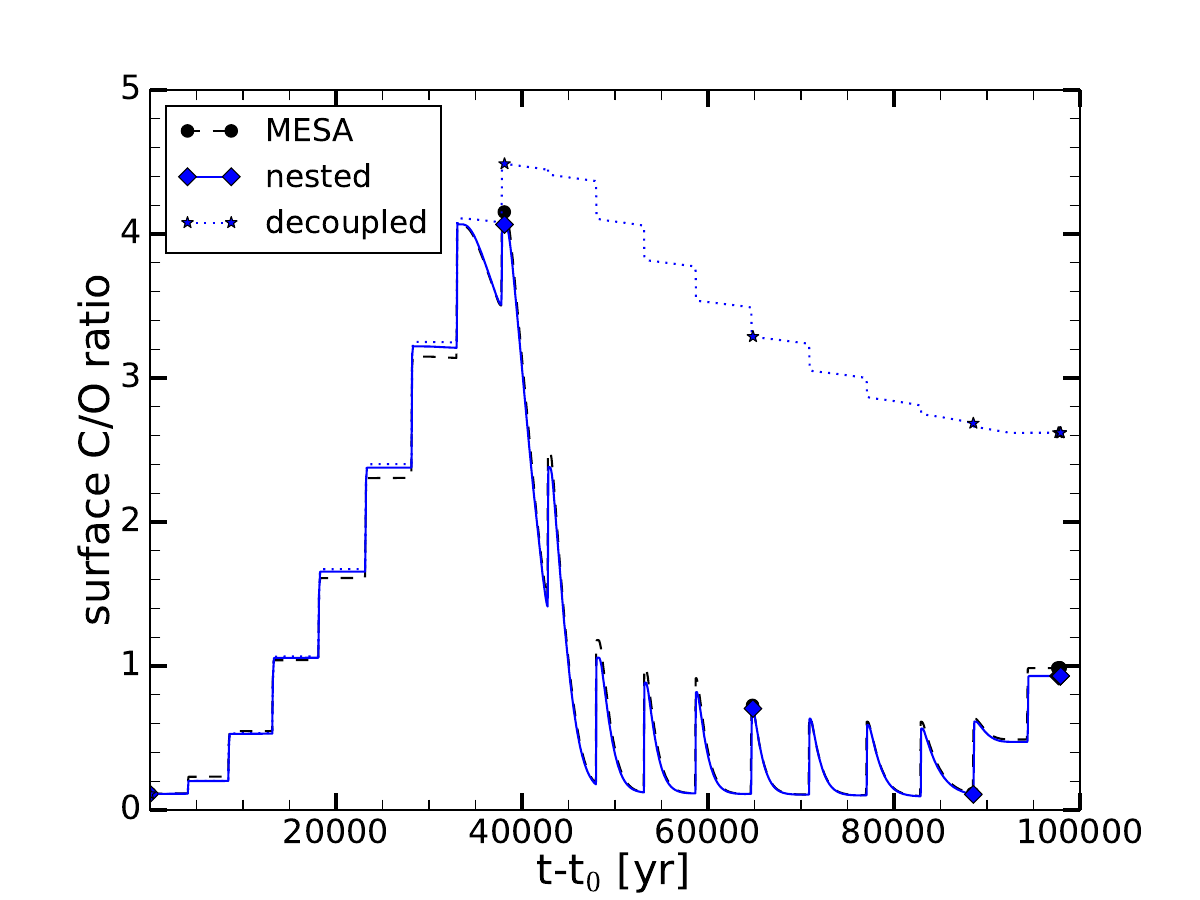}
\includegraphics[width=\columnwidth]{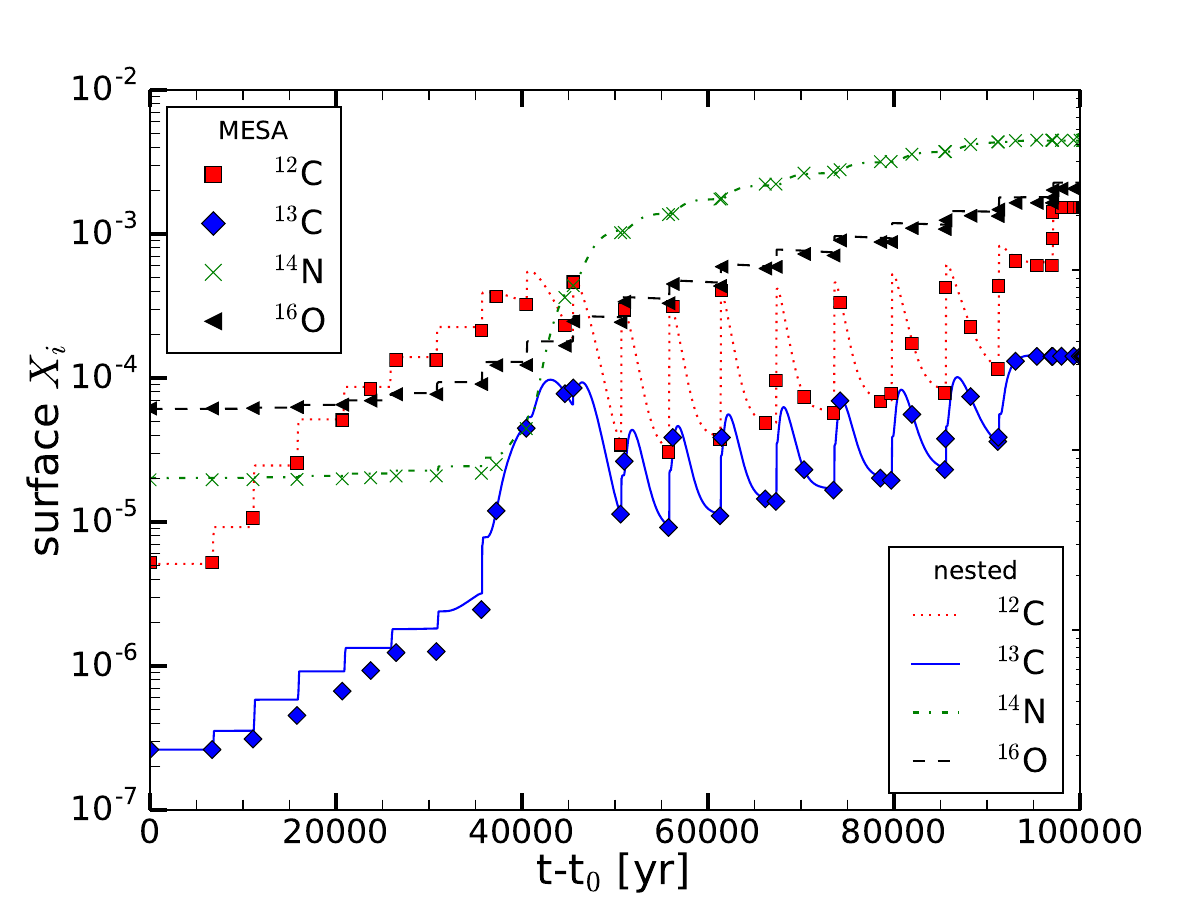}
\caption{Convective turnover timescale $\tau_{\rm conv}$ and CNO reaction timescales $\tau_{p}$ relevant for HBB at the bottom of the convective envelope of the $\mzams=4\msun$, $Z=0.0001$ model (top).
The evolution of the surface C/O number ratio of this stellar model based on the coupled solution of MESA, on the nested-network method and the decoupled method (middle). $t_0$ marks
the beginning of the TP-AGB phase. Surface CNO abundances from the nested-network method in comparison with abundances from MESA (bottom).} 
\label{fig:hbb_timescales}
\end{figure}

We compare of the surface C/O ratio of the $\mzams=4\msun$, $Z=0.0001$ model of the coupled solution with
the nested-network solution and the decoupled solution
(\fig{fig:hbb_timescales}).  Our nested-network method results in the same
evolution of the surface C/O ratio.  The decoupled solution based time steps as
given for the coupled solution of \MESA\ strongly overestimates the surface C/O
ratio compared to the coupled solution from \MESA. 
We find good agreement of the surface abundance of  %\isotope[12]{C}, isotope[13]{C}, \isotope[14]{N},\isotope[16]{O} 
CNO isotopes based on our nested-network method in comparison with predictions from \MESA\ (\fig{fig:hbb_timescales}).

The final stellar yields of CNO isotopes based on the nested-network method are similar 
to \citet[H04]{herwig:04a} and \citet[K10]{karakas:10} who couple mixing and burning (\tab{hbb_yield_table}). 
Neither study includes s-process species, although more recent work by \cite{karakas:16} for $0.007 < Z < 0.03$ does now include heavy elements.
The high \isotope[12]{C}/\isotope[13]{C} ratio of the decoupled solution shows that HBB is not properly resolved.
Even larger values of \citet[C15]{cristallo:15} could be due to resolution issues during HBB with the mix-burn-mix approximation.
%are likely a result of their  mix-burn-mix approximation instead of a coupled network.
The nested-network approach predicts Li production via HBB as well because Cameron-Fowler mechanism is resolved. In summary,  the nested-network method allows to predict Li, CNO isotopes and heavy elements in these HBB  stellar models.

The total stellar yield of element/isotope $i$ of a stellar model with initial mass $m$ includes the yield from stellar winds and the SN explosion as in P16.
The yield ejected by stellar winds ${\rm EM}^{\rm wind}_{\rm im}$ is calculated as
\begin{equation}
{\rm EM}^{\rm wind}_{\rm im} = \int^{\tau(m)}_0 \dot{M}(m,t)\,X^{S}_i(m,t) dt \label{eqn:wind}
\end{equation}
where $ \dot{M}(m,t)$ is the mass loss rate, $X^{S}_i(m,t)$ is the mass fraction of the 
element/isotope $i$ at the surface and $\tau(m)$ is the stellar lifetime.
The yield from the SN ejecta ${\rm EM}^{\rm SN}_{im}$ is derived as 
\begin{equation}
{\rm EM }^{\rm SN}_{\rm im} = \int^{m_{\tau}}_{M_{{\rm rem},m}}\, X_i(m_r) dm_r \label{eqn:sn}
\end{equation}
where $X_i(m_r)$ is the mass fraction of element/isotope $i$ at mass coordinate $m_r$ and $M_{{\rm rem},m}$
is the remnant mass.
Pre-SN yields are calculated as ${\rm EM}^{\rm SN}_{\rm im}$ but without taking into account the nucleosynthesis from the SN shock.
Instead, the ejecta of matter at the point of collapse is considered.
The overproduction factor $OP_{\rm im}$ of element/isotope $i$ of the stellar model with initial mass $m$ is calculated as
\begin{equation}
{\rm OP}_{\rm im} = \frac{{\rm EM}_{\rm im}}{X_i^0 M_{\rm ej}}
\end{equation}
where ${\rm EM}_{\rm im}$ and $X_i^0$ is the total ejected mass and initial mass fraction of element/isotope $i$ respectively.
$M_{\rm ej}$ is the total ejected mass.

%We distinguish between yields only
%from stellar winds, wind yields plus pre-SN yields and the final
%yields which include wind yields plus SN yields (P16). 

%%% Tables for Methods

\begin{table}
\begin{center}
\begin{tabular}{ccc}
\hline
Isotope & [$X_i$/Fe] & $X_i$\\
\hline
\isotope[12]{C} & 0.562 &1.25E-05\\
\isotope[16]{O}  & 0.886 &7.41E-05\\
\isotope[20]{Ne} & 0.5 &5.75E-06\\
\isotope[24]{Mg}& 0.411&1.51E-06\\
\isotope[28]{Si}& 0.307 &1.51E-06\\
\isotope[32]{S}& 0.435 &1.09E-05\\
\isotope[36]{Ar}& 0.3 &1.64E-07\\
\isotope[40]{Ca}& 0.222 &1.21E-07\\
\isotope[48]{Ti}& 0.251 &5.38E-09\\
\hline
\end{tabular}
\end{center}
\caption{Mass fractions of $\alpha$-enhanced isotopes for $Z=0.0001$ derived from \citet{reddy:06} and \citet{kobayashi:06}. The solar normalization based on \citet{grevesse:93} and \citet{lodders:03} as introduced in \sect{sec:iniabu}.}
\label{tab:iniabuset1p5}
\end{table}

\begin{table*}
\begin{center}
%\tabletypesize{\small}

\begin{tabular}{p{0.2\textwidth}p{0.6\textwidth}p{0.1\textwidth}}
\hline
Method  &  Comparison   & Reference   \\ 
\hline
Stellar evolution code  & \MESA\ rev. 3709 is used for AGB models and massive star models. 		
		         P16 uses \MESA\ rev. 3372 for AGB models and GENEC for massive star models. & \sect{sec:stellarevol} \\
Initial abundance & Adoption of $\alpha$-enhancement for stellar models with $Z<0.01$, otherwise solar-scaled abundance as in P16. & \sect{sec:iniabu} \\
\MESA\ network & Same network as in P16 except for massive AGB and S-AGB models which have an extended network for C burning. & \sect{sec:iniabu} \\
Mass loss & Introduction of a Z-dependence of the AGB mass loss. The massloss of massive-star models is as in P16. & \sect{method_massloss} \\
CBM model & Massive and S-AGB models have a reduced CBM efficiency at the bottom of the convective envelope compared& \\
 & to AGB models of P16. & \sect{subsecCBM} \\
CCSN prescription & Same prescription as in P16 except that the $12\msun$ models have the remnant mass of the $15\msun$ models. & \sect{sec:method:exp} \\
HBB  & HBB in AGB models is modeled with a nested-network approach in which burning and mixing are coupled during 
   post-processing in contrast to P16. & \sect{sec:nugrid_codes} \\ 
Post-processing code & Post-processing in this work is done with MPPNP network as in P16. & \sect{sec:nugrid_codes}\\
% \noalign{\smallskip}
\hline
\end{tabular}
%}
\caption{Overview and comparison of stellar model assumptions of this work with P16.}
\label{tab:overview_assump}
\end{center}
\end{table*}

\begin{table}
\begin{center}
%\scalebox{0.0}{
\begin{tabular}{cccc|cccc}
\hline
 \noalign{\smallskip}
    \multicolumn{4}{c}{$\mzams<4\msun$} &  \multicolumn{4}{c}{$\mzams\geq4\msun$}\\
    \cmidrule(lr){1-4}\cmidrule(lr){5-8} 
    \multicolumn{2}{c}{f$_{\rm CE}$} &  \multicolumn{2}{c|}{f$_{\rm PDCZ}$} & \multicolumn{2}{c}{f$_{\rm CE}$} &  \multicolumn{2}{c}{f$_{\rm PDCZ}$} \\  
    \cmidrule(lr){1-2}\cmidrule(lr){3-4}\cmidrule(lr){5-6}\cmidrule(lr){7-8} 
    burn & non-burn & \multicolumn{2}{c|}{burn} & burn & non-burn & \multicolumn{2}{c}{burn} \\  
    \hline  
    0.014 & 0.126 & \multicolumn{2}{c|}{0.008}  & 0.0035 & 0.126 & \multicolumn{2}{c}{0.008} \\
\hline    
\end{tabular}% }
\end{center}
\caption[CBM efficiencies $f$ for the diffusive CBM mechanism in the range of
initial masses $\mzams$ of AGB models.]{CBM efficiencies $f$ for the diffusive CBM mechanism in the range of
initial masses $\mzams$ of AGB models. $f_{\rm CE}$ is adopted at the bottom boundary of the convective
envelope while f$_{\rm PDCZ}$ is adopted at the bottom boundary of the PDCZ. 'burn' or 'non-burn'
stand for burning or no burning at the bottom of the respective convective zone.}
\label{tab:f_table}
\end{table}

\begin{table*}
\begin{center}
%\tabletypesize{\small}

%\scalebox{0.6}{%

\begin{tabular}{ccccccccccc}
\hline
\mzams & \multicolumn{2}{c}{$Z=0.02$}  & \multicolumn{2}{c}{$Z=0.01$}  & \multicolumn{2}{c}{$Z=0.006$} &  \multicolumn{2}{c}{$Z=0.001$} &  \multicolumn{2}{c}{$Z=0.0001$}  \\
\hline
  &  delay       & rapid &  delay       & rapid &  delay       & rapid     &  delay       & rapid &  delay       & rapid \\
\hline
  12     & 1.61 &1.44 & 1.61   &1.44 & 1.62   &1.44 & 1.62   &1.44& 1.62   &1.44\\
  15     & 1.61 &1.44 & 1.61   &1.44 & 1.62   &1.44 & 1.62   &1.44& 1.62   &1.44\\
  20     & 2.73 &2.7  & 2.77   &1.83 & 2.79   &1.77 & 2.81   &1.76& 2.82   &1.76\\
  25     & 5.71 & -   & 6.05   &9.84 & 6.18   &7.84 & 6.35   &5.88& 6.38   &5.61\\

\hline
 \noalign{\smallskip}
\hline
\end{tabular}
%}
\caption{Remnant masses
of massive star models according to \citet{fryer:12} for the delayed and rapid explosion prescriptions. 
The $\mzams=25\msun$, $Z=0.02$ model based on the rapid explosion prescription
collapses directly into a black hole. See text for description of details regarding the prescription. 
}
\label{tab:coo_fallback}
\end{center}
\end{table*}

\begin{table}
\begin{center}
%\tabletypesize{\small}
\begin{tabular}{cccccc}
\hline
  \mzams & Z=0.02 & Z=0.01 & Z=0.006 & Z=0.001 & Z=0.0001     \\
\hline          
       12 & 1.60 & 1.52 & 1.55 & 1.50 & 1.64    \\        
       15 & 1.46 & 1.50 & 1.66 & 1.55 & 1.53   \\       
       20 & 1.68 & 1.32 & 2.02 & 2.08 & 1.65      \\     					
       25 & 1.55 & 1.78 & 1.66 & 1.56 & 1.69   \\
\hline
 \noalign{\smallskip}
\hline
\end{tabular}
\caption[Fe core masses of massive star models presented in this work.]{Fe core mass of massive star models presented in this work. The Fe core boundary is defined where the mass fraction of Fe, Co and Ni falls below $50\, \%$. Units are in $\msun$.}
\label{tab:fe_cores}
\end{center}
\end{table}

 \begin{table*}
%\begin{center}
\centering
%\scalebox{0.75}{%
%\centering
\begin{tabular}{rrrrrr}
\hline
 species & nested & decoupled &H04       & K10       & C15  \\ 
\hline  
         \multicolumn{6}{c}{CNO isotopes} \\
\hline
 C-12 & 1.755E-03  & 9.075E-03 & 2.739E-03 & 5.068E-03 & 1.39E-02\\ 
 C-13 & 2.333E-04  & 3.798E-04 & 2.612E-04 & 4.289E-04 & 5.42E-05\\
 N-14 & 1.019E-02  & 1.230E-03 & 7.110E-03 & 2.634E-02 & 1.17E-04\\
 O-16 & 4.070E-03  & 4.373E-03 & 1.864E-03 & 7.987E-04 & 7.88E-04\\
\hline
        \multicolumn{6}{c}{isotopic ratios} \\
\hline
 C-12/C-13 & 7.52   & 23.89  & 10.48 & 11.82 &   256.46   \\
 C-12/O-16 & 0.43   &  2.08   & 1.47  &  6.35 & 17.64 \\
\hline
\multicolumn{6}{c}{s-process isotopes} \\
\hline
 Sr-88 & 2.240E-09 & 2.240E-09 &        &         &  1.87E-08\\
 Zr-90 & 5.069E-10 & 5.069E-10 &         &         &  3.72E-09\\
 Ba-136& 7.573E-11 & 7.674E-11 &        &         &  4.69E-09\\
 Pb-208& 3.776E-10 & 3.776E-10 &         &         &  4.21E-08\\
\hline
\end{tabular}%
\caption[The final yields for the $\mzams=4\msun$, $Z=0.0001$ model in comparison with yields of H04, K10 and C15.]{The final yields for the $\mzams=4\msun$, $Z=0.0001$ model based on the nested-network approach and the decoupled approach in comparison with yields of H04, K10 and C15. Units are in $\msun$.} 
\label{hbb_yield_table}
%}
%\end{center}

\end{table*}

%%%%%%%%%%%%%%%%%%%%%%%%%

%%%%%%%%%%%%%%%%%%%%%%%%%

%%%%%%%%%%%%%%%%%%%%%%%%%

%stellar evolution
\section{Results of stellar evolution and explosion} \label{stellarevol}

\subsection{General properties}

\subsubsection{The mass and metallicity grid}

The new set of models and stellar yields are all calculated with the same stellar evolution code \MESA.
We calculate massive star models with $\mzams = 15$, $20$ and $25\msun$ at $Z=0.02$ and $Z=0.01$ as an alternative to the massive star GENEC models from P16.
Stellar models with $\mzams = 12\msun$ are added at all metallicities to cover the lower-mass end of the massive star mass range. \cite{cote:16} show that based on our assumption of the remnant mass distribution (cf.\ \sect{sec:method:exp}) adding more masses to the grid would not significantly improve galactic chemical evolution models. \cite{cote:16} find that the metallicity
range covered is more important than the number of metallicities within that range. 
In addition to the $\mzams=5\msun$ models in P16 we are adding intermediate and S-AGB models at all metallicities ($6$ and $7\msun$). We also add a $\mzams=1\msun$ models at all metallicities.

\subsubsection{Stellar evolution tracks}

\noindent \textbf{AGB stars}

\begin{figure}
\centering
\includegraphics[width=\columnwidth]{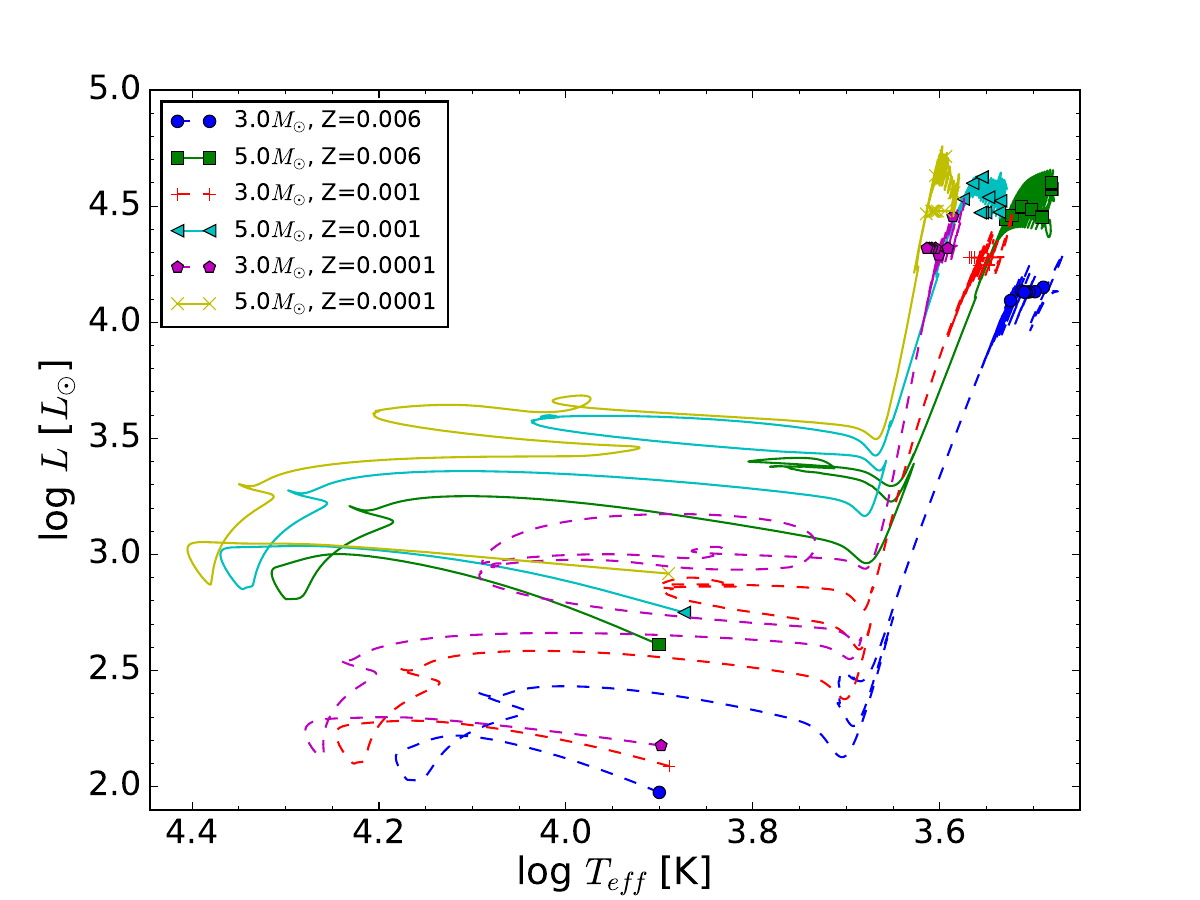}
\includegraphics[width=\columnwidth]{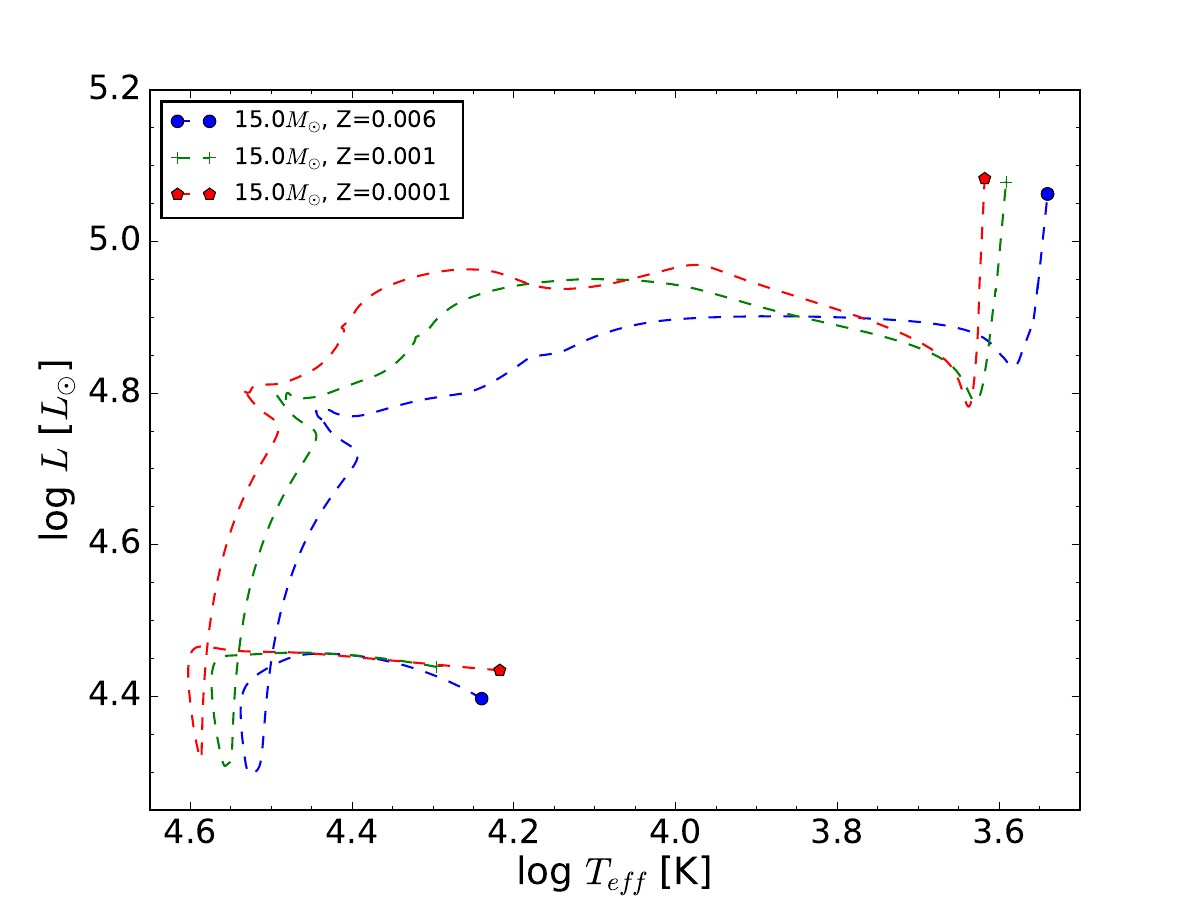}
\includegraphics[width=\columnwidth]{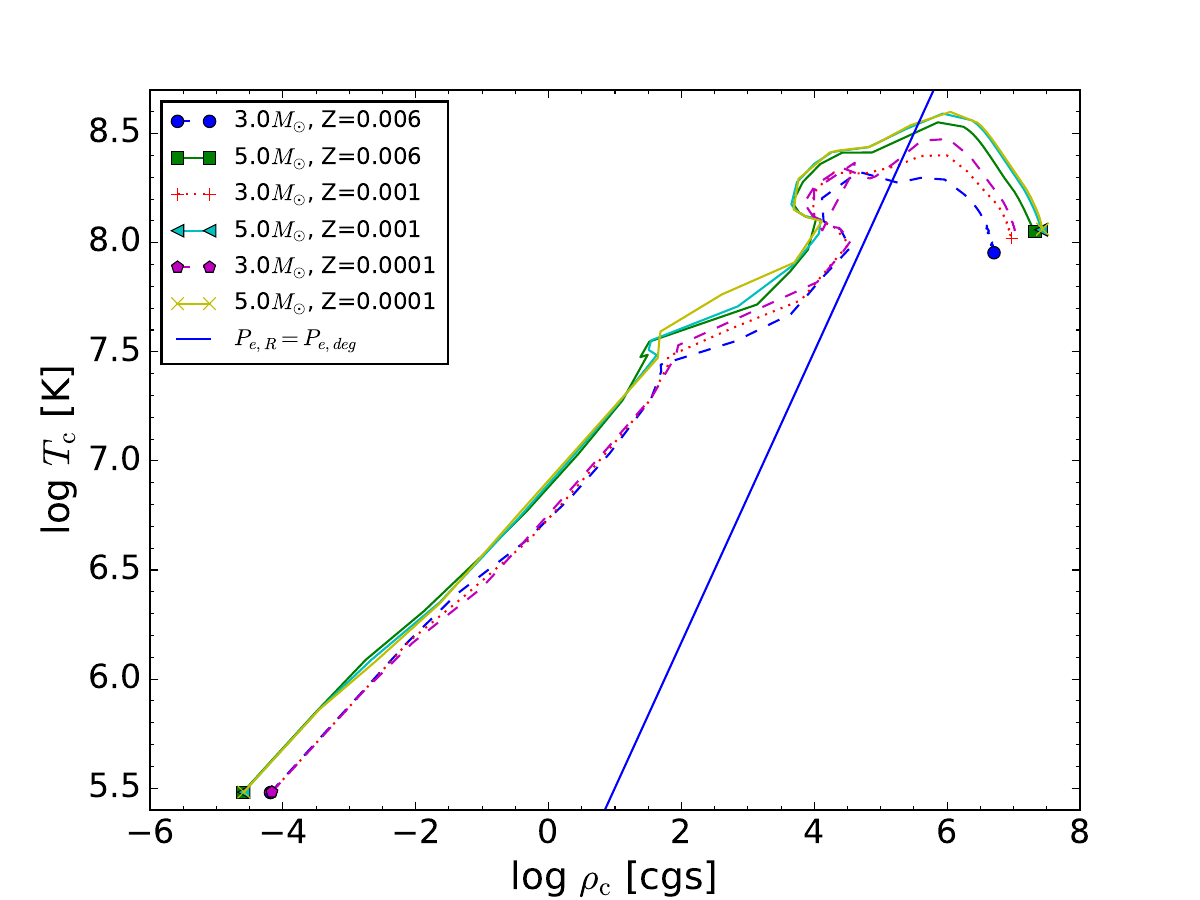}
\caption{Comparison of HRD's for AGB models with $\mzams=3\msun$ and $\mzams=5\msun$ (top) and massive star models with $\mzams=15\msun$ (middle) for $Z=0.006$, $Z=0.001$ and $Z=0.0001$. Central temperatures $T_c$ and densities $\rho_c$ for those AGB models (bottom).} 
\label{fig:hrd_centralT_rho}
\end{figure}

\noindent The influence of metallicity on the stellar evolution
is visible in the Hertzsprung-Russell diagram (HRD) with the stellar models with $\mzams=3\msun$ and $\mzams=5\msun$ shown in \fig{fig:hrd_centralT_rho}.
The shift of the tracks of lower metallicity to higher luminosities and higher surface temperatures is the result
of the larger core masses and lower opacities of the envelopes \citep{herwig:04a}.
The central temperature-density tracks of $\mzams=5\msun$ models
are separated from $\mzams=3\msun$ models. The central densities $\rho_c$ depend on stellar mass $M$ as $\rho_c\propto M^{-2}$ under the assumption of constant temperature during each burning phase. Lower metallicity models behave as models with higher initial masses which is visible in the approach 
of the $\mzams=3\msun$ tracks at low metallicities towards the $\mzams=5\msun$ tracks in the central temperature-density diagram (\fig{fig:hrd_centralT_rho}).

Stellar models with $\mzams\leq1.65\msun$ for $Z=0.006$, $Z=0.001$ and $Z=0.0001$ exhibit He-core flashes.
First dredge-up appears at $Z=0.006$ and $Z=0.001$ in all the AGB models but at $Z=0.0001$ only in stellar models with $\mzams\leq2\msun$.
Second dredge-up  occurs in models with $\mzams\geq4\msun$ and $\mzams\geq3\msun$ at $Z=0.006$ and $Z=0.001$ respectively.
Core flash, first dredge-up and second dredge-up at $Z=0.006$ show the same initial-mass dependence as the AGB models at $Z=0.01$ in P16.

The average luminosity of low-mass, non-HBB stellar models follows the linear core-mass luminosity relation of \cite{bloecker:93b} that was originally derived for $Z=0.02$ models (\fig{fig:corelum_Tceb}).  AGB models with higher initial masses which experience HBB agree with the exponential
core-mass luminosity relationship of \cite{herwig:04a}. 

AGB models with $\mzams\geq 5\msun$ for $Z=0.006$, $Z=0.001$ and $Z=0.0001$ ignite C and reach the S-AGB stage. 
For S-AGB models with initial mass below $\mzams=7\msun$ at $Z=0.006$ and below $\mzams=6\msun$ at $Z=0.001$ and $Z=0.0001$ 
a convective C-burning flame does not appear as in stellar models of higher initial mass. In these models C burning takes place under radiative conditions.
For $\mzams=4\msun$ models the maximum temperatures in the C/O core
do not exceed $T\approx3\times10^{8}\, \mathrm{K}$ which is far below the ignition temperature of $T\approx6\times10^{8}\, \mathrm{K}$ found by \citep{siess:07}.
\cite{farmer:15} has provided a recent, detailed study of the onset of C burning, which also depends sensitively on the still very uncertain \czw+\czw\ reaction rate \citep[see also][]{chen:14}.

\begin{figure*}
\centering
\includegraphics[width=\columnwidth]{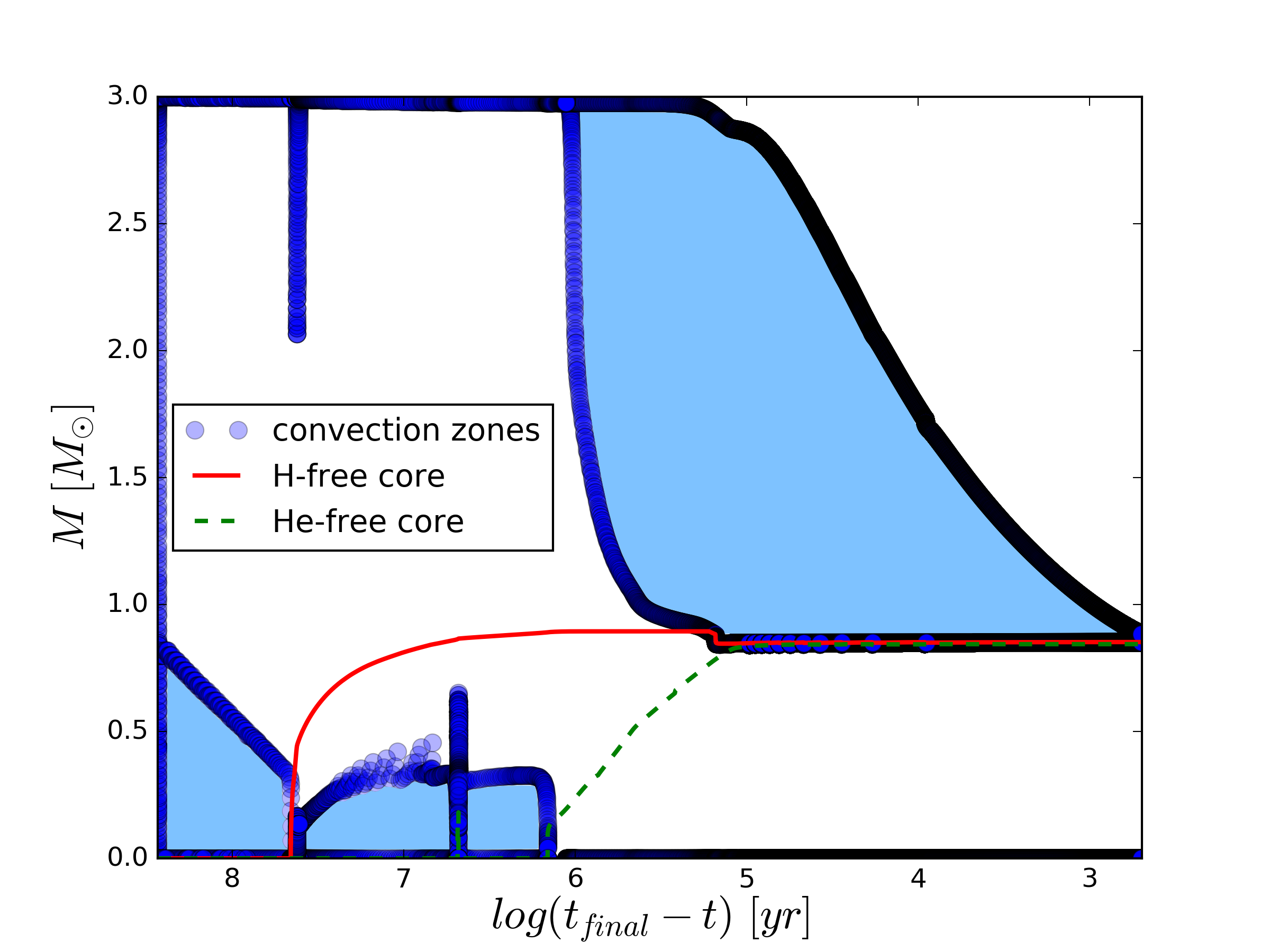} %preAGB
\includegraphics[width=\columnwidth]{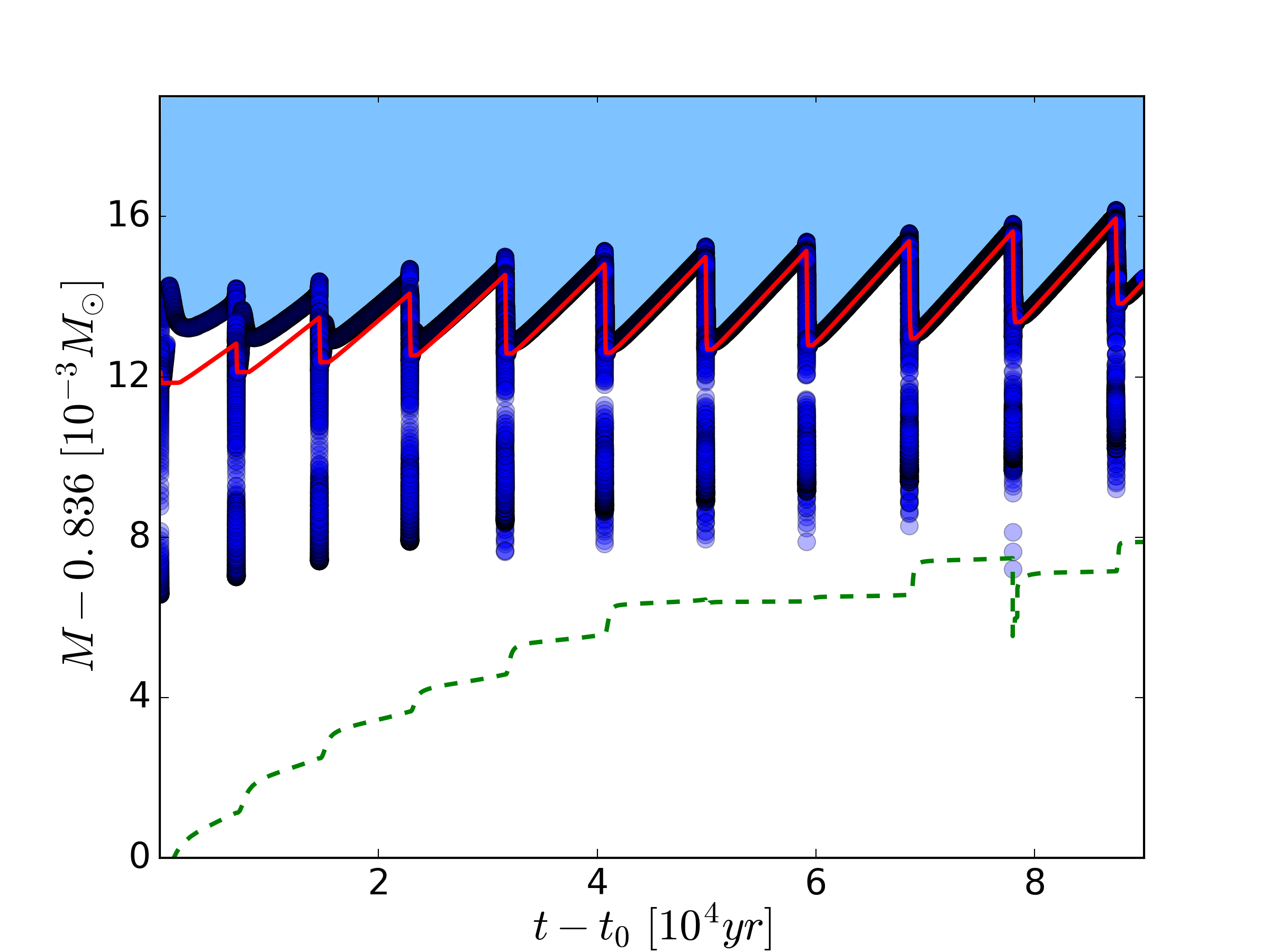}
\includegraphics[width=\columnwidth]{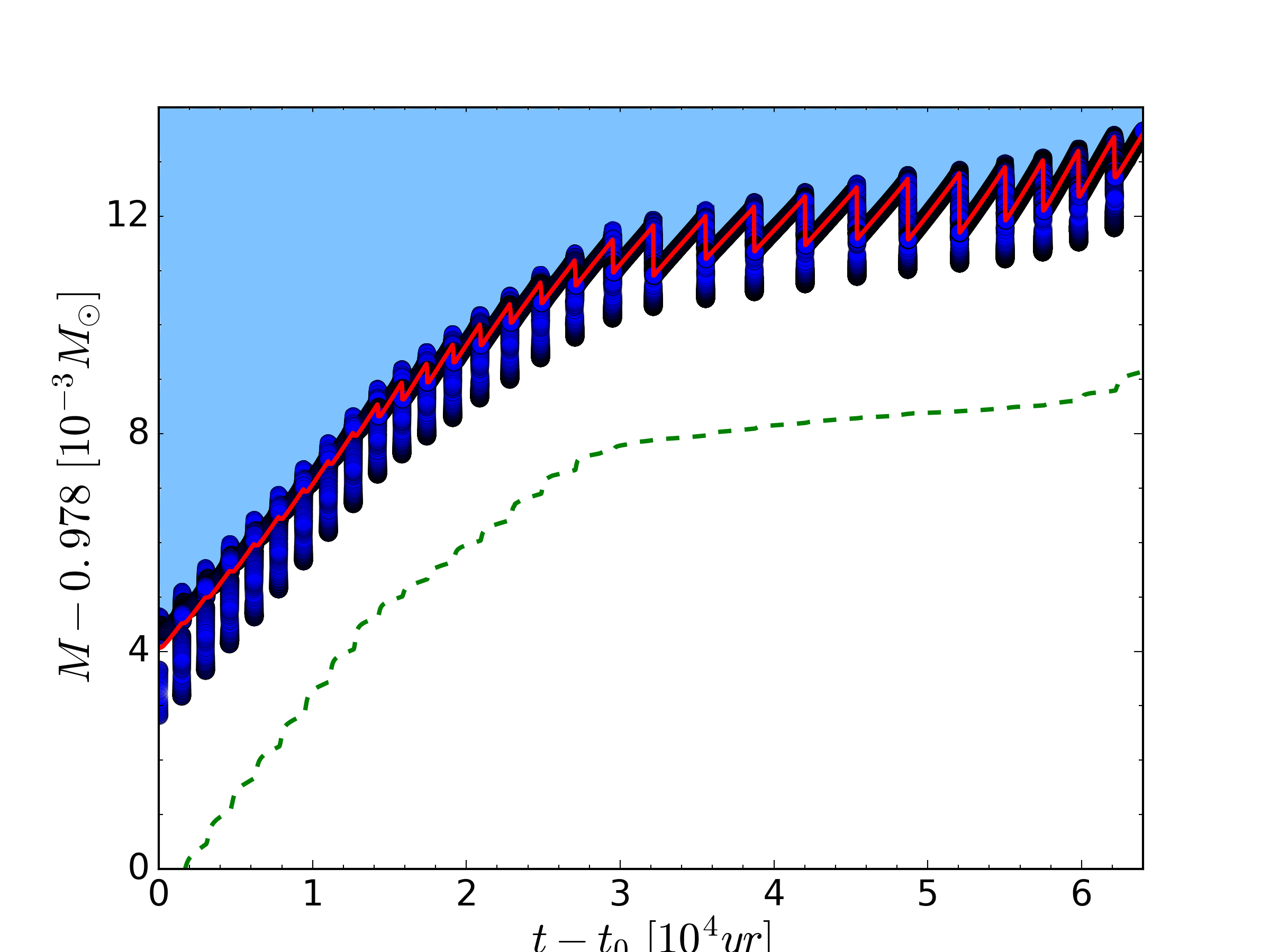}
\includegraphics[width=\columnwidth]{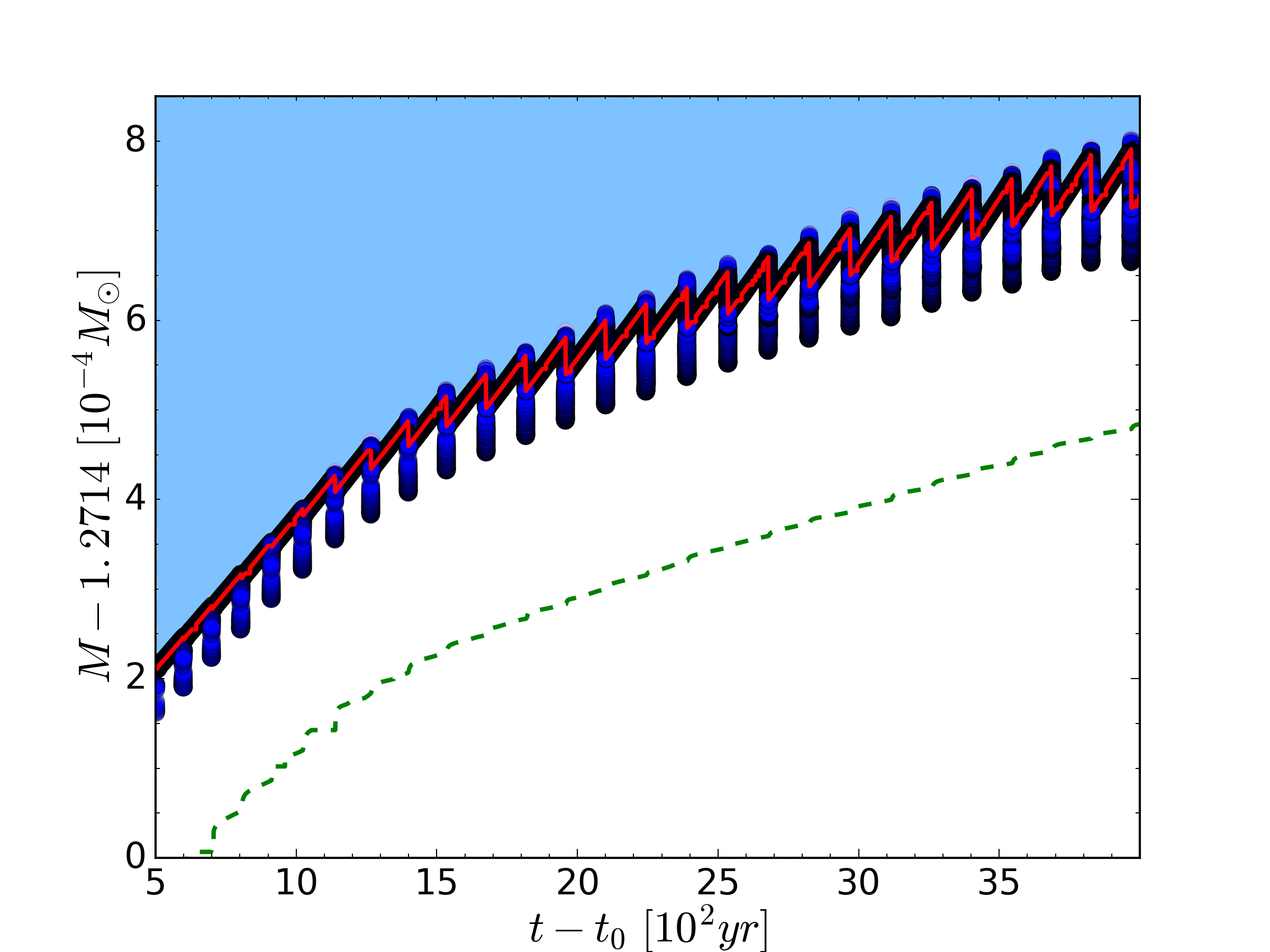}
\
\
\caption{Kippenhahn diagrams of a $\mzams=3\msun$, $Z=0.0001$ model with its pre-AGB phase (top, left) and TP-AGB phase (top, right). 
The H-free and He-free core boundaries are displayed by red solid and green dashed lines. The convective zones are marked in blue. 
$t_0$ and $t_{\rm final}$ are the times at the beginning and the end of the TP-AGB phase respectively.
The TP-AGB phase of a massive AGB model with $\mzams=5\msun$ (bottom, left) and S-AGB model with $\mzams=7\msun$ (bottom, right) at $Z=0.0001$ are shown.
}
\label{fig:kipp_agb}
\end{figure*}

Model properties of the TP-AGB phase for each initial mass and metallicity are shown in \tab{table:agb_properties}.
We present in \tab{agbmodelprop2_1_5a} the detailed TP properties for stellar models of $Z=0.0001$.
The structure evolution of models $\mzams=3\msun$, $\mzams=5\msun$ and $\mzams=7\msun$ at $Z=0.0001$ are shown in the Kippenhahn diagrams in \fig{fig:kipp_agb}.%  and \fig{fig:kipp_agb2}.
The final core mass and lifetimes for AGB models are shown in \tab{tab:cores_agb}.

We compare stellar models with $\mzams=2\msun$ and $\mzams=5\msun$ at $Z=0.001$ with models
of \citet[][K03]{karakas:03b} and \citet[][W09]{weiss:09} who calculated models with $\mzams=1.9\msun$ and $\mzams=5\msun$
based on $\alpha$-enhanced initial abundances and models of $\mzams=2\msun$ and $\mzams=5\msun$ of solar-scaled abundance respectively.
The core mass of these two stellar models at the first TP are $0.63\msun$ and
$0.985\msun$ while K03 obtain $0.548\msun$ and $0.888\msun$ and W09 get $0.494\msun$ and $0.908\msun$.
As P16 we find larger core masses compared
to K03 and W09.  Our number of thermal pulses of the stellar
models are 14 and 32 while K03 have 16 and 83 and W09 have 10 and 38.

The final surface C/O ratio of these stellar models is 3.243 and 3.379 compared
to 8.18 and 4.48 of K03 and 3.449 and 0.772 of W09. The latter value of W09 is taken when the envelope mass is $2.642\msun$ and their simulation stops. It differs from ours
because in our $\mzams=5\msun$ simulation the dominance of the 3DUP over
weakening HBB as described in \citet{frost:98c} increases the C/O
ratio from $<1$ to the large final value $>3$ over the last two thermal pulses when the stellar model loses the last $1\msun$ of envelope mass. The $Z=0.006$, $\mzams=5\msun$ simulation still has $0.75\msun$ of envelope when it stops and the C/O ratio is $\approx 0.13$. This case does not experience the final thermal pulse where TDUP could have significantly increased the C/O ratio. 

\begin{figure}
\centering
\includegraphics[width=\columnwidth]{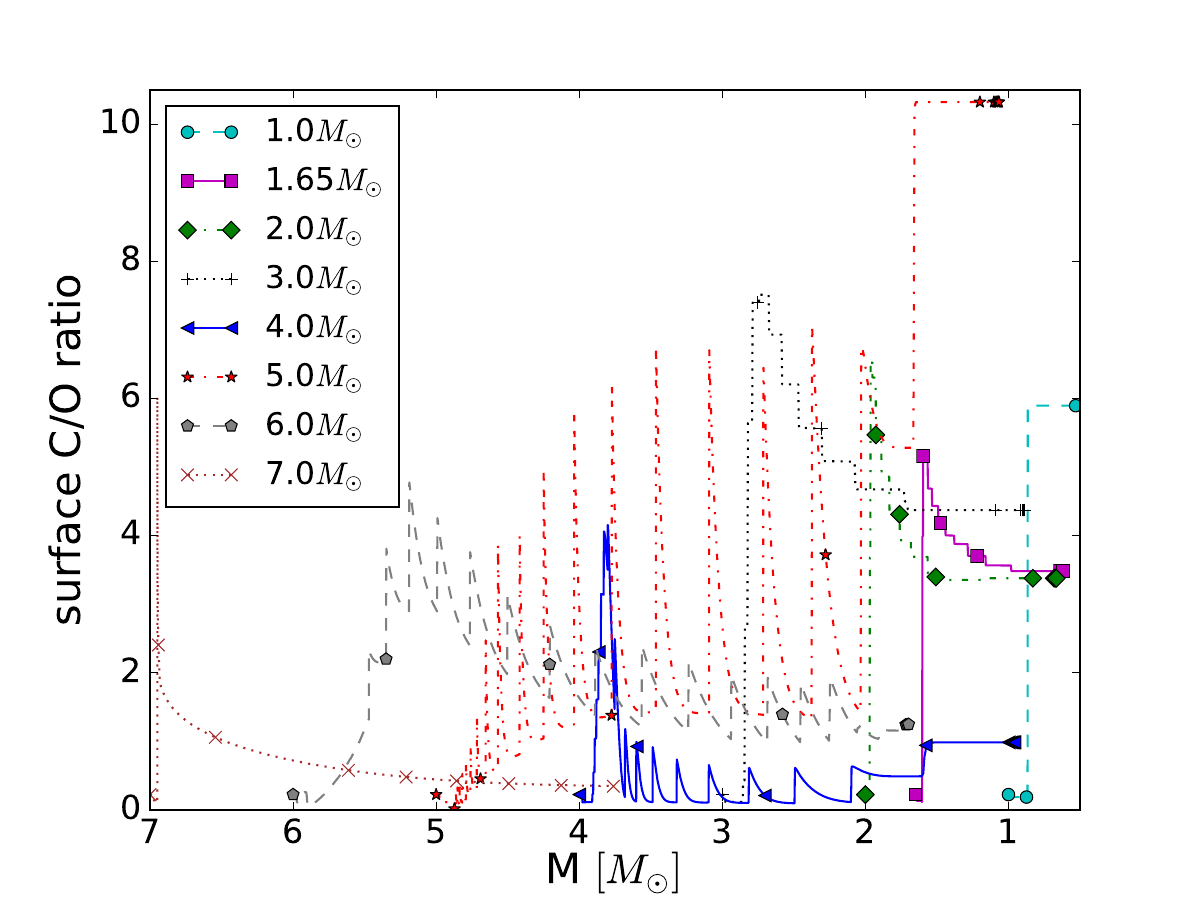}
\includegraphics[width=\columnwidth]{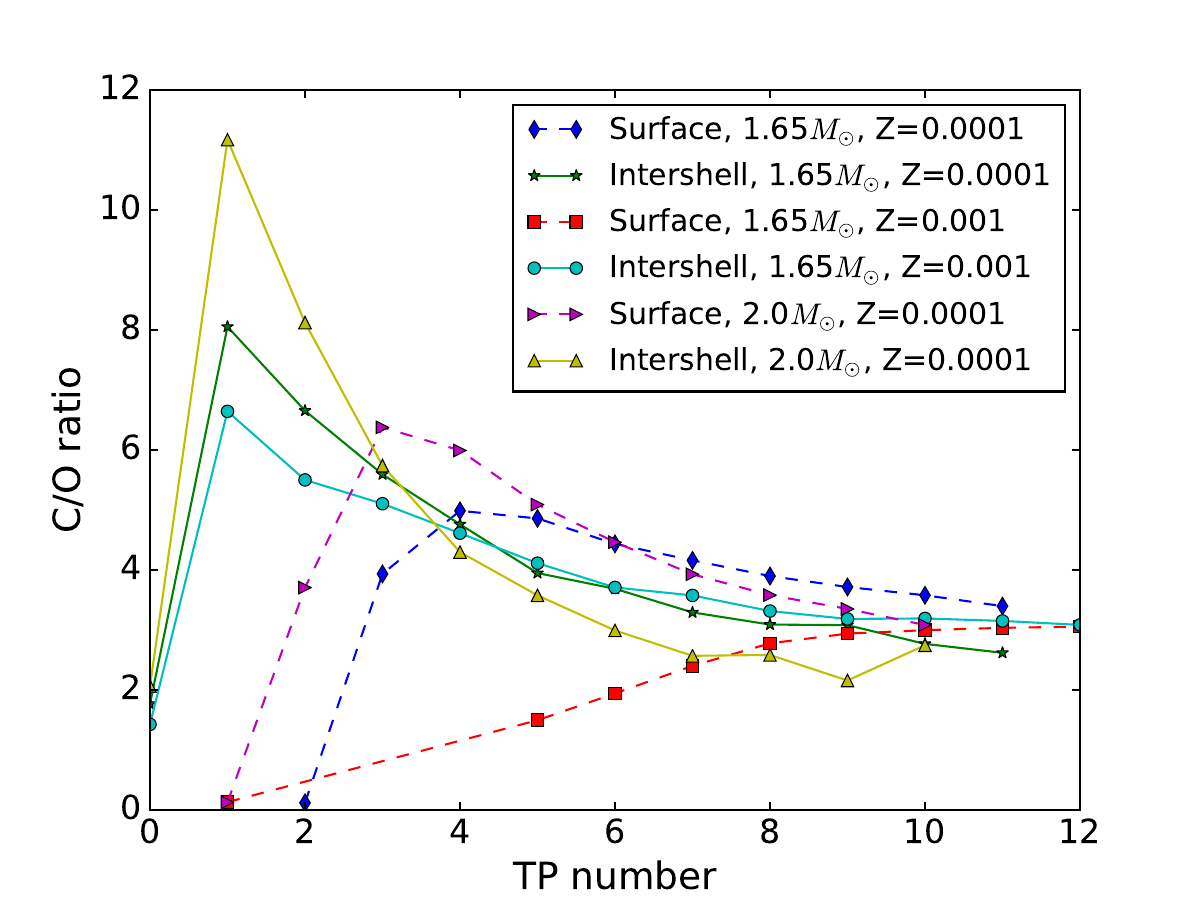}
\caption{Surface C/O ratio versus total stellar mass for $Z=0.0001$ (top). The He intershell and surface 
C/O ratio for each TP of two stellar models with $\mzams=1.65\msun$ and $\mzams=2\msun$ (bottom).}
\label{fig:coratio_set1_5a}
\end{figure}

The surface C/O ratio increases due to TDUP and decreases during the
interpulse HBB in massive AGB models
\citep{lattanzio:96,lattanzio:97a,lattanzio:97b}.  The surface C/O
ratios for stellar models of $Z=0.0001$ presented is complex (\fig{fig:coratio_set1_5a}). %than found in stars at higher
While at $Z=0.02$ low mass stellar models steadily increase their surface C/O ratio (see Fig.\,4 in P16), at low metallicity the first pulses can lead to a
surface enhancement close to or even above the He-intershell C/O ratio
as shown in \fig{fig:coratio_set1_5a}. At low metallicity the envelope C/O ratio quickly represents that of the intershell because the total initial amount of O and C in the envelope is smaller due
to the low initial metallicity. Due to a steady decrease of
the C/O ratio in the He intershell over time the TDUP leads to a
decline in the surface C/O ratio.  Stellar models at higher metallicity such as
the $\mzams=1.65\msun$, $Z=0.001$ model
%do not reach the intershell C/O ratio and hence 
experience only an increase of the surface C/O ratio during their evolution.
%This explains in combination with HBB the C/O surface evolution.
For models with higher initial mass a higher C/O intershell ratio is reached which leads to a higher C/O surface enhancement in the non-HBB models. 
%The fact that less  material is dredged up as well as more envelope mass is available
%leads to a slower... 
%\textbf{TDUP}

\begin{figure}
\centering
\includegraphics[width=\columnwidth]{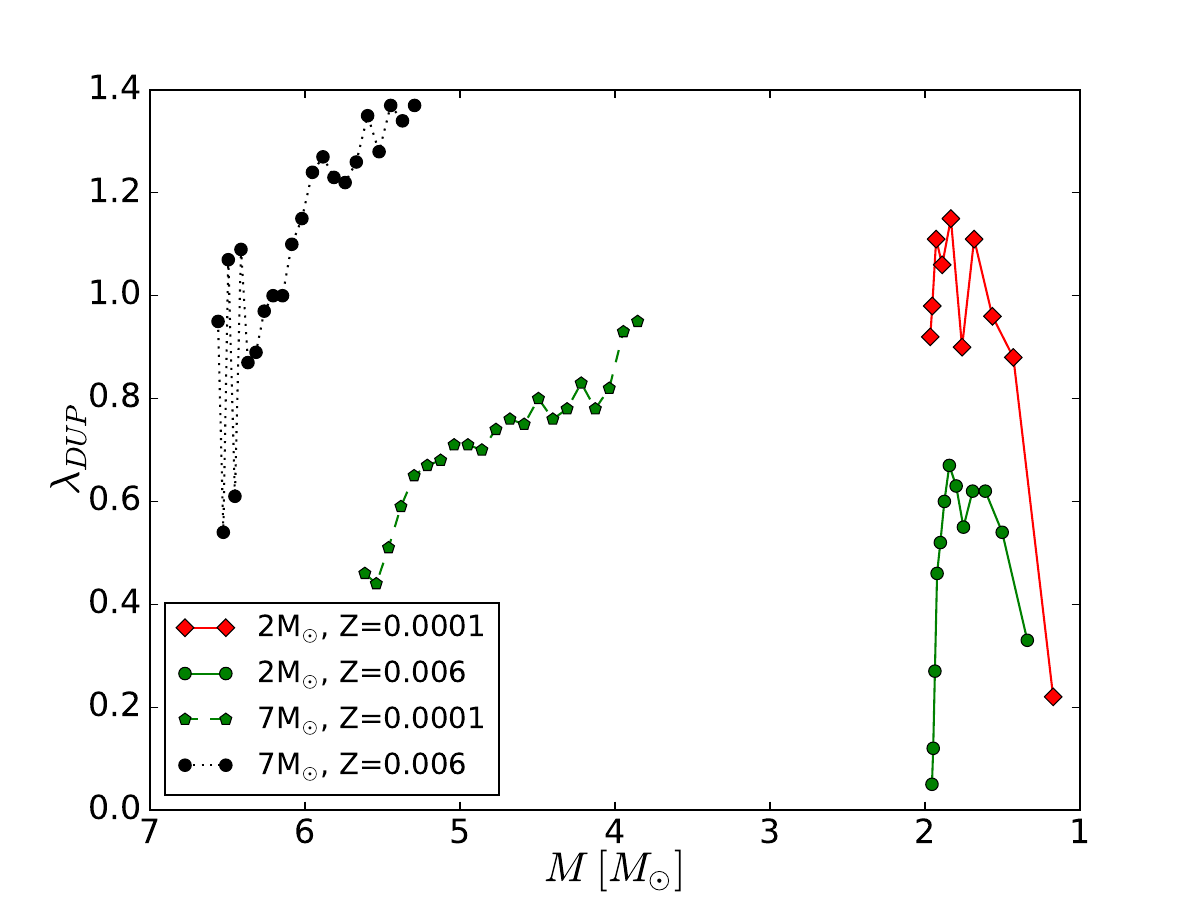}
\caption{Metallicity dependence of the DUP parameter $\lambda$ shown at the example of
low-mass AGB models and a S-AGB models with $\mzams=2\msun$ and $\mzams=7\msun$ for $Z=0.0001$ and $Z=0.006$.}
\label{fig:DUP_lambda_Zcomparison}
\end{figure}

\begin{figure}
\centering
\includegraphics[width=\columnwidth]{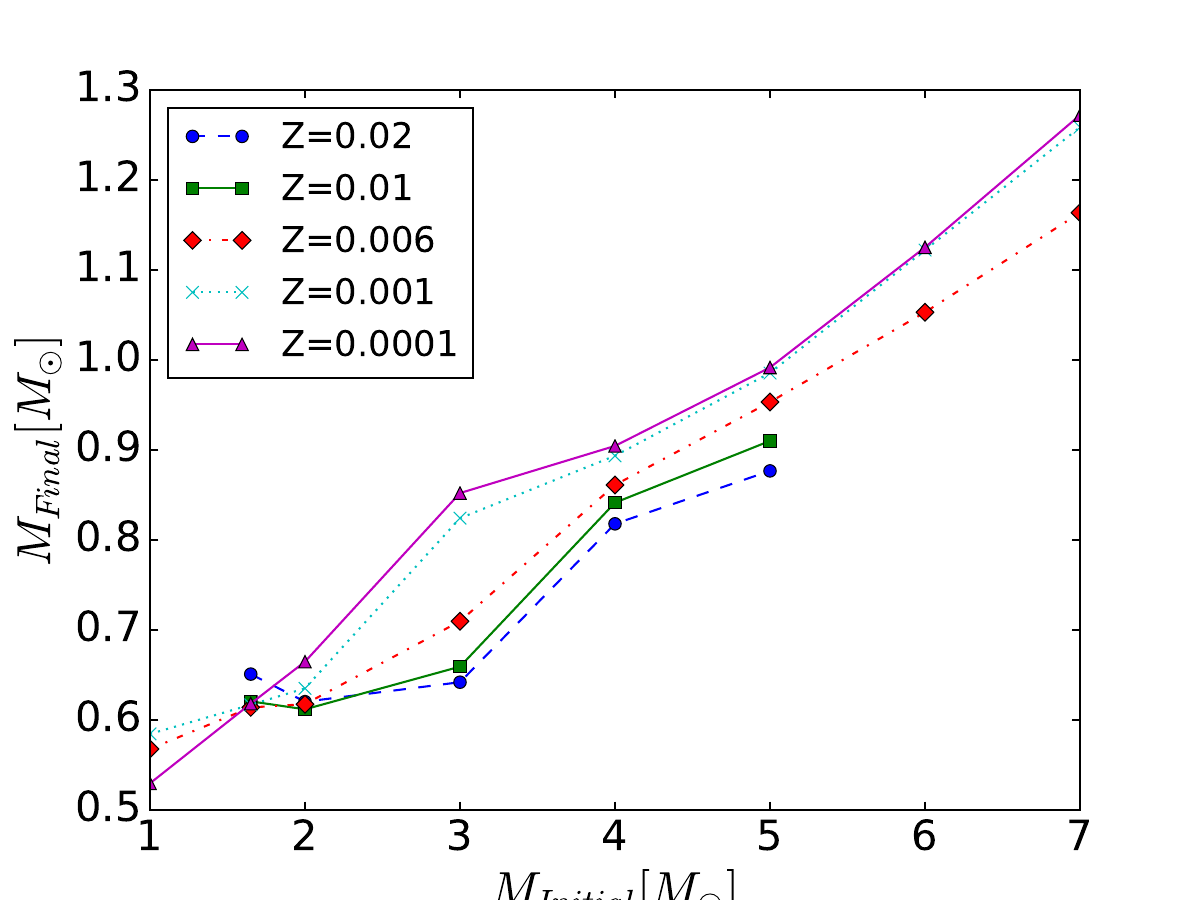}
\caption{Initial-final mass relation for AGB models of this work with AGB models
between $\mzams=1.65\msun$ and $\mzams=5\msun$ at $Z=0.02$ and $Z=0.01$ from P16.}
\label{fig:ifmr}
\end{figure}

The TDUP strength is described by the dredge-up parameter
$\lambda_\mathrm{DUP}$ defined as $\lambda = \Delta M_\mathrm{DUP}/
\Delta M_\mathrm{H}$ where $\Delta M_\mathrm{DUP}$ is the amount of mass
dredged-up into the envelope and $ \Delta M_{\rm H}$ is the increase in mass of the
H-free core during the previous interpulse phase.
$\lambda_\mathrm{DUP}$ shows a strong dependence \sout{behaviour} on metallicity.
In \fig{fig:DUP_lambda_Zcomparison} we compare the stellar models with
$\mzams=2\msun$ and $\mzams=7\msun$ at $Z=0.006$ and $Z=0.0001$.  The
low-mass AGB star model with lower metallicity has higher $\lambda_\mathrm{DUP}$
than the higher metallicity model, while the S-AGB model has higher
$\lambda_\mathrm{DUP}$ at higher metallicity. This can be understood by
considering that the dredge-up efficiency has a maximum for a core
mass $ \approx 0.8\msun$, which at $Z=0.02$ corresponds to an initial
mass of $4\msun$ (P16), and in combination with the metallicity dependence of
the initial to final mass relation (\fig{fig:ifmr}). The lower metallicity
$7\msun$ model has a higher core mass and therefore lower
$\lambda_\mathrm{DUP}$. The low-mass model has also a higher core
mass at lower metallicity, but here this implies larger $\lambda_\mathrm{DUP}$.

While the $\mzams=2\msun$, $Z=0.006$ model is very similar to the $\mzams=2\msun$, $Z=0.01$ model shown in Fig.\, 5 of P16, 
the $\mzams=2\msun$, $Z=0.0001$ model reaches $\lambda_\mathrm{DUP}\approx 1$, similar to the stellar model of the same initial mass and metallicity in \cite{herwig:03c}.
The maximum of total dredged-up mass increase
up to $\mzams=3\msun$ for $Z=0.006$ and up to $\mzams=2\msun$ for $Z=0.001$ and $Z=0.0001$.
For low-mass models both quantities decline towards higher initial masses (\tab{table:agb_properties}). 
For comparison, \cite{fishlock:14} found that the $\mzams=2.5\msun$ and $\mzams=2.75\msun$
models at $Z=0.001$ dredge-up the most material.
In intermediate-mass stellar models we find lower total mass dredged up  compared to \cite{fishlock:14} who reach another maximum at $\mzams=4\msun$.

The final core masses are larger at lower metallicity for most stellar models. This implies a steeper initial-final mass relation (IFMR, \fig{fig:ifmr}). The core masses of models from P16 are added for comparison. 
The IFMRs in \cite{weiss:09} which spans from $\mzams=1\msun$ to $\mzams=6\msun$ and covers $Z=0.02$ down to $Z=0.0005$ 
show in general a smaller final core mass than the present stellar models and those by P16. The spread
in metallicity is more pronounced for these models. Our IFMR covers the upper part of the compiled data of observed open cluster objects
shown in Fig.\ 10 of \cite{weiss:09}. 
The AGB phase of the $\mzams=1\msun$, $Z=0.0001$ model is terminated due to a H-ingestion event 
which prevents further core growth.

\noindent \textbf{Massive stars}

\noindent 

The massive star models used the same \MESA\ code version and input
parameters used by \citet[][J15]{jones:15}. 
%It is thus not surprising that the present models produce final total and core masses extremely similar to those of J15 (\tab{tab:cores_comparison_jones}).
J15 conducted a resolution study of the time steps at the end of core helium burning and
we use the coarsest time step resolution that reproduced the He-free
and C/O core masses. The impact of metallicity in the HRD evolution
(e.g. $\mzams=15\msun$ models, \fig{fig:hrd_centralT_rho}) is similar
to that shown in low-mass models. There is little impact of metallicity
on central temperature and density.

P16 found the final fate of massive stellar models in the mass range $\mzams=15\msun$ to $\mzams=25\msun$ to be the red super giant phase
which is in agreement with other non-rotating models \cite[e.g.][]{hirschi:04}. 
All these massive star models experience the same phase except stellar models of $\mzams=20\msun$ and $\mzams=25\msun$ 
at $Z=0.0001$. The latter move from the blue region of the HRD
into the region of yellow supergiants but not further, similar to models of Pop III stars of \citet{heger:10}. 
Due to their low metallicity these stellar models 
experience negligible mass loss and their intermediate convective zones
are largest among all models. This leads to higher compactness
which favours the blue region of the HRD \citep{hirschi:07,peters:13}.

The stellar models with $\mzams=25\msun$ at all metallicities and the $\mzams=20\msun$, $Z=0.01$ model burn C under radiative conditions consistent
with solar-metallicity and PopIII models \citep[][P16]{heger:10}. The occurrence of convective core C burning in the $\mzams=20\msun$, $Z=0.02$ model 
results from the higher luminosity of C core burning present in stellar models of higher metallicity \citep{hirschi:07,rauscher:02,eleid:04}.
Convective core C burning is present in all massive star models of lower initial mass as in P16.

%Their envelope is not fully developed until the surface.
The lifetimes of the core-burning stages are given in \tab{tab:lifetimesmassive}, using the definition of the lifetimes as in P16.
Most burning stages are shorter for higher initial masses and lower metallicities, as expected.
The final masses and the masses of the He, CO and Si cores are shown
in \fig{fig:cores_massive}, using the definitions of the core masses  as
in P16.
The final mass increases towards lower metallicity at each initial mass.
The He core masses and CO core masses show only a mild metallicity dependence compared to the
clear metallicity dependence of the final mass. For some initial masses the core mass does not increase with decreasing metallicity
such as the CO core masses of the stellar models with $\mzams=25\msun$.
The Si cores do not increase with initial mass as found for the He and CO cores. Instead,
we find large variations in the metallicity of similar magnitude at different initial masses and no clear trend with metallicity (\fig{fig:cores_massive}).
This is due to the non-monotonicity for the Si core \citep[e.g.][]{ugliano:12,sukhbold:14}

We compare the core masses of these stellar models with initial mass of $15\msun$ and $25\msun$ for $Z=0.006$ with those of \citet[][M02]{meynet:02}
at $Z=0.004$ and P16 at $Z=0.01$ in \tab{tab:cores_comparison_meynet}. 
Our He core masses are in better agreement with P16 who got larger values than M02 in spite of the metallicity difference. This is because we adopt a similar convective overshooting strength for the H-burning cores
as P16 while M02 do not adopt any overshooting.
More precisely, for core H and He-burning phases, in \MESA\ models an exponentially-decaying diffusion coefficient with $f=0.022$ is used whereas in GENEC, an instantaneous penetrative overshoot with $\alpha_{\rm ov}= 0.2$ H$_{\rm P}$ is used. The different treatment of convective boundary mixing explains the differences in core masses between this work and P16.
%Our He core masses are $5.09\msun$ and $9.66\msun$ while M02 obtained $4.45\msun$ and $8.44\msun$.
% P16 found $4.811\msun$ and $9.39\msun$. Similar considerations hold as for the He core
The CO core masses show larger differences between this work, P16 and M02 than found for the He core masses.
The mass of the Si core is in better agreement with P16 than the CO core mass (\tab{tab:cores_comparison_meynet}).

The structural differences of stellar models with $\mzams=25\msun$ at $Z=0.001$ and $Z=0.0001$ are
shown in the Kippenhahn diagram in \fig{fig:kip_cont_massive}.
%Both models do not show much differences in core masses but their envelopes
%due to mass loss and convection.
Contacts between convective burning shells occur in different advanced burning stages and can have, 
in particular for a complete shell merger, a profound impact on stellar structure and nucleosynthesis (see \sect{ppnucleo})
The contact between the convective H-burning shell and convective He-burning shell 
leads in the $\mzams=25\msun$, $Z=0.0001$ model to a H-ingestion event. % during O shell burning.
The occurrence of shell merger is affected by considerable
uncertainties \citep{woosley:02} and requires studies with 3D hydrodynamic simulations \citep[e.g.][]{meakin:07a, jones:17}.
This point is discussed in more details below.
%, including
%C, Ne,O,Si shell?.
%In Fig. \ref{fig:kip_cont_massive} shell merging of...
%What effect on nucleosynthesis? 
%compared to Rauscher02 20Msun,
%weak s-process? p process?

%Their definition of the He-free core  is in agreement with ours.

\subsubsection{Core-collapse supernovae}

\begin{figure}
\centering
\includegraphics[width=\columnwidth]{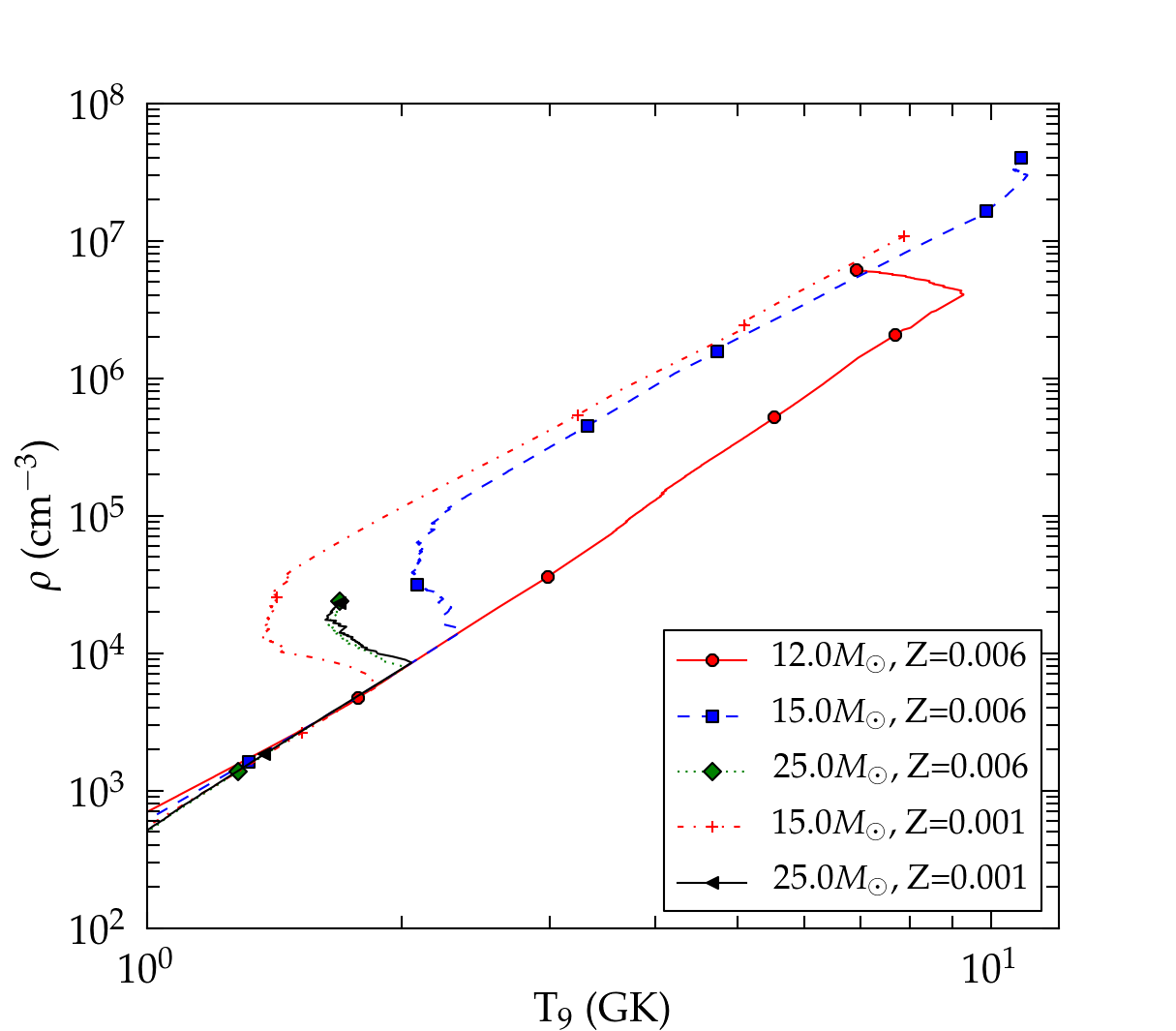}
\caption{Maximum temperature $T_{9}$ and density $\rho$ of each zone during the CCSN explosion based on the delayed explosion prescription for massive star models of different initial masses at $Z=0.006$ and $Z=0.001$.}
\label{fig:ex_characteristic}
\end{figure}

%\noindent The shock conditions are set by the mass cut and the pre-SN structure which which depend on mass and metallicity.
%\cfcom{I would say: ``The explosion energy and remnant mass of a progenitor depends on the pre-SN structure (the compactness 
%parameter represents this in a single parameter but our analysis uses the full core structure\cite{fryer:12}).  The explosion 
%properties determine the layers of the star that are ejected and the shock conditions of that ejecta''}
The explosion energy and remnant mass of a progenitor depends strongly on the pre-SN structure \cite{fryer:99,mueller:16,janka:07}.
The explosion properties determine the layers of the star that are ejected and the shock conditions.
We compare the maximum temperatures and densities reached
during the shock passage for massive star models of $Z=0.006$ and $Z=0.001$ obtained with the delayed explosion prescription (\fig{fig:ex_characteristic}).
The shock temperature for stellar models with $\mzams=12\msun$ and $\mzams=15\msun$ at $Z=0.006$ are the largest of all metallicities. % \textbf{ true for solar,half-solar?}
Up to $Z=0.006$ stellar models with $\mzams=15\msun$ reach the highest shock temperatures and densities followed by the $\mzams=12\msun$ models but
at higher metallicity the trend is reversed.% \crcom{While the density at the zone of the highest shock temperature is about the same }

\begin{figure*}
\centering
\includegraphics[width=2.\columnwidth]{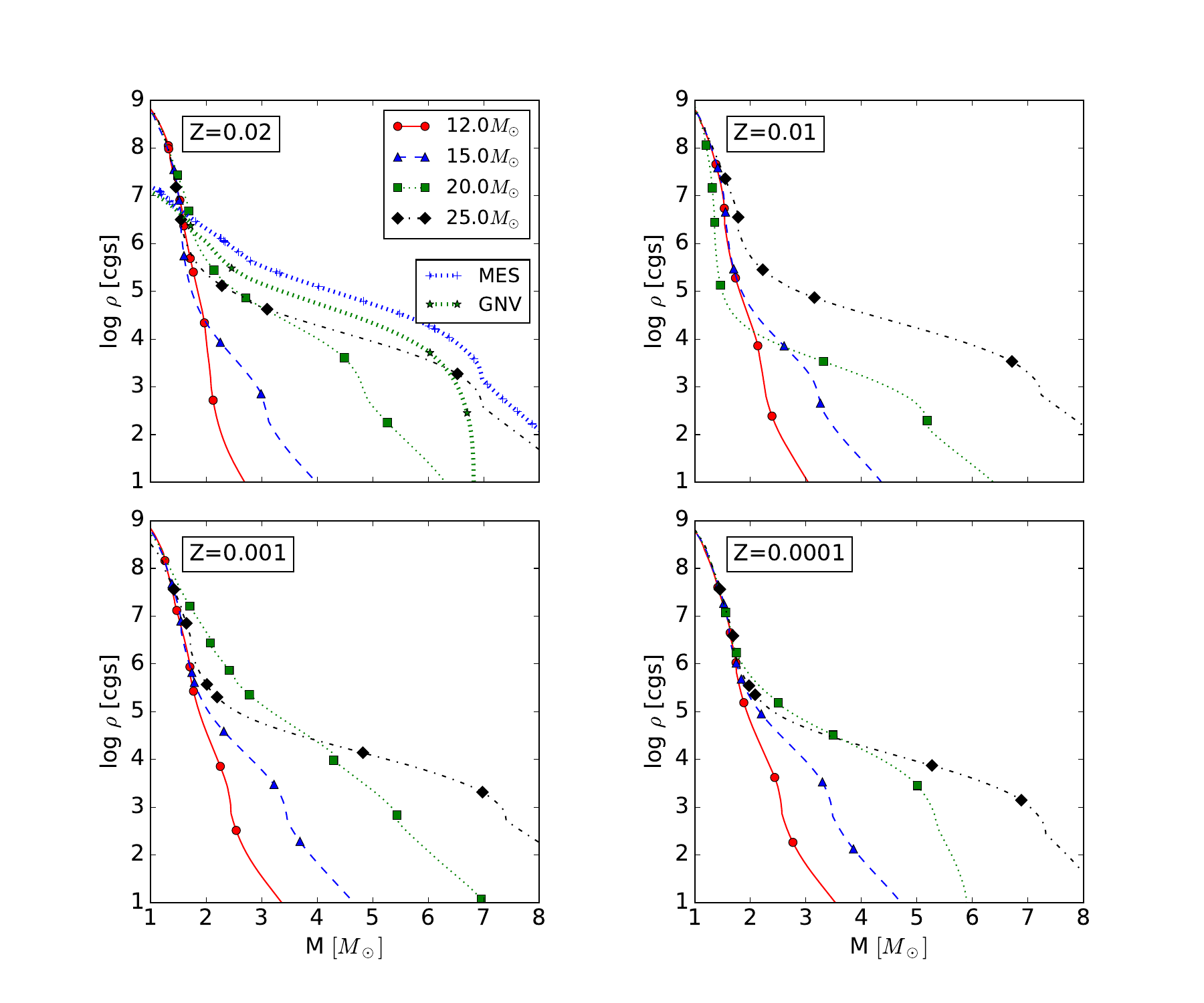}
\caption{Density profiles at core collapse when the infall
    velocity reaches $1000\, \mathrm{km/s}$ for massive star models at
    $Z$=0.02, 0.01, 0.001, 0.0001.  The 1 to 8 Mo range is shown. This
    includes, from left to right, from the outer part of the Fe core
    to at least the end of the O shell.  Comparison of the density
    profiles at the end of Si core burning of our stellar model with
    $\mzams=25\msun$ at $Z=0.02$ computed with the MESA code (MES,
    this work) and the model provided by P16 and calculated with the
    GENEC code (GNV).}
\label{fig:preexpdensityprofile}
\end{figure*}

The pre-SN structure of these stellar models do not always
show trends  with metallicity (\fig{fig:preexpdensityprofile}) and the same counts for the shock temperatures.
There is no trend in the Fe-core mass with
mass and metallicity and instead  %especially at the low-mass end of the models presented here
the stellar models with $\mzams=20\msun$ show the largest Fe core masses (\fig{fig:preexpdensityprofile}). %, with the star of set1.5a having a less metal-rich extended iron core.
Recent studies show that there is no monotonous compactness trend with initial mass and metallicity \citep{ugliano:12,sukhbold:14,sukhbold:16}.
%The remnant masses of all our massive stars is shown in \tab{tab:coo_fallback}.

Under the convective engine paradigm \citep{herant:94}, whether or not the model explodes depends sensitively on the ram pressure of the stellar material falling onto the outer edge of the convective region \citep{fryer:99}. To drive an explosion, the energy in the convective region must overcome this ram pressure and the energy in the convective region when\
 this occurs determines the explosion energy of the supernova. Typically, the energy in the convective region required to overcome an accretion rate of $0.5 \mathrm{\msun \mathrm{s^{-1}}}$ is $1-3 \times 10^{51}\, {\rm erg}$.  The \citet{fryer:12} formalism assumes that the energy in the convective region increases over time (either on rapid or delayed timescales) and assumed that, when the pressure in the convective region exceeded the ram pressure, an explosion was launched.

\begin{figure*}
\centering
\includegraphics[width=2.\columnwidth]{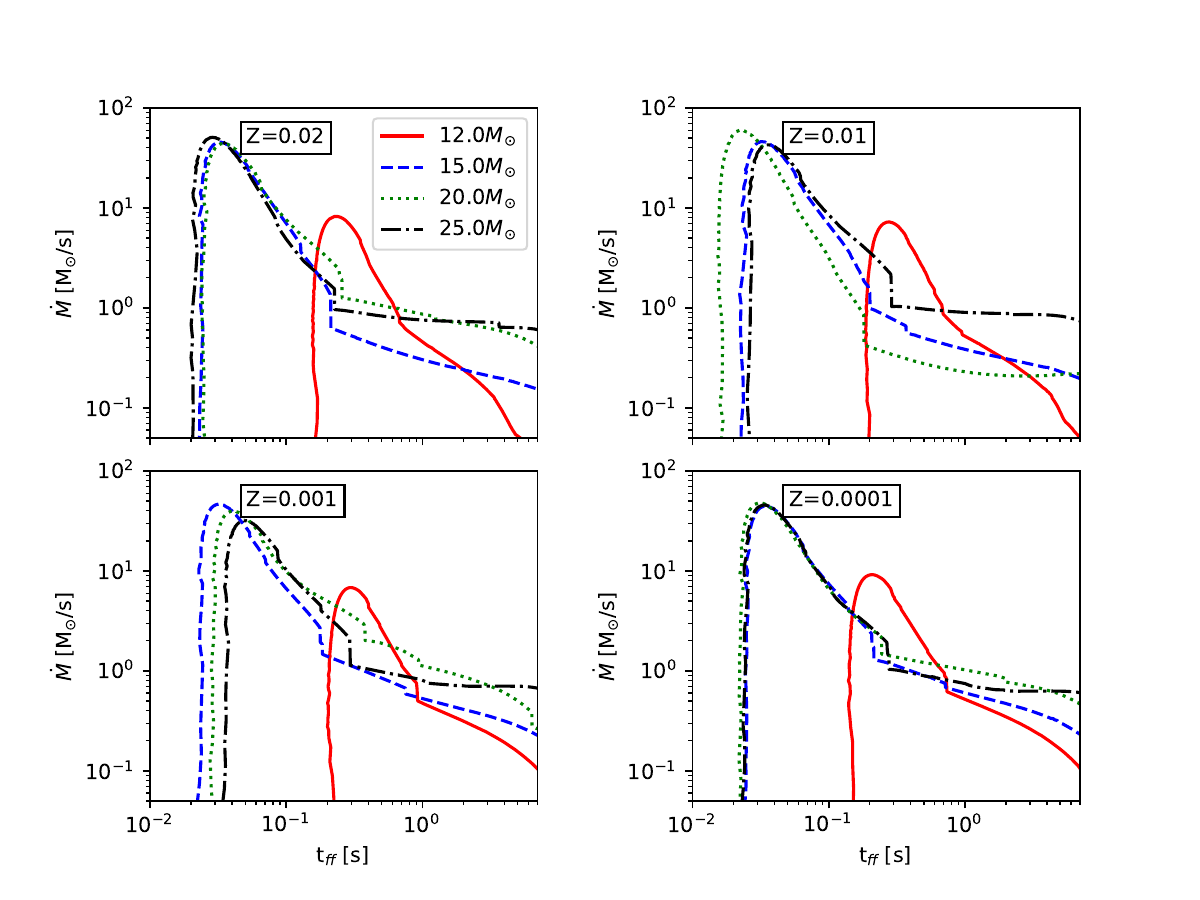}
\caption{Free-fall mass accretion rates for all massive star models for four metallicities at the time of collapse. The time offset for the $12\msun$ models is due to the larger core mass and associated choice of the mass cut (\sect{sec:method:exp}).}
\label{fig:Mdot_tff_collapse}
\end{figure*}

To determine this pressure and, hence, the likelihood of the star exploding, we calculate the accretion rate as a function of time (\fig{fig:Mdot_tff_collapse}).  Just based on these accretion rates, we see that the convective engine is likely to explode both the $\mzams=12$ and $\mzams=15\msun$ progenitors at z=0.02 and z=0.01.  Although the $\mzams=12\msun$ will explode at all metallicities, it becomes increasingly difficult to drive explosions in the $\mzams=15\msun$ at lower metallicities.  Typically, the high accretion rates for the $\mzams=20$ and $\mzams=25\msun$ models make them difficult to explode and we expect no or weak explosions from these models.  The exception is the z=0.01 metallicity $\mzams=20\msun$ star where a shell merger occurred.  This altered the density profile at collapse sufficiently to make this star more-likely to explode, but the high accretion rates at late times is indicative of a large density that may lead to considerable fallback.

The mass at the launch of the explosion can be estimated by looking at the accretion rate as a function of accreted mass.  The unique feature of our $\mzams=12\msun$ \MESA\ progenitor is evident here.  Its core is larger than other progenitors in the literature and it is more likely to make more massive neutron stars.  When the accretion rate falls below $0.5\msun \mathrm{s^{-1}}$, the accreted baryonic mass is $\approx 1.7$ to $1.9\msun$, corresponding to a gravitational mass of $\approx 1.5$ to $1.7\msun$.  In the $Z=0.01$ sequence the $\mzams=12$, $15\msun$ and $20\msun$ models experience O-C shell mergers (\sect{sec:shellmerger}). The consequence is a rapid decline of the mass accretion rate at the location of the bottom of the merged shell. At least at the launch of the explosion, the  $\mzams=20\msun$ stellar model can produce a smaller remnant than our $\mzams=12\msun$ and $\mzams=15\msun$ models.

The maximum temperatures and densities of these delayed explosions at $Z=0.02$ are similar to those shown in Fig.\,31 of P16
for stellar models with $\mzams/\msun$ =  15, 20, 25. We find qualitatively the same increase with initial mass but lower explosion temperatures 
except for the model with $\mzams=25\msun$. We attribute the different explosion conditions to the different pre-SN structures which were
calculated with different stellar evolution codes. The density in Fe core layers at collapse  of our $\mzams=25\msun$, $Z=0.02$ model
is more than $1\, \mathrm{dex}$ larger than in the model of P16 that were calculated with the GENEC code and for which the pre-collapse phase was not modeled (\fig{fig:preexpdensityprofile}).
The densities of the O shell layers are in better agreement.

\subsection{Features at low metallicity}

\subsubsection{H ingestion}
\label{sec:se-hing}
H-ingestion episodes are found in many phases of stellar evolution
particularly in low and zero-metallicity AGB and He-core flash models
\citep[e.g.][]{fujimoto:00,cristallo:09b,campbell:10}, in very late thermal
pulses in models of post-AGB stars \citep{herwig:99c}, in S-AGB stars \citep[cf.\ \sect{sec:hdup},][]{jones:16}. 
The mixing between the H-burning shell and He-burning shell in massive stars have been reported 
for models  at low metallicity in \cite{woosley:95,hirschi:07} and for Pop III models in \cite{heger:10}.

At the first TP of the $\mzams=1\msun$, $Z=0.0001$ model the PDCZ penetrates
slightly into the H-rich envelope. The protons from the envelope are mixed into the PDCZ and react
with \isotope[12]{C} and form \isotope[13]{N}. The latter decays to \isotope[13]{C} which activates the
\isotope[13]{C}($\alpha$,n) neutron source. This leads to the production of heavy elements.
In the following TP the convective He-burning zone penetrates again into the envelope
which leads to the ingestion of much larger amounts of H than previously and stronger surface enrichment of He intershell material.
A H-ingestion flash (HIF) with a peak luminosity of $L_\mathrm{H}\approx10^{10}\, \lsun$ occurs.
This HIF terminates the AGB phase and is shown in the Kippenhahn diagram in
\fig{fig:hif_kipps}.  The conditions are similar to those found in
\cite{iwamoto:04}.

The $\mzams=1\msun$, $Z=0.006$ model experiences a He-shell flash when
it leaves its horizontal post-AGB evolution towards the white dwarf (WD) cooling
track \citep{iben:83a,iben:95b}. This Very-late Thermal Pulse
\citep[VLTP,][]{herwig:00a} causes the PDCZ to reach into the H-rich
envelope, leading to H ingestion, and a born-again phase
\citep{herwig:99c,herwig:11}. The calculation is terminated six years
after the H ingestion due to convergence problems. The H ingestion
leads to the production of heavy elements up to
the first s-process peak in the He intershell which are mixed to the surface.  Due to the
energy release of H burning a connected stable layer forms within the PDCZ and
the convective zone splits into two.  
VLTP events like this are not expected to
influence significantly the composition of the stellar ejecta and the total yields, because the remaining envelope is small and the
cool born-again evolution phase is short. VLTP events have been shown
to posses significant non-radial, global oscillations \citep{herwig:14} which make their
one-dimensional stellar evolution modelling unreliable. 
This applies equally to H-ingestion flashes in low-metallicity AGB stars (Woodward \etal\ in prep.).

In the S-AGB models the time between TP and TDUP becomes shorter for lower metallicity, and this may lead to H-ingestion events \citep{jones:16}. 
Due to the choice of convective boundary mixing parameters this happens only occasionally in these models,  below $Z=0.01$. For example, H ingestion happens during the 29th TP of the $\mzams=7\msun$, $Z=0.001$ model.
For this thermal pulse we obtain neutron densities of up to $N_n=10^{12}\, \mathrm{cm}^{-3}$ in the deepest layers of the PDCZ,
for about five days. The splitting of the PDCZ due to H burning prevents the transport of material from the deep
layers to the surface. 
%Since these events happen so rarely in these models the nucleosynthesis of HIFs does not contribute significantly to the stellar yields. 
Since these events are not frequent in these $7\msun$ models, the nucleosynthesis of HIFs does not contribute significantly to the stellar yields presented here.

\begin{figure}
\centering
\includegraphics[width=\columnwidth]{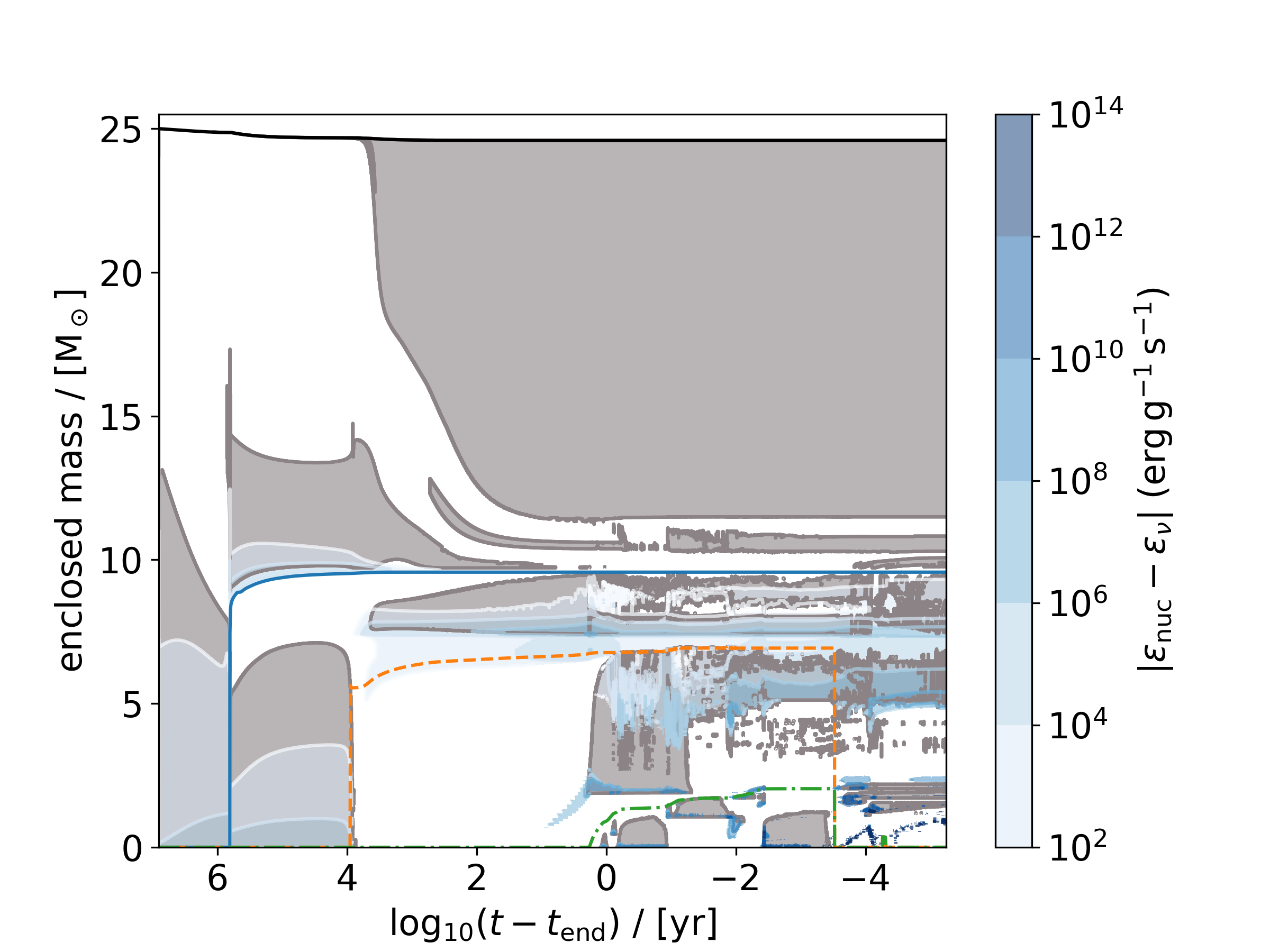}
\includegraphics[width=\columnwidth]{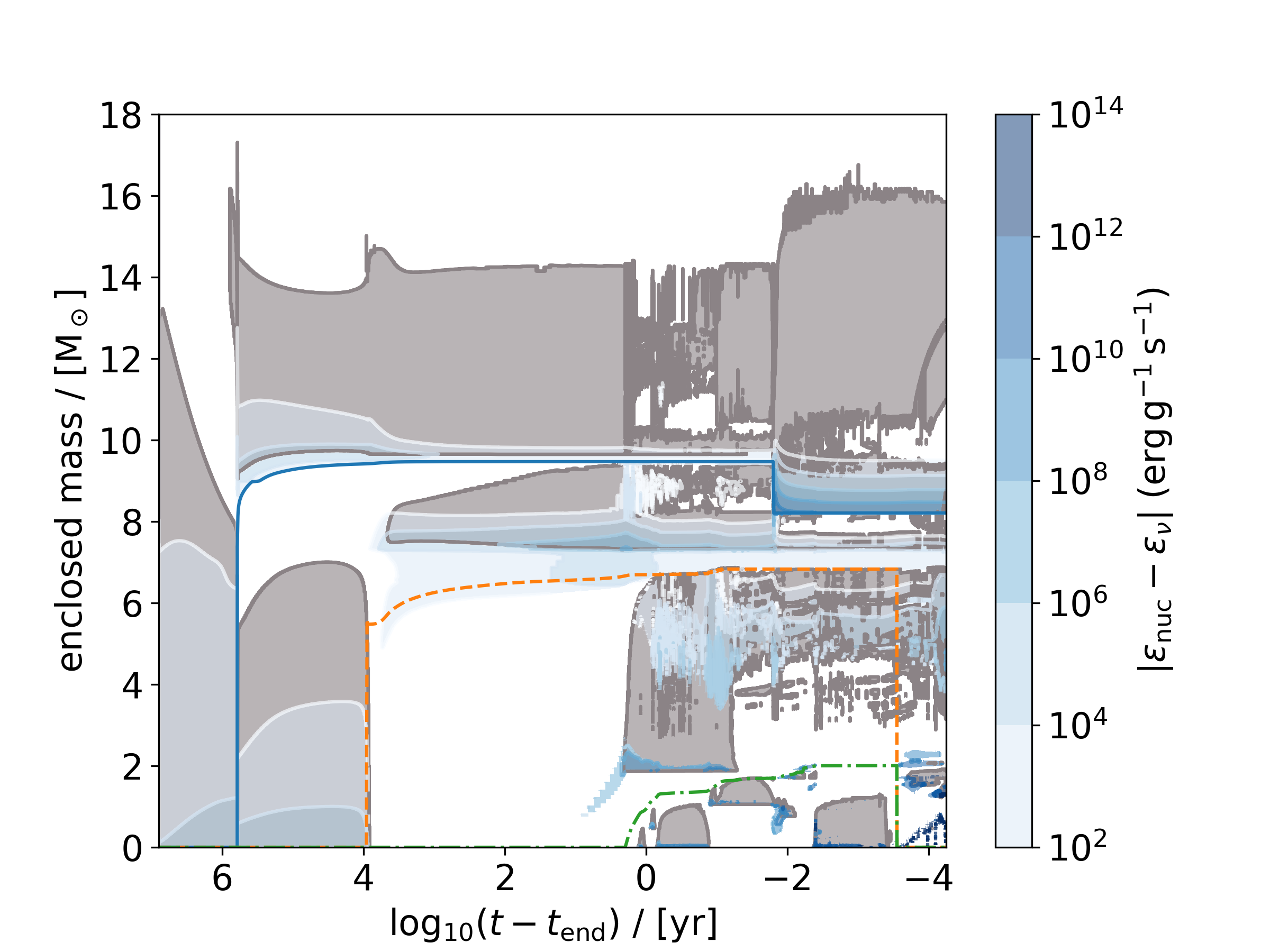}
\
\
\caption{Kippenhahn diagrams for two stellar models with
  $\mzams=25\msun$ at $Z=0.001$ (top) and $Z=0.0001$ (bottom).  Grey
  areas are convective zones. The blue solid line and orange dashed
  line represent H-free and He-free cores respectively.  The green
  dot-dashed line represents the C-free core. The x axis is the
  logarithm of the time until $t_\mathrm{end}$ when the infall
  velocity reaches $1000\, \mathrm{km/s}$. Also shown is the
  nuclear energy generation $\epsilon_{nuc}$ in blue shades. The
  specific energy loss rate due to neutrino production via nuclear
  reaction $\epsilon_{\nu}$ is subtracted and only positive values of
  $|\epsilon_{nuc}- \epsilon_{\nu}|$ are plotted.}
\label{fig:kip_cont_massive}
\end{figure}

Stellar models of $\mzams=20\msun$ and $\mzams=25\msun$ of $Z=0.0001$ experience H ingestion
at the beginning of convective C shell burning and during O shell burning respectively.
At higher metallicity we find H ingestion in the $\mzams=20\msun$, $Z=0.001$ model and in the
$\mzams=12\msun$, $Z=0.006$ model. In both models H ingestion events occur during Si shell burning.
%The ingestion events are independent of the choice of CMB since we do not apply CBM beyond He-core burning.
These H-ingestion, or sometimes H/He-shell mixing events happen without the application of CBM at the boundaries of the convective He shell (as all other convective boundaries post-He core burning). The penetration into the convective He-burning layer is visible
for the $\mzams=25\msun$, $Z=0.0001$ model in \fig{fig:kip_cont_massive}.
The resulting energy release leads to the formation of two extended convective regions which persist until collapse. We find at the bottom of the He-shell convective zone neutron densities
close to $N_n=10^{11}\, \mathrm{cm}^{-3}$ which remain for days until core collapse. There
is only a minor production of heavy elements but lighter elements such as F are effectively produced and
contribute a relevant fraction of the total stellar yields of this stellar model. 
%\mpcom{NOT BEFORE SUBMISSION, BUT CAN YOU CHECK WHAT IS THE CHANNEL TO MAKE THIS EXTRA F? IS O18(NG)O19(B-)F19? YOU CAN ADD THEN A SENTENCE ABOUT IT IN THE NEXT ROUND.
%IT WOULD BE GOOD TO SAY HOW MUCH F IS MADE. YOU SAY "A RELEVANT FRACTION". CAN YOU SAY IF IT IS E.G., 50\% OR 80\% OF THE TOTAL YIELDS?}

Detailed investigations of the nucleosynthesis and 3D stellar hydrodynamics of the H-ingestion event in the post-AGB star Sakurai's object \citep{herwig:11,herwig:14} have shown that the assumption of spherical symmetry and the approximation of mixing via mixing length theory of convection are not appropriate. This is consistent with the failure of such 1D models to reproduce  several key observables, such as light-curve and heavy-element abundance patterns of Sakurai's object. This suggests that the properties of H-ingestion events in this stellar yield grid are indicative at best, and need to be investigated further through  3D hydrodynamics simulations.

%Nucleosynthesis predictions
%need to be derived from 3D hydrodynamic simulations when available.
%in 3D are possible one has to rely on simplifications.

%This limits the predictability
%of nucleosynthesis within the HIF.

%All ingestion events have a strong influence on the final yields and will be addressed
%in Sec. ..

%The
%ingestion event of the 25Msun star is shown in the kippenhahn diagram of Fig. \ref{fig:hif_kipps}. 
%There ..

\subsubsection{Hot bottom burning} \label{sec:hbb}

\noindent 

\begin{figure}
\centering
\includegraphics[width=\columnwidth]{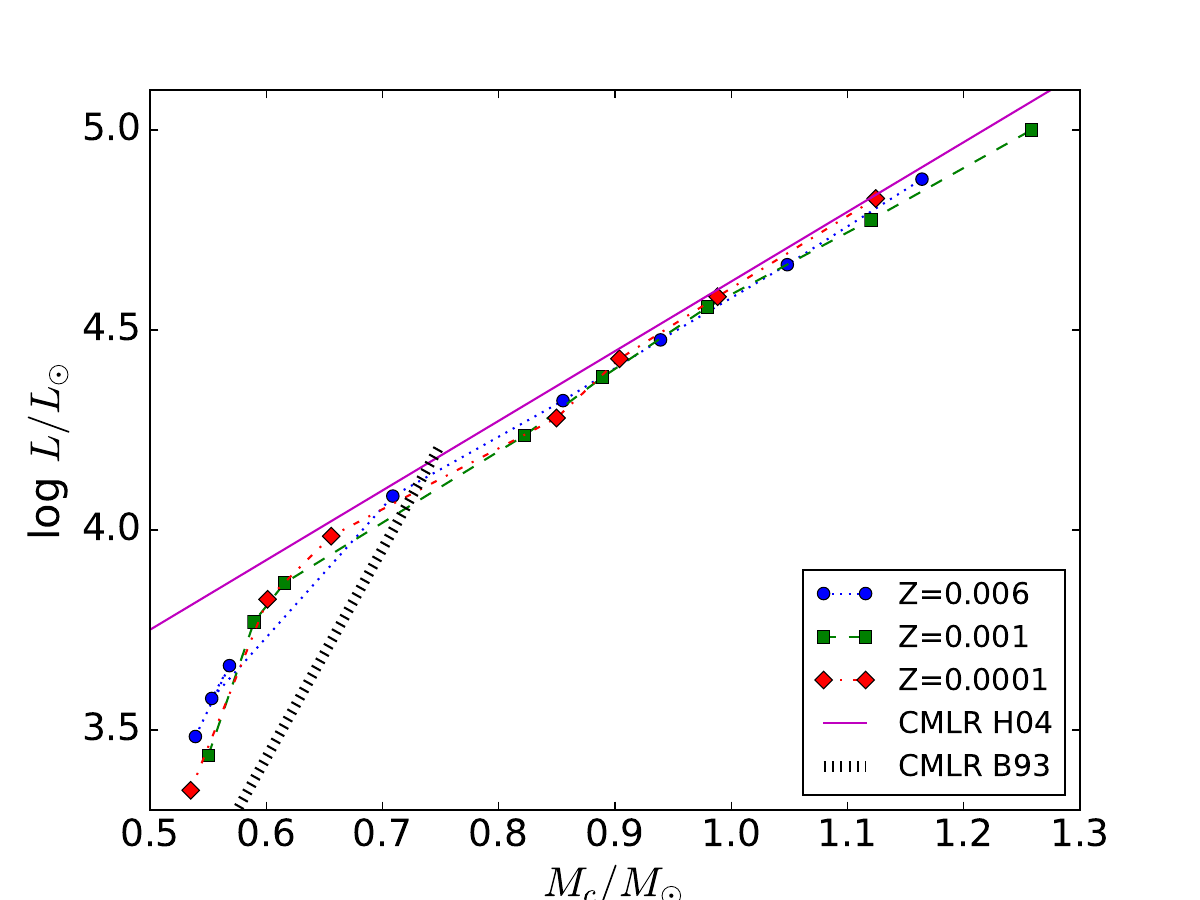}
\includegraphics[width=\columnwidth]{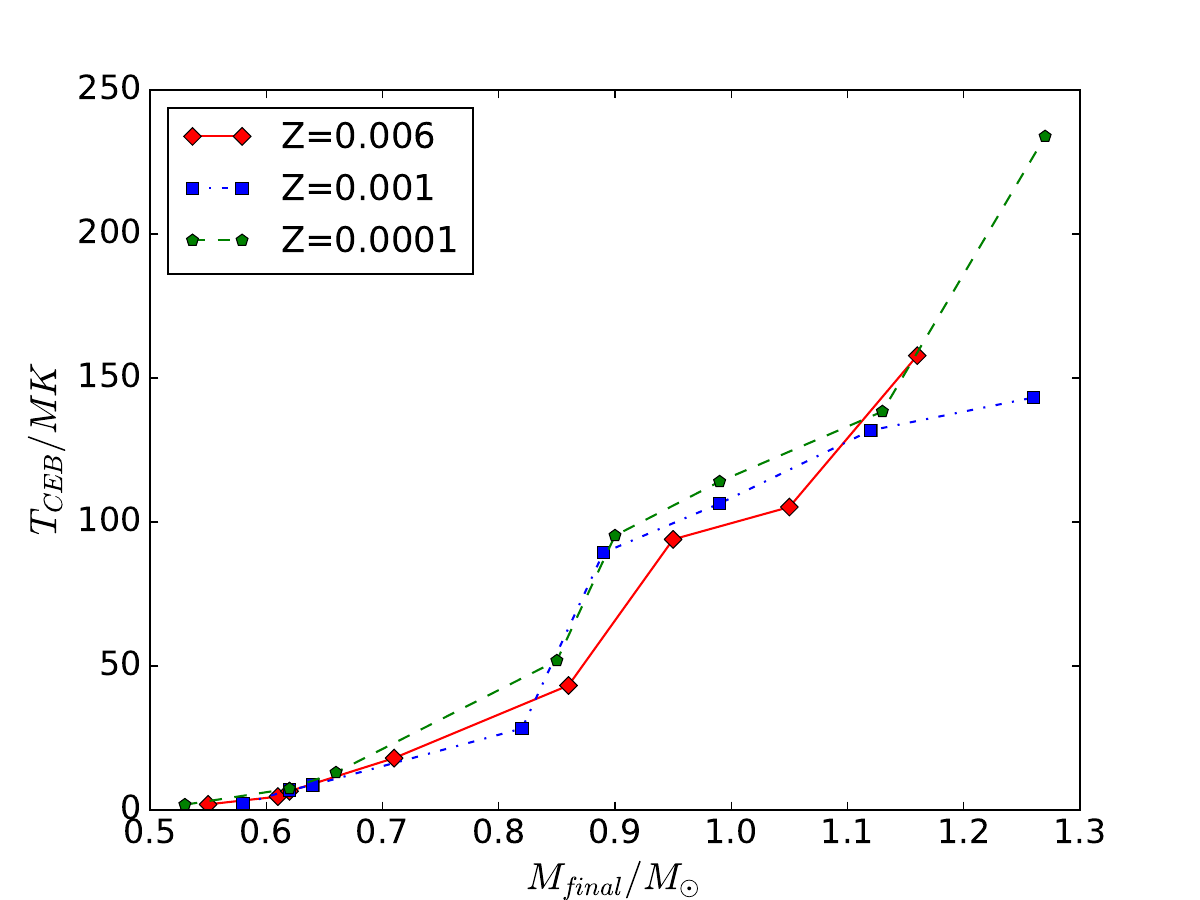}
\caption{Average luminosity versus average core mass of the TP-AGB stage for stellar models at $Z=0.006, 0.001$ and $0.0001$ in comparison
with the linear and exponential core-mass luminosity relations of \citet[][CMLR B93]{bloecker:93b} and \citet[][CMLR H04]{herwig:04a} respectively (top). The luminosities of low-mass models are higher than the the classical CMLR because of third dredge-up \citep{herwig:98b}.
Maximum temperature at the bottom of the convective envelope $T_{\rm CEB}$ versus final core mass during the AGB evolution (bottom).} 
%In case of the stellar model of $7\msun$ we have chosen the maximum temperature found in the H shell burning phases.}
\label{fig:corelum_Tceb}
\end{figure}

\noindent The temperature at the bottom of the convective envelope $T_\mathrm{CEB}$ increases with increasing initial mass
and decreasing metallicity as shown in \fig{fig:corelum_Tceb} and reaches up to $T_\mathrm{CEB}=2.3\times10^8\, \mathrm{K}$.
In S-AGB models with $\mzams=7\msun$ at $Z=0.0001$ and $Z=0.006$ temperatures reach more than $T_\mathrm{CEB}=1.5\times10^8\, \mathrm{K}$ which allows
the activation of the NeNa and MgAl cycles. The $\mzams=3\msun$, $Z=0.0001$ model reaches
$T_\mathrm{CEB}=4\times10^7\, \mathrm{K}$ which leads to HBB. Models of the same mass but of higher metallicity
do not experience HBB (\tab{table:agb_properties}).
The threshold initial mass for HBB in \cite{ventura:13} was found 
to be $\mzams=3\msun$ at $Z=0.0003$ and $\mzams=3.5\msun$ at $Z=0.008$ which is similar to our findings.
%\cite{fishlock:14} found the initial mass for HBB to be $3\msun$ at $Z=0.001$.ÃÂ
HBB is active in stellar models of masses as low as $\mzams=3\msun$ in agreement with models at $Z=0.001$
of \cite{fishlock:14}.

%e.g. Siess has only massive,SAGB stars...

\subsubsection{Effects of hot dredge-up and dredge-out}
\label{sec:hdup}

\cite{herwig:03c} find that hot dredge-up is characterized by extreme H-burning luminosities of $L_\mathrm{H}=2\times10^{6}\, \lsun$ for their  $\mzams=5\msun$, $Z=0.0001$ model.
For stellar models with $\mzams\leq4\msun$ and $Z=0.0001$ $L_\mathrm{H}$ often exceeds the peak He-burning luminosities of the TP.
Under the most extreme conditions in models with $\mzams=6\msun$ and $\mzams=7\msun$ we find $L_\mathrm{H}>10^{9}\, \lsun$. %L_{\odot}$.
At higher metallicities $L_\mathrm{H}$ is lower.
Because of the reduced CBM efficiency $f_\mathrm{CE}$ in massive AGB and S-AGB models (see \sect{subsecCBM}) the size of \cdr\ pocket  decreases substantially with increasing initial mass. Additionally, the pressure scale height at the core-envelope interface decreases with increasing initial mass which leads to a further decrease of the CBM in the parametrized model.
%This is in addition to the decrease of any partial-mixing zone with increasing core mass in any model parameterized with the 
%pressure scale height at the core-envelope interface. % \fig{fig:C13-pocket-sizes} \fhcom{@CR: include C13-pocket size figure}. 
This leads to \cdr\ pockets in S-AGB models below $10^{-7}\, \msun$ at $Z=0.0001$.
%\mpcom{THIS MAY BE A STUPID POINT. WITH MODELS DOING HDU, CAN WE STILL TALK ABOUT C13-POCKETS? WITH ~1E-7 MSUN C13-RICH %REGIONS, THIS IS MORE LIKE THE BOTTOM OF THE H SHELL, AND NO IMPACT ON THE PRODUCTION OF HEAVY ELEMENTS. }
 	
Dredge-out is found in the most massive AGB models during second DUP when the
convective He-burning shell grows in mass and merges with the
convective envelope. This leads to the enrichment of the surface with
products of He-shell burning \citep{ritossa:99}.  H can be entrained
into the He-burning convection zone and ignite as a flash.  This is
another H-ingestion event \citep{gil-pons:10,jones:16}.  We find
dredge-out in S-AGB models with $\mzams=7\msun$ at $Z=0.001$ and
$Z=0.0001$.  The flash at $Z=0.0001$ produces a peak in luminosity of up to
$L_\mathrm{H}\approx10^8\, \lsun$.  The maximum H-burning luminosities
agree well with \cite{jones:16}. 
The initial masses of our stellar models with dredge-out are below the lower initial mass limit of
dredge-out of $\mzams\gtrapprox9\msun$ as reported by
\cite{gil-pons:10}, presumably due to difference in the core overshooting prescription.
%\mpcom{I CANNOT UNDERSTAND WHAT YOU WANT TO SAY IN THIS LAST SENTENCE. CAN YOU MAKE IT SIMPLER?}
%and % which
%is 2  higher then found in \cite{gil-pons} for their $9\msun$ model of $Z=0.00001$.

\subsubsection{Carbon flame quenching in S-AGB stars}

%\textbf{Rework this section!!!}
%CBM also active in C burning conv. zones?
%is it well resolved as in denissenkov?
% look at final profile, centroal c/o  core?
\noindent 
In the  $\mzams=7\msun$, $Z=0.006$ S-AGB model the propagation of the C flame toward the center is
quenched \citep{denissenkov:13}. The C-flame quenching depends sensitively on the assumption of CBM, which is essentially unconstrained. 
If CBM at the bottom of the C-burning shell is efficient enough to quench the flame, then the result is a hybrid core. It consists of a inner C-O core of $\approx0.145\msun$ surrounded by
thicker layers of O, Ne and Mg. 
For stellar models with $\mzams=7\msun$ the first C-burning flash occurs at $Z=0.0001$ closer to the center
than at higher metallicity. 
The C burning moves outward in mass through a series of convective C-shell burning episodes. % (\fig{fig:core_evolution_hybrids}).
The location of the first C ignition is further outwards for models of higher metallicity due to the higher
degeneracy of the lower core masses \citep{garcia-berro:97,siess:07}.
The onset of C burning coincides with the beginning of the second DUP for the $\mzams=7\msun$, $Z=0.006$ model. At higher metallicity
the C burning starts earlier than at lower metallicity. %(\fig{fig:core_evolution_hybrids}).
%show time instead of model number in the figures?!
The difference in metallicity has a qualitatively similar effect on
convective C burning as the difference in initial mass between
$\mzams=7.6\msun$ and $\mzams=9\msun$ shown in Fig.\,3 in \cite{farmer:15}.
%The C-flame quenching takes place closer to the center for the stellar model with initial mass of $6\msun$ at $Z=0.0001$. 
Possible implications of hybrid WDs are discussed in \cite{denissenkov:17}.

%Tables for Results

%\begin{table}
%\rotatebox{90}{
%\resizebox{0.5\textwidth}{!}{
%\begin{minipage}{\textwidth}
%\begin{center}
%\begin{sidewaystable}
\begin{landscape}
\begin{table}
%{ %\tiny
%\fontsize{5}{4}\selectfont
\centering
%\resizebox{0.98\textwidth}{!}{
%\scalebox{0.8}{%
\centering
\begin{tabular}{cccccccccccccccc}
\hline
$\mzams$ &  $m_c$       & $\log L_{\ast}$ & $R_{\ast}$  & $N_{\rm TP}$ & $N_{\rm TDUP}$ & $t_{\rm TPI}$  & $\Delta M_{\rm Dmax}$ & $M_D$              & $t_{ip}$ & $M_{\rm lost}$ & $\log T_{\rm CEB,max}$ & $\log T_{\rm PDCZ,max}$ & $M_{\rm PDCZ}$ & $\log L_{\rm He,max}$ & $\log L_{\rm max}$\\
\hline
{[$M_{\odot}$]}  &  {[$M_{\odot}$]}  &  {[$L_{\odot}$]}   & {[$R_{\odot}$]} &         &                &  {[$10^6\, {\rm yr}$]} & {[$10^{-2}\, \msun$]} & {[$10^{-2}\, \msun$]} & {[${\rm yr}$]}     & {[$M_{\odot}$]} &   {[${\rm K}$]}  &  [${\rm K}$] & {[$10^{-2}\, \msun$]} & {[$L_{\odot}$]} & {[$L_{\odot}$]}\\
\hline
\multicolumn{16}{c}{Z=0.0001} \\
\hline
1.00   & 0.532  & 3.19 & 75&  2      &  1 &  5.726E+03  &  2.485       &  2.485         &  274820  & 0.33       &  6.266  &  8.312  &  3.986  &  6.72  &  3.63\\
1.65   & 0.589  & 3.77 & 161&  12    &  11 &  1.231E+03 &  0.702       &  5.482         &  91155  & 0.99       &  6.870  &  8.461  &  2.956  &  7.53  &  4.05\\
2.00   & 0.655  & 3.97 & 205&  11    &  10 &  7.494E+02 &  0.895       &  6.242         &  56131  & 1.31       &  7.114  &  8.490  &  1.906  &  7.89  &  4.14\\ 
3.00   & 0.848  & 4.29 & 295&  11    &  10 &  2.722E+02 &  0.242       &  1.897         &  8765  & 2.01       &  7.715  &  8.514  &  0.569  &  7.63  &  4.48\\
4.00   & 0.899  & 4.43 & 345&  19    &  18 &  1.414E+02 &  0.246       &  2.506         &  5253  & 3.01       &  7.979  &  8.541  &  0.376  &  8.00  &  4.55\\  
5.00   & 0.982  & 4.59 & 413&  29    &  20 &  8.805E+01 &  0.111       &  1.331         &  2228  & 3.93       &  8.057  &  8.553  &  0.165  &  7.82  &  4.76\\
6.00   & 1.124  & 4.83 & 572&  19    &  16 &  6.115E+01 &  0.038       &  0.428         &  824  & 4.52       &  8.141  &  8.561  &  0.049  &  7.07  &  4.96\\
7.00   & 1.272  & 5.05 & 743&  27    &  21 &  4.557E+01 &  0.007       &  0.091         &  134  & 4.70       &  8.369  &  8.597  &  0.009  &  6.72  &  5.10\\
\hline
\multicolumn{16}{l}{$M_{ini}$: Initial stellar mass.} \\
\multicolumn{16}{l}{  $m_c$: H-free core mass at the first TP.}\\
\multicolumn{16}{l}{ $L_{\ast}$: Approximated mean stellar luminosity.}\\
\multicolumn{16}{l}{ $R_{\ast}$ : Approximated mean stellar radius.}\\
\multicolumn{16}{l}{ $N_{\rm TP}$: Number of TPs.}\\
\multicolumn{16}{l}{ $N_{\rm TDUP}$ : Number of TPs with TDUP.}\\
\multicolumn{16}{l}{ $t_{\rm TPI}$: Time at first TP.} \\
\multicolumn{16}{l}{ $\Delta M_{\rm Dmax}$: Maximum dredged-up mass after a single TP.} \\
\multicolumn{16}{l}{ $M_D$: Total dredged-up mass of all TPs.} \\
\multicolumn{16}{l}{ $t_{ip}$ : Average interpulse duration of TPs.} \\
\multicolumn{16}{l}{$M_{\rm lost}$: Total mass lost during the TP-AGB phase.} \\
\multicolumn{16}{l}{$T_{\rm PDCZ,max}$: Maximum temperature during the TP-AGB phase.}\\
\multicolumn{16}{l}{$M_{\rm PDCZ}$: Maximum size of PDCZ.} \\
\multicolumn{16}{l}{$\log L_{\rm He,max}$: Maximum He luminosity during TP-AGB phase.}\\
\multicolumn{16}{l}{$\log L_{max}$: Maximum total luminosity during TP-AGB phase.}\\
\end{tabular}
%}
\caption{TP-AGB properties for models at $Z=0.0001$. Properties for $Z=0.006$, $0.001$ are available with the online version of the paper (\app{sec:appendix}).}
\label{table:agb_properties}
%\end{sidewaystable}
\end{table}
%\caption{TP-AGB properties for models at $Z=0.0001$. The complete table is available with the online version of the paper.
%We provide tables for other metallicities online (\app{sec:appendix}).}
\end{landscape}

\begin{landscape}
%\begin{sidewaystable}
\begin{table}
{
%\fontsize{2.5}{4}\selectfont
\centering
%\resizebox{0.98\textwidth}{!}{
{ %\tiny
%\fontsize{2.5}{4}\selectfont
%\scalebox{0.8}{%
\centering
\begin{tabular}{cccccccccc}
\hline
 TP & $t_{\rm TP}$ & $\log T_{\rm FBOT}$ & $\log T_{\rm HES}$ & $\log T_{\rm HS}$ & $\log T_{\rm CEB}$ & $m_{\rm FBOT}$ & $m_{\rm HTP}$ & $m_{\rm D,max}$ & $M_{\ast}$\\
\hline
 & [${\rm yr}$] & [${\rm K}$] & [${\rm K}$] & [${\rm K}$] & [${\rm K}$] & [$M_{\odot}$] & [$M_{\odot}$] & [$M_{\odot}$] & [$M_{\odot}$] \\
\hline
\multicolumn{10}{c}{$M=1.0M_{\odot}$} \\
\hline
1  & 0.00E+00   &  8.31 &  8.16 &  7.75     &  5.98  &  0.4926  &  0.5324        &  0.5358          &  0.867 \\
2  & 2.75E+05   &  8.09 &  8.05 &  8.05     &  5.99  &  0.5218  &  0.5374        &  0.0000          &  0.867 \\
\hline
\multicolumn{10}{l}{TP: TP number.} \\
\multicolumn{10}{l}{ $t_{\rm TP}$: Time since the first TP.} \\
\multicolumn{10}{l}{$T_{\rm FBOT}$: Largest temperature at the bottom of the PDCZ.} \\
\multicolumn{10}{l}{$T_{\rm HES}$: Temperature in the He-burning shell during deepest extend of TDUP.} \\
\multicolumn{10}{l}{ $T_{\rm CEB}$: Temperature at the bottom of the convective envelope during deepest extend of TDUP.} \\
\multicolumn{10}{l}{$m_{\rm FBOT}$: Minimum mass coordinate of the bottom of the He-flash convective zone.} \\
\multicolumn{10}{l}{$m_{\rm D,max}$: Mass coordinate of the H-free core at the time of the TP.} \\
\multicolumn{10}{l}{ $M_{\ast}$: Stellar mass at the TP.}
\end{tabular}
}
}
\caption{Model properties of the TP-AGB phase for $Z=0.0001$. Properties for $Z=0.006$, $0.001$ are available with the online version of the paper (\app{sec:appendix}).}
\label{agbmodelprop2_1_5a}
%}
%}
%\end{sidewaystable}
\end{table}
\end{landscape}
%}
 %all in table 

\begin{table}
\center
\begin{tabular}{lrr}
%\hline
%\multicolumn{3}{c}{Z=0.0001} \\
\hline
\mzams & $M_{\mathrm{final}}$ & $\tau_{\mathrm{total}}$ \\
\hline
 [$\msun$]           & [$\msun$] & [${\rm yr}$] \\
\hline
1.0 & 0.592 & 5.670E+09 \\
1.65& 0.637 & 1.211E+09 \\
2.0 & 0.665 & 6.972E+08 \\
3.0 & 0.852 & 2.471E+08 \\
4.0 & 0.905 & 1.347E+08 \\
5.0 & 0.992 & 8.123E+07 \\
6.0 & 1.125 & 5.642E+07 \\
7.0 & 1.272 & 4.217E+07 \\

\hline
\end{tabular}
\caption{Final core masses $M_{\mathrm{final}}$ and total lifetime 
$\tau_{\mathrm{total}}$ for $Z=0.0001$. We provide tables for other
metallicities online (Appendix \ref{sec:appendix}).}

\label{tab:cores_agb}
\end{table}

\begin{table*}
\centering
%\scalebox{0.6}{%
\centering
\begin{tabular}{lrrrrrrr}
\hline
\mzams & $\tau_{\mathrm{H}}$ & $\tau_{\mathrm{He}}$ & $\tau_{\mathrm{C}}$ & $\tau_{\mathrm{Ne}}$ & $\tau_{\mathrm{O}}$ & $\tau_{\mathrm{Si}}$ & $\tau_{\mathrm{total}}$\\
\hline
\multicolumn{8}{c}{Z = 0.02} \\
\hline
12  & 1.742E+07 & 1.669E+06 & 1.046E+04 & 1.046E+01 & 2.973E+00 & 1.895E-01 & 1.935E+07\\
15  & 1.243E+07 & 1.250E+06 & 1.835E+03 & 2.829E+00 & 1.361E+00 & 8.840E-02 & 1.386E+07\\
20  & 8.687E+06 & 8.209E+05 & 1.270E+02 & 1.811E+00 & 7.086E-01 & 5.071E-02 & 9.596E+06\\
25  & 6.873E+06 & 6.426E+05 & 2.525E+02 & 5.303E-01 & 1.390E-01 & 1.385E-02 & 7.585E+06\\
\hline
\end{tabular}
%}
\caption{Lifetimes of major central burning stages of massive star models.
Shown are lifetimes for H burning, $\tau_{\mathrm{H}}$, He burning, $\tau_{\mathrm{He}}$, C burning, $\tau_{\mathrm{C}}$,
Ne burning $\tau_{\mathrm{Ne}}$, O burning, $\tau_{\mathrm{O}}$, Si burning, $\tau_{\mathrm{Si}}$, and the total lifetime
of the stellar models, $\tau_{\mathrm{total}}$. Times in $\mathrm{yr}$. The complete table is available with the online version of the paper.}
\label{tab:lifetimesmassive}
\end{table*}

\begin{table}
\center
\begin{tabular}{lrrrrrr}
\hline
\mzams & \multicolumn{3}{c}{$15\msun$} & \multicolumn{3}{c}{$25\msun$} \\
\hline
  & this work & M02 & P16 & this work & M02 & P16\\
\hline
 $M_{\mathrm{\alpha}}$ & 5.09 & 4.45 & 4.81 &  9.66 & 8.44 & 9.39 \\
 $M_{\mathrm{CO}}$            & 3.27 & 2.27 & 2.84 &  7.26 & 5.35 & 6.45 \\
 $M_{\mathrm{Si}}$         & 2.02  &  & 1.7 &1.99  &  & 1.85 \\
\hline
\end{tabular}
\caption{Comparison of the He core mass ($M_{\mathrm{\alpha}}$), CO core mass ($M_{\mathrm{CO}}$)
and Si core mass $M_{\mathrm{Si}}$ of massive star models at $Z=0.006$ of this work with models at $Z=0.004$ of M02 and models at $Z=0.01$ of P16. 
Core masses are in $\msun$.}
\label{tab:cores_comparison_meynet}
\end{table}

%%%%%%%%%%%%%%%%%%%%%%%%%

%%%%%%%%%%%%%%%%%%%%%%%%%

%%%%%%%%%%%%%%%%%%%%%%%%%

%%%%%%%%%%%%%%%%%%%%%%%%%

%%%%%%%%%%%%%%%%%%%%%%%%%

%%%%%%%%%%%%%%%%%%%%%%%%%

%%%%%%%%%%%%%%%%%%%%%%%%%%%%%%

\begin{figure*}
\centering
\includegraphics[width=2.\columnwidth]{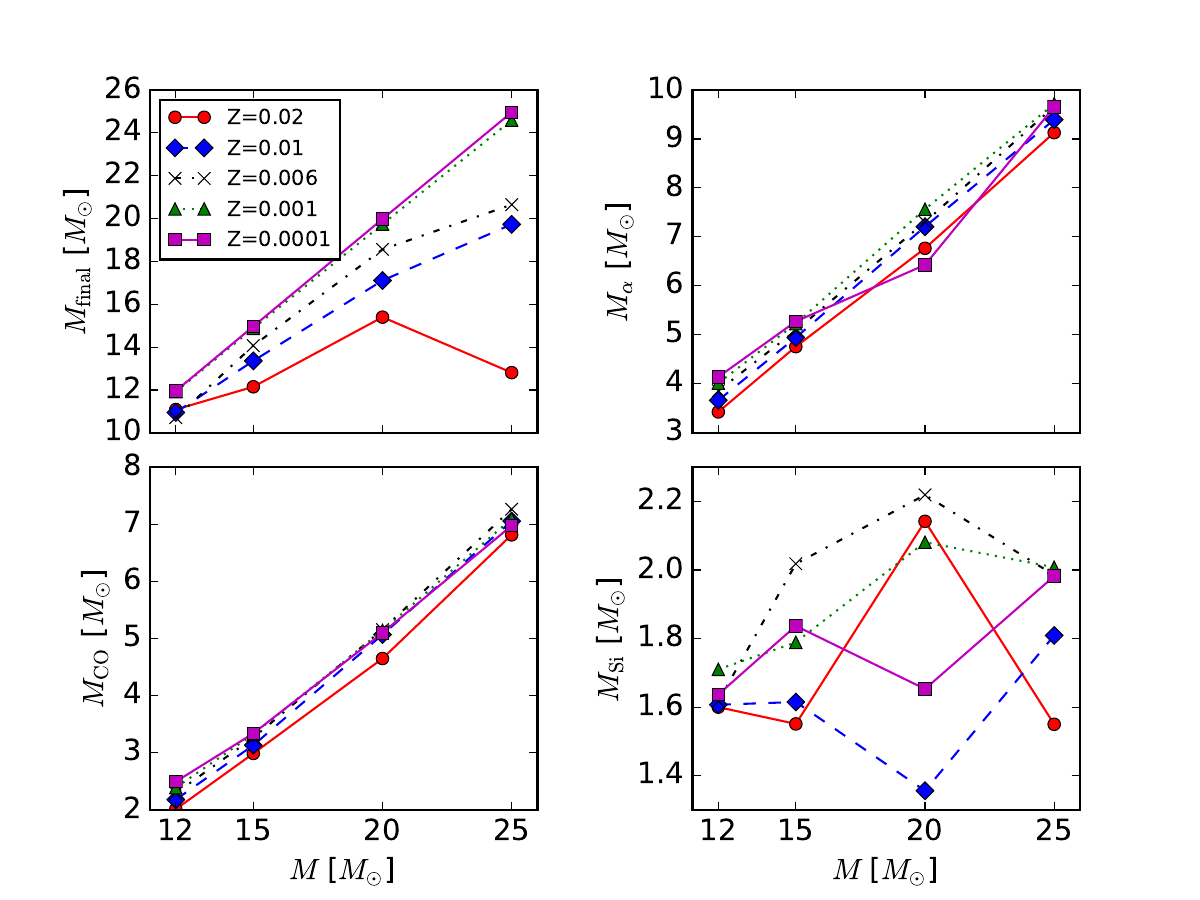}
\caption{Core masses versus initial mass at the time of core collapse for massive star models.
For each model the final mass ($M_{\mathrm{final}}$),
He core mass ($M_{\mathrm{\alpha}}$), CO core mass ($M_{\mathrm{CO}}$) and %M^{75\%}
Si core mass ($M_{\mathrm{Si}}$) are shown.
%Values with $^*$ were taken at maximum of the sum of Si+Ar+Ca+Ti.
}
\label{fig:cores_massive}
\end{figure*}

%%%%%%%%%%%%%%%%%%%%%%%%% PNG for now due to large size of kippenhahn diagrams

%%%%%%%%%%%%%%%%%%%%%%%%%

%%%%%%%%%%%%%%%%%%%%%%%%%

%%%%%%%%%%%%%%%%%%%%%%%%%
%%%%%%%%%%%%%%%%%%%%%%%%%

\begin{figure}
\centering
\includegraphics[width=\columnwidth]{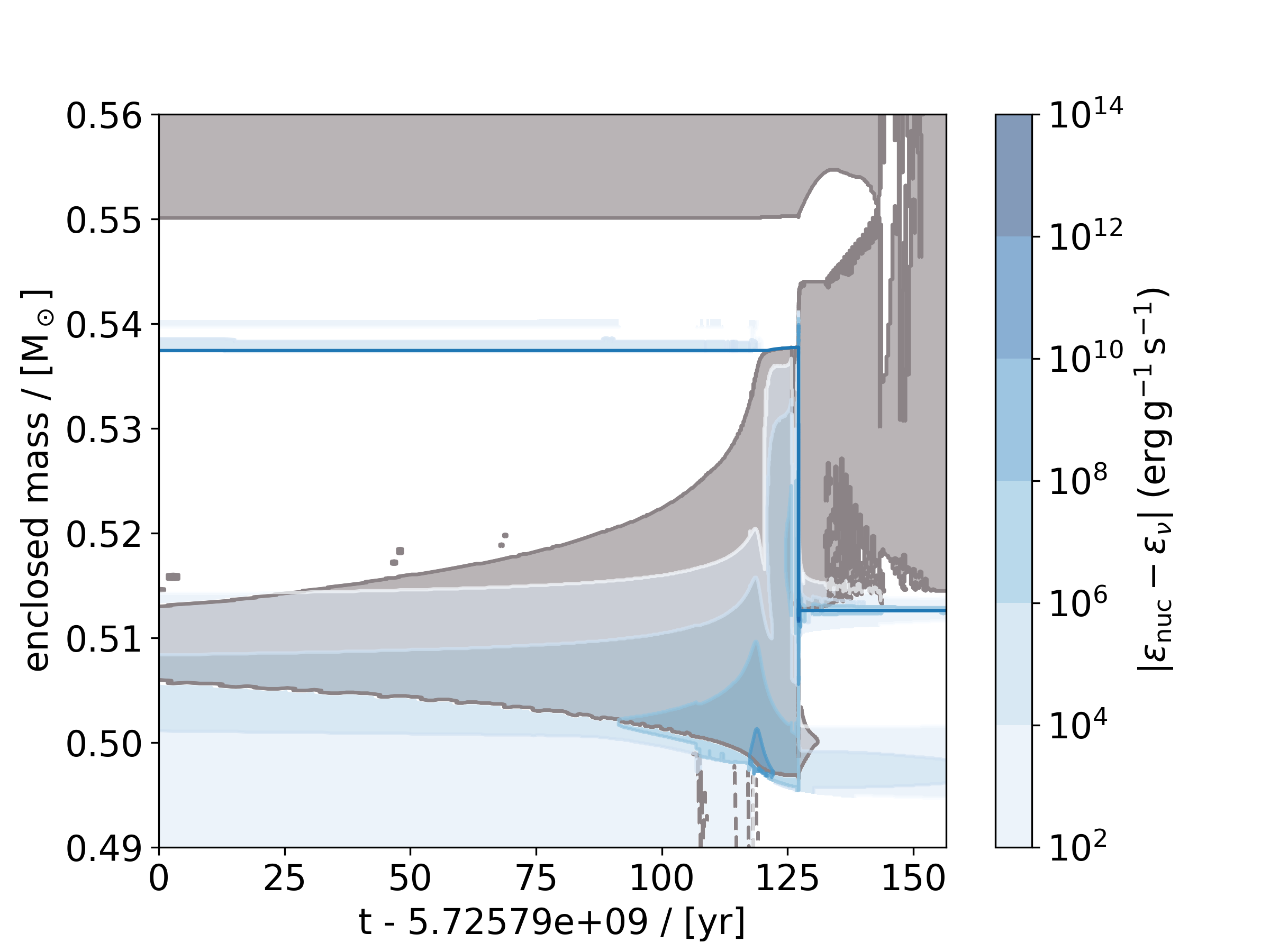}
\caption{H-ingestion in the $\mzams=1\msun$, $Z=0.0001$ model. The
  blue solid line represents the H-free core. The H ingestion during
  the second thermal pulse terminates the TP-AGB phase.}
\label{fig:hif_kipps}
\end{figure}

%nucleosynthesis

\section{Post-processing nucleosynthesis results}
\label{ppnucleo}

This section is complementary to the discussion in P16 ($Z\ge0.01$) and the main focus are results obtained for $Z\le0.006$. 
Processes covered include, among others,  the weak and main s process
\citep{kaeppeler:89,straniero:95,gallino:98,kaeppeler:11}, the
$\alpha$-process \citep{woosley:92,magkotsios:10} and $\gamma$ process
\citep{rayet:95,arnould:03}. Overproduction factors ${\rm OP}_{\rm im}$  (\sect{sec:nugrid_codes})
provide an overview of which stellar models at which metallicity contribute to which
elements/isotopes (Figs.\,\ref{fig:agb_prodfac_set1_2} through \ref{fig:massive_prodfac_set1_5a}).

Final yields with their wind contribution, pre-SN and SN contribution (\sect{sec:nugrid_codes}) are shown for $Z=0.0001$ in \tab{tab:yields_all_delay_set1_5}, and all others are available online (Appendix \ref{sec:appendix}).  In this section we briefly discuss 
the results from our post-process calculations.

\subsection{First dredge-up, second dredge-up and dredge-out}

%%%%%%%%He
%\noindent \textbf{He} \textemdash{ } 
In the AGB models with $\mzams=1.65\msun$, He originates mostly from the first dredge-up.
For higher initial masses the contribution of the second dredge-up increases while the contribution of the first dredge-up decreases. 
Stellar models of the same initial mass experience deeper first dredge-up at higher metallicity.
The initial mass above which the second dredge-up
is responsible for most He  production is $\mzams=2\msun$ at $Z=0.0001$ and $\mzams=3\msun$ at $Z=0.006$.
The largest overproduction of He in AGB models occurs at the highest initial masses.

%%%%%%%%%%%%C
%\noindent \textbf{C} \textemdash{ } 
The C overproduction factors of AGB models peak at $\mzams=2\msun$ for $Z=0.006$ and $Z=0.001$ and
at $\mzams=1.65\msun$ for $Z=0.0001$ (\fig{fig:cno_prodfac_zdep}).
The total amount of dredged-up material reaches a maximum in these three initial stellar models (\tab{table:agb_properties}).
The largest overproduction factors of AGB models are slightly larger than those found in massive star models.
We find dredge-out \citep{ritossa:99,jones:16} 
%\fhcom{add Jones+ 16 ref} 
in stellar models with initial mass of $7\msun$ at $Z=0.001$ and $Z=0.0001$ where it is the main source of 
surface enrichment of C. %\cite{doherty:14}

%%%%%%%%%%%%%%%%O
%\noindent \textbf{O} \textemdash{ } 
In the lowest-metallicity cases, O production factors in AGB stars can reach $10\, \%$ of that in massive stars (\fig{fig:cno_prodfac_zdep}). In AGB stars O is produced in AGB models in the He intershell \citep{herwig:05}. %and during He- and Ne-burning layers of massive star models \citep{woosley:02}. 
CBM   at the bottom of the PDCZ in AGB models leads to an O enhancement
in the He intershell of $X(\isotope[16]{O}) \approx 15\, \%$ compared to $2\, \%$ without CBM \citep{herwig:05,herwig:07a}.  
At $Z=0.006$ and $Z=0.001$ the largest overproduction factors of O of AGB models are from $\mzams=2\msun$ models while
at $Z=0.0001$ it is the $\mzams=1.65\msun$ model.

\subsection{HBB nucleosynthesis}
\label{hbb_nucleos}

\begin{landscape}
\begin{figure}
\centering
\includegraphics[width=0.9\columnwidth]{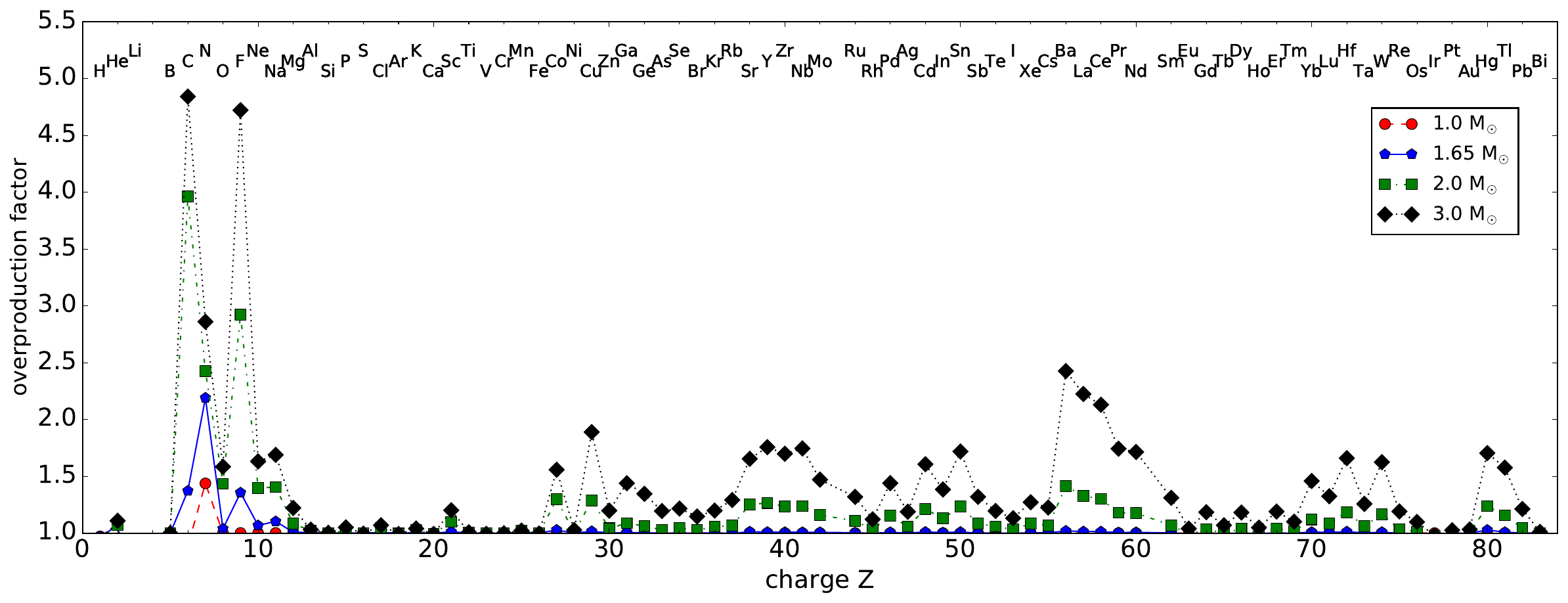}
\includegraphics[width=0.9\columnwidth]{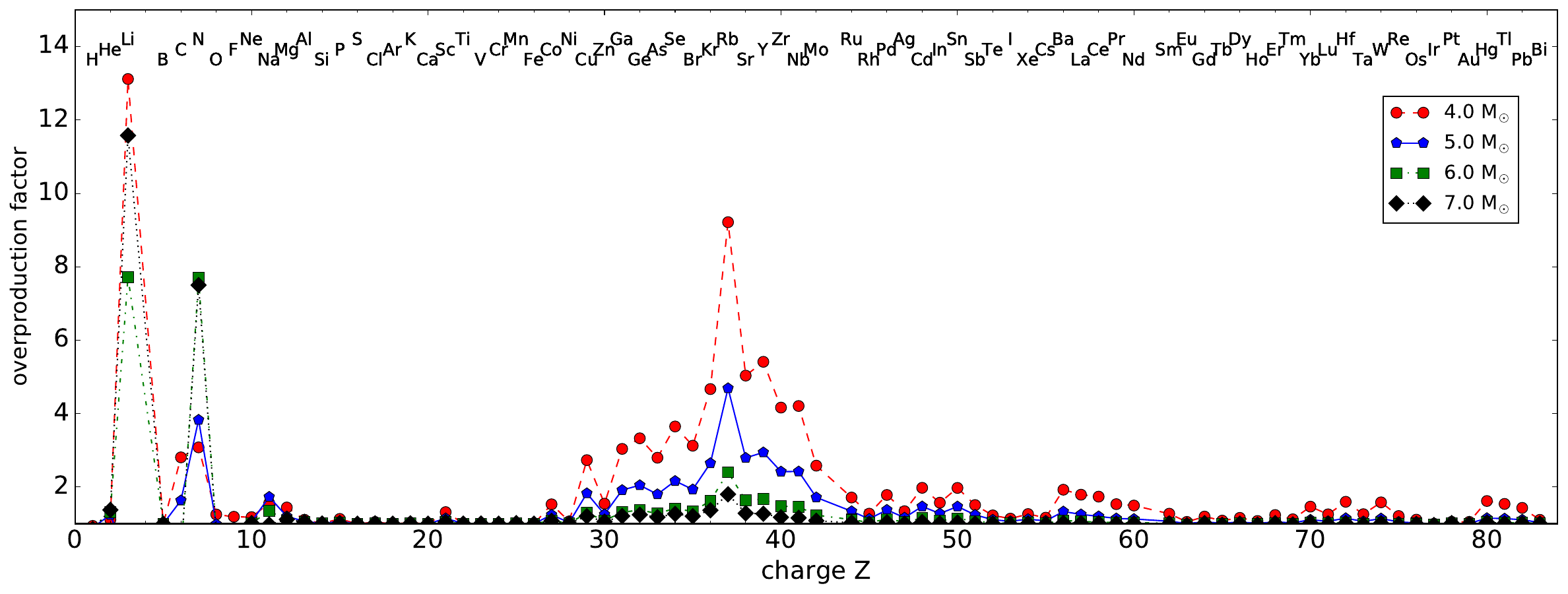}
\caption{Overproduction factors versus charge number of final yields of AGB models at $Z=0.02$ with models of $\mzams/\msun$ = 1.65, 2, 3, 4, 5 of P16.
%\mpcom{FOR NOW IT IS FINE AS IT IS, BUT I FIND A BIT STRANGE TO USE "CHARGE NUMBER", INSTEAD OF PROTON NUMBER Z. ALSO IN THE PLOTS, WOULD NOT BE BETTER TO USE SIMPLY Z INSTEAD OF "CHARGE Z"?}
}
\label{fig:agb_prodfac_set1_2}
\end{figure}
\end{landscape}

\begin{landscape}

\begin{figure}
\centering
\includegraphics[width=0.9\columnwidth]{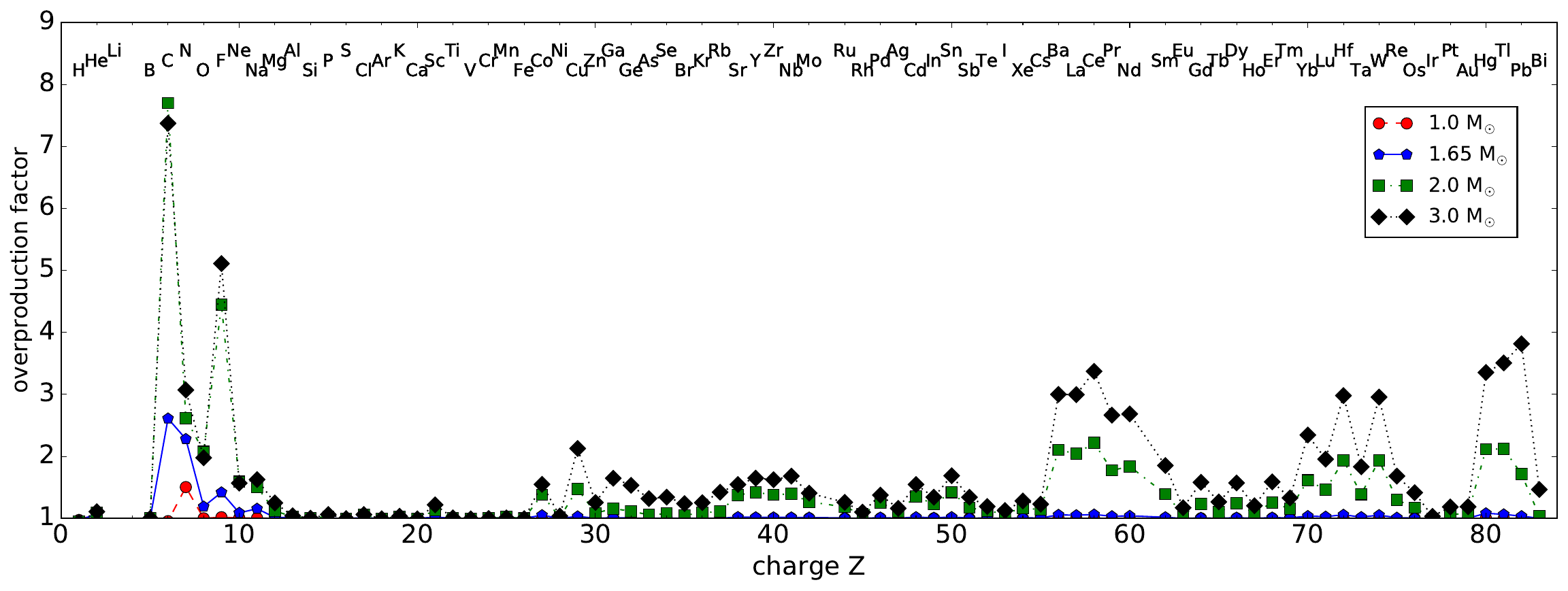}
\includegraphics[width=0.9\columnwidth]{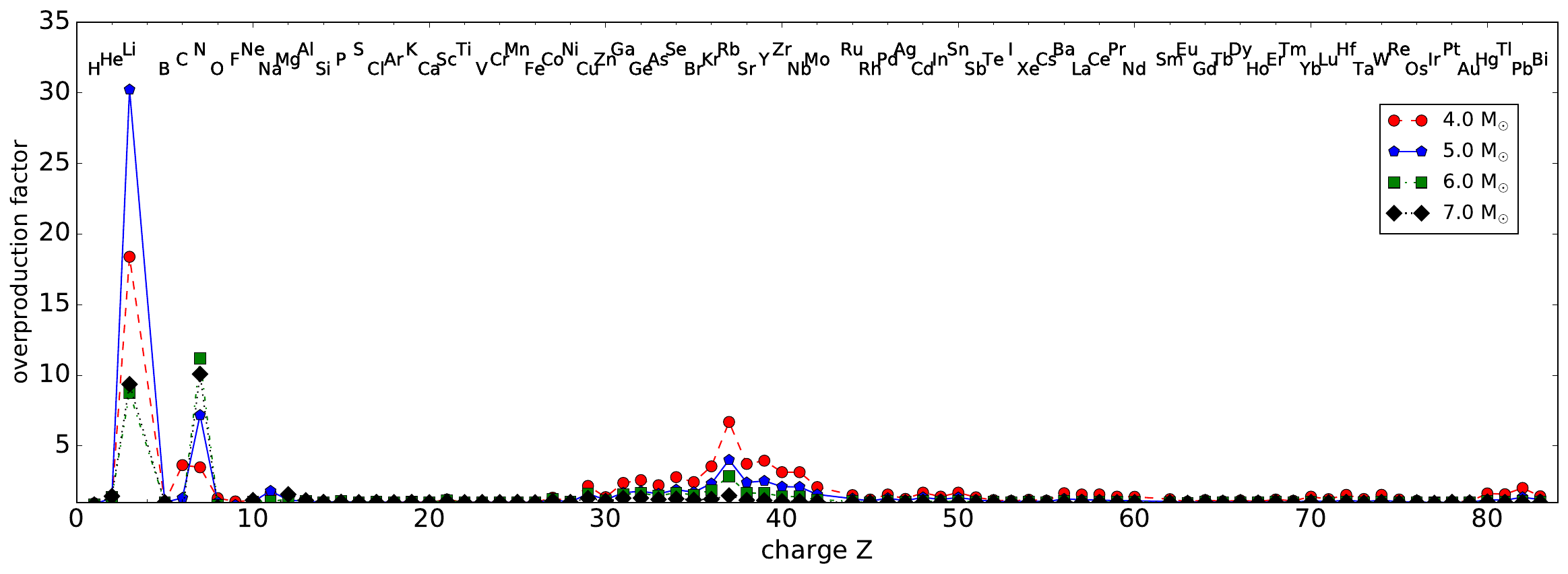}
\caption{Overproduction factors versus charge number of final yields of AGB models at $Z=0.01$ with stellar models of $\mzams/\msun$ = 1.65, 2, 3, 4, 5 of P16.}
\label{fig:agb_prodfac_set1_1}
\end{figure}
\end{landscape}

%\noindent \textbf{Li} \textemdash{ }  
Li is produced during HBB in massive AGB models through the Cameron-Fowler mechanism via \isotope[3]{He}($\alpha$,$\gamma$)\isotope[7]{Be} at the hot bottom of the convective envelope and the decay of \isotope[7]{Be} into \isotope[7]{Li} in cooler outer layers \citep[$T_{\rm CEB}\ge3\times10^7\, \mathrm{K}$, ][]{cameron:71,sackmann:92}.
We improved over the approach of P16 and resolve the simultaneous burning and mixing of CNO isotopes
while still including all heavy species in the calculation (\sect{sec:hbb}).
Li is effectively produced in all these massive AGB models and the
largest yields for each metallicity result from the most massive AGB models (\fig{fig:agb_prodfac_set1_2} to \fig{fig:agb_prodfac_set1_5a}).%\citep[e.g.][]{ventura:13}.

%%%%%%%%%%%%%%%N
HBB in AGB models synthesizes large amounts of primary N in the form of \isotope[14]{N}.
The overproduction factors of N increase with initial mass above $3\msun$ at
$Z=0.001$ and $Z=0.0001$ due to HBB for these stellar models (\fig{fig:cno_prodfac_zdep}). 
The production of N increases in stellar models at lower metallicity due to the larger temperatures at the bottom of the convective envelope $T_{\rm CEB}$ (\tab{table:agb_properties}).

In these most massive AGB models the activation of the complete CNO cycle at $T_{\rm CEB}\approx8\times10^7\, \mathrm{K}$ 
owing to HBB leads to effective O destruction (\fig{fig:cno_prodfac_zdep}) as in \cite{ventura:13}. 
More efficient destruction of O occurs at lower metallicity due to higher $T_{\rm CEB}$.

\subsection{C/Si zone and n process}

During explosive nucleosynthesis of massive star models O is transformed through $\alpha$ captures into heavier isotopes including \isotope[28]{Si} 
at the bottom of the He shell which leads to the formation of a C/Si zone \citep{pignatari:13b}. 
The presence of \isotope[4]{He} is crucial to activate explosive He-burning and to form
the C/Si zone, for which temperatures in excess of $10^9\, \mathrm{K}$ are required. The $\alpha$-capture chain can produce isotopes up to \isotope[44]{Ti}, 
which are observed in C-rich presolar stellar dust together with \isotope[28]{Si} \citep[][]{pignatari:13b,zinner:14}. 
We find the C/Si zone in all our massive star models where the most abundant isotopes are \isotope[20]{Ne}, \isotope[24]{Mg} 
and \isotope[28]{Si}.

\begin{figure}
\centering
\includegraphics[width=\columnwidth]{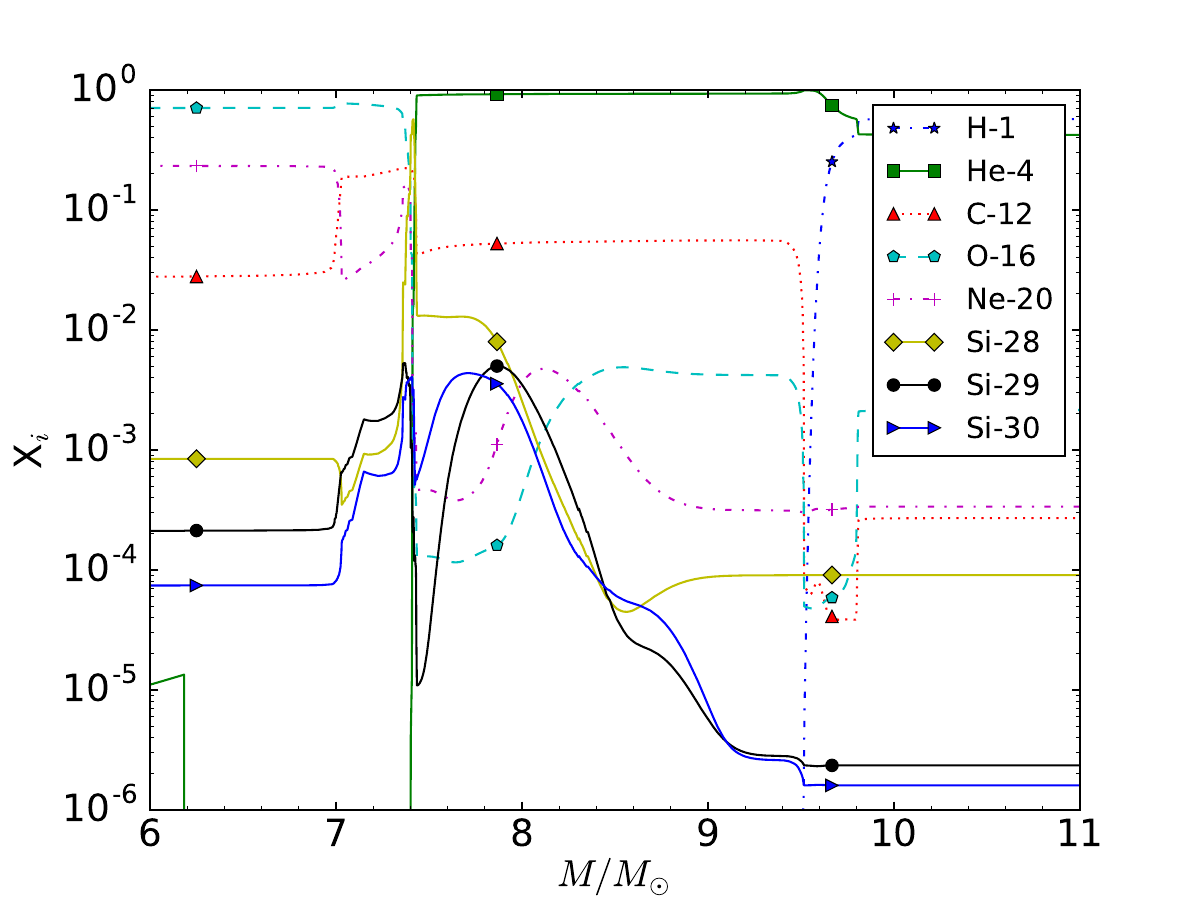}
\includegraphics[width=\columnwidth]{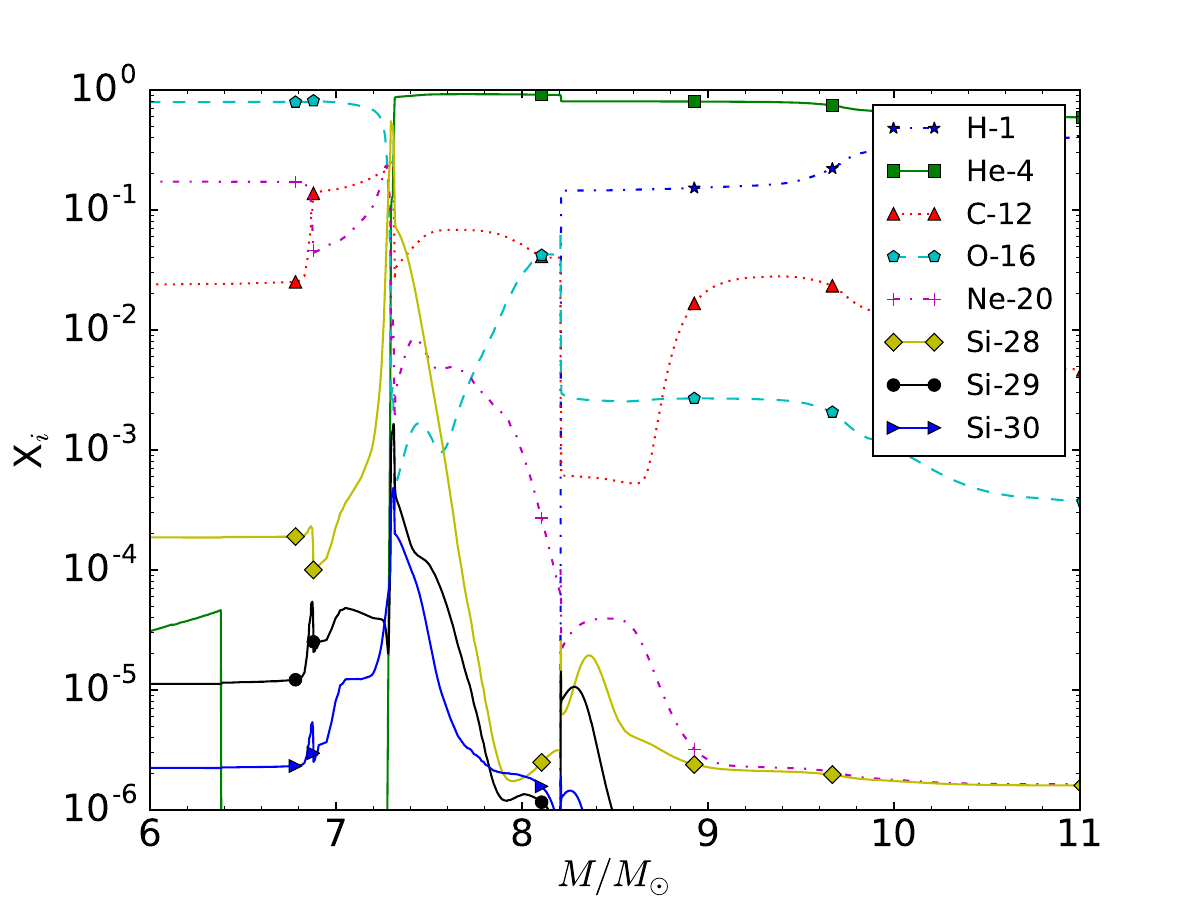}
\caption{Abundance profiles of the C/Si zones after the passage of the SN shock for stellar models with $\mzams=25\msun$ at $Z=0.006$ (top) and $Z=0.0001$ (bottom). 
Shown are the $\alpha$-chain isotope \isotope[28]{Si} and the n-process isotopes \isotope[29,30]{Si}.}
\label{fig:npr_profiles}
\end{figure}

The C/Si zone in the stellar models with higher metallicity is more extended as shown
in the comparison of the stellar models with $\mzams=25\msun$ at $Z=0.006$ and $Z=0.0001$ in \fig{fig:npr_profiles}.
As discussed in \cite{pignatari:13c}, $\alpha$-captures on \isotope[16]{O} and \isotope[20]{Ne} are in competition with the 
nucleosynthesis channel (n,$\gamma$)($\alpha$,n), leading to the production of the same species as the 
($\alpha$,$\gamma$) reactions. The ($\alpha$,p) reactions are in balance with their reverse reactions.
As a consequence, the nucleosynthesis in the C/Si zone is not much affected by metallicity and the observed C/Si zone size
is due to the metallicity-dependence of the pre-SN evolution and the SN shock temperature.

Neutron-rich isotopes are produced %and upper part of the C shell 
via the neutron source \isotope[22]{Ne}($\alpha$,n)\isotope[25]{Mg} of the n process in the He/C zone of the He shell during the explosive nucleosynthesis of massive star models \citep{thielemann:79,meyer:00,rauscher:02,pignatari:17}.
As fallback in the most massive stellar models with  $\mzams=25\msun$ prevents the ejection of deeper layers
the more externally located C/Si zone and n-process enriched He/C zone become more relevant for the total yields. 
In the $\mzams=25\msun$, $Z=0.006$ stellar model the largest contribution to the n-rich \isotope[40]{Ar} originates from the n process inside the C/Si zone. % due to fallback. % \citep{blake:76}.
The efficiency of the n-process production decreases with metallicity as indicated in the decrease of the yields of its tracer \isotope[30]{Si} (\fig{fig:npr_profiles}). This is due to the secondary nature of \isotope[22]{Ne} which abundance is made by the initial CNO abundances \citep[e.g.][]{peters:68}.

\subsection{Shell merger nucleosynthesis}
\label{sec:shellmerger}
%P,S,Cl,Ar,K,Ca,Sc
%no Ca-48

\begin{landscape}

\begin{figure}
\centering
\includegraphics[width=0.9\columnwidth]{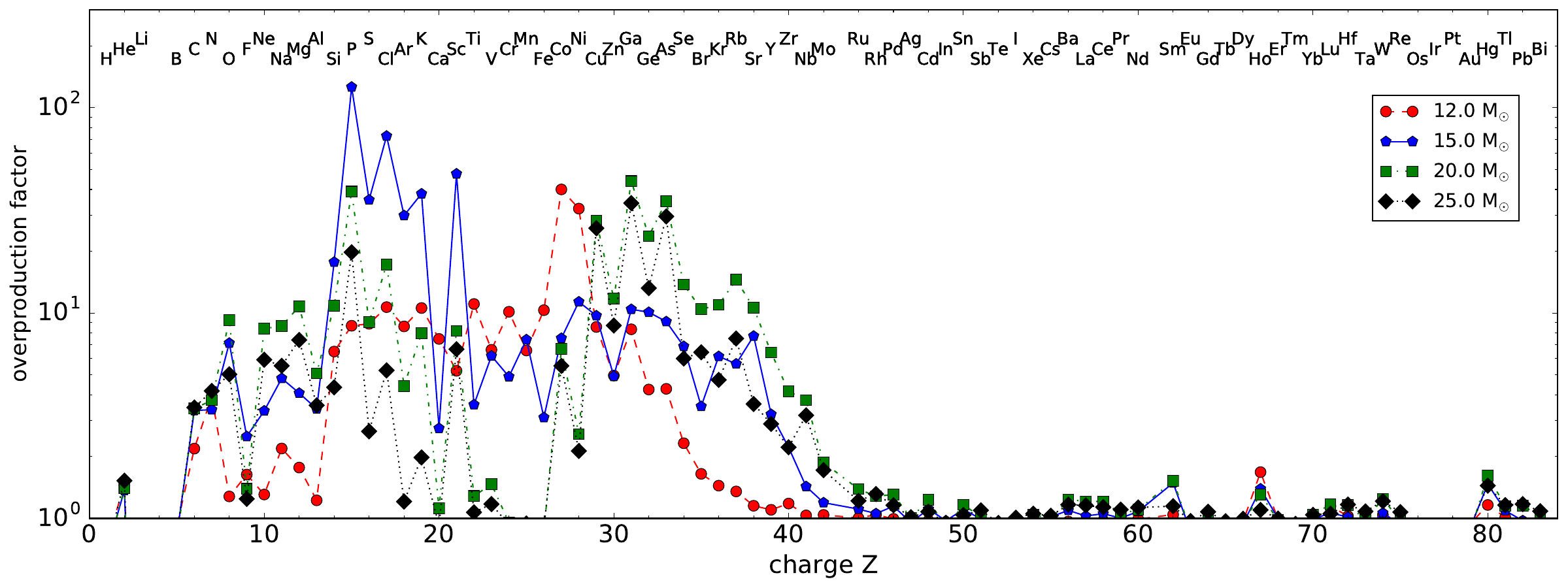}
\includegraphics[width=0.9\columnwidth]{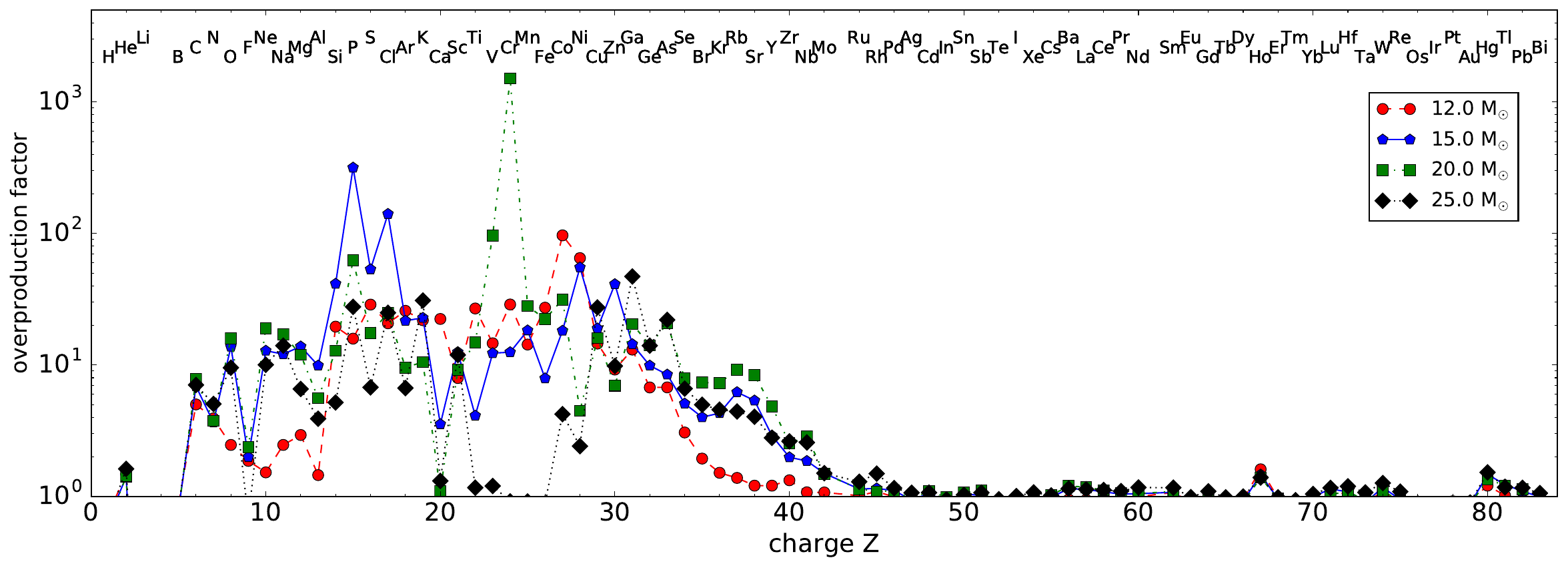}

\caption{Overproduction factors of final yields massive star models at $Z=0.02$ (top) and $Z=0.01$ (bottom).}
\label{fig:massive_prodfac_set1_21}
\end{figure}
\end{landscape}

\noindent During Si shell burning convective O-C shell mergers occur in the massive star models with $\mzams/\msun$ = 12, 15, 20 at $Z=0.01$ and $\mzams=15\msun$ at $Z=0.02$.
In these models the convective O shell increases in mass and touches the C-shell. C-shell material is mixed into the O shell until both convective shells fully merge.
Burning of the ingested Ne results in large overproduction factors of the odd-Z elements P, Cl, K and Sc in \fig{fig:massive_prodfac_set1_21} \citep{ritter:17b}.
These shell merger may harbour significant additional production of p-process nuclei such as \isotope[130,132]{Ba}. The amount of p-process nuclei produced depends on initial mass and metallicity.

 In the stellar model with initial mass of $20\msun$ at $Z=0.01$ the convective Si burning shell grows in mass until
it reaches the C shell. In the following merger of the convective Si-O shell and convective C shell Fe-peak elements are transported out of the deeper layers which fall back onto the remnant during CCSN. This boosts the production of Fe peak elements, in particular Cr and leads to large overproduction factors (\fig{fig:massive_prodfac_set1_21}). 
The overproduction factor of Cr of the $\mzams=25\msun$, $Z=0.01$ model is more than $1.7\, \mathrm{dex}$ larger than found in other stellar models at the same metallicity and the Cr production in our stellar models is already too high compared to observations \citep{cote:17}. 

Stellar evolution simulations based on the mixing-length theory describe convection through time and spherically-symmetric averages. This approach can not describe the interaction of convective C, O and Si burning shells \citep{meakin:06a,arnett:11}. Results from 1D stellar evolution are therefore mostly qualitative (Andrassy et al., in prep.).  3D hydrodynamic simulations are required to analyze in which situations O-C shell merger happen, and the dynamics of the convective shells when they happen \citep{ritter:17b}.

\subsection{Fe-peak elements}

The nucleosynthesis of the Fe-peak elements with even number of protons in massive stars is primary, and therefore does not depend on the initial metallicity \citep[e.g.][]{prantzos:00,woosley:02}. %FOR INSTANCE, MN AND CO PRODUCTION ARE NOT PRIMARY
However, the supernova progenitor evolution and the amount of fallback do depend on the initial metallicity and hence the total yields of these primary Fe-peak elements depend in some cases strongly on the initial metallicity (\tab{tab:yields_all_delay_set1_5} and online yield tables).

\begin{landscape}

\begin{figure}
\centering
\includegraphics[width=0.9\columnwidth]{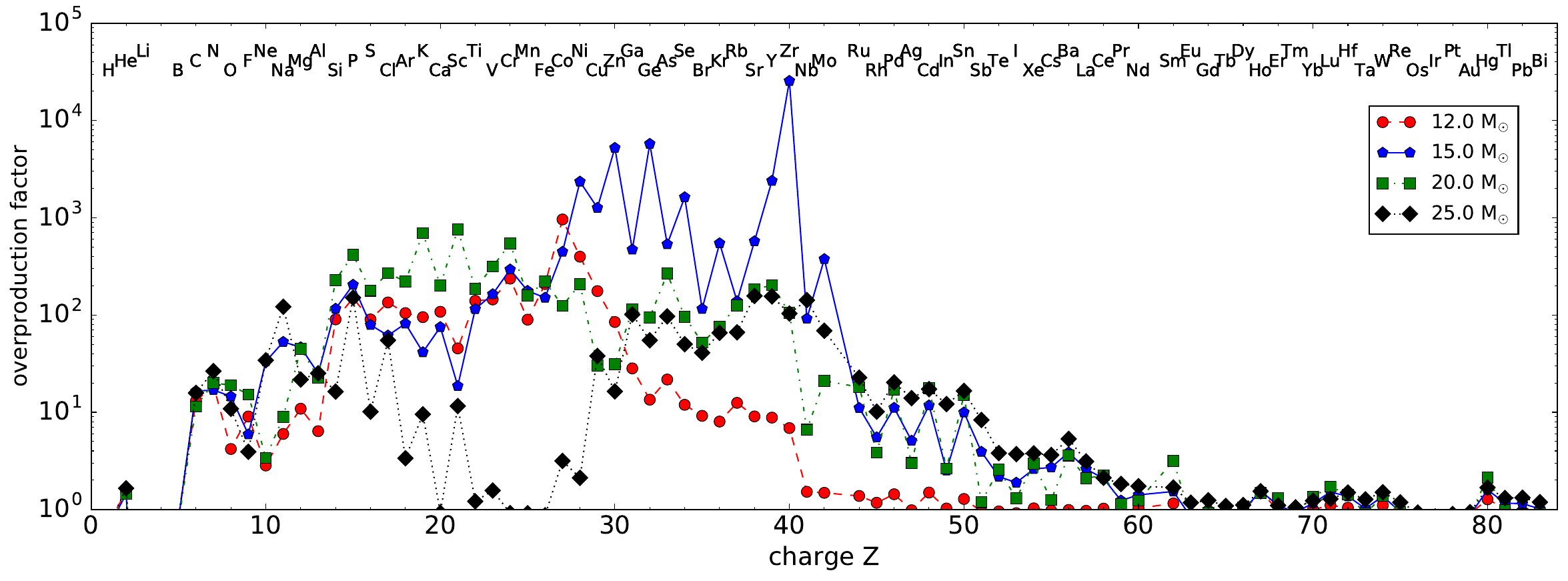}
\includegraphics[width=0.9\columnwidth]{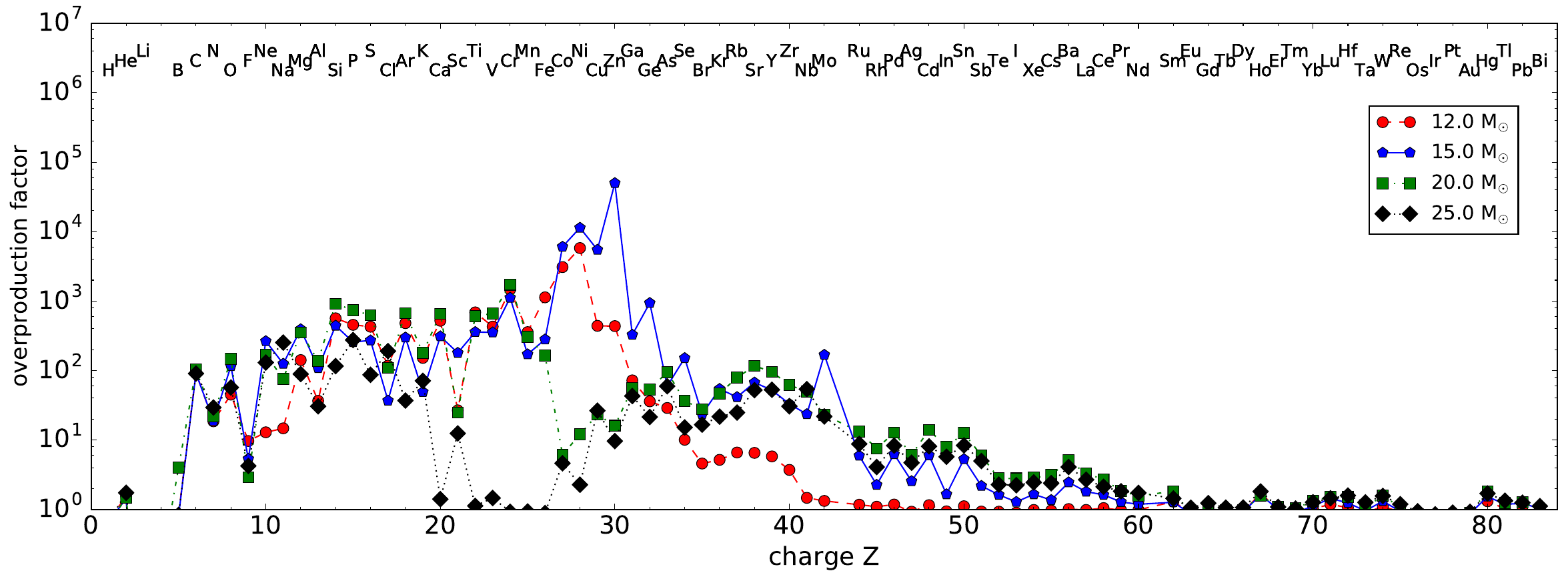}
\caption{Overproduction factors versus charge number of massive star models at $Z=0.006$ (top) and $Z=0.001$ (bottom).}
\label{fig:massive_prodfac_set1_34a}
\end{figure}
\end{landscape}

If not mentioned otherwise we discuss the delayed explosions (see \sect{sec:method:exp}). Fallback limits the ejection of Fe-peak elements which becomes important in $\mzams\ge20\msun$ models, but less so at lower initial mass. Fallback prevents any Fe ejection in stellar models with $\mzams=25\msun$ which results in low overproduction factors of Fe (\fig{fig:massive_prodfac_set1_21} to \fig{fig:massive_prodfac_set1_5a}).
In the stellar models with $\mzams=15\msun$ the ratio of explosive production to pre-SN production (see pre-SN yield definition in \sect{sec:nugrid_codes}) of Fe peak elements is much larger 
at $Z=0.0001$ compared to $Z=0.006$ due to a contribution of Fe-core layers to the pre-SN production at the latter metallicity (\fig{fig:ratio_presn_sn}). This is due to a lower explosive Fe-peak production in stellar model of higher metallicity.

Of all stellar models with $\mzams=20\msun$ only the $Z=0.006$ model produces Fe peak elements during SN shock nucleosynthesis.
Consequently this model has the largest ratio of Fe peak elements produced during SN to the pre-SN production.
In stellar models with $\mzams=12$ and $15\msun$ additional production and ejection of Fe-peak elements originates
from the  $\alpha$-rich freeze-out layer which falls back in stellar models of higher initial mass (\sect{alphaproc}). 
The interplay of the core masses at collapse (\fig{fig:cores_massive}) and the effect of fallback (\tab{tab:coo_fallback}) results in
 much larger variations of the Fe-peak elements ejection with initial mass and metallicity, compared to other yield sets for massive stars \citep[e.g.][]{woosley:95,nomoto:06}.

\subsection{H-ingestion nucleosynthesis}
\label{hingnucleo}
While Li is produced through HBB in AGB models (\sect{hbb_nucleos}) it is also effectively produced by H ingestion events (\sect{sec:se-hing}) in the second thermal pulse of the 
$1\msun$, $Z=0.0001$ model and the post-AGB thermal pulse of the $1.0\msun$, $Z=0.006$ model as decayed \isotope[7]{Be} via \isotope[3]{He}($\alpha$,g)\isotope[7]{Be} where
 \isotope[3]{He} is ingested with H \citep{herwig:00f,iwamoto:04}. The post-AGB model production does not contribute to the yields as the enriched mass ejected into the interstellar medium is too small. The $Z=0.0001$ model loses \isotope[7]{Li}-enriched mass efficiently leading to large Li overproduction factors (\fig{fig:agb_prodfac_set1_3a}). 

H-ingestion events are also present in massive star models. They involve a ingestion of protons into the He-convection shell and reduce the He-core core mass by about $1\msun$. The nucleosynthetic effect of H-ingestion events becomes apparent when the SN shock reaches the He shell which results in explosive He-burning with a small amount of added H. 
The exact amount and nature and amount of the H ingestion would depend on the 3D hydrodynamic nature of convection in such conditions. 

H ingestion in massive stars (see \sect{sec:se-hing}) can lead to the production of \isotope[7]{Be} during the explosion. H that reaches to the bottom of the He-shell just before the collapse produces \isotope[3]{He} under explosive conditions and then \isotope[7]{Be} via the reaction mentioned above.
This \isotope[7]{Be} would be ejected without the possibility to capture an electron to produce \isotope[7]{Li}, and thus SN with previous H ingestions would be \isotope[7]{Be} producers. Other Li production might occur by $\nu$-induced production in CCSN or via galactic cosmic rays \citep[e.g.][]{prantzos:12,banerjee:13}, which are both not considered in this work.

\begin{landscape}

\begin{figure}
\centering
\includegraphics[width=0.9\columnwidth]{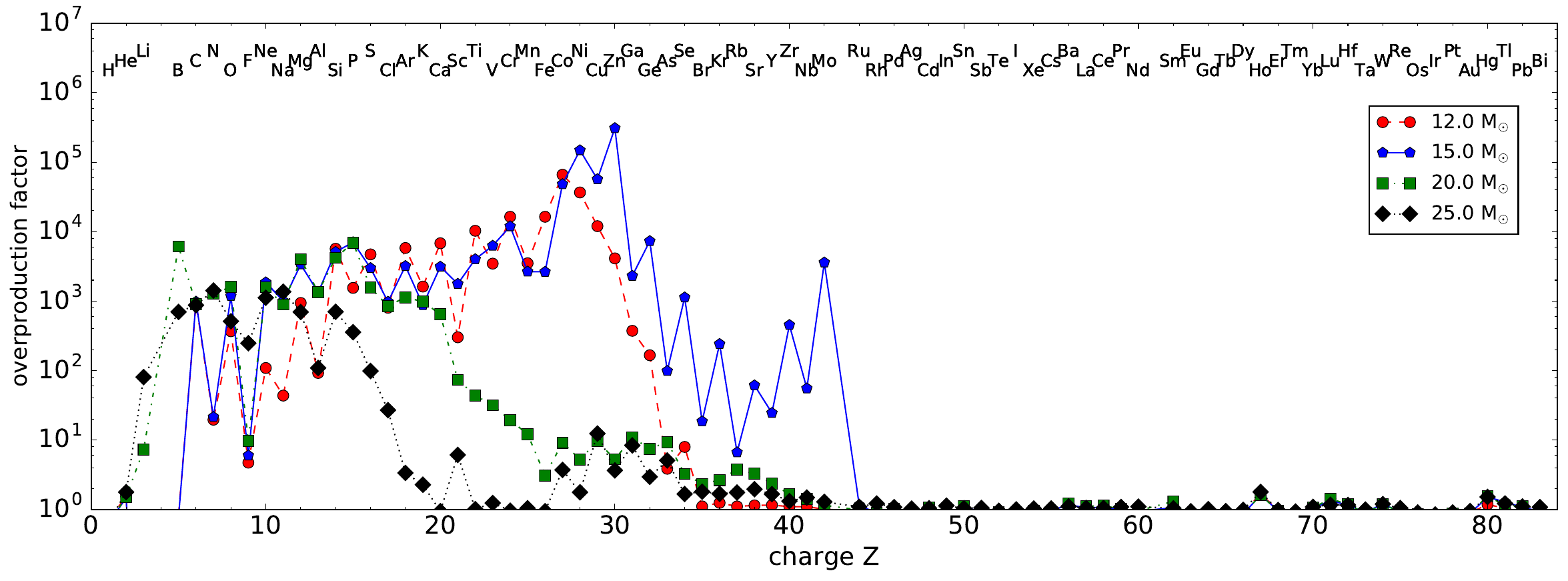}
\caption{Overproduction factors versus charge number of final yields of massive star models at $Z=0.0001$.}
\label{fig:massive_prodfac_set1_5a}
\end{figure}
\end{landscape}

\begin{figure}
\centering
\includegraphics[width=\columnwidth]{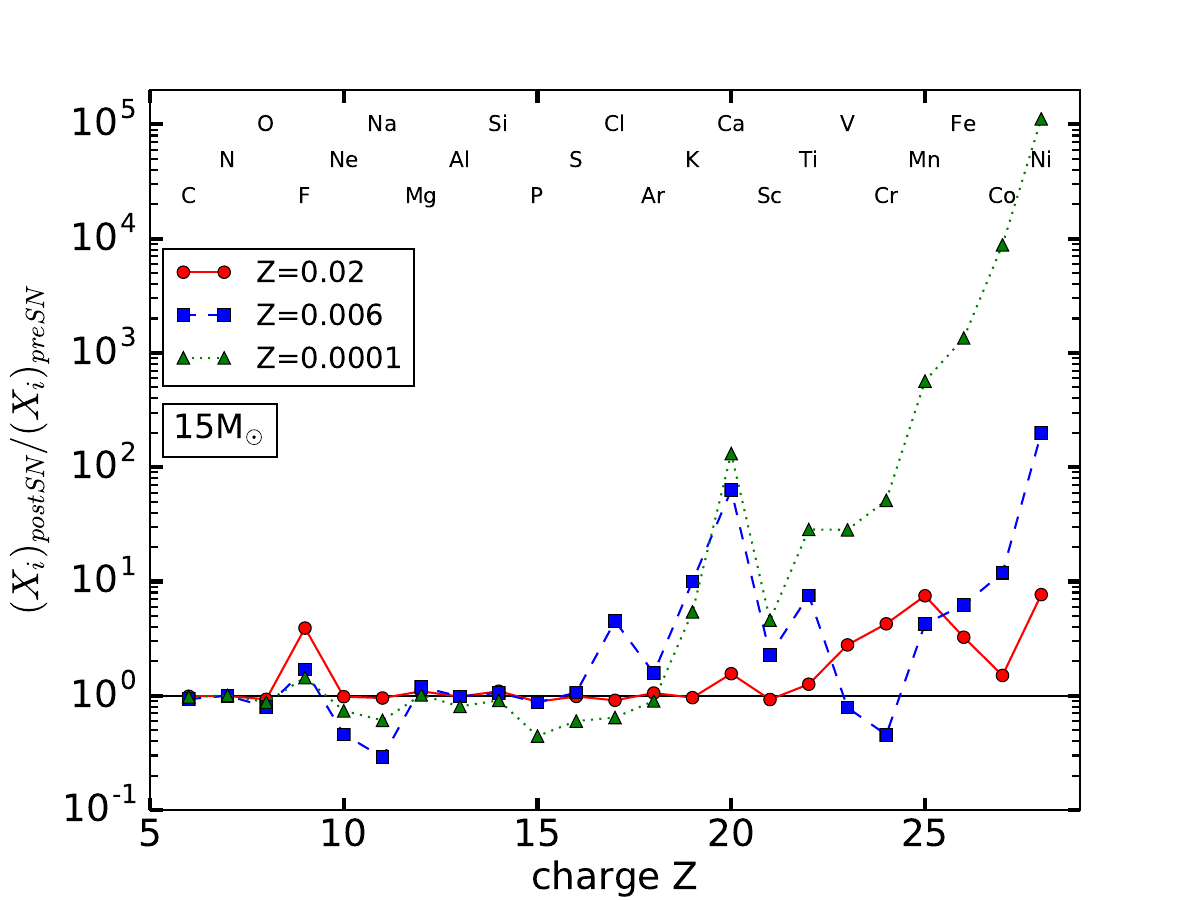}
\includegraphics[width=\columnwidth]{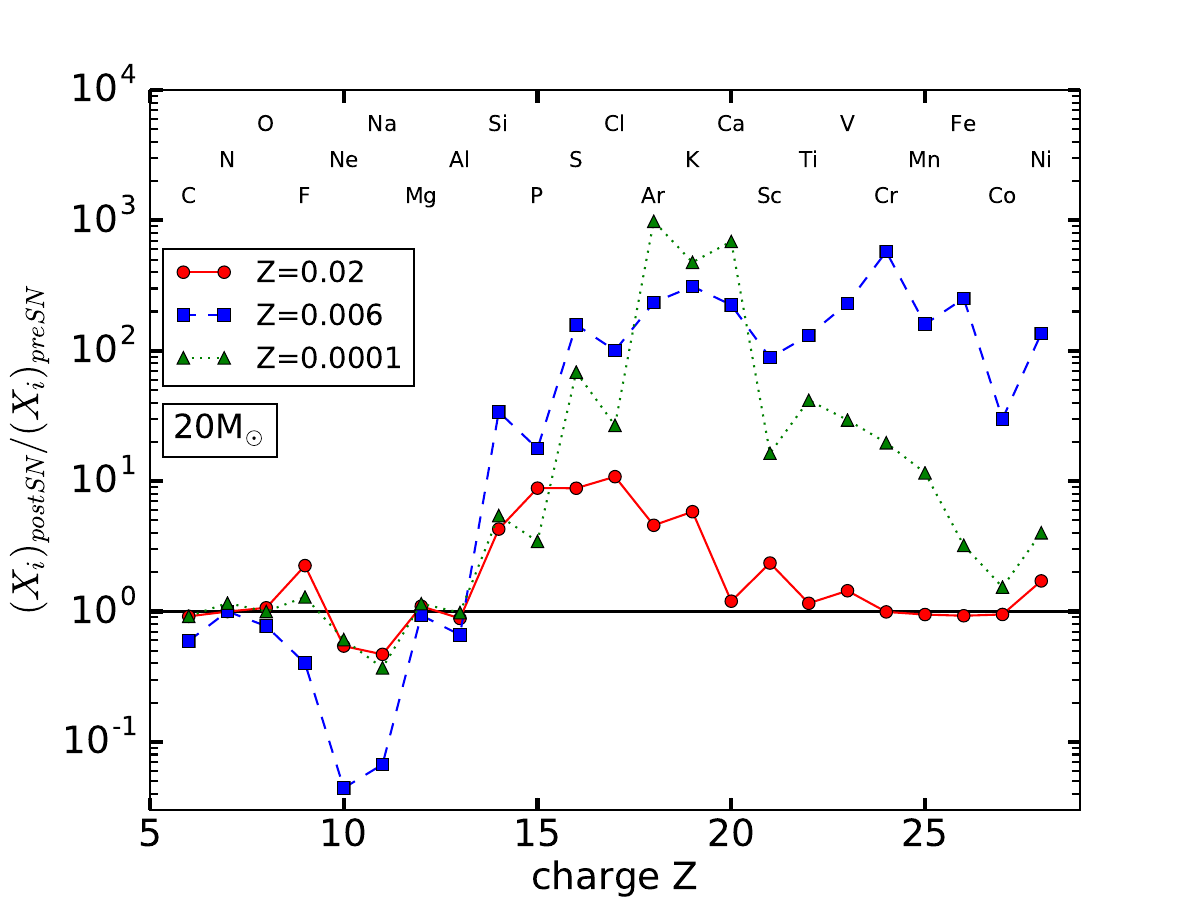}
\includegraphics[width=\columnwidth]{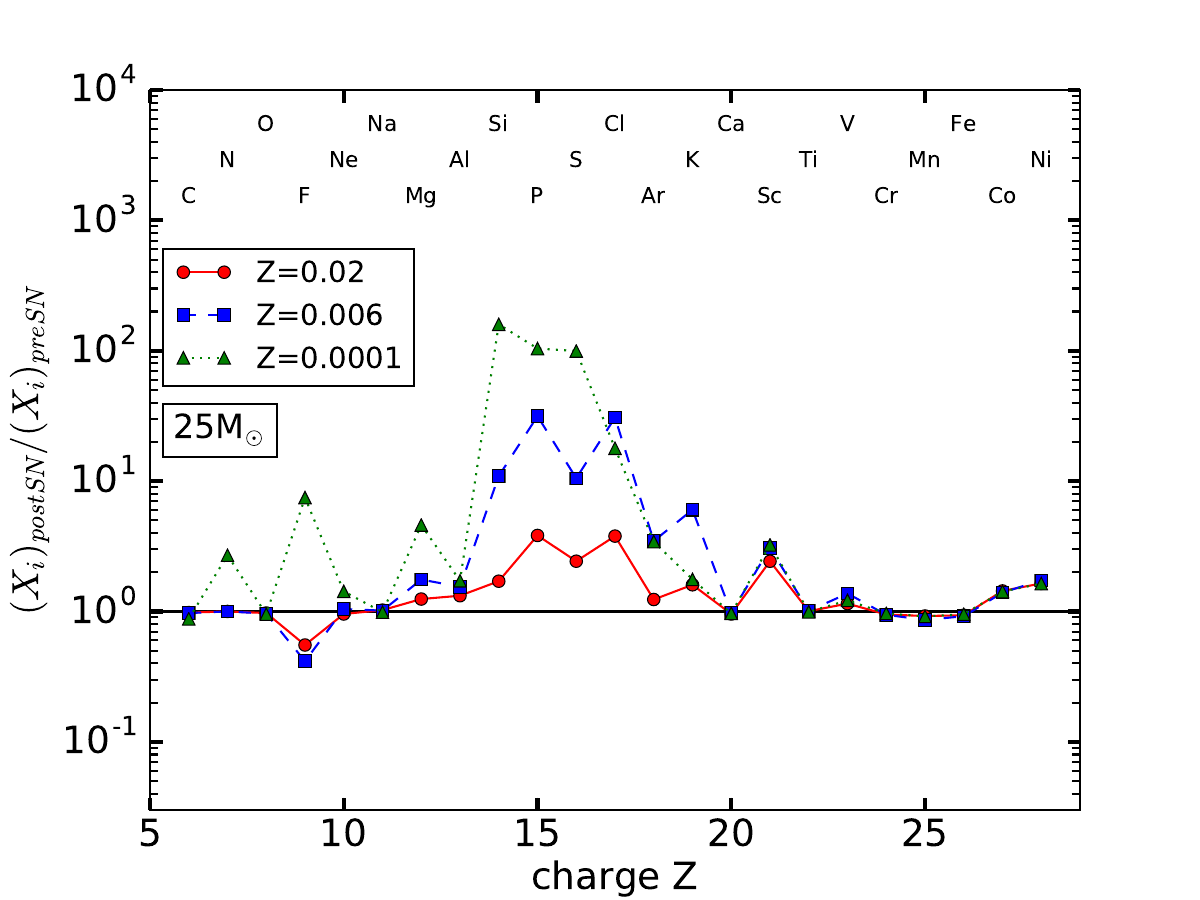}
\caption{Ratio of SN to pre-SN yields versus charge number of stellar models with $\mzams=15$, $20$, $25\msun$ for $Z=0.02$, $Z=0.006$ and $Z=0.0001$.
The pre-SN yields are the ejected pre-SN composition above the mass cut (\sect{sec:nugrid_codes}).}
\label{fig:ratio_presn_sn}
\end{figure}

H-ingestion leads to significant production of light elements such as Li and N in the $20$ and $25\msun$, $Z=0.0001$ models (\fig{fig:massive_prodfac_set1_5a}). \fig{fig:ratio_presn_sn} shows that there is however no explosive contribution to N from the $20\msun$, $Z=0.0001$ model, while the 25\msun\ explosion adds approximately the same amount of N compared to the the pre-SN evolution. N production has been seen in massive star models at low-metallicity previously \citep[e.g.][]{woosley:95,ekstroem:08} and Pop III models \citep{heger:10}. 

\begin{figure}
\includegraphics[width=\columnwidth]{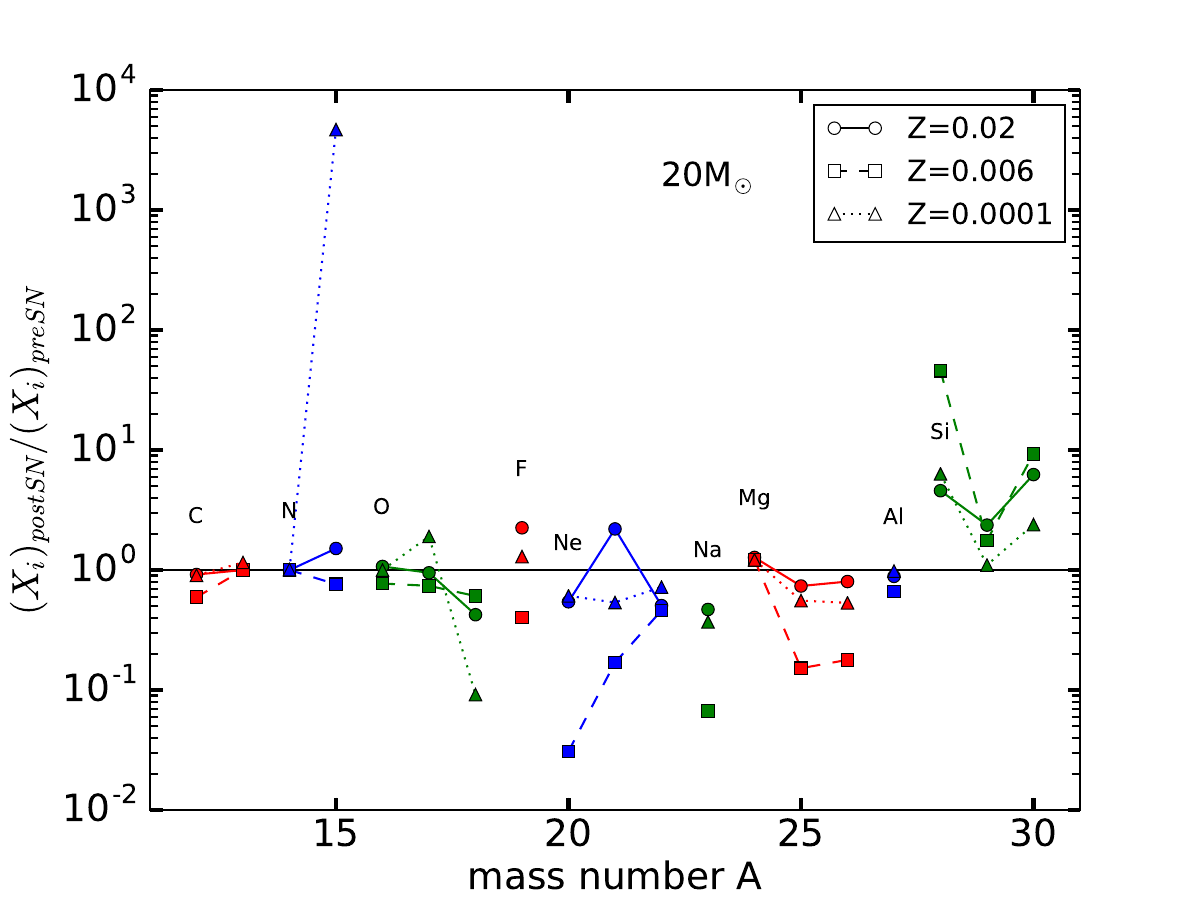}
\caption{Ratio of SN to pre-SN yields versus mass number of stellar
  models with $\mzams=20\msun$ for $Z=0.02$, $Z=0.006$ and $Z=0.0001$.
  The pre-SN yields are the ejected pre-SN composition above the mass
  cut (\sect{sec:nugrid_codes}).}
\label{fig:ratio_presn_sn_iso}
\end{figure}

As previously reported by \cite{pignatari:15}, \isotope[15]{N} is effectively produced in the region of pre-SN H ingestion during the explosive nucleosynthesis in our models. 
In the SN explosion \isotope[15]{N} is relative to its pre-SN abundance orders of magnitude more produced than \isotope[14]{N} as
visible in the ratio of SN yields to pre-SN yields of the stellar model with $\mzams=20\msun$ in \fig{fig:ratio_presn_sn_iso}.
The ingestion events might be a relevant source of primary production of \isotope[14]{N} and \isotope[15]{N} at low metallicity
in contrast to the pre-explosive production in rotating massive star models \citep[e.g.][]{hirschi:07} which do not predict the low \isotope[14]{N}/\isotope[15]{N} ratio 
observed at high redshift and the isotopic ratio of the Sun \citep{pignatari:15}. 
\isotope[19]{F} is also produced efficiently through \isotope[15]{N}($\alpha$,$\gamma$)\isotope[19]{F} in these stellar explosion with $\mzams=25\msun$ (\fig{fig:massive_prodfac_set1_5a}). 

During the first TP of the $\mzams=1\msun$, $Z=0.0001$ model the PDCZ reaches into the radiative H-rich envelope and small amounts of H are ingested similar to H-ingestion events reported previously \citep[e.g.][and references within]{fujimoto:00,cristallo:09b}. % in their  $1.5\mzams$, $Z=0.00005$ model.
%The following (p,$\gamma$)\isotope[13]{N} reaction, \isotope[13]{N}()\isotope[13]{C} 
Most ingested H is absorbed by \isotope[12]{C} to produce
\isotope[13]{C} which produces neutron densities $N_n\approx10^{7}\,
\mathrm{cm}^{-3}$ via the \isotope[13]{C} ($\alpha$,n) neutron source
and synthesizes heavy elements up to Pb.  During the second TP the
PDCZ reaches out into the convective envelope (\fig{fig:hif_kipps})
and large amounts of H are mixed into the PDCZ which leads to the
convective-reactive production of \isotope[13]{C} as in the $2\msun$,
${\rm [Fe/H]} = - 2.7$ model of \cite{iwamoto:04}.  The energy
generation due to proton burning leads to a split of the convective
zone and its bottom part reaches a neutron density of $N_n\approx
5\times10^{13}\, \mathrm{cm}^{-3}$ which leads to additional
production of heavy elements with large overproduction factors
(\fig{fig:agb_prodfac_set1_5a}). The process of neutron release is as
in \cite{iwamoto:04}. \cite{iwamoto:04} and \cite{cristallo:09b}
report higher neutron densities of $N_n\approx 10^{14}\,
\mathrm{cm}^{-3}$ and $N_n\approx 10^{15}\, \mathrm{cm}^{-3}$
respectively. This is the heavy-element production through
\iprn\ which is poorly described in stellar evolution models. As
discussed in \sect{sec:se-hing} it has been shown by \citet{herwig:11}
and \citet{herwig:14} that the convective-reactive
\ipr\ nucleosynthesis can not be modelled correctly by present versions
of MLT based convective mixing in 1D stellar
evolution simulations. We do therefore not make any effort to ensure
numerical convergence of a demonstrably insufficient modeling
approximation, and defer more reliable \iprn\ predictions to a time
when better modeling approaches have been developed for this
particular regime found in our models.

\subsection{$\mathbf{\alpha}$ process}
\label{alphaproc}

\begin{landscape}
\begin{figure}
\centering
\includegraphics[width=0.8\columnwidth]{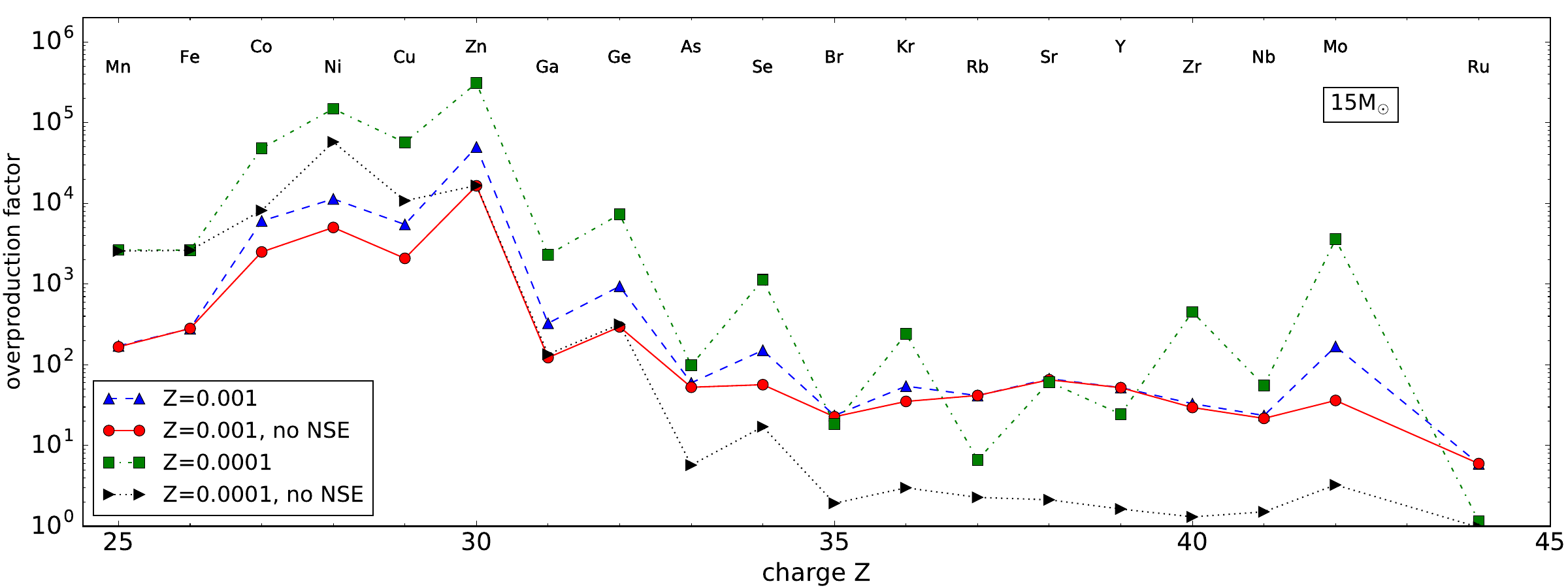}
\includegraphics[width=0.8\columnwidth]{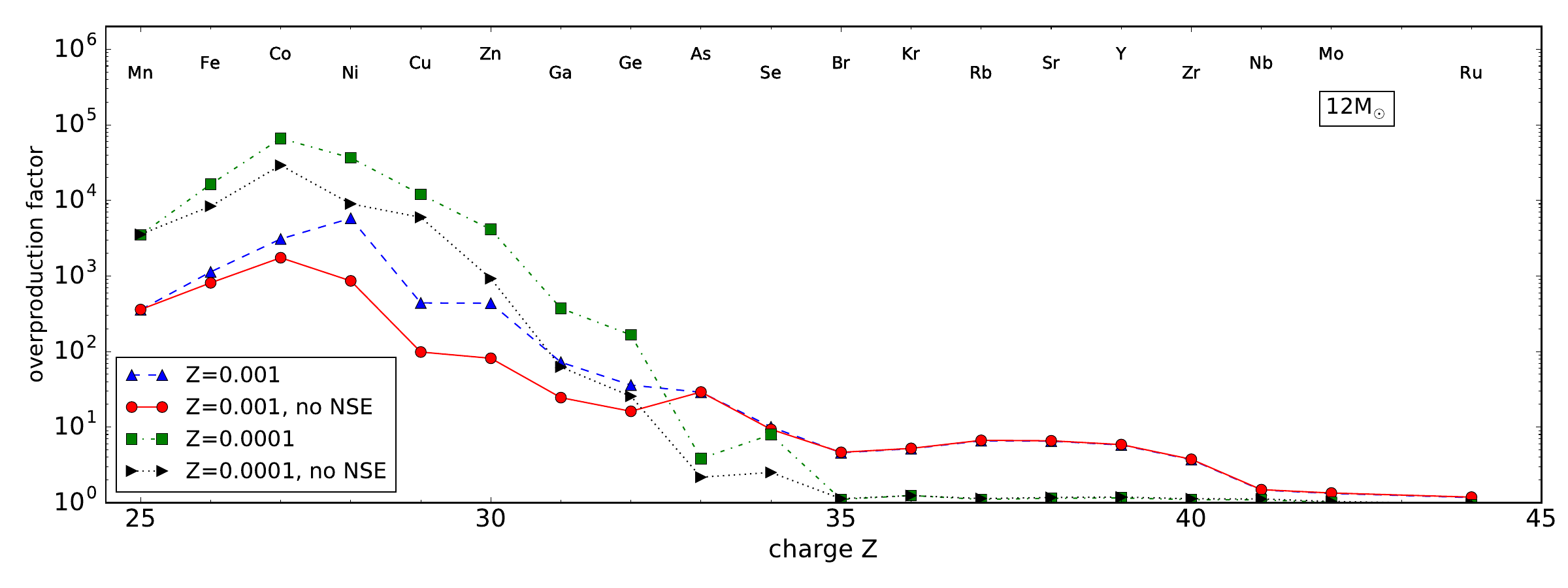}
\caption{Overproduction factors versus charge number for stellar models with $\mzams=12\msun$ and $\mzams=15\msun$ at $Z=0.001$ and $Z=0.0001$
which reach $\alpha$-rich freeze out conditions during the CCSN explosion.
Shown are the production factors of final yields under the assumption of fallback of the layers which went into NSE (no NSE).
In the latter case layers which experience NSE conditions during the explosion were assumed not to be ejected.}
\label{fig:apr_prod_facs}
\end{figure}
\end{landscape}

%\noindent \textbf{} \textemdash{ }
Matter in nuclear statistical equilibrium (NSE) during the CCSN explosion
which later on cools and expands can experience an $\alpha$-rich freeze-out \citep{woosley:73,woosley:92}.
%$\alpha$ nuclei merge back into the Fe peak via the $\alpha$ process and produce
%Fe-peak elements and heavier elements up to Ag 
Such $\alpha$-rich freeze out conditions are reached in all our $\mzams=12\msun$ and
$\mzams=15\msun$ models (\fig{fig:apr_prod_facs}). % with the exception of the $15\msun$ at solar Z.
%for all metallicities and indicate a independence from the initial composition.
%The $12\msun$ stars experience the highest shock temperatures reaching above $2\times10^{10} K$.
A larger $\alpha$-rich freeze-out layer formed during the explosive nucleosynthesis of the stellar models with $\mzams=15\msun$ compared to the stellar models 
with $\mzams=12\msun$ leads to a larger production of Fe-peak elements compared to the production in explosive Si burning. The $\alpha$-rich freeze out layers in the stellar models with $\mzams=15\msun$ produce elements up to Mo in agreement with P16 (their Fig.\ 24).
The massive star models of lower initial mass produce only elements up to Ge and Br at $Z=0.001$ and $Z=0.001$ respectively as indicated by their overproduction factors (\fig{fig:apr_prod_facs}). At lower metallicity heavier elements are produced in the NSE region than in stellar models of higher metallicity (\fig{fig:apr_prod_facs}).

\subsection{Weak s process}

\begin{landscape}
\begin{figure}
\centering
\includegraphics[width=0.6\columnwidth]{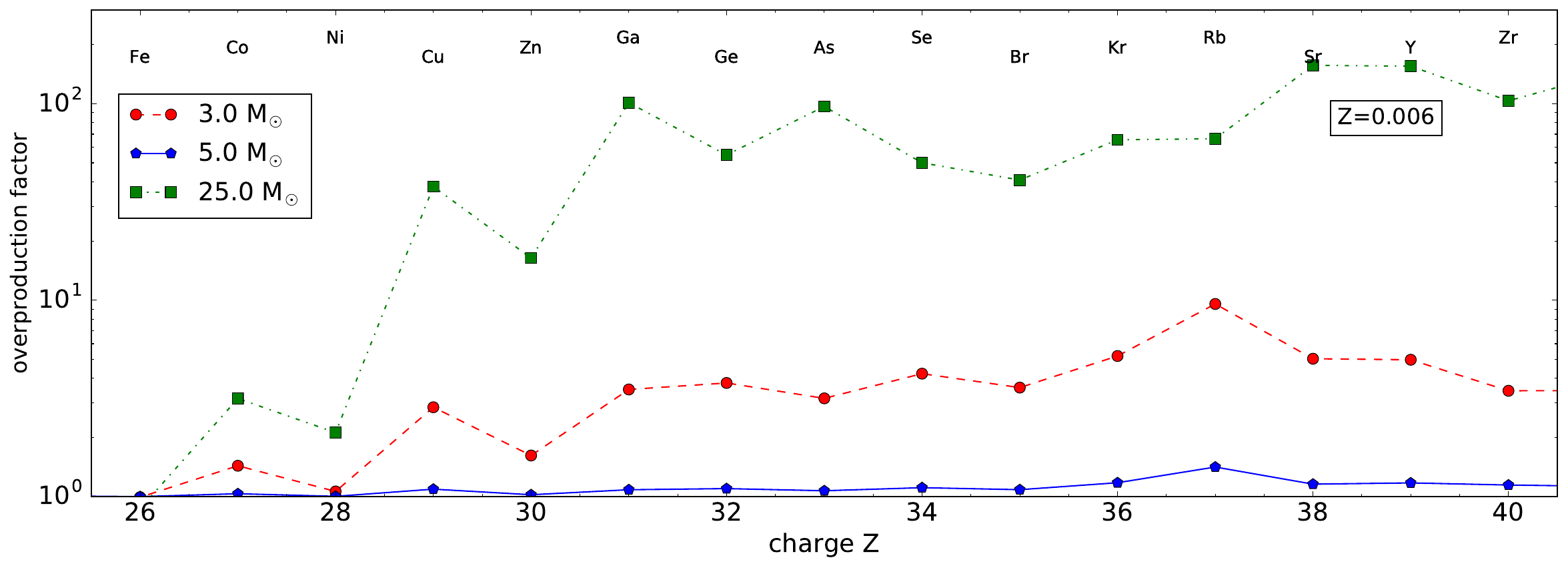}
\includegraphics[width=0.6\columnwidth]{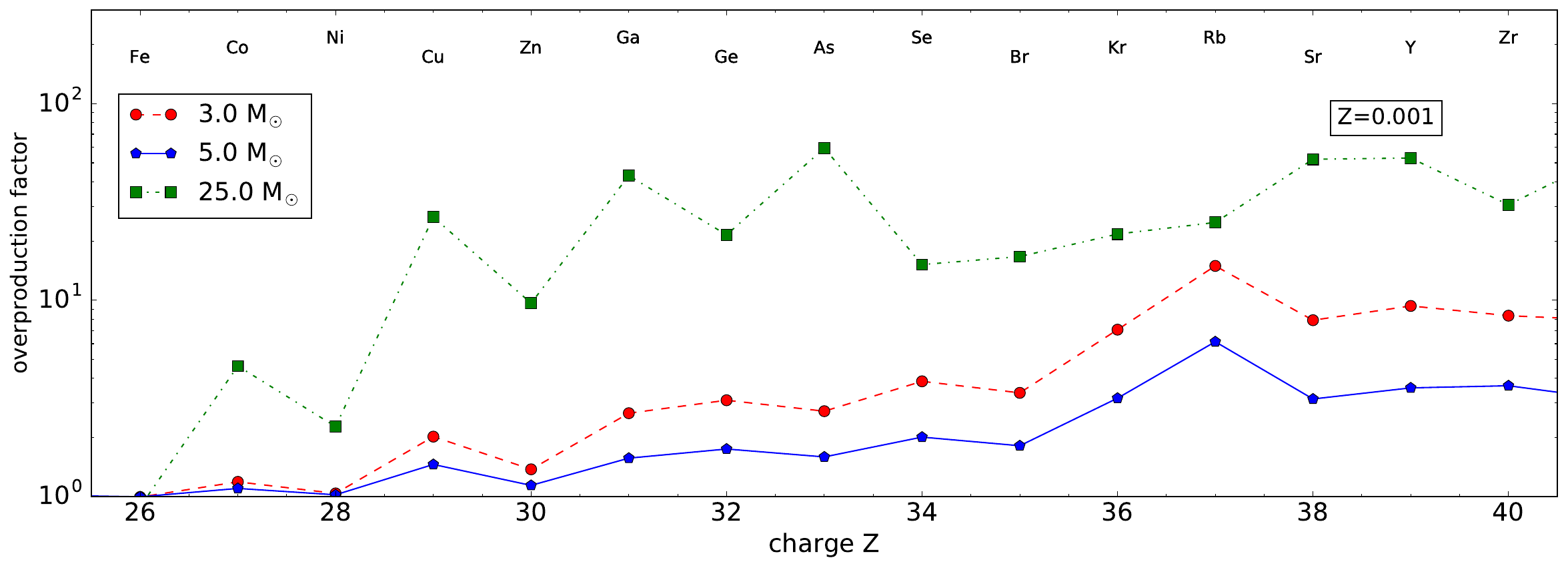}
\includegraphics[width=0.6\columnwidth]{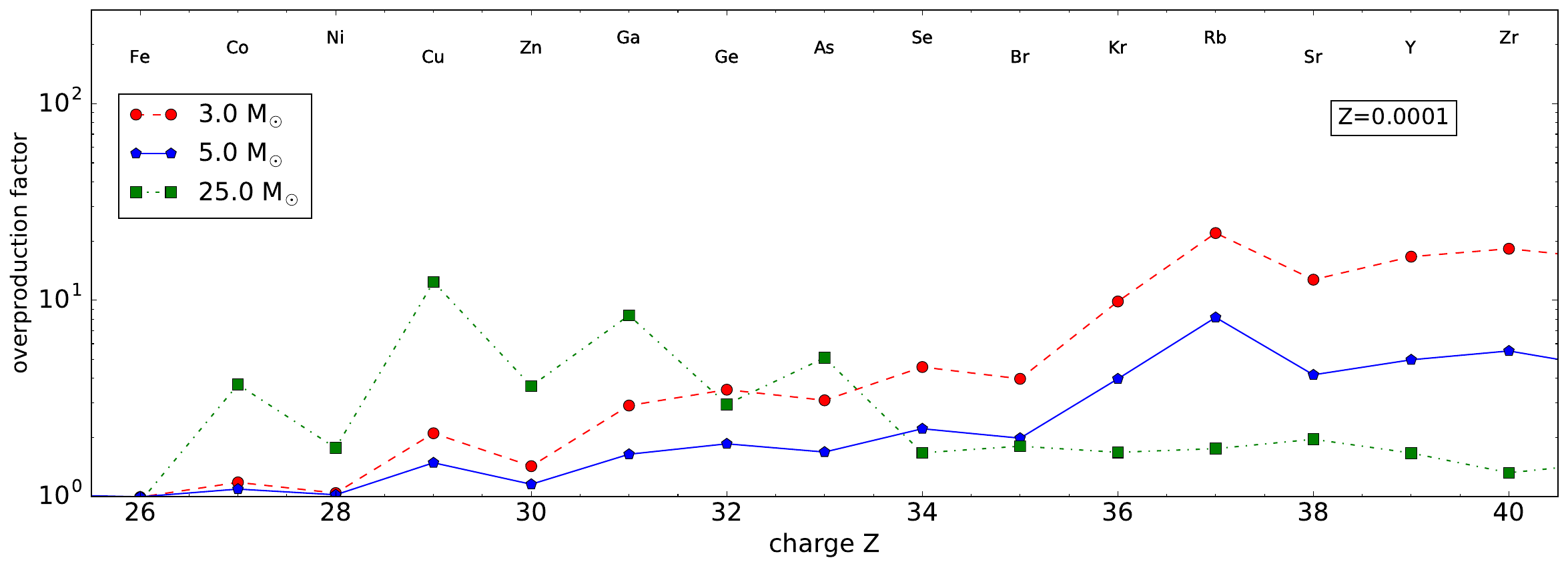}
\caption{Comparison of overproduction factors versus charge of stellar models with $\mzams=3\msun$ and $\mzams=5\msun$ (main s-process) and of stellar models with $\mzams=25\msun$ at $Z=0.006, 0.001$ and $0.0001$ (weak s process).}
\label{fig:m_w_component_vsZ}
\end{figure}
\end{landscape}

The weak s process takes place at the end of core He-burning and
during convective C shell burning in massive star models and 
is metallicity dependent. The process depends on the initial abundance of Fe seeds, and on the initial abundance of CNO nuclei that will make most of the \isotope[22]{Ne} available as a neutron source \citep[e.g.][]{kaeppeler:89,prantzos:90,raiteri:92,the:07,pignatari:10,kaeppeler:11,frischknecht:16,sukhbold:16}.
 We compare the heavy element production up to the first s-process peak originating from the weak s process in these stellar models with $\mzams=25\msun$ with element production from the main s process in these models with $\mzams=3\msun$ and $\mzams=5\msun$ for $Z=0.006$, $Z=0.001$ and $Z=0.0001$ in \fig{fig:m_w_component_vsZ}. % in the mass range $26\leq Z\leq 38$).
The weak s-process efficiency is overall the largest at $Z=0.006$, also more than in models at higher metallicities. 
Because of the secondary nature of the weak s-process, this could appear as a surprising result. However, as already discussed in e.g. \cite{pignatari:07}, this is mostly due to the $\alpha$-enhancement on \isotope[16]{O} at low metallicity, causing a smaller decrease of \isotope[22]{Ne} with respect to the Fe seeds, that are instead decreasing linearly with the metallicity. As a consequence, the s-process distribution is also partially modified, showing a high production up to the Sr neutron-magic peak. For lower metallicities, also by taking into account $\alpha$-enhancement the resulting abundance of \isotope[22]{Ne} becomes too low and the weak s-process contribution to the stellar yields becomes marginal. The fewer neutrons made by the \isotope[22]{Ne}($\alpha$,n)\isotope[25]{Mg} reaction are captured by primary neutron poisons like \isotope[16]{O} \citep{baraffe:92,pignatari:07}.
The overproduction factors of elements above As even decrease in the massive star models at $Z=0.0001$ below those of the AGB models (\fig{fig:m_w_component_vsZ}).

The overproduction factors of the s-only isotopes
\isotope[70]{Ge}, \isotope[76]{Se}, \isotope[80,82]{Kr} and \isotope[86,87]{Sr} of the stellar models with $\mzams=25\msun$ at $Z=0.006$, $Z=0.001$ and $Z=0.0001$ 
show a decrease in the s-process efficiency below $Z=0.001$ (\fig{fig:wspr_prodfacs}). Most
production of \isotope[70]{Ge} takes place in the pre-explosive nucleosynthesis as indicated by the overproduction
factors of the pre-SN ejecta compared to the SN ejecta for the model at $Z=0.006$ (\fig{fig:wspr_prodfacs}). 
In stellar models of lower initial mass the explosive nucleosynthesis produces further \isotope[70]{Ge} which increases
the overproduction factors of the SN ejecta over that of the pre-SN ejecta. The high production in the 
stellar model with $\mzams=15\msun$ at $Z=0.006$ originates from a thin shocked Fe core layer.

\subsection{Main s process}
\label{sec:mspr}

\begin{landscape}
\begin{figure}
%\resizebox{0.9\textwidth}{!}{
%\begin{minipage}{\textwidth}
\centering
\includegraphics[width=0.9\columnwidth]{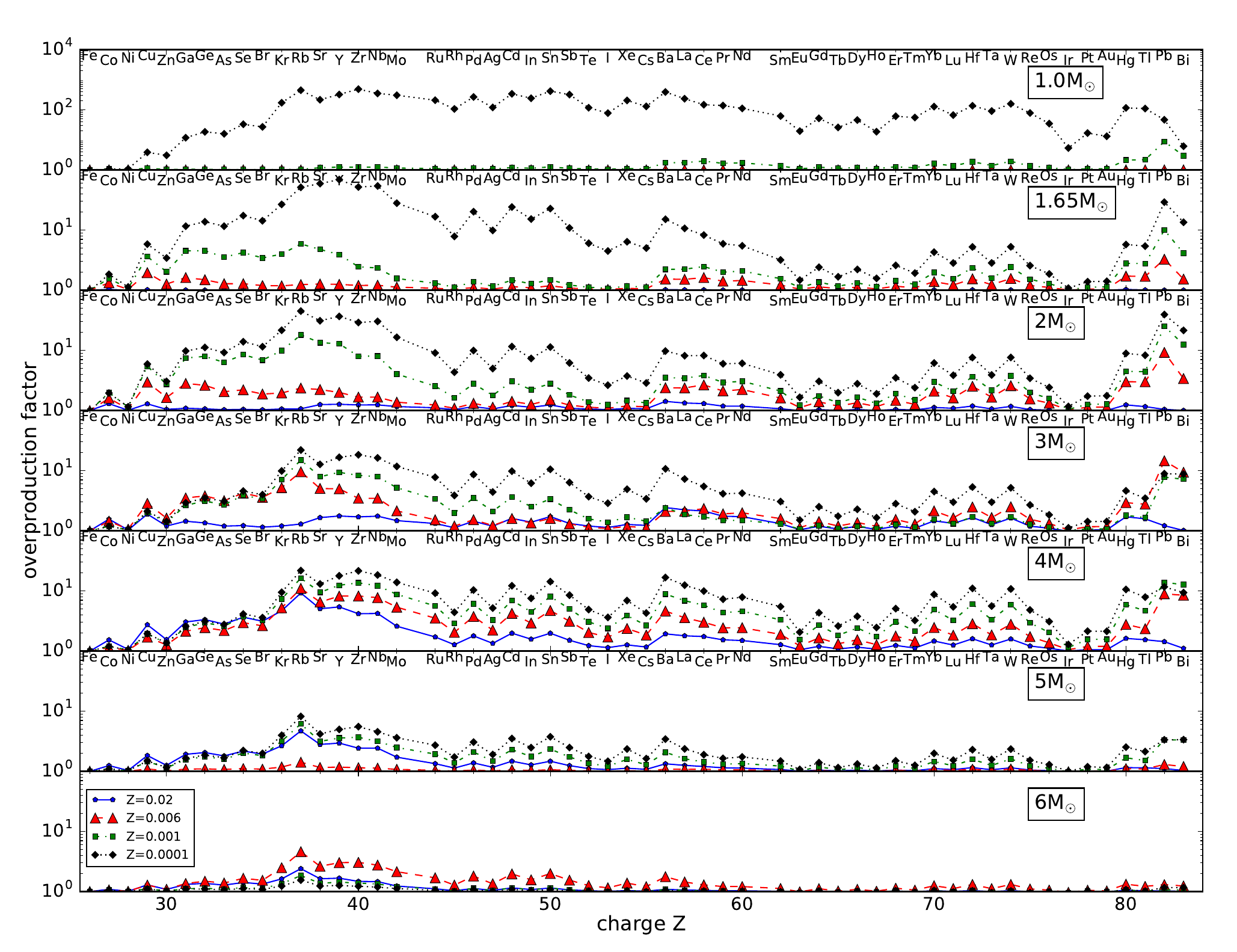}
\caption{Overproduction factors of heavy elements versus charge number of low-mass, massive
and SAGB models. Included are AGB models at $Z=0.02$ of P16.}
\label{fig:mspr_prodfac_vsZ_agb_all}
%\end{minipage}} 
%\end{sidewaysfigure}
%}
\end{figure}
\end{landscape}

The main s process takes place in the \cdr-pocket of low-mass AGB stars and in much smaller amounts in the PDCZ
of massive AGB stars. The s-process abundance distribution depends on the metallicity of the star, because of the 
combined effect of the primary neutron source \isotope[13]{C} and of the secondary nature of the Fe seeds \citep{gallino:98,busso:99}.
%The secondary nature of the process allows
%significant production in low-mass low-Z AGB stars which contributes to $\approx$50\% of solar \isotope[208]{Pb}.
%The strong s-process component which is related to
%to contribute to about 50\%  of solar \isotope[208]{Pb}.
In these AGB models the \cdr-pocket size $M_{\cdr}$ depends on the efficiency of the CBM and decreases
at $Z=0.0001$ from the $M_{\cdr}\approx10^{-4}\msun$ in the model with
$\mzams=1.65\msun$ to $M_{\cdr}\approx10^{-8}\msun$ in the model with $\mzams=7\msun$ (\fig{fig:mspr_prodfac_vsZ_agb_all}). 
This is to a large extent due to the drastic reduction of $f_\mathrm{\rm CE}$ during the dredge-up in AGB stars with $\mzams\geq 4\msun$ (\tab{tab:f_table}). $M_{\cdr}$ in the 2\msun\ model is similar to $3.7\times10^{-5}\msun$ in stellar models at solar metallicity of \cite{lugaro:03a} and $2-3\times10^{-5}\msun$ in models at $Z=0.02$ in P16.
The decreasing pocket size with initial mass leads to a drastic decrease of s-process production in massive AGB and S-AGB models.

We compare the overproduction factors of heavy elements of  the low-mass AGB models, massive AGB models and S-AGB models with
AGB models of $Z=0.02$ from P16 in \fig{fig:mspr_prodfac_vsZ_agb_all}.
%There is an trend of increasing overproduction factors with decreasing metallicity.
In stellar models with initial mass of $\mzams=1\msun$ at $Z=0.006$ and $Z=0.001$ inefficient TDUP leads to little surface enrichment
except for the model at $Z=0.0001$ which experiences H ingestion (\sect{hingnucleo}).
The total dredged-up mass $M_D$ of AGB models increases in initial mass up to $\mzams=2\msun$ (\tab{table:agb_properties}) which leads
to an increase of the overproduction factors of heavy elements with initial mass (\fig{fig:mspr_prodfac_vsZ_agb_all}).
For larger initial masses the overproduction factors of peak s-process elements tend to decrease because of the larger envelope masses dilute the heavy elements,
a decrease of $M_D$ and smaller \cdr\ pockets. %s (\tab{agbmodelprop2_1_5a}).

With decreasing metallicity lower initial masses have the largest overproduction factors (\fig{fig:mspr_prodfac_vsZ_agb_all}).
The largest overproduction factors of Sr and Pb are present in low-mass AGB models with initial masses below $4\msun$.
%The overproduction factors are the largest around Zr at the first s-process peak
%due to efficient production of \isotope[96]{Zr} \fhcom{n: this is odd - Zr96 is only 2-3\% of all Zr in the sun. Most of the Zr mus then not come from these AGBs? This needs to be checked.} in the TP of AGB models with $\mzams/\msun$ = 3, 4, 5 (\fig{fig:mspr_prodfac_vsZ_agb_all}).
Rb is efficiently produced in the TP of massive AGB stars and its ratio to Sr, which is mostly produced in low-mass AGB models,
increases from low-mass AGB stars to massive AGB stars (\fig{fig:mspr_prodfac_vsZ_agb_all}) in agreement with the observed high Rb/Sr ratio of massive AGB stars \cite[e.g.][]{garcia-hernandez:13}.
At lower metallicity the higher pulse temperature $T_{\rm PDCZ}$ results in a larger Rb/Sr ratio in the stellar models with $\mzams=2\msun$ (\fig{fig:mspr_prodfac_vsZ_agb_all}).

%and its production relative to  %as \isotope[87]{Rb}
% shows 
%the relative importance of the \isotope[22]{Ne}($\alpha$,n)\isotope[25]{Mg} neutron source compared to the \isotope[13]{C}($\alpha$,n) neutron source

\isotope[87]{Rb} and \isotope[88]{Sr} have the largest overproduction
factors of all AGB models at $Z=0.006$ in the $\mzams=4\msun$, $Z=0.006$ model in agreement with models of the same initial mass at $Z=0.01$ of P16.
%The highest overproduction factors of \isotope[88]{Sr} are found for models of $4\msun$, $2\msun$ and $1\msun$ at $Z=0.006$, $Z=0.001$ and 
%$Z=0.0001$ respectively. 
%\isotope[138]{Ba} has the highest overproduction factors at the initial mass of $4\msun$ for $Z=0.006$ and $Z=0.001$ and
%for $Z=0.0001$ at $1\msun$. P16 find the highest factors at $Z=0.01$ in the stellar model with initial mass of $3\msun$. 
\isotope[208]{Pb} has the highest overproduction factor of all AGB models at $Z=0.006$  in the $\mzams=3\msun$, $Z=0.006$ model
in agreement with AGB models at $Z=0.01$ of P16. 
A comparison between these results and other models available in the literature is provided in \sect{discussion}.

\subsection{$\gamma$ process}

\begin{figure}
\centering
\includegraphics[width=\columnwidth]{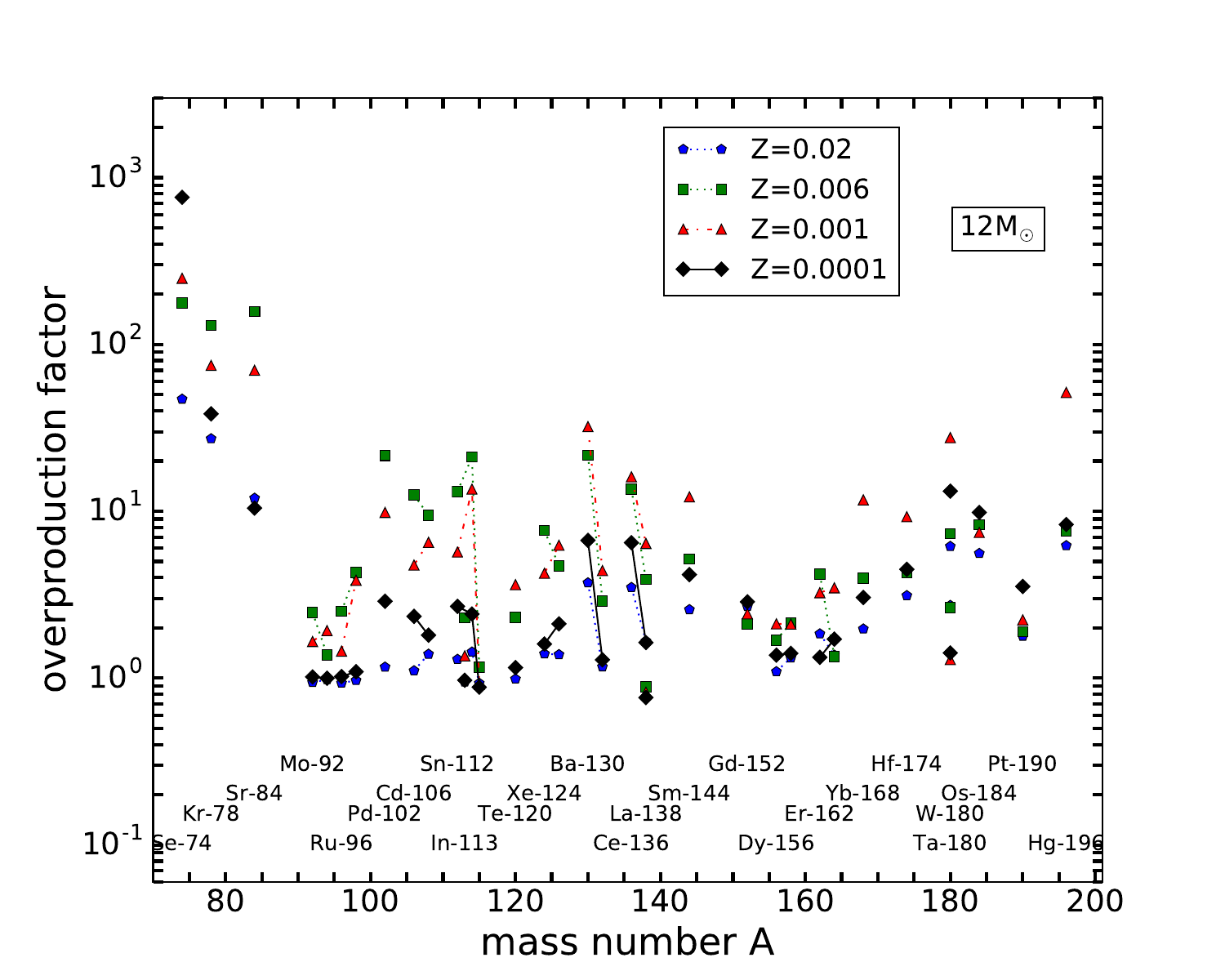}
\includegraphics[width=\columnwidth]{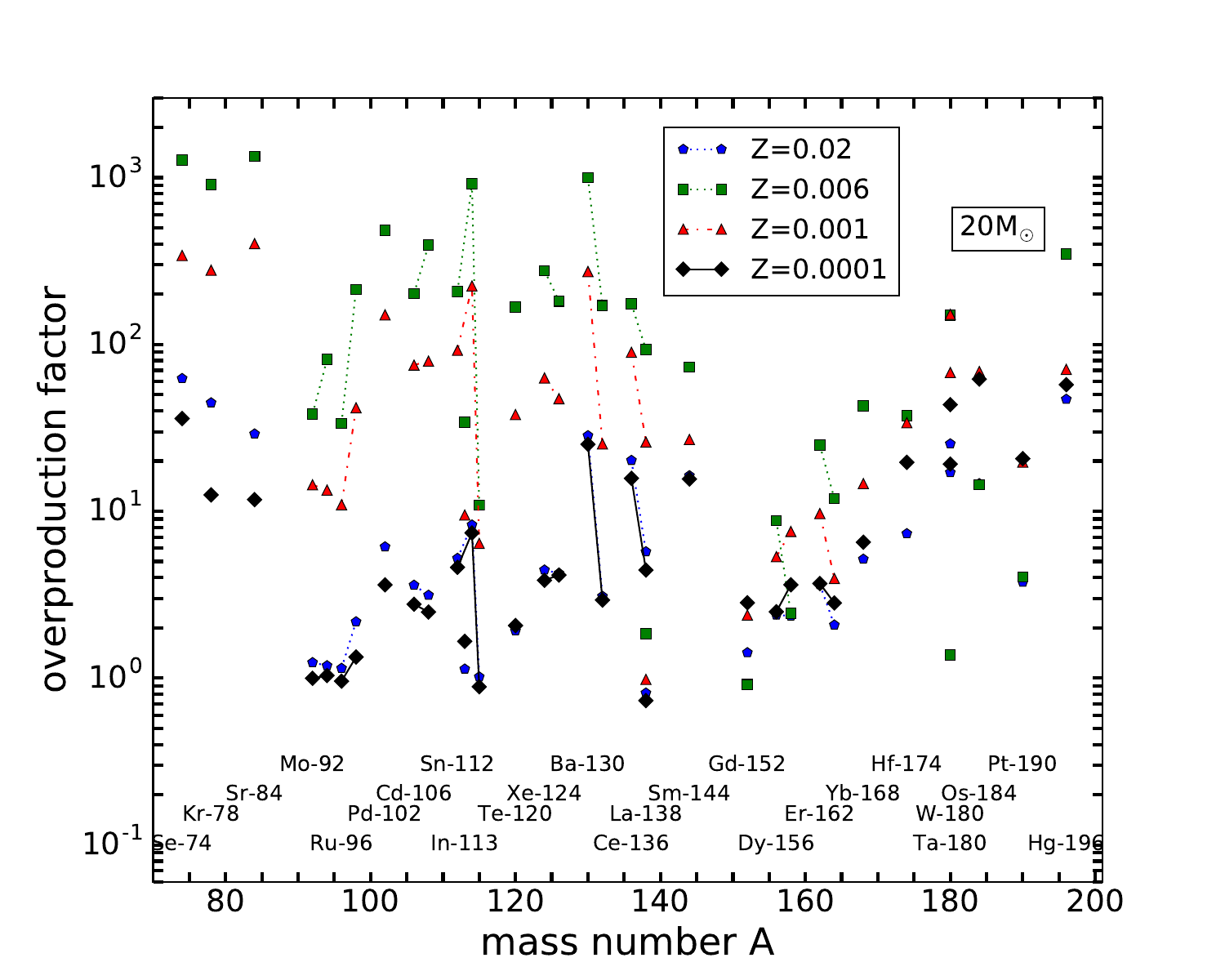}

\caption{Overproduction factors versus mass number of p nuclei and their metallicity-dependence in massive star models with $\mzams=12\msun$ (top) and $\mzams=20\msun$ (bottom).}
\label{fig:ppr_prodfacs}
\end{figure}

The $\gamma$ process produces proton-rich (p) nuclei in explosive Ne- and O-burning layers of CCSN models,
where heavy seed nuclei are destroyed through photo-disintegration and proton capture \citep{woosley:78}.
For a review of the $\gamma$-process production and its uncertainties we refer to e.g. \cite{arnould:03,rauscher:13,pignatari:16,rauscher:16}.
In the massive star models presented here, the lightest p-process nuclei \isotope[74]{Se}, \isotope[78]{Kr} and \isotope[84]{Sr}
are more produced than most heavier $\gamma$-process isotopes in stellar models with initial mass up to $\mzams=20\msun$ (\fig{fig:ppr_prodfacs}). % initial mass of $12\msun$ and $25\msun$ in
These isotopes are formed in the deepest layers of explosive
O burning owing to their light masses and strong fallback prevents any production of \isotope[74]{Se}, \isotope[78]{Kr} and \isotope[84]{Sr}
in the massive star models with $\mzams=25\msun$. In the latter models only the heaviest p-process nuclei such as \isotope[180]{Ta} and \isotope[180]{W} are ejected.

Models with $\mzams=15\msun$ produce the majority of $\gamma$-process isotopes from the $\alpha$-rich freeze-out layers.
%in contrast to stellar models with initial mass of $12\msun$.
For increasing metallicity the relative contribution of the $\alpha$-rich freeze-out material
to the total amount of produced \isotope[74]{Se}, \isotope[78]{Kr} and \isotope[84]{Sr} decreases.
At $Z=0.006$ the production in $\alpha$-rich freeze-out layers of the stellar model with $\mzams=15\msun$ would become negligible.
But in this model an additional production of light p-process nuclei takes place in a shocked and ejected thin Fe core layer.

The dominant production of \isotope[92,94]{Mo}, including contributions to \isotope[96,98]{Ru},
occurs in the same $\alpha$-rich freeze-out layers as \isotope[74]{Se}, \isotope[78]{Kr} and \isotope[84]{Sr}. %(P16).
Heavier $\gamma$-process isotopes are mostly produced in O and Ne shell burning of these massive star models. %see NSE comparison
Both burning sites are the only $\gamma$-process sites in stellar models with $\mzams=20\msun$ and $\mzams=25\msun$ because of the
lack of ejected $\alpha$-rich freeze-out layers. In stellar models with $\mzams=20\msun$ we find larger overproduction factors
than in those with $\mzams=12\msun$ (\fig{fig:ppr_prodfacs}). 
Nucleosynthesis in O-C shell mergers involves the $\gamma$ process (\sect{sec:shellmerger}).

The $\gamma$-process production in massive stars is considered to be dominated by the SN explosive component \citep[e.g.][and references therein]{arnould:03}. However, in case of O-C shell mergers the pre-SN production is increased by orders of magnitudes, and it may become more relevant than the explosive $\gamma$-process component \citep[][]{ritter:17b}. In our stellar model set, the $\mzams=15\msun$, $Z=0.02$ model and $\mzams=12\msun$, $15\msun$,  and $20\msun$  models at $Z=0.01$ include O-C shell mergers, and carry this anomalous pre-SN signature.

%\begin{sidewaystable}
%\resizebox{0.80\textwidth}{!}{
%\begin{minipage}{\textwidth}
\begin{landscape}
\begin{table}
\centering
\begin{center}
\begin{tabular}{ccccccccccccc}
\hline
\multicolumn{13}{|c|}{wind}  \\
\hline
 species & $1\msun$  & $1.65\msun$ & $2\msun$  & $3\msun$  & $4\msun$  & $5\msun$  & $6\msun$  & $7\msun$  & $12\msun$ & $15\msun$ & $20\msun$ & $25\msun$  \\
\hline
 C & 8.708E-03 & 2.147E-02 & 2.356E-02 & 8.884E-03 & 1.988E-03 & 7.856E-04 & 2.937E-04 & 1.797E-03 & 4.031E-07 & 2.472E-07 & 2.913E-07 & 4.980E-07 \\
 N & 7.763E-05 & 5.710E-05 & 3.870E-05 & 3.408E-05 & 1.019E-02 & 4.692E-03 & 2.360E-03 & 8.018E-03 & 4.301E-07 & 2.674E-07 & 2.471E-08 & 4.223E-08 \\
 O & 1.835E-03 & 8.626E-03 & 9.952E-03 & 2.781E-03 & 4.080E-03 & 1.833E-04 & 2.239E-04 & 3.249E-03 & 2.939E-06 & 1.671E-06 & 1.721E-06 & 2.942E-06 \\
 F & 3.522E-08 & 2.432E-06 & 2.565E-06 & 3.756E-08 & 7.060E-09 & 1.065E-09 & 6.403E-10 & 9.822E-09 & 2.166E-11 & 1.246E-11 & 1.304E-11 & 2.229E-11 \\
 Ne & 1.487E-05 & 8.758E-04 & 9.089E-04 & 1.138E-04 & 2.528E-04 & 4.369E-05 & 5.373E-05 & 1.268E-04 & 2.461E-07 & 1.412E-07 & 1.371E-07 & 2.344E-07 \\
 Na & 1.903E-07 & 6.951E-06 & 6.375E-06 & 1.020E-06 & 1.500E-06 & 3.222E-07 & 1.138E-07 & 5.084E-07 & 8.330E-09 & 4.204E-09 & 9.296E-10 & 1.589E-09 \\
 Mg & 1.005E-06 & 5.403E-05 & 9.078E-05 & 3.960E-05 & 1.193E-04 & 2.429E-05 & 1.937E-05 & 4.907E-05 & 7.500E-08 & 4.634E-08 & 3.896E-08 & 6.660E-08 \\
 Al & 3.506E-08 & 3.905E-07 & 8.674E-07 & 6.366E-07 & 2.057E-06 & 2.275E-06 & 1.841E-06 & 1.037E-06 & 3.202E-09 & 1.937E-09 & 1.506E-09 & 2.575E-09 \\
 Si & 7.645E-07 & 1.791E-06 & 2.559E-06 & 3.970E-06 & 6.714E-06 & 7.136E-06 & 1.495E-05 & 9.418E-06 & 6.802E-08 & 3.953E-08 & 3.666E-08 & 6.266E-08 \\
 S & 5.121E-07 & 1.134E-06 & 1.469E-06 & 2.391E-06 & 3.455E-06 & 4.471E-06 & 5.436E-06 & 6.386E-06 & 4.806E-08 & 2.793E-08 & 2.590E-08 & 4.427E-08 \\
 Ar & 8.228E-08 & 1.820E-07 & 2.359E-07 & 3.843E-07 & 5.537E-07 & 7.182E-07 & 8.738E-07 & 1.027E-06 & 7.738E-09 & 4.498E-09 & 4.170E-09 & 7.129E-09 \\
 Ca & 5.590E-08 & 1.240E-07 & 1.607E-07 & 2.630E-07 & 3.790E-07 & 4.920E-07 & 5.993E-07 & 7.038E-07 & 5.303E-09 & 3.082E-09 & 2.858E-09 & 4.885E-09 \\
 Fe & 6.453E-07 & 1.451E-06 & 1.881E-06 & 3.073E-06 & 4.429E-06 & 5.754E-06 & 7.012E-06 & 8.238E-06 & 6.202E-08 & 3.605E-08 & 3.342E-08 & 5.714E-08 \\
 Sr & 6.005E-09 & 3.708E-09 & 2.515E-09 & 1.657E-09 & 2.446E-09 & 1.014E-09 & 3.665E-10 & 3.752E-10 & 2.619E-12 & 1.522E-12 & 1.411E-12 & 2.412E-12 \\
 Ba & 3.250E-09 & 2.820E-10 & 2.351E-10 & 4.156E-10 & 9.354E-10 & 2.492E-10 & 9.297E-11 & 1.114E-10 & 7.818E-13 & 4.544E-13 & 4.213E-13 & 7.202E-13 \\
 Eu & 3.857E-12 & 6.504E-13 & 9.461E-13 & 1.386E-12 & 2.736E-12 & 1.848E-12 & 1.967E-12 & 2.290E-12 & 1.849E-14 & 1.075E-14 & 9.967E-15 & 1.704E-14 \\
 Pb & 3.221E-10 & 4.503E-10 & 7.906E-10 & 2.868E-10 & 5.420E-10 & 2.010E-10 & 8.544E-11 & 9.068E-11 & 6.482E-13 & 3.768E-13 & 3.493E-13 & 5.972E-13 \\
\\
\hline
\noalign{\smallskip}
\end{tabular}
%\centering
%\begin{center}
\begin{tabular}{ccccc}
\hline
\multicolumn{5}{|c|}{wind + pre-SN ejecta}  \\
\hline
species & $12\msun$ & $15\msun$ & $20\msun$ & $25\msun$  \\
\hline
 C & 1.294E-01 & 1.686E-01 & 2.128E-01 & 2.320E-01 \\
 N & 2.159E-04 & 3.034E-04 & 2.001E-02 & 1.041E-02 \\
 O & 7.383E-01 & 1.339E+00 & 2.044E+00 & 7.407E-01 \\
 F & 2.482E-08 & 3.120E-08 & 7.212E-08 & 3.463E-07 \\
 Ne & 6.624E-02 & 1.989E-01 & 2.638E-01 & 8.609E-02 \\
 Na & 2.475E-04 & 8.799E-04 & 1.662E-03 & 1.031E-03 \\
 Mg & 6.435E-02 & 7.448E-02 & 1.011E-01 & 4.728E-03 \\
 Al & 1.640E-03 & 1.436E-03 & 1.536E-03 & 7.644E-05 \\
 Si & 4.380E-02 & 1.183E-01 & 2.093E-02 & 1.294E-04 \\
 S & 6.015E-03 & 7.430E-02 & 4.377E-04 & 2.055E-05 \\
 Ar & 3.365E-04 & 8.531E-03 & 3.551E-06 & 3.270E-06 \\
 Ca & 3.565E-06 & 3.916E-05 & 1.977E-06 & 2.233E-06 \\
 Fe & 1.204E-02 & 3.753E-05 & 2.378E-05 & 2.648E-05 \\
 Sr & 9.471E-10 & 1.915E-09 & 3.577E-09 & 2.253E-09 \\
 Ba & 2.181E-10 & 3.039E-10 & 4.202E-10 & 3.869E-10 \\
 Eu & 3.695E-12 & 4.508E-12 & 5.702E-12 & 6.867E-12 \\
 Pb & 1.697E-10 & 2.311E-10 & 3.149E-10 & 3.012E-10 \\
\\
\hline
\noalign{\smallskip}
\hline
\end{tabular}
\begin{tabular}{cccc}
\hline
\multicolumn{4}{|c|}{Wind + SN ejecta}  \\
\hline
      $12\msun$ & $15\msun$ & $20\msun$ & $25\msun$  \\
\hline
 1.226E-01 & 1.646E-01 & 1.956E-01 & 2.037E-01 \\
 2.164E-04 & 3.050E-04 & 2.320E-02 & 2.813E-02 \\
 2.828E-01 & 1.176E+00 & 2.043E+00 & 7.062E-01 \\
 2.752E-08 & 4.500E-08 & 9.317E-08 & 2.585E-06 \\
 6.669E-03 & 1.469E-01 & 1.607E-01 & 1.227E-01 \\
 1.812E-05 & 5.359E-04 & 6.145E-04 & 1.018E-03 \\
 1.638E-02 & 7.562E-02 & 1.160E-01 & 2.178E-02 \\
 6.219E-05 & 1.157E-03 & 1.505E-03 & 1.315E-04 \\
 9.289E-02 & 1.078E-01 & 1.138E-01 & 2.063E-02 \\
 5.446E-02 & 4.459E-02 & 2.999E-02 & 2.046E-03 \\
 1.083E-02 & 7.684E-03 & 3.495E-03 & 1.122E-05 \\
 8.686E-03 & 5.152E-03 & 1.363E-03 & 2.159E-06 \\
 2.443E-01 & 5.073E-02 & 7.649E-05 & 2.524E-05 \\
 7.219E-10 & 4.976E-08 & 3.417E-09 & 2.215E-09 \\
 1.771E-10 & 2.812E-10 & 3.836E-10 & 3.739E-10 \\
 3.874E-12 & 4.845E-12 & 5.932E-12 & 7.541E-12 \\
 1.450E-10 & 2.163E-10 & 2.869E-10 & 3.093E-10 \\
\\
\hline
\noalign{\smallskip}
\hline
\end{tabular}
\end{center}
\caption{Yields derived from stellar winds, pre-SN and SN ejecta for $Z=0.0001$. The SN ejecta is produced with the delayed explosion prescription.
We provide tables for other metallicities online (Appendix \ref{sec:appendix}).}
%\end{minipage}
%}
%\end{sidewaystable}
\label{tab:yields_all_delay_set1_5}
\end{table}
\end{landscape}

%%%%%%%%%%%%%%%%%%%%%%%%%

%%%%%%%%%%%%%%%%%%%%%%%%%

%%%%%%%%%%%%%%%%%%%%%%%%%
\begin{landscape}
\begin{figure}
\centering
\includegraphics[width=0.9\columnwidth]{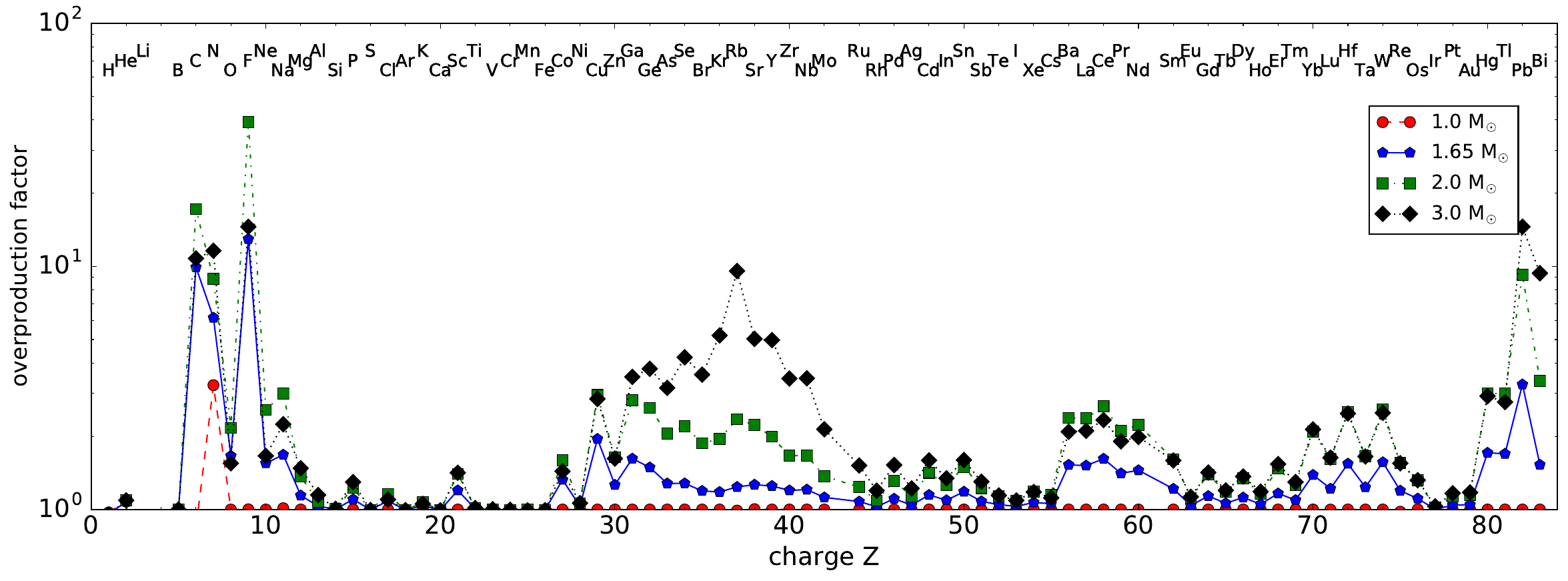}
\includegraphics[width=0.9\columnwidth]{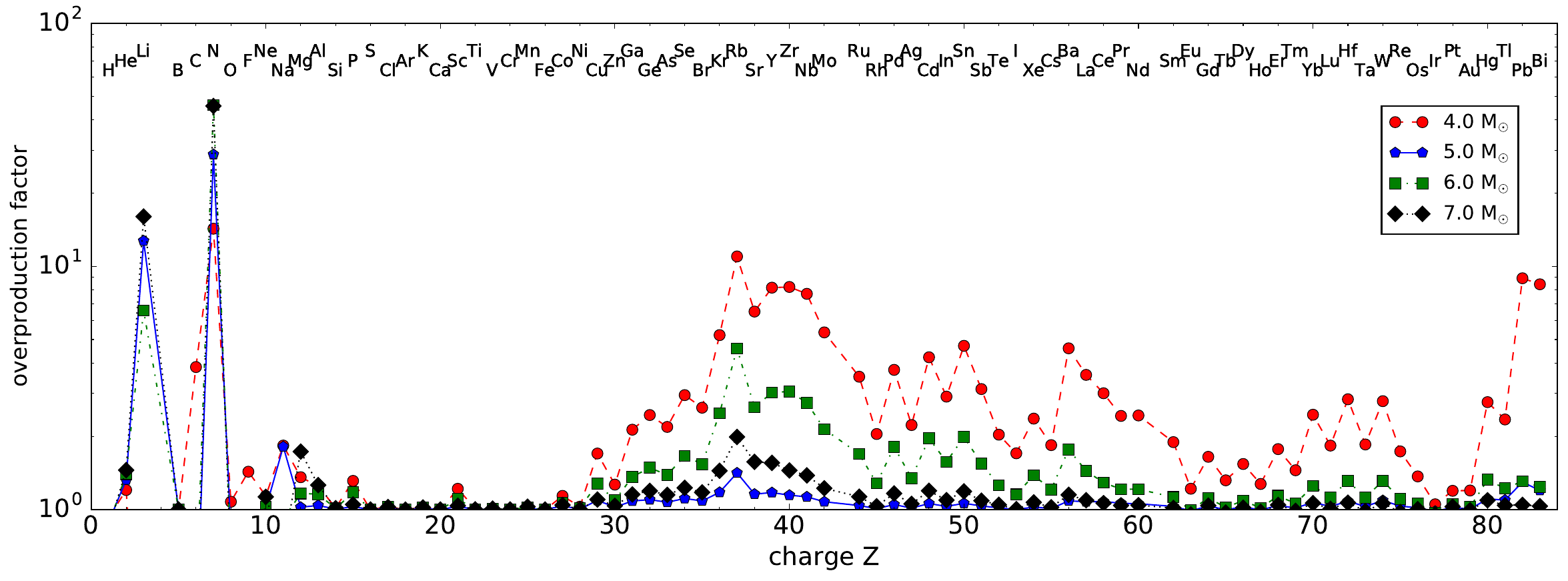}
\caption{Overproduction factors versus charge number of final yields of AGB models at $Z=0.006$.}
\label{fig:agb_prodfac_set1_3a}
\end{figure}
\end{landscape}

%%%%%%%%%%%%%%%%%%%%%%%%%
\begin{landscape}

\begin{figure}
\centering
\includegraphics[width=0.9\columnwidth]{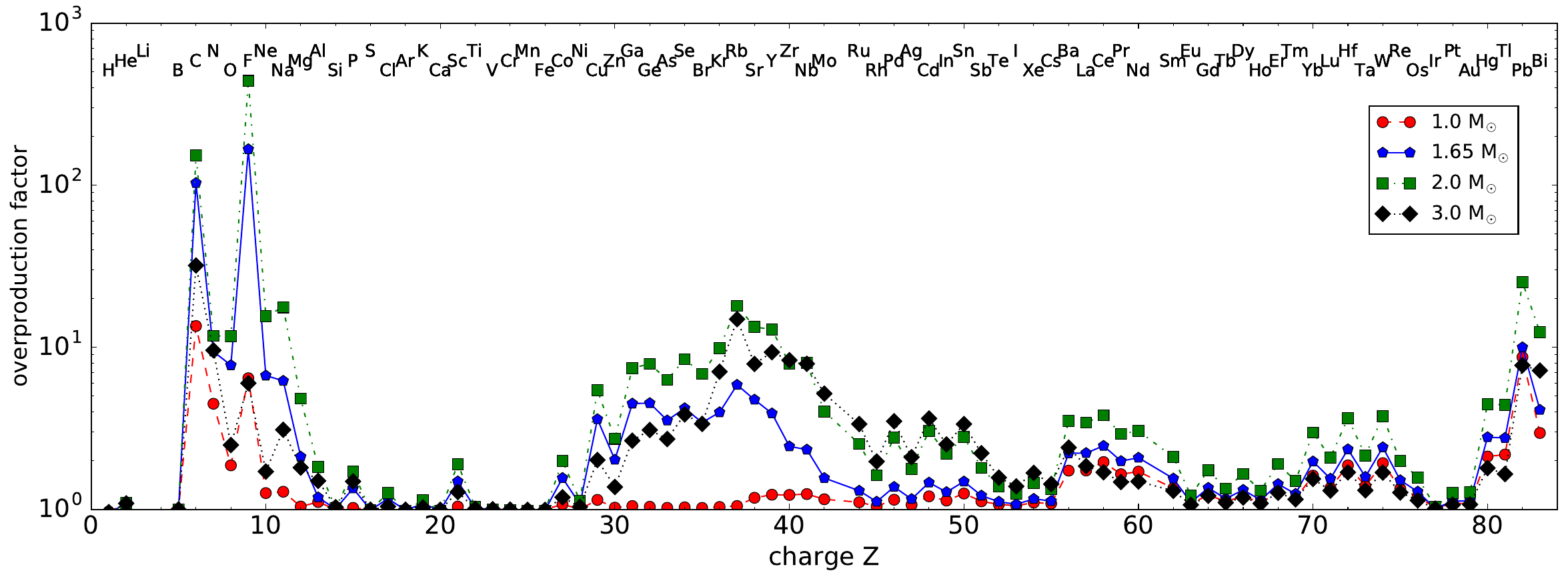}
\includegraphics[width=0.9\columnwidth]{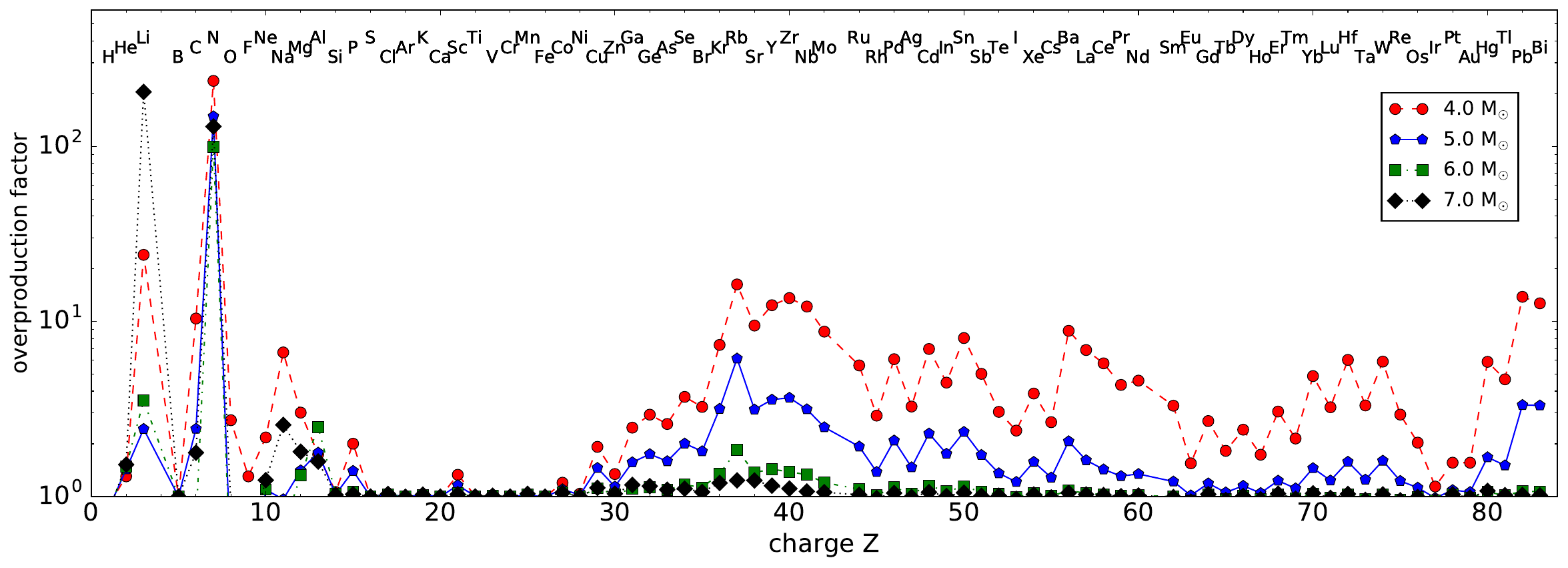}
\caption{Overproduction factors versus charge number of final yields of AGB models at $Z=0.001$.}
\label{fig:agb_prodfac_set1_4a}
\end{figure}
\end{landscape}

%%%%%%%%%%%%%%%%%%%%%%%%%
\begin{landscape}

\begin{figure}
\centering
\includegraphics[width=0.9\columnwidth]{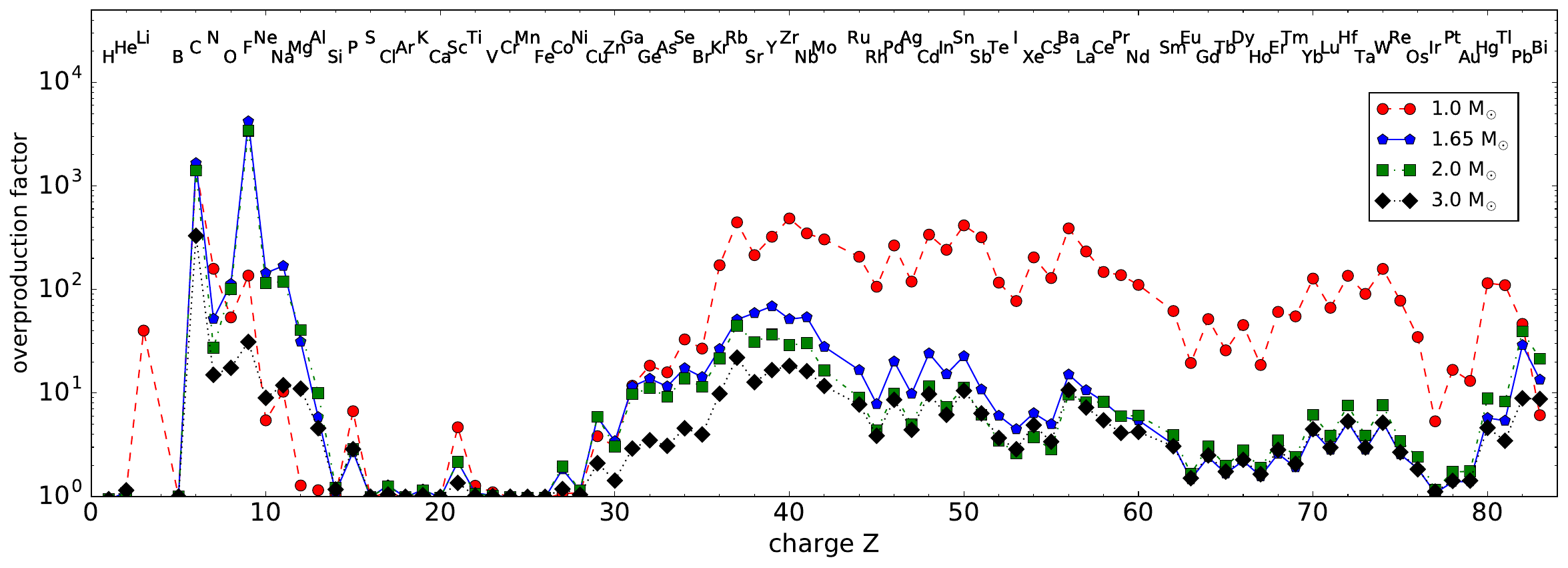}
\includegraphics[width=0.9\columnwidth]{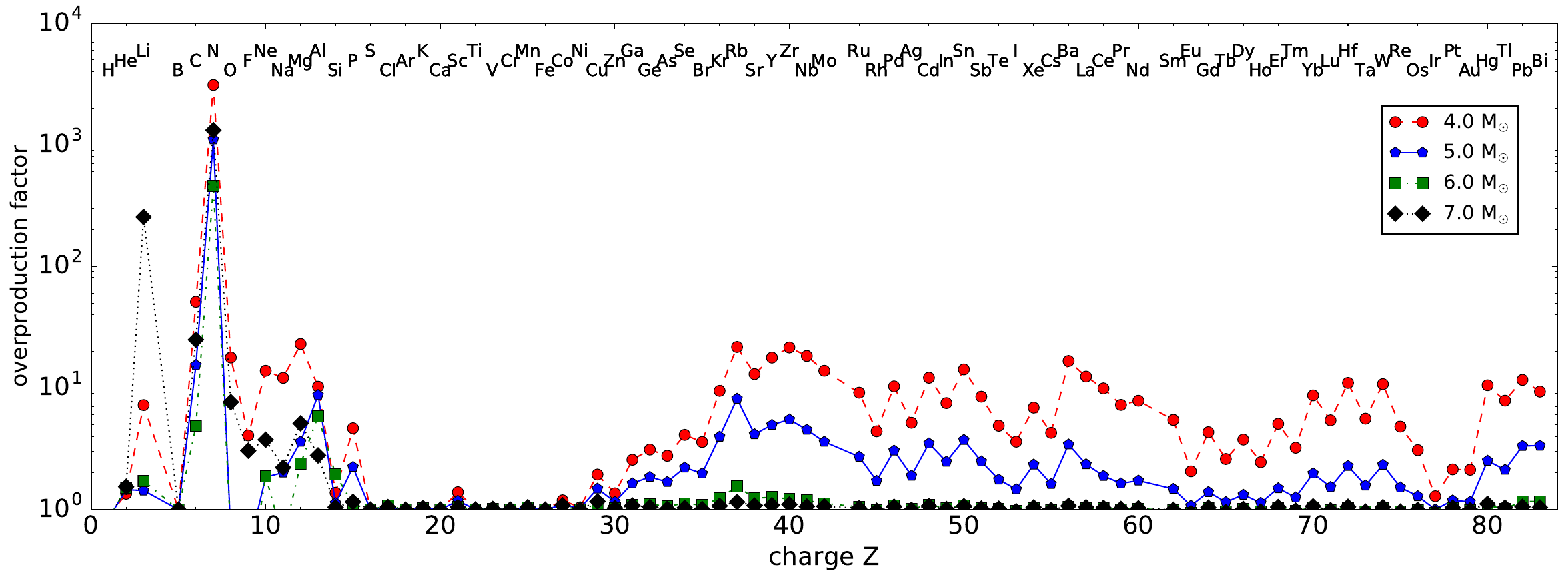}
\caption{Overproduction factors versus charge number of final yields of AGB models at $Z=0.0001$.}
\label{fig:agb_prodfac_set1_5a}
\end{figure}
\end{landscape}

%%%%%%%%%%%%%%%%%%%%%%%%%

%%%%%%%%%%%%%%%%%%%%%%%%%

%%%%%%%%%%%%%%%%%%%%%%%%%
\begin{figure}
\centering
\includegraphics[width=\columnwidth]{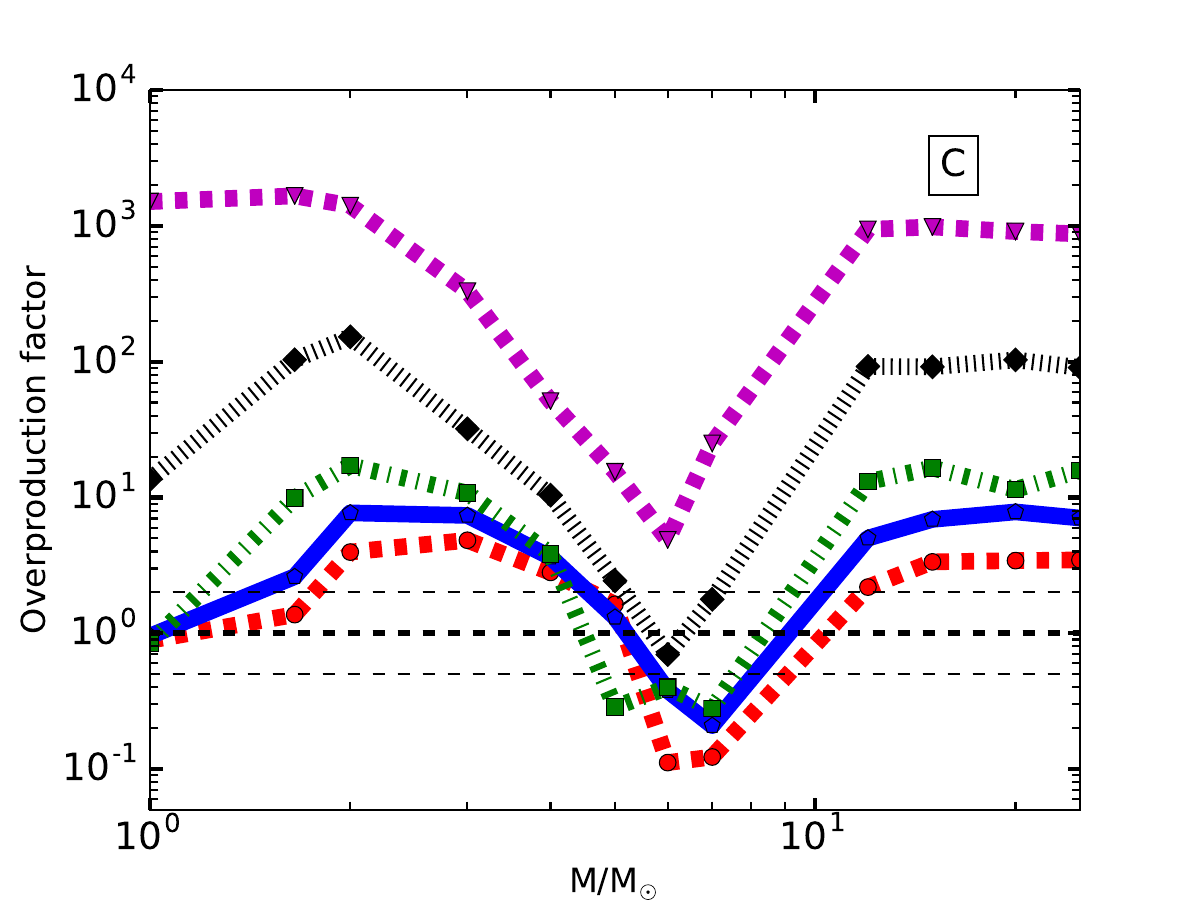}
\includegraphics[width=\columnwidth]{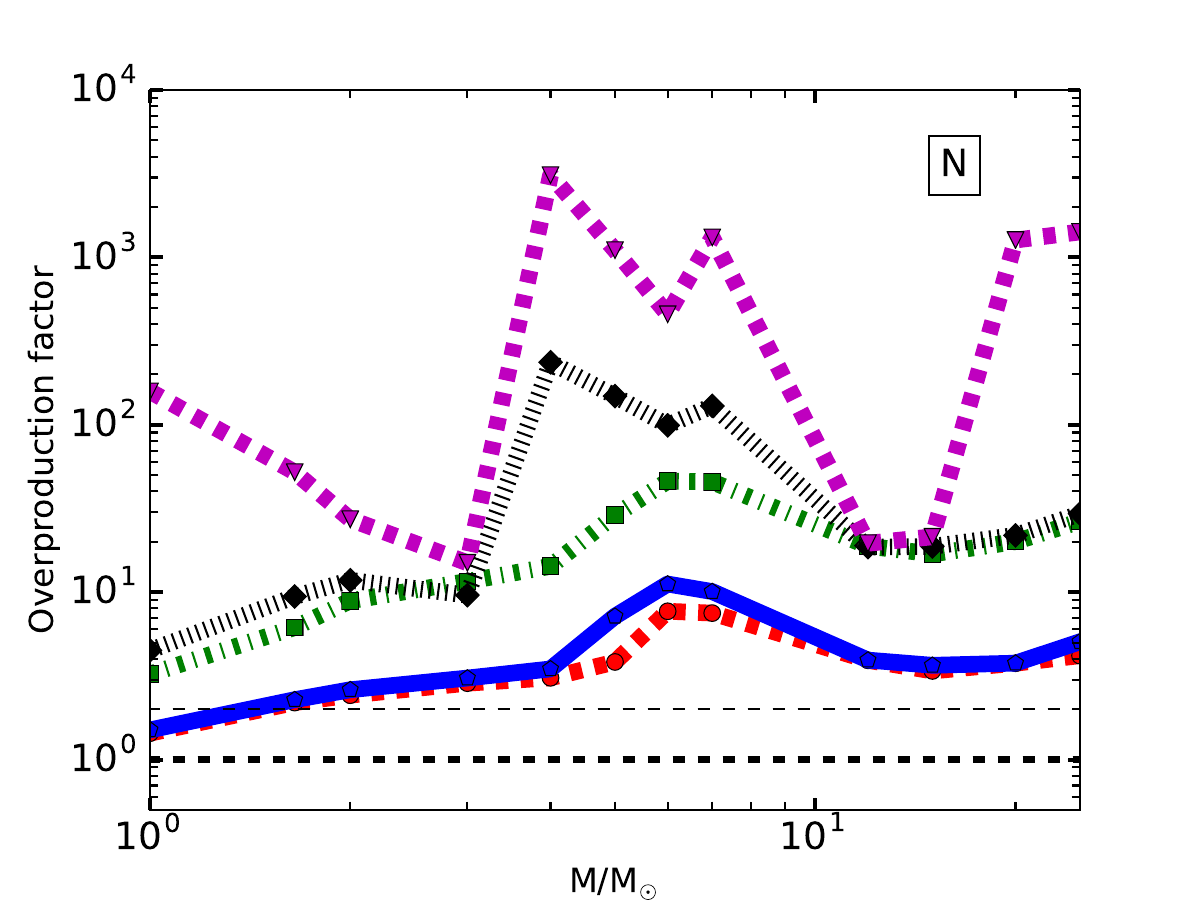}
\includegraphics[width=\columnwidth]{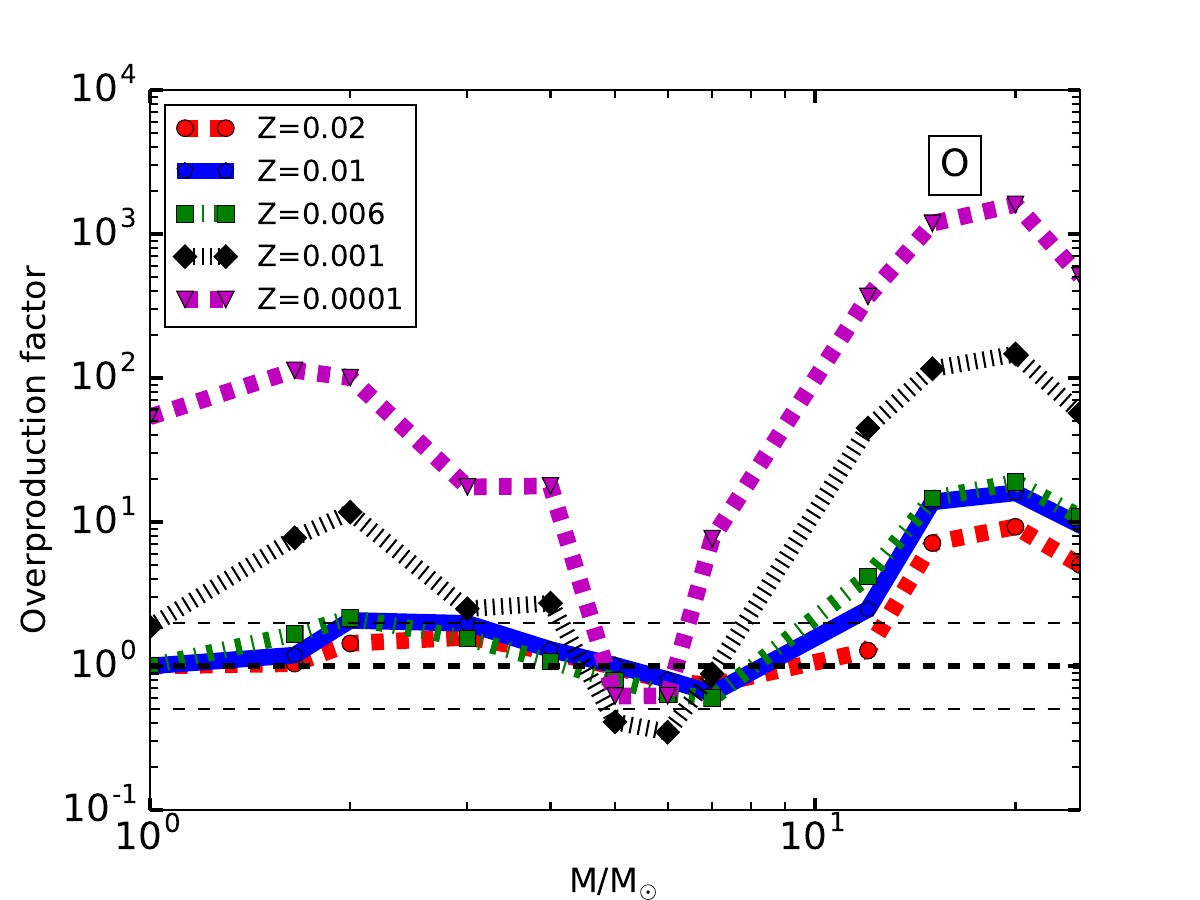}
\caption{Overproduction factors of CNO isotopes versus initial mass of final yields. The dashed horizontal lines indicate
the values 0.5, 1 and 2. 
Plots for all stable elements and many isotopes at all metallicities presented in this work are available online at 
\protect\url{http://nugridstars.org/data-and-software/yields/set-1}.} 
\label{fig:cno_prodfac_zdep}
\end{figure}

%%%%%%%%%%%%%%%%%%%%%%%%

%%%%%%%%%%%%%%%%%%%%%%%%%
%%%%%%%%%%%%%%%%%%%%%%%%%

%%%%%%%%%%%%%%%%%%%%%%%%%

%%%%%%%%%%%%%%%%%%%%%%%%%

%%%%%%%%%%%%%%%%%%%%%%%%%
\begin{figure}
\centering
\includegraphics[width=\columnwidth]{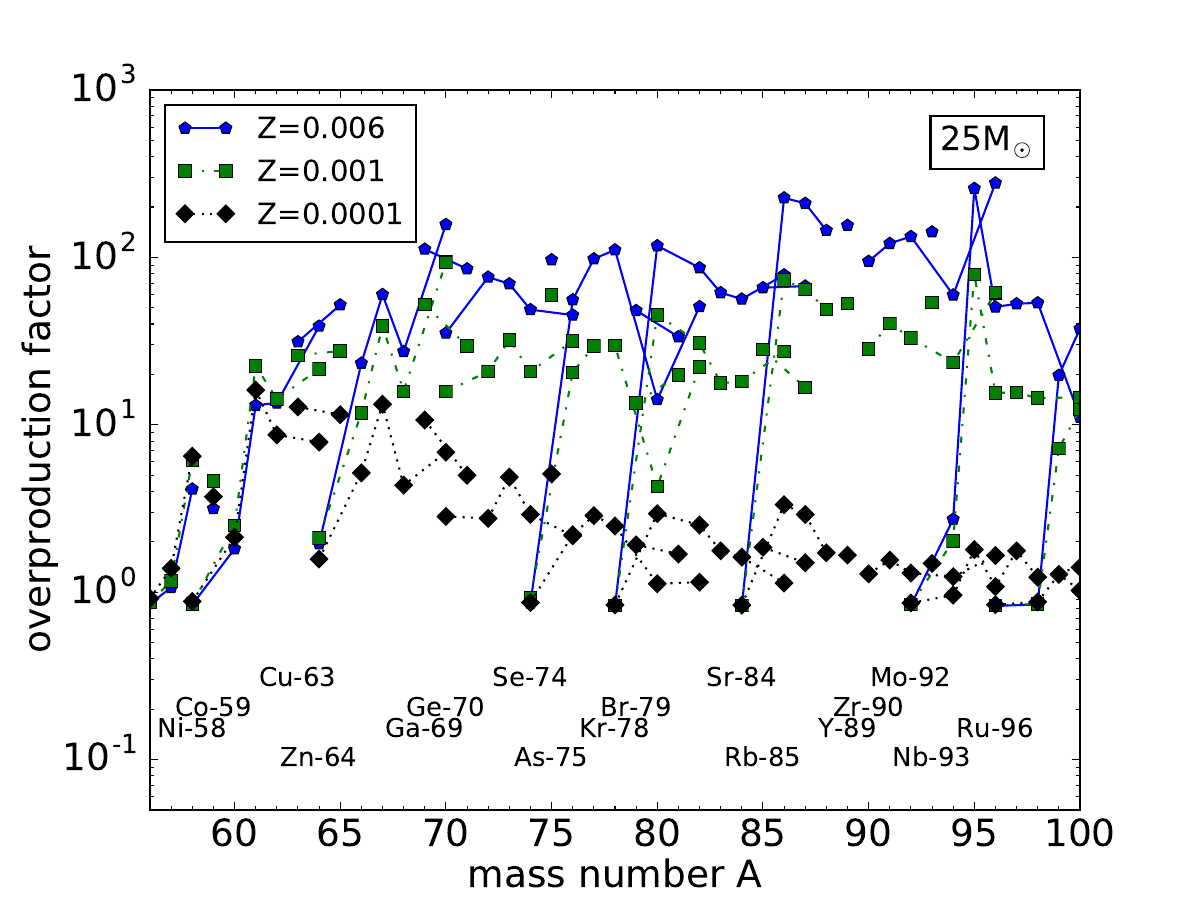}
\includegraphics[width=\columnwidth]{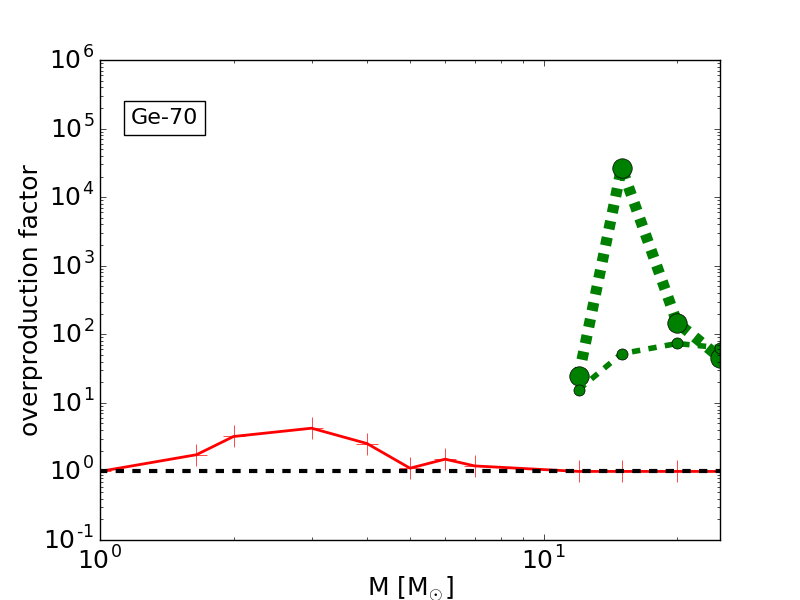}
\caption{Overproduction factors versus mass number of final yields of stars with $\mzams=25\msun$ 
at $Z=0.006$, $Z=0.001$ and $Z=0.0001$ with focus on the
weak s-only isotopes \isotope[70]{Ge}, \isotope[76]{Se},\isotope[80,82]{Kr} and \isotope[86,87]{Sr} (top).
The overproduction factors versus initial mass of \isotope[70]{Ge} at $Z=0.006$ for stellar wind ejecta (solid line),
pre-explosive ejecta (small circles) and explosive ejecta (large circles) with delayed explosion prescription (bottom).
Plots for all stable elements and many isotopes at all metallicities including delayed and rapid explosion prescriptions are available online at 
\protect\url{http://nugridstars.org/data-and-software/yields/set-1}.
}
\label{fig:wspr_prodfacs}
\end{figure}

%%%%%%%%%%%%%%%%%%%%%%%%%

%\begin{sidewaysfigure}

%%%%%%%%%%%%%%%%%%%%%%%%%%

%discussion
\section{Discussion} \label{discussion}

In this section we first address model limitations arising from the choice of resolution in AGB models and massive star models. Afterwards, the resulting stellar yields of AGB models and massive star models are compared with 
previous works.

\subsection{Resolution of AGB models}
\label{resagb}

\noindent In AGB models the H and He shells become hotter and thinner
with increasing initial mass and decreasing metallicity.
This makes it challenging to model the bottom of the convective envelope and its boundary in massive AGB stars and S-AGB stars \citep{siess:10}. 
To model the effect of HDUP (see \sect{subsecCBM}) a high temporal and spatial resolution is required at the bottom of the
convective envelope.

To assess the sensitivity of resolution on the final yields a $\mzams=4$, $Z=0.0001$ model was calculated with increased resolution at the bottom boundary of the convective envelope and at the location of the \cdr\ pocket. We compare this high-resolution calculation with up to $1.8\times10^{4}$ zones with the
calculation with moderate resolution of below $2\times10^3$ zones. The latter resolution is similar to the resolution of the stellar models for which yields are calculated. No efforts have been made to reach convergence based on the resolution.
Among light elements the biggest difference is in F which is overproduced by factor four. Mg and Al that are underproduced by less than a factor of two in the high-resolution runs compared to the lower-resolution run. 
%\mpcom{World-WideHAT IS THE RESOLUTION IN THE MODELS USED TO CALCULATE THE YIELDS PROVIDED IN THIS PAPER? YOU MAY WANT TO SPECIFY THIS MAYBE?} 
Furthermore, the lower-resolution model produces two times as many elements at $Z \approx 50$ than the high-resolution model.
For most elements beyond $Z \approx 50$ this production relative to the high-resolution model increases with a maximum of 6.7 for Bi and Pb  (\fig{fig:agb_4Msun_set1_4a_comp_highlowres}).
%\mpcom{IF THE C13-POCKET IS NEGLIGIBLE, WHY WE GET AN EFFECT ON THE ABUNDANCES OF BI AND PB? IS THIS THE NE22 ACTIVATION? I DO NOT FULLY UNDERSTOOD THIS RESULT. }

In order to resolve the  \cdr\ pocket down to the size of $M_{\cdr}\approx10^{-8}\msun$ that would be found, for example, in the 7\msun\ models owing
to the decreased CBM efficiency $f_\mathrm{CE}$ in the most massive S-AGB models
further resolution refinement below the convective boundary would be  necessary. However, this is not required because at this level the \cdr\ pocket does not contribute significantly to 
the \spr\ production of the more massive low-resolution AGB models.

The heavy element production in the \cdr\ pocket for stellar models with  $\mzams\geq4\msun$ decreases strongly due to a rapidly decreasing \cdr-pocket size (\fig{fig:agb_prodfac_set1_3a} to \fig{fig:agb_prodfac_set1_5a}).
For all AGB models with $\mzams>3\msun$ we find $T_{\rm PDCZ}>3\times10^8\, \mathrm{K}$ (\tab{table:agb_properties}) which is high enough to activate efficiently the \isotope[22]{Ne}($\alpha$,n)\isotope[25]{Mg} reaction. In these stellar models with $\mzams>4\msun$ the PDCZ becomes the main production site of first-peak s-process elements (\fig{fig:agb_prodfac_set1_3a} to \fig{fig:agb_prodfac_set1_5a}). The impact of the \cdr-pocket resolution on the yields of s-process elements in these stellar models is low. 
%removed this because Sam feels its contradicting: Another factor is the total dredged-up mass $M_D$, which at $Z=0.0001$ is larger in the $\mzams=4\msun$ model than in the $\mzams=3\msun$ model.
The influence of the resolution on the bottom boundary is not
  relevant for the chemical imprint of the AGB models with
  $\mzams>4\msun$ and for GCE because there is no relevant element
  production in the \cdr\ pocket. F is strongly affected by the
  resolution but is only produced in small amounts compared to AGB
  models of lower initial mass. We have therefore not improved the
  resolution for these models.

\subsection{Resolution of massive star models} \label{sec:sprscaling} %Weak s-process

\noindent In these models the C/O ratio from the post-processing at the end of the core He-burning differs from  the stellar evolution calculations by a factor of $\approx 30\, \%$ 
at $Z=0.02$ and $\approx$ 50\, \% at $Z=0.0001$. This discrepancy can be reduced in future calculations by reducing the time step, as done in \citet{jones:15}. 
%\fhcom{@SJ: is that what you really did in that paper? Did you check that the time-steps were small enough so that the post-processing yields a reasonably close answer compared to the MESA run?}. 
The largest absolute differences are found at $Z=0.02$ because the the C/O ratio decreases with metallicity.

The weak s-process at the end of core He-burning depends on the He-burning conditions for which the the ratio of the He-burning products C and O is an indicator.
%such as the abundance of available $\alpha$'s for which the ratio of the He-burning products C and O ratio.}
To analyze the impact of the C/O ratio difference  on the weak s process we have  
selected a $\mzams=15\msun$, $Z=0.02$ model which shows stronger s-process production compared to stars at lower metallicity. 
%\fhcom{@CR: Do we discuss the questionable 12Msun production of heavy elements, and if so where?}
Using a sub-time stepping method we generate a resolved post-processing simulation that fully agrees with the stellar evolution simulation in the He-core C/O ratios. First-peak s-process element yields are lower by up to a factor of three in the the converged post-processed model compared to the lower-resolution model.  (\fig{fig:co_core_M15Z0p02_yield_exp_comparison}).
%largest differences 
%are for weak s-process elements with up to a factor of three. 
These differences in final yields is within the uncertainty which results from the method which is applied to artificially explode the models: the difference between delay and rapid explosion prescription.
The difference in yields between the resolved and lower-resolution models will impact GCE simulation only to the extent that the elements have a significant weak s-process contribution. In particular, this needs to be taken into account for the GCE of Cu, Ga and Ge \citep[][]{pignatari:10}.
%The differences are similar to the difference between these yields and those of P16 
%where the core He burning is resolved (see \sect{sect_yield_comparison}).
%The heavier elements not receive a positive contribution from explosive nucleosynthesis!!!!??
%The factor of three might be reduced due to the impact of the 
%Converging the advanced stage of stellar evolution is a challenge \cite{}.

\begin{figure}
\centering
\includegraphics[width=\columnwidth]{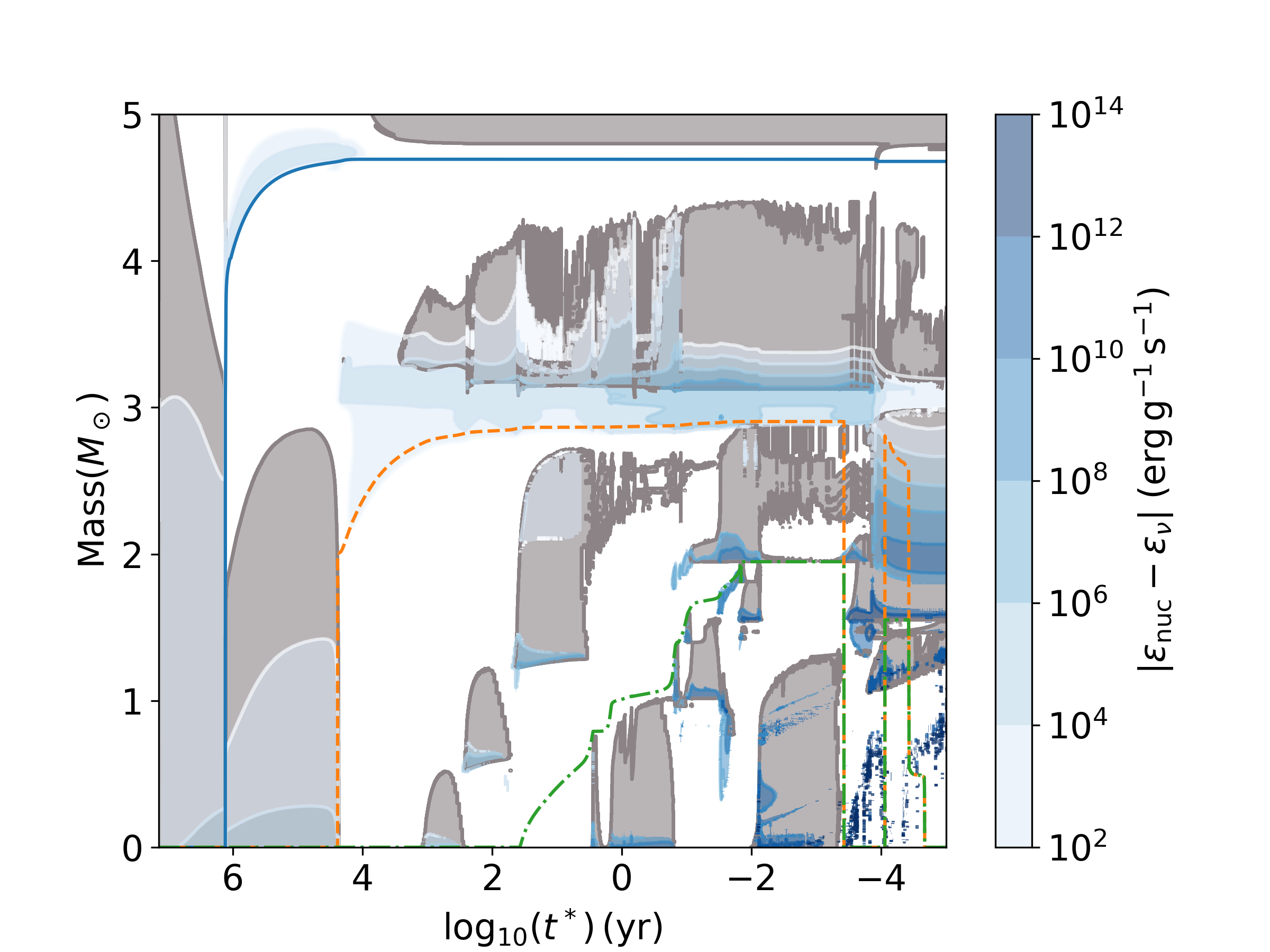}
\includegraphics[width=\columnwidth]{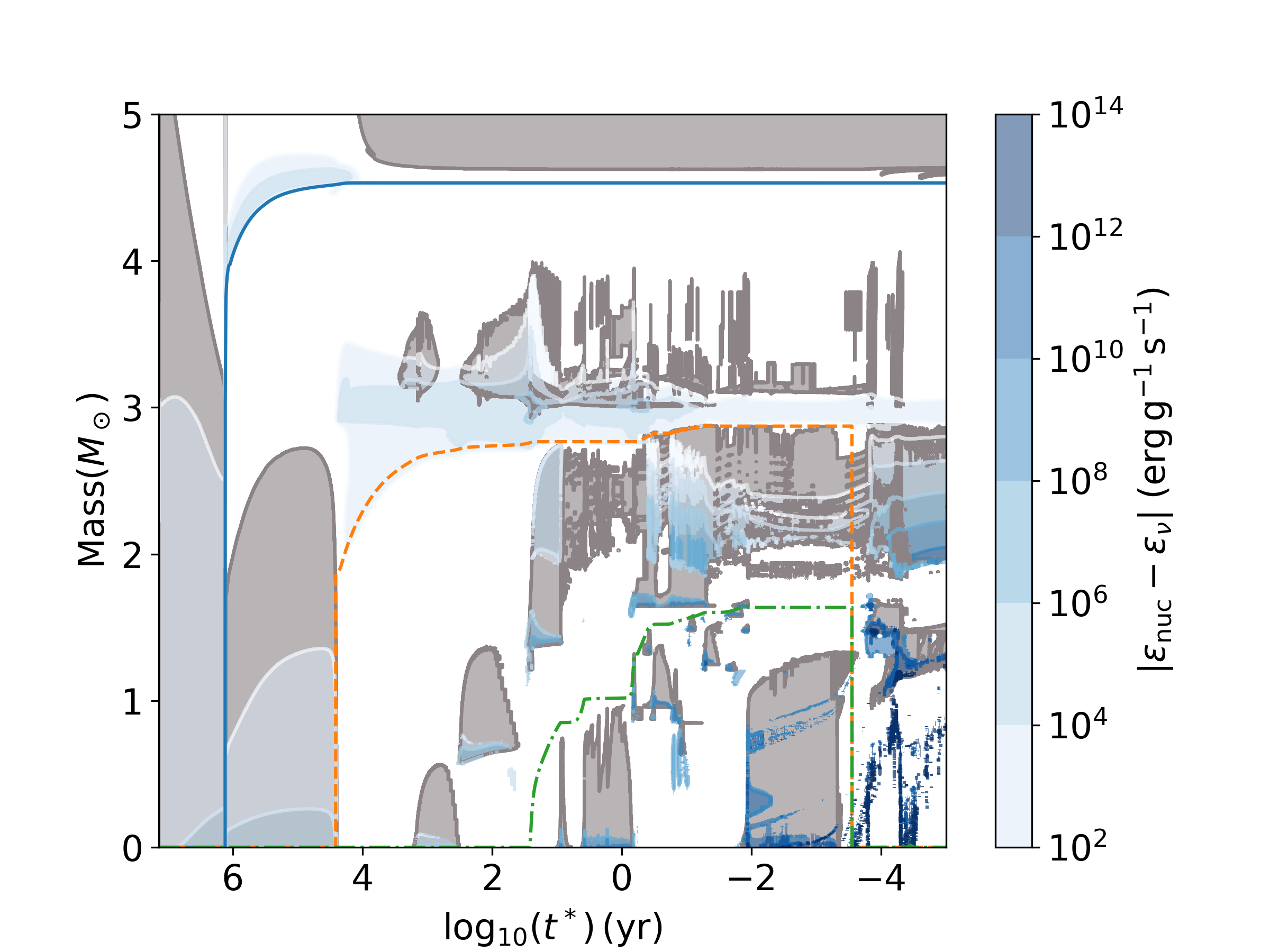}

\caption{Kippenhahn diagrams as in \fig{fig:kip_cont_massive} of two
  $\mzams=15\msun$, $Z=0.02$ models with the default resolution (top)
  and with an increased resolution during core He-burning
  (bottom).}
\label{fig:kip_cont_res_massive}
\end{figure}

To analyze the effect of resolution on the stellar evolution simulation of massive stars we calculate a $\mzams=15\msun$, $Z=0.02$ model
with a factor between about two and ten higher time resolution of the He core-burning phase 
compared to the default-resolution model. We find that the He core is slightly smaller at higher resolution (\fig{fig:kip_cont_res_massive}).
 A major O-C shell merger which is present in the default-resolution model disappears at higher resolution. The occurrence of O-C shell mergers is ultimately dependent on the 3D hydrodynamic properties of  of convection
and CBM in the late stages of massive star models, which require multi-D hydrodynamic simulations \citep[][Andrassy et al., in preparation]{meakin:06a,herwig:14,jones:17}. Numerical resolution of massive star models adds to the numerous other sources of uncertainty (Petermann et al., in preparation), and we plan to reduce this error in future updates of theses data sets. 

\subsection{Comparison with stellar yields in literature} \label{sect_yield_comparison}

\subsubsection{AGB models}

\noindent 
Yield sets of of AGB and S-AGB models have been presented by  \cite{karakas:10}, \cite{siess:10}, the FRUITY database \citep{cristallo:11,cristallo:15},
 \cite{lugaro:12}, \cite{ventura:13}, \cite{fishlock:14}, \cite{doherty:14b} and \cite{karakas:16}.
Others have published AGB and massive star yields such as the 
Padova group \citep{portinari:98,marigo:01b}. %alongi:93b,

\noindent \textbf{Low-mass AGB models}

We compare yields of the $\mzams=2\msun$, $Z=0.0001$ model of this work with
yields of models of the same initial mass and metallicity from \cite{herwig:04a}, \cite{karakas:10} and \citet[S14]{straniero:14}
in \tab{compareOtherWork_M2_set1_5a}.
For the isotopes \isotope[12]{C} and \isotope[14]{N} we find yields in-between those of K10 and S14 and within
a factor two of those of H04. The larger production of \isotope[16]{O} compared to K10 and S14 is %\fhcom{n: might? aren't we sure it is?}
due to the choice of the CBM applied in the He intershell \citep{herwig:05}. H04 get about 2.5 times lower O yield
with the application of the same CBM efficiency.
In this work s-process isotopes are less produced than in S14. The yield of \isotope[88]{Sr} from the AGB model of this work
is roughly 50\% lower than of S14. The yields of \isotope[138]{Ba} and \isotope[208]{Pb} are by more than  $1\, \mathrm{dex}$ and  $2\, \mathrm{dex}$ lower than those of S14, respectively.
We attribute the differences %in part \fhcom{n: what are the other parts?} 
to less convective boundary mixing and smaller \cdr\ pockets than in S14. 
The small \cdr\ pocket sizes of these models correspond to lower [hs/ls] abundances of the observed s-process spread of C stars (see P16, \sect{sec:mspr}). An updated model of \cdr-pocket formation through CBM induced by gravity waves
has been recently proposed by \cite{battino:16}. It is able to explain the larger  observed neutron-exposure signatures.  The physics process that causes the spread of observed s-process is still matter of debate \citep[e.g.][]{iben:82a,herwig:97,cristallo:01,denissenkov:03t,herwig:05,busso:07,denissenkov:09,battino:16,trippella:16}.

\noindent \textbf{Massive AGB models}

Yields of the $\mzams=5\msun$, $Z=0.0001$ model of this work and yields 
from models of the same initial mass and metallicity from H04, K10 and C15
are shown in \tab{compareOtherWork_M5_set1_5a}. The same trends as in the discussion on the nested-network for HBB can be observed here (\sect{sec:nugrid_codes}).
Differences between the yields from this work, H04 and K10 are similar within a factor 2-3, while showing the same trends in isotopic ratios. 
Our $\mzams=5\msun$, $Z=0.0001$ model  has lower s-process yields than our $\mzams=2\msun$, $Z=0.0001$ model, %possibly 
because of the HDUP limit to the mixing at the bottom of the convective envelope CBM parameter (\sect{sec:hdup}, \sect{subsecCBM}) which leads to a lower \cdr\ pocket contribution. 
The $\mzams=5\msun$ s-process yields from this work are more than  $1\, \mathrm{dex}$ below those of C15, which is due to the difference of CBM assumptions in these models.

\subsubsection{Massive star models}

Groups have published massive star yields at various metallicities, among others \cite{woosley:95}, the Geneva group \citep{hirschi:05b,frischknecht:16},
\cite{chieffi:04}, \cite{heger:10} and P16.
We choose the yields of P16, \citet[CL04]{chieffi:04} and \citet[K06]{kobayashi:06} for 
a comparison with models of the same initial mass and metallicity of this work.

%Note that \cite{woosley:95} (W95) and  which are without magnetic fields and rotation 
The metallicity dependence of the mass loss has a significant impact on the final yields.
K06 apply a metallicity-dependent mass loss while CL04 do not include any mass loss. The total mass lost is $0.13\msun$ and $0.41\msun$ for the $\mzams=15\msun$ model and the $\mzams=25\msun$ model at $Z=0.001$ while K06 finds $0.08\msun$ and $0.58\msun$. 
The reduced mass loss at lower metallicity results in larger core masses when considering a similar mass cut and hence 
the ejection of larger amounts of O compared to models at higher metallicity.
%Both use expl;
%Both use scaled solar abundances.
%CL uses no NSE approximation
%Both use more sophisticated explosion prescription.
%CL chooses masss cutso that ejecta is $0.1\msun$ of \isotope[56]{Ni}.

Yields of the $\mzams=15\msun$, $Z=0.02$ model of this work are compared with yields of the same initial mass
and metallicity based on the models of P16 in \tab{compareOtherWork_M15_set1_2}.
The yields of \isotope[12]{C} and \isotope[14]{N} are close to those of P16 while \isotope[16]{O} yields are by about a factor
three larger than in P16. We find only low sensitivity of the yields on the amount of fallback.
For \isotope[56]{Fe} the difference in yields to P16 increases and we find a factor of 3.2 and 4.5 lower Fe yields than in P16
because of less production of Fe in the explosion.  \isotope[88]{Sr} yields are more sensitive to
the amount of fallback than the CNO species as in P16. \isotope[88]{Sr} yield based on the delayed
explosion prescription are by about a factor two larger than those of P16 which is
within the expected difference due to the under-resolved He burning (\sect{sec:sprscaling}).

At low metallicity we show yields of the $\mzams=15\msun$, $Z=0.001$ model of this work
and models of the same initial mass and metallicity by CL04 and K06 in
\tab{compareOtherWork_M15_set1_4a}.
Yields of \isotope[12]{C} and \isotope[14]{N} are in between those of CL04 and K06 while \isotope[16]{O}
yields are larger than both works and roughly a factor of two larger than those of CL04.
The range of \isotope[56]{Fe} yields given through the delayed and rapid explosion prescriptions includes
the yields of CL04 and K06. 
%\fhcom{n: why is our Sr88 not bracketing  that of CL04 and K06 while the Fe yield does? There is more then just fall-back?!?} 
In contrast to this work CL04 and K06 fix the ejecta of Ni and Fe respectively and then adjust the amount of fallback. 
The yields of \isotope[88]{Sr} in these massive star models are considerably
larger than found in K06. Little fallback due to the rapid explosion prescription leads to about $3\, \mathrm{dex}$ more production of
\isotope[88]{Sr} compared to CL04, K06 and the corresponding model in our set calculated using the delay prescription.  
Most of the large production of \isotope[88]{Sr} shown in \tab{compareOtherWork_M15_set1_4a} originates from the innermost $\sim$ 0.1M$_{\odot}$ layers, due to the activation of the $\alpha$-rich freeze out. %\fhcom{but even our delayed case has 1dex more Sr88 than CL04?!}

We compare yields of the $\mzams=25\msun$, $Z=0.001$ model of this work with yields of models of the same initial mass and metallicity from  CL04 and K06 in \tab{compareOtherWork_M25_set1_4a}.
\isotope[12]{C} yields agree well with K06 yields and \isotope[14]{N} yields with CL04 yields.
We find lower  \isotope[16]{O} yields than CL04 and K06 which might be due to the fallback of larger parts of the O shell in these models.
Fallback strongly reduces \isotope[56]{Fe} ejection in stellar models presented here while it does not affect \isotope[56]{Fe} in CL04 and K06 because
of their fixed Ni and Fe ejecta. This leads to more than  $2\, \mathrm{dex}$ lower \isotope[56]{Fe} yields of the presented models than CL04 and K06.
\isotope[88]{Sr} yields of these massive star models are larger than those of CL04 similar to the $\mzams=15\msun$, $Z=0.001$ models.
This requires more efficient weak s-process production during He-core burning in these stellar models than in CL04 as most of the formerly He-core burning layers
fall back onto the remnant.

\begin{table}
\begin{center}
\caption{Comparison of the final yields of $\mzams=2\msun$, $Z=0.0001$ models from this work
with H04, K10 and S14. Units in $\msun$.
}
\begin{tabular}{ccccc}
\hline
\hline
species & this work         & H04       & K10       & S14     \\ 
\hline
 C-12   &  2.356E-02  & 1.834E-02 & 3.274E-02 & 1.424E-02    \\ 
% C-13   &  9.692E-07  & 5.379E-07 & 1.655E-06 & 8.593E-07    \\ 
 N-14   &  3.870E-05  & 2.767E-05 & 7.458E-05 & 4.110E-05    \\ 
% N-15   &  5.275E-09  & 6.114E-09 & 3.936E-08 & 1.616E-08    \\ 
 O-16   &  9.951E-03  & 3.830E-03 & 1.015E-03 & 5.031E-04    \\ 
 Sr-88  &  2.161E-09  &           &           & 3.528E-09    \\ 
% Y-89   &  6.017E-10  &           &           & 9.747E-10    \\ 
% Zr-90  &  4.159E-10  &           &           & 8.964E-10    \\ 
% Ba-136 &  2.079E-11  &           &           & 2.418E-10    \\ 
 Ba-138 &  1.678E-10  &           &           & 3.901E-09    \\ 
% La-139 &  1.942E-11  &           &           & 4.839E-10    \\ 
 Pb-208 &  6.656E-10  &           &           & 1.084E-07    \\ 
\hline
\noalign{\smallskip}
\hline
\end{tabular}
\label{compareOtherWork_M2_set1_5a}
\end{center}
\end{table}

 \begin{table}
\begin{center}
\caption{Comparison of the final yields of the $\mzams=5\msun$, $Z=0.0001$ models from this work
with H04, K10 and C15. Units in $\msun$.}
\begin{tabular}{ccccc} 
\hline
\hline
species & this work & H04       & K10       & C15  \\ 
\hline  
 C-12   &  6.948E-04  & 1.830E-04 & 2.787E-03 & 1.274E-02    \\ 
% C-13   &  9.086E-05  & 4.372E-05 & 4.059E-04 & 1.856E-04    \\ 
 N-14   &  4.692E-03  & 6.703E-03 & 2.405E-02 & 3.405E-04    \\ 
% N-15   &  2.194E-07  & 4.431E-08 & 6.923E-07 & 2.607E-08    \\ 
 O-16   &  1.824E-04  & 1.200E-03 & 6.094E-04 & 9.350E-04    \\ 
 Sr-88  &  8.969E-10  &           &           & 2.238E-08    \\ 
% Y-89   &  2.447E-10  &           &           & 5.853E-09    \\ 
% Zr-90  &  1.986E-10  &           &           & 4.399E-09    \\ 
% Ba-136 &  2.236E-11  &           &           & 1.029E-09    \\ 
 Ba-138 &  1.450E-10  &           &           & 5.523E-09    \\ 
% La-139 &  1.693E-11  &           &           & 5.654E-10    \\ 
 Pb-208 &  1.465E-10  &           &           & 1.284E-08    \\ 
\hline
\noalign{\smallskip}
\hline
\end{tabular}
\label{compareOtherWork_M5_set1_5a}
\end{center}
\end{table}

\begin{table}
\begin{center}
\caption{Comparison of the final yields of $\mzams=15\msun$, $Z=0.02$ models 
of this work (delay, rapid) with those of P16. Units in $\msun$.} %, CL04 and K06.}
\begin{tabular}{ccccc}
\hline  
\hline  
species & delay & rapid & P16 (delay) & P16 (rapid) \\ %& CL04      & K06        \\  
\hline  
 C-12   &  1.543E-01  & 1.528E-01  &1.761E-01	 & 1.785E-01	\\% & 2.050E-01 & 6.510E-02  \\ 
% C-13   &  1.000E-03  & 1.001E-03  &9.805E-04	 & 9.813E-04	\\ %& 1.310E-03 & 1.150E-03  \\ 
 N-14   &  4.965E-02  & 4.989E-02  &4.967E-02	 & 4.973E-02	\\% & 5.070E-02 & 6.150E-02  \\ 
% N-15   &  8.457E-05  & 9.629E-05  &5.308E-05	 & 5.457E-05	\\% & 2.190E-05 & 5.970E-05  \\ 
 O-16   &  9.162E-01  & 8.137E-01  &2.986E-01	 & 3.011E-01	\\% & 5.160E-01 & 1.620E-01  \\ 
% O-17   &  6.673E-05  & 6.110E-05  &7.713E-05	 & 7.736E-05	\\% & 1.550E-04 & 8.120E-04  \\ 
% O-18   &  5.592E-03  & 6.444E-03  &4.882E-03	 & 5.004E-03	\\% & 4.190E-03 & 2.540E-03  \\ 
% F-19   &  1.876E-05  & 1.537E-05  &1.518E-05	 & 1.483E-05	\\% & 4.190E-06 & 1.640E-05  \\ 
% Ne-20  &  7.256E-02  & 6.391E-02  &3.151E-02	 & 3.141E-02	\\% & 3.600E-02 & 3.390E-02  \\ 
% Na-23  &  2.597E-03  & 2.215E-03  &1.299E-03	 & 1.301E-03	\\% & 1.380E-03 & 1.060E-03  \\ 
% Mg-24  &  2.842E-02  & 2.523E-02  &1.548E-02	 & 1.505E-02	\\% & 3.390E-02 & 3.790E-02  \\ 
% Al-27  &  2.965E-03  & 2.510E-03  &1.259E-03	 & 1.264E-03	\\% & 1.730E-03 & 2.440E-03  \\ 
% Si-28  &  1.494E-01  & 1.442E-01  &9.677E-02	 & 8.910E-02	\\% & 1.310E-01 & 8.380E-02  \\ 
% S-32   &  5.666E-02  & 6.022E-02  &6.575E-02	 & 6.385E-02	\\% & 6.060E-02 & 3.470E-02  \\ 
% Ar-36  &  5.762E-03  & 7.807E-03  &2.651E-02	 & 2.651E-02	\\% & 9.990E-03 & 4.900E-03  \\ 
% Ca-40  &  1.970E-03  & 4.160E-03  &1.971E-02	 & 1.967E-02	\\% & 8.230E-03 & 4.010E-03  \\ 
% Ti-48  &  8.435E-05  & 2.085E-04  &5.717E-04	 & 4.802E-04	\\% & 2.060E-04 & 1.070E-04  \\ 
% V-51   &  3.294E-05  & 8.094E-05  &1.024E-04	 & 9.875E-05	\\% & 3.350E-05 & 9.960E-06  \\ 
% Cr-52  &  8.825E-04  & 1.280E-03  &3.766E-03	 & 3.564E-03	\\% & 2.040E-03 & 1.270E-03  \\ 
% Mn-55  &  1.092E-03  & 1.853E-03  &2.124E-03	 & 2.159E-03	\\%& 1.360E-03 & 3.800E-04  \\ 
 Fe-56  &  4.306E-02  & 5.395E-02  &1.915E-01	 & 1.681E-01	\\% & 1.150E-01 & 8.520E-02  \\ 
% Co-59  &  4.033E-04  & 3.047E-03  &1.023E-02	 & 1.148E-02	\\% & 2.980E-04 & 9.000E-05  \\ 
% Ni-58  &  1.058E-02  & 6.879E-02  &1.823E-01	 & 1.131E-01	\\% & 5.020E-03 & 1.150E-03  \\ 
% Zn-70  &  1.327E-06  & 8.782E-07  &2.820E-06	 & 2.846E-06	\\% & 9.130E-07 & 5.590E-07  \\ 
% Ge-70  &  5.866E-06  & 7.985E-04  &8.356E-04	 & 3.634E-03	\\% & 1.130E-06 & 8.380E-07  \\ 
% Se-76  &  1.499E-06  & 8.123E-05  &3.013E-05	 & 1.847E-04	\\% & 2.900E-07 &            \\ 
% Kr-80  &  3.052E-07  & 2.279E-05  &1.555E-05	 & 5.410E-05	\\% & 8.460E-08 &            \\ 
% Kr-82  &  1.332E-06  & 1.011E-05  &4.350E-06	 & 3.938E-05	\\% & 2.140E-07 &            \\ 
% Sr-86  &  4.457E-07  & 6.409E-06  &3.747E-06	 & 1.756E-05	\\% & 1.080E-07 &            \\ 
% Sr-87  &  2.883E-07  & 1.397E-06  &6.093E-07	 & 3.172E-06	\\% & 5.650E-08 &            \\ 
 Sr-88  &  5.537E-06  & 1.752E-05  &2.648E-06	 & 4.056E-05	\\%& 7.880E-07 &            \\ 
\hline  
\noalign{\smallskip}          
\hline  
\end{tabular}
\label{compareOtherWork_M15_set1_2}
\end{center}
\end{table}

%\FloatBarrier
%\vspace{-20cm}
\begin{table}
 \begin{center}
\caption{Comparison of the final yields of $\mzams=15\msun$, $Z=0.001$ models from this work (delay, rapid)
with CL04 and K06. Units in $\msun$.}
\begin{tabular}{ccccc}
\hline
\hline
species & delay & rapid & CL04      & K06        \\ 
\hline
 C-12   &  1.537E-01  & 1.538E-01  & 1.840E-01 & 8.500E-02  \\ 
% C-13   &  3.379E-05  & 3.387E-05  & 5.730E-05 & 5.380E-05  \\ 
 N-14   &  2.675E-03  & 2.677E-03  & 2.990E-03 & 3.580E-03  \\ 
% N-15   &  9.472E-07  & 9.650E-07  & 8.550E-07 & 8.590E-07  \\ 
 O-16   &  1.148E+00  & 1.022E+00  & 5.270E-01 & 2.940E-01  \\ 
% O-17   &  9.748E-06  & 8.529E-06  & 1.610E-05 & 2.560E-05  \\ 
% O-18   &  3.680E-05  & 4.122E-05  & 8.300E-05 & 3.660E-04  \\ 
% F-19   &  4.026E-07  & 3.887E-07  & 2.280E-07 & 1.980E-07  \\ 
% Ne-20  &  2.079E-01  & 1.494E-01  & 1.210E-01 & 1.900E-01  \\ 
% Na-23  &  6.705E-04  & 4.477E-04  & 1.050E-03 & 1.960E-03  \\ 
% Mg-24  &  8.508E-02  & 7.286E-02  & 6.370E-02 & 6.370E-02  \\ 
% Al-27  &  9.428E-04  & 7.628E-04  & 1.520E-03 & 2.350E-03  \\ 
% Si-28  &  9.166E-02  & 1.043E-01  & 1.040E-01 & 4.290E-02  \\ 
% S-32   &  3.976E-02  & 5.319E-02  & 4.600E-02 & 1.640E-02  \\ 
% Ar-36  &  7.013E-03  & 1.056E-02  & 8.650E-03 & 2.460E-03  \\ 
% Ca-40  &  5.097E-03  & 8.190E-03  & 7.650E-03 & 1.730E-03  \\ 
% Ti-48  &  2.709E-04  & 3.698E-04  & 1.990E-04 & 5.190E-05  \\ 
% V-51   &  1.966E-05  & 2.632E-05  & 1.090E-05 & 1.620E-05  \\ 
% Cr-52  &  2.815E-03  & 3.181E-03  & 1.660E-03 & 3.030E-04  \\ 
% Mn-55  &  2.520E-04  & 3.972E-04  & 5.770E-04 & 8.010E-05  \\ 
 Fe-56  &  5.280E-02  & 1.294E-01  & 1.000E-01 & 7.080E-02  \\ 
% Co-59  &  3.237E-03  & 3.552E-03  & 1.500E-04 & 2.320E-04  \\ 
% Ni-58  &  8.744E-02  & 1.725E-02  & 2.030E-03 & 9.110E-04  \\ 
% Zn-70  &  4.330E-08  & 4.287E-08  & 8.920E-09 & 1.050E-08  \\ 
% Ge-70  &  9.422E-06  & 3.756E-03  & 3.980E-08 & 1.010E-07  \\ 
% Se-76  &  6.016E-07  & 3.788E-04  & 9.810E-09 &            \\ 
% Kr-80  &  3.748E-08  & 1.186E-04  & 2.690E-09 &            \\ 
% Kr-82  &  1.405E-07  & 4.234E-05  & 9.900E-09 &            \\ 
% Sr-86  &  8.381E-08  & 3.164E-05  & 5.420E-09 &            \\ 
% Sr-87  &  5.220E-08  & 5.954E-06  & 3.100E-09 &            \\ 
 Sr-88  &  3.935E-07  & 5.845E-05  & 3.230E-08 &            \\ 
\hline
\noalign{\smallskip}
\hline
\end{tabular}
\label{compareOtherWork_M15_set1_4a}
\end{center}
\end{table}

\begin{table}
\begin{center}
\caption{Comparison of the final yields of $\mzams=25\msun$, $Z=0.001$ models from this work (delay, rapid)
with CL04 and K06. Units in $\msun$.}
\begin{tabular}{ccccc}
\hline
\hline
 species & delay & rapid & CL04      & K06        \\ 
\hline
 C-12   &  2.115E-01  & 2.242E-01  & 5.300E-01 & 2.150E-01  \\ 
% C-13   &  5.655E-05  & 5.655E-05  & 8.860E-05 & 9.810E-05  \\ 
 N-14   &  5.825E-03  & 5.833E-03  & 4.560E-03 & 9.200E-03  \\ 
% N-15   &  1.703E-05  & 1.465E-05  & 1.330E-06 & 7.240E-06  \\ 
 O-16   &  7.878E-01  & 1.151E+00  & 2.280E+00 & 3.820E+00  \\ 
% O-17   &  2.872E-06  & 2.933E-06  & 1.430E-05 & 2.790E-05  \\ 
% O-18   &  7.849E-06  & 8.169E-06  & 2.240E-05 & 7.050E-05  \\ 
% F-19   &  4.441E-07  & 4.536E-07  & 9.460E-07 & 6.120E-07  \\ 
% Ne-20  &  1.424E-01  & 2.263E-01  & 5.280E-01 & 1.220E+00  \\ 
% Na-23  &  1.888E-03  & 3.178E-03  & 3.610E-03 & 8.090E-03  \\ 
% Mg-24  &  2.527E-02  & 2.993E-02  & 1.520E-01 & 1.790E-01  \\ 
% Al-27  &  3.698E-04  & 4.907E-04  & 3.700E-03 & 5.050E-03  \\ 
% Si-28  &  3.300E-02  & 3.200E-02  & 2.410E-01 & 1.200E-01  \\ 
% S-32   &  1.779E-02  & 1.496E-02  & 1.130E-01 & 5.510E-02  \\ 
% Ar-36  &  1.212E-03  & 8.563E-04  & 2.140E-02 & 9.290E-03  \\ 
% Ca-40  &  3.056E-05  & 2.602E-05  & 1.860E-02 & 7.940E-03  \\ 
% Ti-48  &  8.861E-07  & 8.951E-07  & 2.050E-04 & 1.170E-04  \\ 
% V-51   &  1.119E-07  & 1.158E-07  & 1.720E-05 & 7.520E-06  \\ 
% Cr-52  &  2.747E-06  & 2.775E-06  & 3.610E-03 & 2.180E-03  \\ 
% Mn-55  &  1.919E-06  & 1.933E-06  & 8.790E-04 & 2.990E-04  \\ 
 Fe-56  &  2.140E-04  & 2.152E-04  & 1.010E-01 & 7.110E-02  \\ 
% Co-59  &  3.443E-06  & 3.927E-06  & 3.540E-05 & 3.420E-05  \\ 
% Ni-58  &  9.057E-06  & 9.068E-06  & 5.050E-04 & 2.510E-04  \\ 
% Zn-70  &  2.753E-07  & 3.007E-07  & 1.380E-08 & 2.350E-08  \\ 
% Ge-70  &  1.520E-07  & 2.549E-07  & 1.600E-07 & 3.780E-06  \\ 
% Se-76  &  4.982E-08  & 8.510E-08  & 2.950E-08 &            \\ 
% Kr-80  &  2.183E-08  & 3.805E-08  & 8.820E-09 &            \\ 
% Kr-82  &  7.574E-08  & 1.291E-07  & 2.870E-08 &            \\ 
% Sr-86  &  8.012E-08  & 1.363E-07  & 1.910E-08 &            \\ 
% Sr-87  &  5.329E-08  & 8.762E-08  & 1.130E-08 &            \\ 
 Sr-88  &  4.575E-07  & 6.784E-07  & 8.010E-08 &            \\ 
\hline
\noalign{\smallskip}
\hline
\end{tabular}
\label{compareOtherWork_M25_set1_4a}
\end{center}
\end{table}

%\FloatBarrier
%\input{figures4a}

%%%%%%%%%%%%%%%%%%%%%%% stuff for discussion

\begin{landscape}
\begin{figure}
\centering
\includegraphics[width=0.8\columnwidth]{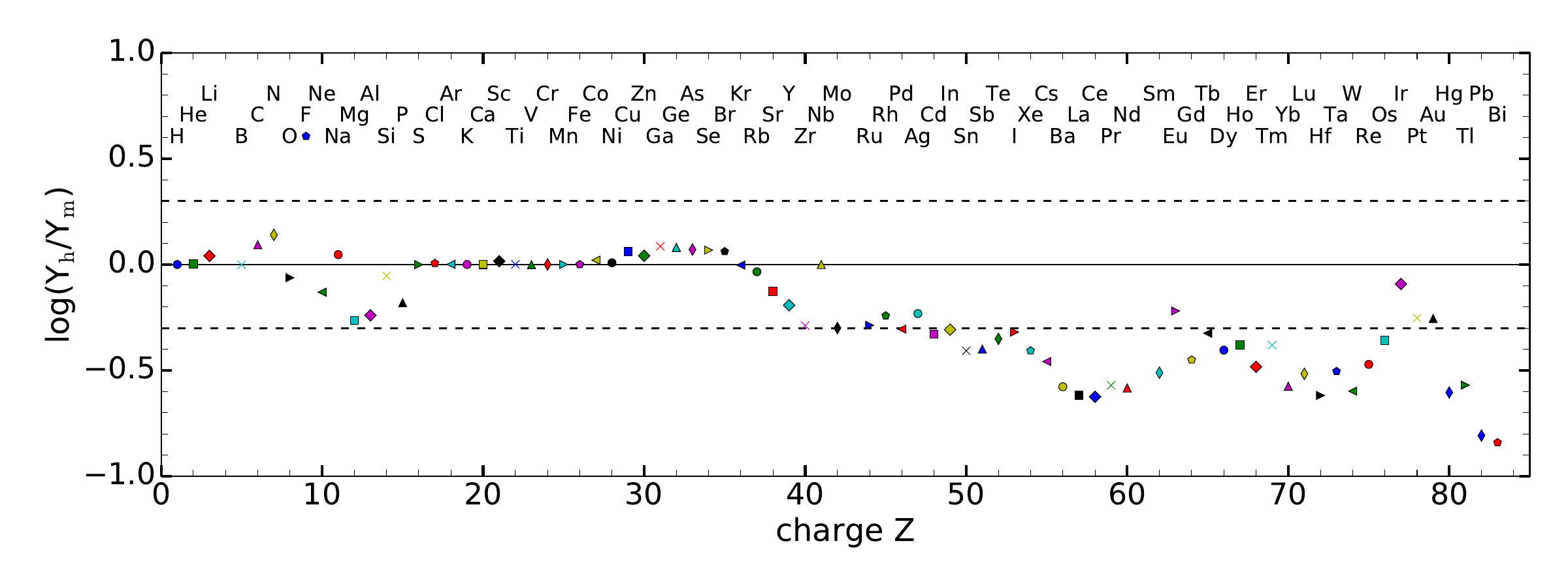}
\caption{Ratio of final yields versus charge number based on a medium resolution ($Y_{m}$) and a high resolution ($Y_{h}$) of the $\mzams=4\msun$, $Z=0.0001$ model.
The dashed lines indicate a factor of 0.5 and 2.}
\label{fig:agb_4Msun_set1_4a_comp_highlowres}
\end{figure}
%\end{landscape}

%%%%%%%%%%%%%%%%%%%%%%%%%

%\begin{landscape}
\begin{figure}
\centering
\includegraphics[width=0.8\columnwidth]{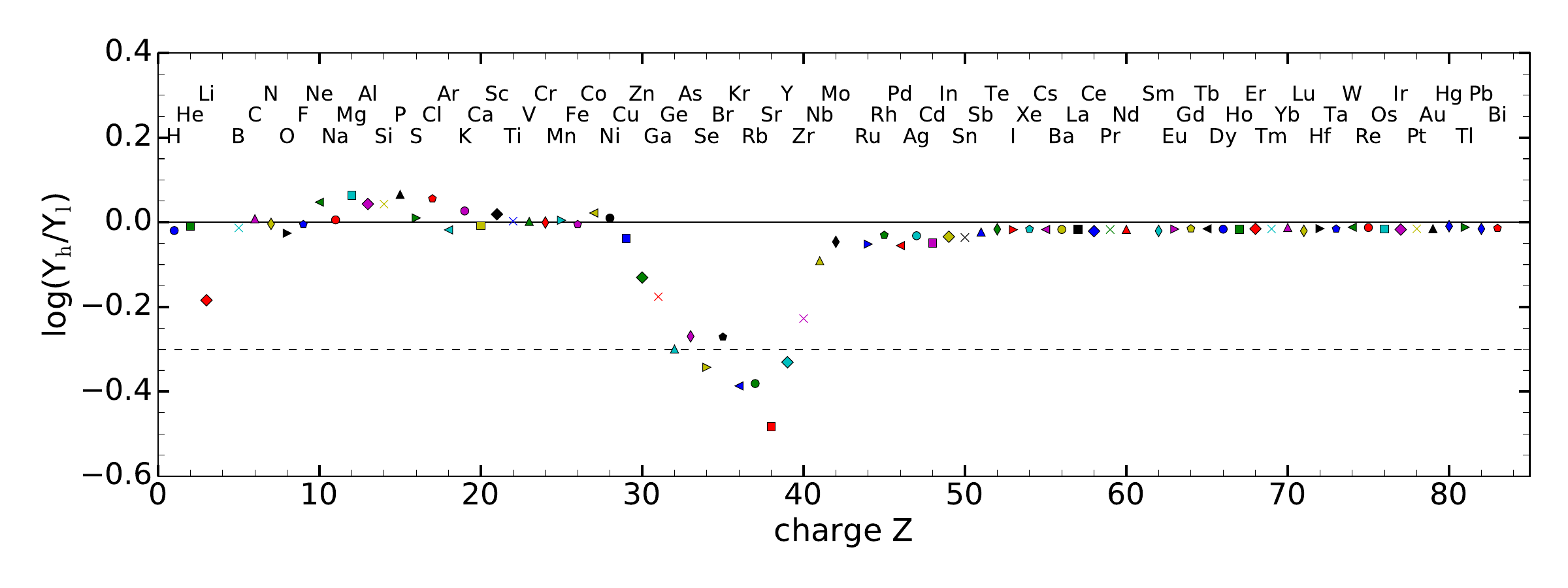}
\caption{Ratios of yields versus charge number based on the $15\msun$, $Z=0.02$ models computed with highly resolved core He-burning ($Y_{h}$)
and with the default resolution ($Y_{l}$). The dashed line indicates a factor
of 0.5.}
\label{fig:co_core_M15Z0p02_yield_exp_comparison}
\end{figure}
\end{landscape}

%%%%%%%%%%%%%%%%%%%%%%%%%%PNG FOR NOW due to size

\section{Summary} \label{sec:concl}

\noindent Stellar models and complete yields from H to Bi for $\mzams/\msun$ = 1, 1.65, 2, 3, 4, 5, 6, 7, 12, 15, 20, 25
at each metallicity $Z=0.006$, $Z=0.001$ and $Z=0.0001$ are calculated. Further stellar evolution tracks
with initial masses of $\mzams/\msun$ = 1, 6, 7, 12, 15, 20, 25 for the metallicities $Z=0.02$ and $Z=0.01$
are added to the models of $\mzams/\msun$ = 1.65, 2, 3,4,5 of P16 to complete the NuGrid yield grid Set 1. Set 1 models have at all five metallicities the same initial masses
and all its stellar evolution and post-processing data is available online at the CADC\footnote{The Canadian Astronomical Data Center, http://www.cadc-ccda.hia-iha.nrc-cnrc.gc.ca/vosui/\#nugrid} and can be analyzed interactively through the web interface WENDI at wendi.nugridstars.org.

We provide explosive yields for each massive star model based on two 1D semi-analytic CCSN prescriptions. %sensitivity of yiedls on explosion properties
%which create the largest mass-metallicity grid of CCSN models with strong shocks currently available.
Predictions of elements and isotopes up to Bi are available for all stellar models which make the largest number of elements available for the considered mass-metallicity space.
These yields are based on the same nuclear reaction rates and are calculated with 
the same stellar evolution code (\MESA) and post-processing code (\mppnp) which provides consistency for chemical evolution simulations.

AGB models include the effect of convective boundary mixing at all boundaries which results in
hot dredge-up in the most massive models. We determine the strong production of N and Li in the massive AGB and S-AGB stars and heavy elements through the application
of a nested network approach which resolves HBB in the post-processing code.
All AGB yields show s-process enhancements based on a  self-consistent
\cdr\ pocket which strongly decreases in massive AGB and S-AGB models. AGB mass loss is reduced towards higher initial masses and lower metallicity
using a mass-and metallicity dependent mass loss prescription.
H ingestion events in the $\mzams=1\msun$, $Z=0.0001$ model lead to heavy-element production up to the third s-process peak.
S-AGB models at the lowest metallicity experience H ingestion events which are a potential i-process site.

Fallback strongly reduces the s-process and $\gamma$-process yields in our most massive stellar models at all metallicities.
In stellar models with $\mzams=12\msun$ and $\mzams=15\msun$ $\alpha$-rich freeze-out layers are ejected which produce most of the 
Ni and significant amounts of proton-rich nuclei up to the Mo mass region. %p nuclei.
The non-monotonic behaviour of the core masses with initial stellar mass together with the mass- and metallicity dependent fallback lead to 
variations of the yields of Fe-peak elements with initial mass and metallicity by orders of magnitude.
We find convective O-C shell merger in the stellar models with $\mzams/\msun$ = 12, 15, 20 at $Z=0.01$ and $\mzams=15\msun$ at $Z=0.02$ which
lead to a boost of odd-Z elements P, Cl, K and Sc and overproduction factors of up to $\approx1\, \mathrm{dex}$. The massive star yields of stellar models with $\mzams=20\msun$ and $\mzams=25\msun$ include additional amounts of N and F owing to H ingestion events.

\section*{Acknowledgements}
We have used the Compute Canada/Westgrid computing resources for most
of the computations and analysis presented in this paper.  We have
also used the computer cluster at Keele University.  We would like to
thank Belaid Moa (WestGrid/UVic) for his dedicated research computing
support.  We greatly appreciate the efforts and time commitment of
Stephenson Yang for local research computing systems support. NuGrid
has been supported over the years by the Joint Institute for Nuclear
Astrophysics through NSF grants PHY 02-16783 and PHY 09-22648 (Joint
Institute for Nuclear Astrophysics, JINA), NSF grant PHY-1430152 (JINA
Center for the Evolution of the Elements) and EU MIRG-CT-2006-046520.
We acknowledge the Canadian Advanced Network for Astronomical Research
cloud service which hosts the WENDI web interface.  RH was supported
by the World Premier International Research Center Initiative (WPI
Initiative), MEXT, Japan.  RH acknowledges support from the "ChETEC"
COST Action (CA16117), supported by COST (European Cooperation in
Science and Technology). MP thanks support from STFC (UK, through the
University of Hull Consolidated Grant ST/R000840/1). FH acknowledges
funding through an NSERC Discovery Grant. We thanks Ondrea Clarkson
for help with redoing some of the figures during the reviewing
process.

%%%%%%%%%%%%%%%%%%%%%%%%%%%%%%%%%%%%%%%%%%%%%%%%%%



\bibliographystyle{mnras}
%\bibliography{astro}

\begin{thebibliography}{}
\makeatletter
\relax
\def\mn@urlcharsother{\let\do\@makeother \do\$\do\&\do\#\do\^\do\_\do\%\do\~}
\def\mn@doi{\begingroup\mn@urlcharsother \@ifnextchar [ {\mn@doi@}
  {\mn@doi@[]}}
\def\mn@doi@[#1]#2{\def\@tempa{#1}\ifx\@tempa\@empty \href
  {http://dx.doi.org/#2} {doi:#2}\else \href {http://dx.doi.org/#2} {#1}\fi
  \endgroup}
\def\mn@eprint#1#2{\mn@eprint@#1:#2::\@nil}
\def\mn@eprint@arXiv#1{\href {http://arxiv.org/abs/#1} {{\tt arXiv:#1}}}
\def\mn@eprint@dblp#1{\href {http://dblp.uni-trier.de/rec/bibtex/#1.xml}
  {dblp:#1}}
\def\mn@eprint@#1:#2:#3:#4\@nil{\def\@tempa {#1}\def\@tempb {#2}\def\@tempc
  {#3}\ifx \@tempc \@empty \let \@tempc \@tempb \let \@tempb \@tempa \fi \ifx
  \@tempb \@empty \def\@tempb {arXiv}\fi \@ifundefined
  {mn@eprint@\@tempb}{\@tempb:\@tempc}{\expandafter \expandafter \csname
  mn@eprint@\@tempb\endcsname \expandafter{\@tempc}}}

\bibitem[\protect\citeauthoryear{{Arnett} \& {Meakin}}{{Arnett} \&
  {Meakin}}{2011}]{arnett:11}
{Arnett} W.~D.,  {Meakin} C.,  2011, \mn@doi [\apj]
  {10.1088/0004-637X/733/2/78}, \href
  {http://adsabs.harvard.edu/abs/2011ApJ...733...78A} {733, 78}

\bibitem[\protect\citeauthoryear{{Arnould} \& {Goriely}}{{Arnould} \&
  {Goriely}}{2003}]{arnould:03}
{Arnould} M.,  {Goriely} S.,  2003, \mn@doi [\physrep]
  {10.1016/S0370-1573(03)00242-4}, \href
  {http://adsabs.harvard.edu/abs/2003PhR...384....1A} {384, 1}

\bibitem[\protect\citeauthoryear{{Banerjee}, {Qian}, {Haxton}  \&
  {Heger}}{{Banerjee} et~al.}{2013}]{banerjee:13}
{Banerjee} P.,  {Qian} Y.-Z.,  {Haxton} W.~C.,   {Heger} A.,  2013, \mn@doi
  [Physical Review Letters] {10.1103/PhysRevLett.110.141101}, \href
  {http://adsabs.harvard.edu/abs/2013PhRvL.110n1101B} {110, 141101}

\bibitem[\protect\citeauthoryear{{Baraffe}, {El Eid}  \& {Prantzos}}{{Baraffe}
  et~al.}{1992}]{baraffe:92}
{Baraffe} I.,  {El Eid} M.~F.,   {Prantzos} N.,  1992, \aap, \href
  {http://adsabs.harvard.edu/abs/1992A%26A...258..357B} {258, 357}

\bibitem[\protect\citeauthoryear{{Battino} et~al.,}{{Battino}
  et~al.}{2016}]{battino:16}
{Battino} U.,  et~al., 2016, \mn@doi [\apj] {10.3847/0004-637X/827/1/30}, \href
  {http://adsabs.harvard.edu/abs/2016ApJ...827...30B} {827, 30}

\bibitem[\protect\citeauthoryear{{Belczynski}, {Wiktorowicz}, {Fryer}, {Holz}
  \& {Kalogera}}{{Belczynski} et~al.}{2012}]{belczynski:12}
{Belczynski} K.,  {Wiktorowicz} G.,  {Fryer} C.~L.,  {Holz} D.~E.,   {Kalogera}
  V.,  2012, \mn@doi [\apj] {10.1088/0004-637X/757/1/91}, \href
  {http://adsabs.harvard.edu/abs/2012ApJ...757...91B} {757, 91}

\bibitem[\protect\citeauthoryear{{Bl{\"o}cker}}{{Bl{\"o}cker}}{1993}]{bloecker:93b}
{Bl{\"o}cker} T.,  1993, PhD thesis, , Universit{\"a}t Kiel, Germany, (1993)

\bibitem[\protect\citeauthoryear{Bl\"ocker \& Sch\"onberner}{Bl\"ocker \&
  Sch\"onberner}{1995}]{bloecker:95}
Bl\"ocker T.,  Sch\"onberner D.,  1995, in 2nd Int. Coll. on Hydrogen-Deficient
  Stars.

\bibitem[\protect\citeauthoryear{{Busso}, {Gallino}  \& {Wasserburg}}{{Busso}
  et~al.}{1999}]{busso:99}
{Busso} M.,  {Gallino} R.,   {Wasserburg} G.~J.,  1999, ARA\&A, 37, 239

\bibitem[\protect\citeauthoryear{{Busso}, {Wasserburg}, {Nollett}  \&
  {Calandra}}{{Busso} et~al.}{2007}]{busso:07}
{Busso} M.,  {Wasserburg} G.~J.,  {Nollett} K.~M.,   {Calandra} A.,  2007,
  \mn@doi [ApJ] {10.1086/522616}, \href
  {http://adsabs.harvard.edu/abs/2007ApJ...671..802B} {671, 802}

\bibitem[\protect\citeauthoryear{{Cameron} \& {Fowler}}{{Cameron} \&
  {Fowler}}{1971}]{cameron:71}
{Cameron} A.~G.~W.,  {Fowler} W.~A.,  1971, ApJ, \href
  {http://adsabs.harvard.edu/cgi-bin/nph-bib_query?bibcode=1971ApJ...164..111C&db_key=AST}
  {164, 111}

\bibitem[\protect\citeauthoryear{{Campbell}, {Lugaro}  \& {Karakas}}{{Campbell}
  et~al.}{2010}]{campbell:10}
{Campbell} S.~W.,  {Lugaro} M.,   {Karakas} A.~I.,  2010, \mn@doi [\aap]
  {10.1051/0004-6361/201015428}, \href
  {http://adsabs.harvard.edu/abs/2010A%26A...522L...6C} {522, L6}

\bibitem[\protect\citeauthoryear{{Chen}, {Herwig}, {Denissenkov}  \&
  {Paxton}}{{Chen} et~al.}{2014}]{chen:14}
{Chen} M.~C.,  {Herwig} F.,  {Denissenkov} P.~A.,   {Paxton} B.,  2014, \mn@doi
  [\mnras] {10.1093/mnras/stu108}, \href
  {http://adsabs.harvard.edu/abs/2014MNRAS.440.1274C} {440, 1274}

\bibitem[\protect\citeauthoryear{{Chieffi} \& {Limongi}}{{Chieffi} \&
  {Limongi}}{2004}]{chieffi:04}
{Chieffi} A.,  {Limongi} M.,  2004, \mn@doi [\apj] {10.1086/392523}, \href
  {http://adsabs.harvard.edu/abs/2004ApJ...608..405C} {608, 405}

\bibitem[\protect\citeauthoryear{{C{\^o}t{\'e}}, {Martel}  \&
  {Drissen}}{{C{\^o}t{\'e}} et~al.}{2013}]{cote:13}
{C{\^o}t{\'e}} B.,  {Martel} H.,   {Drissen} L.,  2013, \mn@doi [\apj]
  {10.1088/0004-637X/777/2/107}, \href
  {http://adsabs.harvard.edu/abs/2013ApJ...777..107C} {777, 107}

\bibitem[\protect\citeauthoryear{{C{\^o}t{\'e}}, {West}, {Heger}, {Ritter},
  {O'Shea}, {Herwig}, {Travaglio}  \& {Bisterzo}}{{C{\^o}t{\'e}}
  et~al.}{2016a}]{cote:16}
{C{\^o}t{\'e}} B.,  {West} C.,  {Heger} A.,  {Ritter} C.,  {O'Shea} B.~W.,
  {Herwig} F.,  {Travaglio} C.,   {Bisterzo} S.,  2016a, \mn@doi [\mnras]
  {10.1093/mnras/stw2244}, \href
  {http://ukads.nottingham.ac.uk/abs/2016MNRAS.463.3755C} {463, 3755}

\bibitem[\protect\citeauthoryear{C{\^o}t{\'e}, Ritter, O'Shea, Herwig,
  Pignatari, Jones  \& Fryer}{C{\^o}t{\'e} et~al.}{2016b}]{Cote:2016iia}
C{\^o}t{\'e} B.,  Ritter C.,  O'Shea B.~W.,  Herwig F.,  Pignatari M.,  Jones
  S.,   Fryer C.~L.,  2016b, ApJ, 824, 82

\bibitem[\protect\citeauthoryear{{C{\^o}t{\'e}}, {O'Shea}, {Ritter}, {Herwig}
  \& {Venn}}{{C{\^o}t{\'e}} et~al.}{2017}]{cote:17}
{C{\^o}t{\'e}} B.,  {O'Shea} B.~W.,  {Ritter} C.,  {Herwig} F.,   {Venn} K.~A.,
   2017, \mn@doi [\apj] {10.3847/1538-4357/835/2/128}, \href
  {http://adsabs.harvard.edu/abs/2017ApJ...835..128C} {835, 128}

\bibitem[\protect\citeauthoryear{{Cristallo}, {Straniero}, {Gallino}, {Herwig},
  {Chieffi}, {Limongi}  \& {Busso}}{{Cristallo} et~al.}{2001}]{cristallo:01}
{Cristallo} S.,  {Straniero} O.,  {Gallino} R.,  {Herwig} F.,  {Chieffi} A.,
  {Limongi} M.,   {Busso} M.,  2001, \mn@doi [Nuclear Physics A]
  {10.1016/S0375-9474(01)00701-1}, \href
  {http://adsabs.harvard.edu/abs/2001NuPhA.688..217C} {688, 217}

\bibitem[\protect\citeauthoryear{{Cristallo}, {Piersanti}, {Straniero},
  {Gallino}, {Dom{\'{\i}}nguez}  \& {K{\"a}ppeler}}{{Cristallo}
  et~al.}{2009}]{cristallo:09b}
{Cristallo} S.,  {Piersanti} L.,  {Straniero} O.,  {Gallino} R.,
  {Dom{\'{\i}}nguez} I.,   {K{\"a}ppeler} F.,  2009, \mn@doi [\pasa]
  {10.1071/AS09003}, \href {http://adsabs.harvard.edu/abs/2009PASA...26..139C}
  {26, 139}

\bibitem[\protect\citeauthoryear{{Cristallo} et~al.,}{{Cristallo}
  et~al.}{2011}]{cristallo:11}
{Cristallo} S.,  et~al., 2011, \mn@doi [\apjs] {10.1088/0067-0049/197/2/17},
  \href {http://adsabs.harvard.edu/abs/2011ApJS..197...17C} {197, 17}

\bibitem[\protect\citeauthoryear{{Cristallo}, {Straniero}, {Piersanti}  \&
  {Gobrecht}}{{Cristallo} et~al.}{2015}]{cristallo:15}
{Cristallo} S.,  {Straniero} O.,  {Piersanti} L.,   {Gobrecht} D.,  2015,
  preprint, \href {http://adsabs.harvard.edu/abs/2015arXiv150707338C} {}
  (\mn@eprint {arXiv} {1507.07338})

\bibitem[\protect\citeauthoryear{{Cristini}, {Meakin}, {Hirschi}, {Arnett},
  {Georgy}  \& {Viallet}}{{Cristini} et~al.}{2016}]{cristini:16}
{Cristini} A.,  {Meakin} C.,  {Hirschi} R.,  {Arnett} D.,  {Georgy} C.,
  {Viallet} M.,  2016, preprint, \href
  {http://ukads.nottingham.ac.uk/abs/2016arXiv161005173C} {} (\mn@eprint
  {arXiv} {1610.05173})

\bibitem[\protect\citeauthoryear{{Denissenkov} \& {Tout}}{{Denissenkov} \&
  {Tout}}{2003}]{denissenkov:03t}
{Denissenkov} P.~A.,  {Tout} C.~A.,  2003, \mn@doi [\mnras]
  {10.1046/j.1365-8711.2003.06284.x}, \href
  {http://adsabs.harvard.edu/abs/2003MNRAS.340..722D} {340, 722}

\bibitem[\protect\citeauthoryear{{Denissenkov}, {Pinsonneault}  \&
  {MacGregor}}{{Denissenkov} et~al.}{2009}]{denissenkov:09}
{Denissenkov} P.~A.,  {Pinsonneault} M.,   {MacGregor} K.~B.,  2009, \mn@doi
  [\apj] {10.1088/0004-637X/696/2/1823}, \href
  {http://adsabs.harvard.edu/abs/2009ApJ...696.1823D} {696, 1823}

\bibitem[\protect\citeauthoryear{{Denissenkov}, {Herwig}, {Truran}  \&
  {Paxton}}{{Denissenkov} et~al.}{2013}]{denissenkov:13}
{Denissenkov} P.~A.,  {Herwig} F.,  {Truran} J.~W.,   {Paxton} B.,  2013,
  \mn@doi [\apj] {10.1088/0004-637X/772/1/37}, \href
  {http://ukads.nottingham.ac.uk/abs/2013ApJ...772...37D} {772, 37}

\bibitem[\protect\citeauthoryear{{Denissenkov}, {Herwig}, {Battino}, {Ritter},
  {Pignatari}, {Jones}  \& {Paxton}}{{Denissenkov}
  et~al.}{2017}]{denissenkov:17}
{Denissenkov} P.~A.,  {Herwig} F.,  {Battino} U.,  {Ritter} C.,  {Pignatari}
  M.,  {Jones} S.,   {Paxton} B.,  2017, \mn@doi [\apjl]
  {10.3847/2041-8213/834/2/L10}, \href
  {http://ukads.nottingham.ac.uk/abs/2017ApJ...834L..10D} {834, L10}

\bibitem[\protect\citeauthoryear{{Doherty}, {Siess}, {Lattanzio}  \&
  {Gil-Pons}}{{Doherty} et~al.}{2010}]{doherty:10}
{Doherty} C.~L.,  {Siess} L.,  {Lattanzio} J.~C.,   {Gil-Pons} P.,  2010,
  \mn@doi [\mnras] {10.1111/j.1365-2966.2009.15772.x}, \href
  {http://adsabs.harvard.edu/abs/2010MNRAS.401.1453D} {401, 1453}

\bibitem[\protect\citeauthoryear{{Doherty}, {Gil-Pons}, {Lau}, {Lattanzio},
  {Siess}  \& {Campbell}}{{Doherty} et~al.}{2014}]{doherty:14b}
{Doherty} C.~L.,  {Gil-Pons} P.,  {Lau} H.~H.~B.,  {Lattanzio} J.~C.,  {Siess}
  L.,   {Campbell} S.~W.,  2014, \mn@doi [\mnras] {10.1093/mnras/stu571}, \href
  {http://adsabs.harvard.edu/abs/2014MNRAS.441..582D} {441, 582}

\bibitem[\protect\citeauthoryear{{Eggenberger}, {Meynet}, {Maeder}, {Hirschi},
  {Charbonnel}, {Talon}  \& {Ekstr{\"o}m}}{{Eggenberger}
  et~al.}{2008}]{eggenberger:08}
{Eggenberger} P.,  {Meynet} G.,  {Maeder} A.,  {Hirschi} R.,  {Charbonnel} C.,
  {Talon} S.,   {Ekstr{\"o}m} S.,  2008, \mn@doi [\apss]
  {10.1007/s10509-007-9511-y}, \href
  {http://adsabs.harvard.edu/abs/2008Ap%26SS.316...43E} {316, 43}

\bibitem[\protect\citeauthoryear{{Ekstr{\"o}m}, {Meynet}, {Chiappini},
  {Hirschi}  \& {Maeder}}{{Ekstr{\"o}m} et~al.}{2008}]{ekstroem:08}
{Ekstr{\"o}m} S.,  {Meynet} G.,  {Chiappini} C.,  {Hirschi} R.,   {Maeder} A.,
  2008, \mn@doi [A\&A] {10.1051/0004-6361:200809633}, \href
  {http://adsabs.harvard.edu/abs/2008A%26A...489..685E} {489, 685}

\bibitem[\protect\citeauthoryear{{El Eid}, {Meyer}  \& {The}}{{El Eid}
  et~al.}{2004}]{eleid:04}
{El Eid} M.~F.,  {Meyer} B.~S.,   {The} L.-S.,  2004, \mn@doi [\apj]
  {10.1086/422162}, \href
  {http://ukads.nottingham.ac.uk/abs/2004ApJ...611..452E} {611, 452}

\bibitem[\protect\citeauthoryear{{Farmer}, {Fields}  \& {Timmes}}{{Farmer}
  et~al.}{2015}]{farmer:15}
{Farmer} R.,  {Fields} C.~E.,   {Timmes} F.~X.,  2015, \mn@doi [\apj]
  {10.1088/0004-637X/807/2/184}, \href
  {http://adsabs.harvard.edu/abs/2015ApJ...807..184F} {807, 184}

\bibitem[\protect\citeauthoryear{{Few}, {Courty}, {Gibson}, {Kawata}, {Calura}
  \& {Teyssier}}{{Few} et~al.}{2012}]{few:12}
{Few} C.~G.,  {Courty} S.,  {Gibson} B.~K.,  {Kawata} D.,  {Calura} F.,
  {Teyssier} R.,  2012, \mn@doi [\mnras] {10.1111/j.1745-3933.2012.01275.x},
  \href {http://adsabs.harvard.edu/abs/2012MNRAS.424L..11F} {424, L11}

\bibitem[\protect\citeauthoryear{{Fishlock}, {Karakas}, {Lugaro}  \&
  {Yong}}{{Fishlock} et~al.}{2014}]{fishlock:14}
{Fishlock} C.~K.,  {Karakas} A.~I.,  {Lugaro} M.,   {Yong} D.,  2014, \mn@doi
  [\apj] {10.1088/0004-637X/797/1/44}, \href
  {http://ukads.nottingham.ac.uk/abs/2014ApJ...797...44F} {797, 44}

\bibitem[\protect\citeauthoryear{Freytag, Ludwig  \& Steffen}{Freytag
  et~al.}{1996}]{freytag:96}
Freytag B.,  Ludwig H.-G.,   Steffen M.,  1996, A\&A, 313, 497

\bibitem[\protect\citeauthoryear{{Frischknecht} et~al.,}{{Frischknecht}
  et~al.}{2016}]{frischknecht:16}
{Frischknecht} U.,  et~al., 2016, \mn@doi [\mnras] {10.1093/mnras/stv2723},
  \href {http://adsabs.harvard.edu/abs/2016MNRAS.456.1803F} {456, 1803}

\bibitem[\protect\citeauthoryear{{Frost}, {Cannon}, {Lattanzio}, {Wood}  \&
  {Forestini}}{{Frost} et~al.}{1998}]{frost:98c}
{Frost} C.~A.,  {Cannon} R.~C.,  {Lattanzio} J.~C.,  {Wood} P.~R.,
  {Forestini} M.,  1998, \aap, \href
  {http://adsabs.harvard.edu/abs/1998A%26A...332L..17F} {332, L17}

\bibitem[\protect\citeauthoryear{{Fryer}}{{Fryer}}{1999}]{fryer:99}
{Fryer} C.~L.,  1999, \mn@doi [\apj] {10.1086/307647}, \href
  {http://adsabs.harvard.edu/abs/1999ApJ...522..413F} {522, 413}

\bibitem[\protect\citeauthoryear{{Fryer}, {Belczynski}, {Wiktorowicz},
  {Dominik}, {Kalogera}  \& {Holz}}{{Fryer} et~al.}{2012}]{fryer:12}
{Fryer} C.~L.,  {Belczynski} K.,  {Wiktorowicz} G.,  {Dominik} M.,  {Kalogera}
  V.,   {Holz} D.~E.,  2012, \mn@doi [\apj] {10.1088/0004-637X/749/1/91}, \href
  {http://adsabs.harvard.edu/abs/2012ApJ...749...91F} {749, 91}

\bibitem[\protect\citeauthoryear{{Fujimoto}, {Ikeda}  \& {Iben}}{{Fujimoto}
  et~al.}{2000}]{fujimoto:00}
{Fujimoto} M.~Y.,  {Ikeda} Y.,   {Iben} I. J.,  2000, ApJ Lett., 529, L25

\bibitem[\protect\citeauthoryear{{Gallino}, {Arlandini}, {Busso}, {Lugaro},
  {Travaglio}, {Straniero}, {Chieffi}  \& {Limongi}}{{Gallino}
  et~al.}{1998}]{gallino:98}
{Gallino} R.,  {Arlandini} C.,  {Busso} M.,  {Lugaro} M.,  {Travaglio} C.,
  {Straniero} O.,  {Chieffi} A.,   {Limongi} M.,  1998, \mn@doi [\apj]
  {10.1086/305437}, \href {http://adsabs.harvard.edu/abs/1998ApJ...497..388G}
  {497, 388}

\bibitem[\protect\citeauthoryear{{Garc{\'{\i}}a-Berro}, {Ritossa}  \&
  {Iben}}{{Garc{\'{\i}}a-Berro} et~al.}{1997}]{garcia-berro:97}
{Garc{\'{\i}}a-Berro} E.,  {Ritossa} C.,   {Iben} Jr. I.,  1997, \apj, \href
  {http://ukads.nottingham.ac.uk/abs/1997ApJ...485..765G} {485, 765}

\bibitem[\protect\citeauthoryear{{Garc{\'{\i}}a-Hern{\'a}ndez}, {Zamora},
  {Yag{\"u}e}, {Uttenthaler}, {Karakas}, {Lugaro}, {Ventura}  \&
  {Lambert}}{{Garc{\'{\i}}a-Hern{\'a}ndez} et~al.}{2013}]{garcia-hernandez:13}
{Garc{\'{\i}}a-Hern{\'a}ndez} D.~A.,  {Zamora} O.,  {Yag{\"u}e} A.,
  {Uttenthaler} S.,  {Karakas} A.~I.,  {Lugaro} M.,  {Ventura} P.,   {Lambert}
  D.~L.,  2013, \mn@doi [\aap] {10.1051/0004-6361/201321818}, \href
  {http://adsabs.harvard.edu/abs/2013A%26A...555L...3G} {555, L3}

\bibitem[\protect\citeauthoryear{{Gibson}}{{Gibson}}{2002}]{gibson:02}
{Gibson} B.~K.,  2002, in {Nomoto} K.,  {Truran} J.~W.,  eds,  IAU Symposium
  Vol. 187, Cosmic Chemical Evolution. pp 159--163

\bibitem[\protect\citeauthoryear{{Gil-Pons} \& {Doherty}}{{Gil-Pons} \&
  {Doherty}}{2010}]{gil-pons:10}
{Gil-Pons} P.,  {Doherty} C.~L.,  2010, \memsai, \href
  {http://adsabs.harvard.edu/abs/2010MmSAI..81..974G} {81, 974}

\bibitem[\protect\citeauthoryear{{Gil-Pons}, {Doherty}, {Lau}, {Campbell},
  {Suda}, {Guilani}, {Guti{\'e}rrez}  \& {Lattanzio}}{{Gil-Pons}
  et~al.}{2013}]{gil-pons:13}
{Gil-Pons} P.,  {Doherty} C.~L.,  {Lau} H.,  {Campbell} S.~W.,  {Suda} T.,
  {Guilani} S.,  {Guti{\'e}rrez} J.,   {Lattanzio} J.~C.,  2013, \mn@doi [\aap]
  {10.1051/0004-6361/201321127}, \href
  {http://adsabs.harvard.edu/abs/2013A%26A...557A.106G} {557, A106}

\bibitem[\protect\citeauthoryear{{Glebbeek}, {Gaburov}, {de Mink}, {Pols}  \&
  {Portegies Zwart}}{{Glebbeek} et~al.}{2009}]{glebbeek:09}
{Glebbeek} E.,  {Gaburov} E.,  {de Mink} S.~E.,  {Pols} O.~R.,   {Portegies
  Zwart} S.~F.,  2009, \mn@doi [\aap] {10.1051/0004-6361/200810425}, \href
  {http://adsabs.harvard.edu/abs/2009A%26A...497..255G} {497, 255}

\bibitem[\protect\citeauthoryear{{Goriely} \& {Siess}}{{Goriely} \&
  {Siess}}{2004}]{goriely:04}
{Goriely} S.,  {Siess} L.,  2004, A\&A, \href
  {http://adsabs.harvard.edu/cgi-bin/nph-bib_query?bibcode=2004A%26A...421L..25G&db_key=AST}
  {421, L25}

\bibitem[\protect\citeauthoryear{{Grevesse} \& {Noels}}{{Grevesse} \&
  {Noels}}{1993}]{grevesse:93}
{Grevesse} N.,  {Noels} A.,  1993, in {N.~Prantzos, E.~Vangioni-Flam, \&
  M.~Casse} ed., Origin and Evolution of the Elements. pp 15--25

\bibitem[\protect\citeauthoryear{{Gutierrez}, {Garcia-Berro}, {Iben}, {Isern},
  {Labay}  \& {Canal}}{{Gutierrez} et~al.}{1996}]{1996ApJ...459..701G}
{Gutierrez} J.,  {Garcia-Berro} E.,  {Iben} Jr. I.,  {Isern} J.,  {Labay} J.,
  {Canal} R.,  1996, \mn@doi [\apj] {10.1086/176934}, \href
  {http://adsabs.harvard.edu/abs/1996ApJ...459..701G} {459, 701}

\bibitem[\protect\citeauthoryear{{Heger} \& {Woosley}}{{Heger} \&
  {Woosley}}{2010}]{heger:10}
{Heger} A.,  {Woosley} S.~E.,  2010, \mn@doi [\apj]
  {10.1088/0004-637X/724/1/341}, \href
  {http://adsabs.harvard.edu/abs/2010ApJ...724..341H} {724, 341}

\bibitem[\protect\citeauthoryear{Herant, Benz, Hix, Fryer  \& Colgate}{Herant
  et~al.}{1994}]{herant:94}
Herant M.,  Benz W.,  Hix W.~R.,  Fryer C.~L.,   Colgate S.~A.,  1994, ApJS

\bibitem[\protect\citeauthoryear{{Herwig}}{{Herwig}}{2000}]{herwig:00}
{Herwig} F.,  2000, \aap, \href
  {http://adsabs.harvard.edu/abs/2000A%26A...360..952H} {360, 952}

\bibitem[\protect\citeauthoryear{{Herwig}}{{Herwig}}{2001a}]{herwig:00a}
{Herwig} F.,  2001a, APSS, \href
  {http://adsabs.harvard.edu/cgi-bin/nph-bib_query?bibcode=2001Ap%26SS.275...15H&db_key=AST}
  {275, 15}

\bibitem[\protect\citeauthoryear{Herwig}{Herwig}{2001b}]{herwig:01a}
Herwig F.,  2001b, ApJ Lett., 554, L71

\bibitem[\protect\citeauthoryear{{Herwig}}{{Herwig}}{2004a}]{herwig:04a}
{Herwig} F.,  2004a, ApJS, \href
  {http://cdsads.u-strasbg.fr/cgi-bin/nph-bib_query?bibcode=2004ApJS..155..651H&db_key=AST}
  {155, 651}

\bibitem[\protect\citeauthoryear{Herwig}{Herwig}{2004b}]{herwig:03c}
Herwig F.,  2004b, ApJ, 605, 425

\bibitem[\protect\citeauthoryear{{Herwig}}{{Herwig}}{2005}]{herwig:05}
{Herwig} F.,  2005, \mn@doi [\araa] {10.1146/annurev.astro.43.072103.150600},
  \href {http://adsabs.harvard.edu/abs/2005ARA%26A..43..435H} {43, 435}

\bibitem[\protect\citeauthoryear{{Herwig} \& {Langer}}{{Herwig} \&
  {Langer}}{2001}]{herwig:00f}
{Herwig} F.,  {Langer} N.,  2001, Nuclear Physics A, \href
  {http://adsabs.harvard.edu/cgi-bin/nph-bib_query?bibcode=2001NuPhA.688..221H&db_key=PHY}
  {688, 221}

\bibitem[\protect\citeauthoryear{Herwig, Bl\"ocker, Sch\"onberner  \& {El
  Eid}}{Herwig et~al.}{1997}]{herwig:97}
Herwig F.,  Bl\"ocker T.,  Sch\"onberner D.,   {El Eid} M.~F.,  1997, A\&A,
  324, L81

\bibitem[\protect\citeauthoryear{Herwig, Sch\"onberner  \& Bl\"ocker}{Herwig
  et~al.}{1998}]{herwig:98b}
Herwig F.,  Sch\"onberner D.,   Bl\"ocker T.,  1998, A\&A, 340, L43

\bibitem[\protect\citeauthoryear{Herwig, Bl\"ocker, Langer  \& Driebe}{Herwig
  et~al.}{1999}]{herwig:99c}
Herwig F.,  Bl\"ocker T.,  Langer N.,   Driebe T.,  1999, A\&A, 349, L5

\bibitem[\protect\citeauthoryear{{Herwig}, {Langer}  \& {Lugaro}}{{Herwig}
  et~al.}{2003}]{herwig:03}
{Herwig} F.,  {Langer} N.,   {Lugaro} M.,  2003, \mn@doi [\apj]
  {10.1086/376726}, \href {http://adsabs.harvard.edu/abs/2003ApJ...593.1056H}
  {593, 1056}

\bibitem[\protect\citeauthoryear{{Herwig}, {Freytag}, {Fuchs}, {Hansen},
  {Hueckstaedt}, {Porter}, {Timmes}  \& {Woodward}}{{Herwig}
  et~al.}{2007}]{herwig:07a}
{Herwig} F.,  {Freytag} B.,  {Fuchs} T.,  {Hansen} J.~P.,  {Hueckstaedt} R.~M.,
   {Porter} D.~H.,  {Timmes} F.~X.,   {Woodward} P.~R.,  2007, in {Kerschbaum}
  F.,  {Charbonnel} C.,   {Wing} R.~F.,  eds,  Astronomical Society of the
  Pacific Conference Series Vol. 378, Why Galaxies Care About AGB Stars: Their
  Importance as Actors and Probes. p.~43 (\mn@eprint {} {arXiv:0709.0197})

\bibitem[\protect\citeauthoryear{{Herwig}, {Pignatari}, {Woodward}, {Porter},
  {Rockefeller}, {Fryer}, {Bennett}  \& {Hirschi}}{{Herwig}
  et~al.}{2011}]{herwig:11}
{Herwig} F.,  {Pignatari} M.,  {Woodward} P.~R.,  {Porter} D.~H.,
  {Rockefeller} G.,  {Fryer} C.~L.,  {Bennett} M.,   {Hirschi} R.,  2011,
  \mn@doi [\apj] {10.1088/0004-637X/727/2/89}, \href
  {http://adsabs.harvard.edu/abs/2011ApJ...727...89H} {727, 89}

\bibitem[\protect\citeauthoryear{{Herwig}, {Woodward}, {Lin}, {Knox}  \&
  {Fryer}}{{Herwig} et~al.}{2014}]{herwig:14}
{Herwig} F.,  {Woodward} P.~R.,  {Lin} P.-H.,  {Knox} M.,   {Fryer} C.,  2014,
  \mn@doi [\apjl] {10.1088/2041-8205/792/1/L3}, \href
  {http://adsabs.harvard.edu/abs/2014ApJ...792L...3H} {792, L3}

\bibitem[\protect\citeauthoryear{Herwig et~al.,}{Herwig
  et~al.}{2018}]{Herwig:2018gx}
Herwig F.,  et~al., 2018, ASTROPHYS J SUPPL S, 236, 2

\bibitem[\protect\citeauthoryear{{Hirschi}}{{Hirschi}}{2007}]{hirschi:07}
{Hirschi} R.,  2007, \mn@doi [A\&A] {10.1051/0004-6361:20065356}, \href
  {http://adsabs.harvard.edu/abs/2007A%26A...461..571H} {461, 571}

\bibitem[\protect\citeauthoryear{{Hirschi}, {Meynet}  \& {Maeder}}{{Hirschi}
  et~al.}{2004}]{hirschi:04}
{Hirschi} R.,  {Meynet} G.,   {Maeder} A.,  2004, \mn@doi [\aap]
  {10.1051/0004-6361:20041095}, \href
  {http://adsabs.harvard.edu/abs/2004A%26A...425..649H} {425, 649}

\bibitem[\protect\citeauthoryear{{Hirschi}, {Meynet}  \& {Maeder}}{{Hirschi}
  et~al.}{2005}]{hirschi:05b}
{Hirschi} R.,  {Meynet} G.,   {Maeder} A.,  2005, \mn@doi [\aap]
  {10.1051/0004-6361:20041554}, \href
  {http://adsabs.harvard.edu/abs/2005A%26A...433.1013H} {433, 1013}

\bibitem[\protect\citeauthoryear{Iben \& MacDonald}{Iben \&
  MacDonald}{1995}]{iben:95b}
Iben Jr. I.,  MacDonald J.,  1995, in Koester D.,  Werner K.,  eds, White
  Dwarfs. No.~443 in LNP.
Springer, Heidelberg, p.~48

\bibitem[\protect\citeauthoryear{Iben \& Renzini}{Iben \&
  Renzini}{1982}]{iben:82a}
Iben Jr. I.,  Renzini A.,  1982, ApJ, 259, L79

\bibitem[\protect\citeauthoryear{Iben, Kaler, Truran  \& Renzini}{Iben
  et~al.}{1983}]{iben:83a}
Iben Jr. I.,  Kaler J.~B.,  Truran J.~W.,   Renzini A.,  1983, ApJ, 264, 605

\bibitem[\protect\citeauthoryear{{Iwamoto}, {Kajino}, {Mathews}, {Fujimoto}  \&
  {Aoki}}{{Iwamoto} et~al.}{2004}]{iwamoto:04}
{Iwamoto} N.,  {Kajino} T.,  {Mathews} G.~J.,  {Fujimoto} M.~Y.,   {Aoki} W.,
  2004, ApJ, \href
  {http://adsabs.harvard.edu/cgi-bin/nph-bib_query?bibcode=2004ApJ...602..378I&amp;db_key=AST}
  {602, 378}

\bibitem[\protect\citeauthoryear{{Janka}, {Langanke}, {Marek},
  {Mart{\'{\i}}nez-Pinedo}  \& {M{\"u}ller}}{{Janka} et~al.}{2007}]{janka:07}
{Janka} H.-T.,  {Langanke} K.,  {Marek} A.,  {Mart{\'{\i}}nez-Pinedo} G.,
  {M{\"u}ller} B.,  2007, \mn@doi [\physrep] {10.1016/j.physrep.2007.02.002},
  \href {http://ukads.nottingham.ac.uk/abs/2007PhR...442...38J} {442, 38}

\bibitem[\protect\citeauthoryear{{Jones} et~al.,}{{Jones}
  et~al.}{2013}]{2013ApJ...772..150J}
{Jones} S.,  et~al., 2013, \mn@doi [\apj] {10.1088/0004-637X/772/2/150}, \href
  {http://adsabs.harvard.edu/abs/2013ApJ...772..150J} {772, 150}

\bibitem[\protect\citeauthoryear{{Jones}, {Hirschi}  \& {Nomoto}}{{Jones}
  et~al.}{2014}]{2014ApJ...797...83J}
{Jones} S.,  {Hirschi} R.,   {Nomoto} K.,  2014, \mn@doi [\apj]
  {10.1088/0004-637X/797/2/83}, \href
  {http://adsabs.harvard.edu/abs/2014ApJ...797...83J} {797, 83}

\bibitem[\protect\citeauthoryear{{Jones}, {Hirschi}, {Pignatari}, {Heger},
  {Georgy}, {Nishimura}, {Fryer}  \& {Herwig}}{{Jones} et~al.}{2015}]{jones:15}
{Jones} S.,  {Hirschi} R.,  {Pignatari} M.,  {Heger} A.,  {Georgy} C.,
  {Nishimura} N.,  {Fryer} C.,   {Herwig} F.,  2015, \mn@doi [\mnras]
  {10.1093/mnras/stu2657}, \href
  {http://ukads.nottingham.ac.uk/abs/2015MNRAS.447.3115J} {447, 3115}

\bibitem[\protect\citeauthoryear{{Jones}, {Ritter}, {Herwig}, {Fryer},
  {Pignatari}, {Bertolli}  \& {Paxton}}{{Jones} et~al.}{2016}]{jones:16}
{Jones} S.,  {Ritter} C.,  {Herwig} F.,  {Fryer} C.,  {Pignatari} M.,
  {Bertolli} M.~G.,   {Paxton} B.,  2016, \mn@doi [\mnras]
  {10.1093/mnras/stv2488}, \href
  {http://ukads.nottingham.ac.uk/abs/2016MNRAS.455.3848J} {455, 3848}

\bibitem[\protect\citeauthoryear{{Jones}, {Andrassy}, {Sandalski}, {Davis},
  {Woodward}  \& {Herwig}}{{Jones} et~al.}{2017}]{jones:17}
{Jones} S.,  {Andrassy} R.,  {Sandalski} S.,  {Davis} A.,  {Woodward} P.,
  {Herwig} F.,  2017, \mn@doi [\mnras] {10.1093/mnras/stw2783}, \href
  {http://ukads.nottingham.ac.uk/abs/2017MNRAS.465.2991J} {465, 2991}

\bibitem[\protect\citeauthoryear{{K{\"a}ppeler}, {Beer}  \&
  {Wisshak}}{{K{\"a}ppeler} et~al.}{1989}]{kaeppeler:89}
{K{\"a}ppeler} F.,  {Beer} H.,   {Wisshak} K.,  1989, Rep. Prog. Phys., \href
  {http://adsabs.harvard.edu/abs/1989RPPh...52..945K} {52, 945}

\bibitem[\protect\citeauthoryear{{K{\"a}ppeler}, {Gallino}, {Bisterzo}  \&
  {Aoki}}{{K{\"a}ppeler} et~al.}{2011}]{kaeppeler:11}
{K{\"a}ppeler} F.,  {Gallino} R.,  {Bisterzo} S.,   {Aoki} W.,  2011, \mn@doi
  [Reviews of Modern Physics] {10.1103/RevModPhys.83.157}, \href
  {http://adsabs.harvard.edu/abs/2011RvMP...83..157K} {83, 157}

\bibitem[\protect\citeauthoryear{{Karakas}}{{Karakas}}{2003}]{karakas:03b}
{Karakas} A.,  2003, Ph.D.~Thesis, \href
  {http://www.mso.anu.edu.au/~akarakas/download/thesis.pdf} {}

\bibitem[\protect\citeauthoryear{{Karakas}}{{Karakas}}{2010}]{karakas:10}
{Karakas} A.~I.,  2010, VizieR Online Data Catalog, \href
  {http://adsabs.harvard.edu/abs/2010yCat..74031413K} {740, 31413}

\bibitem[\protect\citeauthoryear{{Karakas} \& {Lugaro}}{{Karakas} \&
  {Lugaro}}{2016}]{karakas:16}
{Karakas} A.~I.,  {Lugaro} M.,  2016, \mn@doi [\apj]
  {10.3847/0004-637X/825/1/26}, \href
  {http://ukads.nottingham.ac.uk/abs/2016ApJ...825...26K} {825, 26}

\bibitem[\protect\citeauthoryear{{Karakas}, {Garc{\'{\i}}a-Hern{\'a}ndez}  \&
  {Lugaro}}{{Karakas} et~al.}{2012}]{karakas:12}
{Karakas} A.~I.,  {Garc{\'{\i}}a-Hern{\'a}ndez} D.~A.,   {Lugaro} M.,  2012,
  \mn@doi [\apj] {10.1088/0004-637X/751/1/8}, \href
  {http://adsabs.harvard.edu/abs/2012ApJ...751....8K} {751, 8}

\bibitem[\protect\citeauthoryear{{Kobayashi}, {Umeda}, {Nomoto}, {Tominaga}  \&
  {Ohkubo}}{{Kobayashi} et~al.}{2006}]{kobayashi:06}
{Kobayashi} C.,  {Umeda} H.,  {Nomoto} K.,  {Tominaga} N.,   {Ohkubo} T.,
  2006, \mn@doi [\apj] {10.1086/508914}, \href
  {http://adsabs.harvard.edu/abs/2006ApJ...653.1145K} {653, 1145}

\bibitem[\protect\citeauthoryear{Lattanzio \& Boothroyd}{Lattanzio \&
  Boothroyd}{1997}]{lattanzio:97b}
Lattanzio J.~C.,  Boothroyd A.~I.,  1997, in Bernatowitz T.,  Zinner E.,  eds,
  Astrophysical Implications of the Laboratory Study of Presolar Materials. AIP
  Conf.\ Ser., p.~85

\bibitem[\protect\citeauthoryear{{Lattanzio}, {Frost}, {Cannon}  \&
  {Wood}}{{Lattanzio} et~al.}{1996}]{lattanzio:96}
{Lattanzio} J.,  {Frost} C.,  {Cannon} R.,   {Wood} P.~R.,  1996, \memsai,
  \href {http://adsabs.harvard.edu/abs/1996MmSAI..67..729L} {67, 729}

\bibitem[\protect\citeauthoryear{{Lattanzio}, {Frost}, {Cannon}  \&
  {Wood}}{{Lattanzio} et~al.}{1997}]{lattanzio:97a}
{Lattanzio} J.~C.,  {Frost} C.~A.,  {Cannon} R.~C.,   {Wood} P.~R.,  1997,
  \mn@doi [Nuclear Physics A] {10.1016/S0375-9474(97)00286-8}, \href
  {http://adsabs.harvard.edu/abs/1997NuPhA.621..435L} {621, 435}

\bibitem[\protect\citeauthoryear{{Limongi} \& {Chieffi}}{{Limongi} \&
  {Chieffi}}{2006}]{limongi:06}
{Limongi} M.,  {Chieffi} A.,  2006, \mn@doi [\apj] {10.1086/505164}, \href
  {http://adsabs.harvard.edu/abs/2006ApJ...647..483L} {647, 483}

\bibitem[\protect\citeauthoryear{{Lodders}}{{Lodders}}{2003}]{lodders:03}
{Lodders} K.,  2003, \apj

\bibitem[\protect\citeauthoryear{{Lugaro}, {Herwig}, {Lattanzio}, {Gallino}  \&
  {Straniero}}{{Lugaro} et~al.}{2003}]{lugaro:03a}
{Lugaro} M.,  {Herwig} F.,  {Lattanzio} J.~C.,  {Gallino} R.,   {Straniero} O.,
   2003, \mn@doi [\apj] {10.1086/367887}, \href
  {http://adsabs.harvard.edu/abs/2003ApJ...586.1305L} {586, 1305}

\bibitem[\protect\citeauthoryear{{Lugaro}, {Karakas}, {Stancliffe}  \&
  {Rijs}}{{Lugaro} et~al.}{2012}]{lugaro:12}
{Lugaro} M.,  {Karakas} A.~I.,  {Stancliffe} R.~J.,   {Rijs} C.,  2012, \mn@doi
  [\apj] {10.1088/0004-637X/747/1/2}, \href
  {http://adsabs.harvard.edu/abs/2012ApJ...747....2L} {747, 2}

\bibitem[\protect\citeauthoryear{{Maeder} \& {Meynet}}{{Maeder} \&
  {Meynet}}{2001}]{maeder:01}
{Maeder} A.,  {Meynet} G.,  2001, \mn@doi [\aap] {10.1051/0004-6361:20010596},
  \href {http://adsabs.harvard.edu/abs/2001A%26A...373..555M} {373, 555}

\bibitem[\protect\citeauthoryear{{Magkotsios}, {Timmes}, {Hungerford}, {Fryer},
  {Young}  \& {Wiescher}}{{Magkotsios} et~al.}{2010}]{magkotsios:10}
{Magkotsios} G.,  {Timmes} F.,  {Hungerford} A.,  {Fryer} C.,  {Young} P.,
  {Wiescher} M.,  2010, in APS Division of Nuclear Physics Meeting Abstracts.
  p.~H5

\bibitem[\protect\citeauthoryear{{Marigo}}{{Marigo}}{2001}]{marigo:01b}
{Marigo} P.,  2001, A\&A, 370, 194

\bibitem[\protect\citeauthoryear{{Mattsson}, {Wahlin}  \&
  {H{\"o}fner}}{{Mattsson} et~al.}{2010}]{mattsson:10}
{Mattsson} L.,  {Wahlin} R.,   {H{\"o}fner} S.,  2010, \mn@doi [\aap]
  {10.1051/0004-6361/200912084}, \href
  {http://adsabs.harvard.edu/abs/2010A%26A...509A..14M} {509, A14}

\bibitem[\protect\citeauthoryear{{Meakin} \& {Arnett}}{{Meakin} \&
  {Arnett}}{2006}]{meakin:06a}
{Meakin} C.~A.,  {Arnett} D.,  2006, \mn@doi [ApJ Lett.] {10.1086/500544},
  \href {http://adsabs.harvard.edu/abs/2006ApJ...637L..53M} {637, L53}

\bibitem[\protect\citeauthoryear{{Meakin} \& {Arnett}}{{Meakin} \&
  {Arnett}}{2007}]{meakin:07a}
{Meakin} C.~A.,  {Arnett} D.,  2007, \mn@doi [ApJ] {10.1086/519372}, \href
  {http://adsabs.harvard.edu/abs/2007ApJ...665..690M} {665, 690}

\bibitem[\protect\citeauthoryear{{Meyer}, {Clayton}  \& {The}}{{Meyer}
  et~al.}{2000}]{meyer:00}
{Meyer} B.~S.,  {Clayton} D.~D.,   {The} L.-S.,  2000, \mn@doi [\apjl]
  {10.1086/312865}, \href {http://adsabs.harvard.edu/abs/2000ApJ...540L..49M}
  {540, L49}

\bibitem[\protect\citeauthoryear{{Meynet} \& {Maeder}}{{Meynet} \&
  {Maeder}}{2002}]{meynet:02}
{Meynet} G.,  {Maeder} A.,  2002, A\&A, 390, 561

\bibitem[\protect\citeauthoryear{{Moll{\'a}}, {Cavichia}, {Gavil{\'a}n}  \&
  {Gibson}}{{Moll{\'a}} et~al.}{2015}]{molla:15}
{Moll{\'a}} M.,  {Cavichia} O.,  {Gavil{\'a}n} M.,   {Gibson} B.~K.,  2015,
  \mn@doi [\mnras] {10.1093/mnras/stv1102}, \href
  {http://adsabs.harvard.edu/abs/2015MNRAS.451.3693M} {451, 3693}

\bibitem[\protect\citeauthoryear{{M{\"u}ller}}{{M{\"u}ller}}{2016}]{mueller:16}
{M{\"u}ller} B.,  2016, \mn@doi [\pasa] {10.1017/pasa.2016.40}, \href
  {http://ukads.nottingham.ac.uk/abs/2016PASA...33...48M} {33, e048}

\bibitem[\protect\citeauthoryear{{Nomoto}, {Tominaga}, {Umeda}, {Kobayashi}  \&
  {Maeda}}{{Nomoto} et~al.}{2006}]{nomoto:06}
{Nomoto} K.,  {Tominaga} N.,  {Umeda} H.,  {Kobayashi} C.,   {Maeda} K.,  2006,
  \mn@doi [Nuclear Physics A] {10.1016/j.nuclphysa.2006.05.008}, \href
  {http://adsabs.harvard.edu/abs/2006NuPhA.777..424N} {777, 424}

\bibitem[\protect\citeauthoryear{{Nomoto}, {Kobayashi}  \& {Tominaga}}{{Nomoto}
  et~al.}{2013}]{nomoto:13}
{Nomoto} K.,  {Kobayashi} C.,   {Tominaga} N.,  2013, \mn@doi [\araa]
  {10.1146/annurev-astro-082812-140956}, \href
  {http://adsabs.harvard.edu/abs/2013ARA%26A..51..457N} {51, 457}

\bibitem[\protect\citeauthoryear{{Nugis} \& {Lamers}}{{Nugis} \&
  {Lamers}}{2000}]{nugis:00}
{Nugis} T.,  {Lamers} H.~J.~G.~L.~M.,  2000, \aap, \href
  {http://adsabs.harvard.edu/abs/2000A%26A...360..227N} {360, 227}

\bibitem[\protect\citeauthoryear{{Paxton}, {Bildsten}, {Dotter}, {Herwig},
  {Lesaffre}  \& {Timmes}}{{Paxton} et~al.}{2011}]{paxton:11}
{Paxton} B.,  {Bildsten} L.,  {Dotter} A.,  {Herwig} F.,  {Lesaffre} P.,
  {Timmes} F.,  2011, \mn@doi [\apjs] {10.1088/0067-0049/192/1/3}, \href
  {http://adsabs.harvard.edu/abs/2011ApJS..192....3P} {192, 3}

\bibitem[\protect\citeauthoryear{{Peters}}{{Peters}}{1968}]{peters:68}
{Peters} J.~G.,  1968, \mn@doi [\apj] {10.1086/149753}, \href
  {http://adsabs.harvard.edu/abs/1968ApJ...154..225P} {154, 225}

\bibitem[\protect\citeauthoryear{{Peters} \& {Hirschi}}{{Peters} \&
  {Hirschi}}{2013}]{peters:13}
{Peters} G.~J.,  {Hirschi} R.,  2013, {The Evolution of High-Mass Stars}.
p.~447, \mn@doi{10.1007/978-94-007-5615-1_9}

\bibitem[\protect\citeauthoryear{{Pignatari} \& {Gallino}}{{Pignatari} \&
  {Gallino}}{2007}]{pignatari:07}
{Pignatari} M.,  {Gallino} R.,  2007, \memsai, \href
  {http://adsabs.harvard.edu/abs/2007MmSAI..78..543P} {78, 543}

\bibitem[\protect\citeauthoryear{{Pignatari}, {Gallino}, {Heil}, {Wiescher},
  {K{\"a}ppeler}, {Herwig}  \& {Bisterzo}}{{Pignatari}
  et~al.}{2010}]{pignatari:10}
{Pignatari} M.,  {Gallino} R.,  {Heil} M.,  {Wiescher} M.,  {K{\"a}ppeler} F.,
  {Herwig} F.,   {Bisterzo} S.,  2010, \mn@doi [\apj]
  {10.1088/0004-637X/710/2/1557}, \href
  {http://adsabs.harvard.edu/abs/2010ApJ...710.1557P} {710, 1557}

\bibitem[\protect\citeauthoryear{{Pignatari} et~al.,}{{Pignatari}
  et~al.}{2013a}]{pignatari:13b}
{Pignatari} M.,  et~al., 2013a, \mn@doi [\apjl] {10.1088/2041-8205/767/2/L22},
  \href {http://adsabs.harvard.edu/abs/2013ApJ...767L..22P} {767, L22}

\bibitem[\protect\citeauthoryear{{Pignatari} et~al.,}{{Pignatari}
  et~al.}{2013b}]{pignatari:13c}
{Pignatari} M.,  et~al., 2013b, \mn@doi [\apjl] {10.1088/2041-8205/771/1/L7},
  \href {http://adsabs.harvard.edu/abs/2013ApJ...771L...7P} {771, L7}

\bibitem[\protect\citeauthoryear{{Pignatari} et~al.,}{{Pignatari}
  et~al.}{2015}]{pignatari:15}
{Pignatari} M.,  et~al., 2015, preprint, \href
  {http://adsabs.harvard.edu/abs/2015arXiv150609056P} {} (\mn@eprint {arXiv}
  {1506.09056})

\bibitem[\protect\citeauthoryear{{Pignatari}, {G{\"o}bel}, {Reifarth}  \&
  {Travaglio}}{{Pignatari} et~al.}{2016a}]{pignatari:16}
{Pignatari} M.,  {G{\"o}bel} K.,  {Reifarth} R.,   {Travaglio} C.,  2016a,
  \mn@doi [International Journal of Modern Physics E]
  {10.1142/S0218301316300034}, \href
  {http://adsabs.harvard.edu/abs/2016IJMPE..2530003P} {25, 1630003}

\bibitem[\protect\citeauthoryear{Pignatari et~al.,}{Pignatari
  et~al.}{2016b}]{pignatari:13a}
Pignatari M.,  et~al., 2016b, The Astrophysical Journal Supplement Series, 225,
  24

\bibitem[\protect\citeauthoryear{{Pignatari}, {Hoppe}, {Trappitsch}, {Fryer},
  {Timmes}, {Herwig}  \& {Hirschi}}{{Pignatari} et~al.}{2017}]{pignatari:17}
{Pignatari} M.,  {Hoppe} P.,  {Trappitsch} T.,  {Fryer} C.,  {Timmes} F.,
  {Herwig} F.,   {Hirschi} R.,  2017, Geochim.\ Cosmochim.\ Acta, 5, 6

\bibitem[\protect\citeauthoryear{{Portinari}, {Chiosi}  \&
  {Bressan}}{{Portinari} et~al.}{1998}]{portinari:98}
{Portinari} L.,  {Chiosi} C.,   {Bressan} A.,  1998, \aap, \href
  {http://adsabs.harvard.edu/abs/1998A%26A...334..505P} {334, 505}

\bibitem[\protect\citeauthoryear{{Prantzos}}{{Prantzos}}{2000}]{prantzos:00}
{Prantzos} N.,  2000, \mn@doi [\nar] {10.1016/S1387-6473(00)00058-0}, \href
  {http://adsabs.harvard.edu/abs/2000NewAR..44..303P} {44, 303}

\bibitem[\protect\citeauthoryear{{Prantzos}}{{Prantzos}}{2012}]{prantzos:12}
{Prantzos} N.,  2012, \mn@doi [\aap] {10.1051/0004-6361/201219043}, \href
  {http://adsabs.harvard.edu/abs/2012A%26A...542A..67P} {542, A67}

\bibitem[\protect\citeauthoryear{{Prantzos}, {Hashimoto}  \&
  {Nomoto}}{{Prantzos} et~al.}{1990}]{prantzos:90}
{Prantzos} N.,  {Hashimoto} M.,   {Nomoto} K.,  1990, \aap, \href
  {http://adsabs.harvard.edu/abs/1990A%26A...234..211P} {234, 211}

\bibitem[\protect\citeauthoryear{{Raiteri}, {Gallino}  \& {Busso}}{{Raiteri}
  et~al.}{1992}]{raiteri:92}
{Raiteri} C.~M.,  {Gallino} R.,   {Busso} M.,  1992, \mn@doi [\apj]
  {10.1086/171078}, \href {http://adsabs.harvard.edu/abs/1992ApJ...387..263R}
  {387, 263}

\bibitem[\protect\citeauthoryear{{Rauscher}, {Heger}, {Hoffman}  \&
  {Woosley}}{{Rauscher} et~al.}{2002}]{rauscher:02}
{Rauscher} T.,  {Heger} A.,  {Hoffman} R.~D.,   {Woosley} S.~E.,  2002, \mn@doi
  [\apj] {10.1086/341728}, \href
  {http://adsabs.harvard.edu/abs/2002ApJ...576..323R} {576, 323}

\bibitem[\protect\citeauthoryear{{Rauscher}, {Dauphas}, {Dillmann},
  {Fr{\"o}hlich}, {F{\"u}l{\"o}p}  \& {Gy{\"u}rky}}{{Rauscher}
  et~al.}{2013}]{rauscher:13}
{Rauscher} T.,  {Dauphas} N.,  {Dillmann} I.,  {Fr{\"o}hlich} C.,
  {F{\"u}l{\"o}p} Z.,   {Gy{\"u}rky} G.,  2013, \mn@doi [Reports on Progress in
  Physics] {10.1088/0034-4885/76/6/066201}, \href
  {http://adsabs.harvard.edu/abs/2013RPPh...76f6201R} {76, 066201}

\bibitem[\protect\citeauthoryear{{Rauscher}, {Nishimura}, {Hirschi},
  {Cescutti}, {Murphy}  \& {Heger}}{{Rauscher} et~al.}{2016}]{rauscher:16}
{Rauscher} T.,  {Nishimura} N.,  {Hirschi} R.,  {Cescutti} G.,  {Murphy}
  A.~S.~J.,   {Heger} A.,  2016, \mn@doi [\mnras] {10.1093/mnras/stw2266},
  \href {http://adsabs.harvard.edu/abs/2016MNRAS.463.4153R} {463, 4153}

\bibitem[\protect\citeauthoryear{{Rayet}, {Arnould}, {Hashimoto}, {Prantzos}
  \& {Nomoto}}{{Rayet} et~al.}{1995}]{rayet:95}
{Rayet} M.,  {Arnould} M.,  {Hashimoto} M.,  {Prantzos} N.,   {Nomoto} K.,
  1995, \aap, \href {http://adsabs.harvard.edu/abs/1995A%26A...298..517R} {298,
  517}

\bibitem[\protect\citeauthoryear{{Reddy}, {Lambert}  \& {Allende
  Prieto}}{{Reddy} et~al.}{2006}]{reddy:06}
{Reddy} B.~E.,  {Lambert} D.~L.,   {Allende Prieto} C.,  2006, VizieR Online
  Data Catalog, \href {http://adsabs.harvard.edu/abs/2006yCat..73671329R} {736,
  71329}

\bibitem[\protect\citeauthoryear{Reimers}{Reimers}{1975}]{reimers:75}
Reimers D.,  1975, Mem. Soc. Sci. Liege, 8, 369

\bibitem[\protect\citeauthoryear{{Ritossa}, {Garc{\' i}a-Berro}  \&
  {Iben}}{{Ritossa} et~al.}{1999}]{ritossa:99}
{Ritossa} C.,  {Garc{\' i}a-Berro} E.,   {Iben} I.~J.,  1999, ApJ, 515, 381

\bibitem[\protect\citeauthoryear{{Ritter}, {Andrassy}, {C{\^o}t{\'e}},
  {Herwig}, {Woodward}, {Pignatari}  \& {Jones}}{{Ritter}
  et~al.}{2017a}]{ritter:17b}
{Ritter} C.,  {Andrassy} R.,  {C{\^o}t{\'e}} B.,  {Herwig} F.,  {Woodward}
  P.~R.,  {Pignatari} M.,   {Jones} S.,  2017a, \mnras, \href
  {http://adsabs.harvard.edu/abs/2017arXiv170405985R} {}

\bibitem[\protect\citeauthoryear{Ritter, C{\^o}t{\'e}, Herwig, Navarro  \&
  Fryer}{Ritter et~al.}{2017b}]{Ritter:2017uy}
Ritter C.,  C{\^o}t{\'e} B.,  Herwig F.,  Navarro J.~F.,   Fryer C.,  2017b,
  ApJS

\bibitem[\protect\citeauthoryear{{Romano}, {Karakas}, {Tosi}  \&
  {Matteucci}}{{Romano} et~al.}{2010}]{romano:10}
{Romano} D.,  {Karakas} A.~I.,  {Tosi} M.,   {Matteucci} F.,  2010, \mn@doi
  [\aap] {10.1051/0004-6361/201014483}, \href
  {http://adsabs.harvard.edu/abs/2010A%26A...522A..32R} {522, A32}

\bibitem[\protect\citeauthoryear{{Rosenfield} et~al.,}{{Rosenfield}
  et~al.}{2014}]{rosenfield:14}
{Rosenfield} P.,  et~al., 2014, \mn@doi [\apj] {10.1088/0004-637X/790/1/22},
  \href {http://adsabs.harvard.edu/abs/2014ApJ...790...22R} {790, 22}

\bibitem[\protect\citeauthoryear{Sackmann \& Boothroyd}{Sackmann \&
  Boothroyd}{1992}]{sackmann:92}
Sackmann I.-J.,  Boothroyd A.~I.,  1992, ApJ, 392, L71

\bibitem[\protect\citeauthoryear{{Sbordone} et~al.,}{{Sbordone}
  et~al.}{2010}]{sbordone:10}
{Sbordone} L.,  et~al., 2010, \mn@doi [\aap] {10.1051/0004-6361/200913282},
  \href {http://adsabs.harvard.edu/abs/2010A%26A...522A..26S} {522, A26}

\bibitem[\protect\citeauthoryear{{Scalo}, {Despain}  \& {Ulrich}}{{Scalo}
  et~al.}{1975}]{scalo:75}
{Scalo} J.~M.,  {Despain} K.~H.,   {Ulrich} R.~K.,  1975, ApJ, \href
  {http://adsabs.harvard.edu/cgi-bin/nph-bib_query?bibcode=1975ApJ...196..805S&db_key=AST}
  {196, 805}

\bibitem[\protect\citeauthoryear{{Scannapieco}, {Tissera}, {White}  \&
  {Springel}}{{Scannapieco} et~al.}{2005}]{scannapieco:05}
{Scannapieco} C.,  {Tissera} P.~B.,  {White} S.~D.~M.,   {Springel} V.,  2005,
  \mn@doi [\mnras] {10.1111/j.1365-2966.2005.09574.x}, \href
  {http://adsabs.harvard.edu/abs/2005MNRAS.364..552S} {364, 552}

\bibitem[\protect\citeauthoryear{{Schaye} et~al.,}{{Schaye}
  et~al.}{2015}]{schaye:15}
{Schaye} J.,  et~al., 2015, \mn@doi [\mnras] {10.1093/mnras/stu2058}, \href
  {http://adsabs.harvard.edu/abs/2015MNRAS.446..521S} {446, 521}

\bibitem[\protect\citeauthoryear{{Siess}}{{Siess}}{2007}]{siess:07}
{Siess} L.,  2007, \mn@doi [\aap] {10.1051/0004-6361:20078132}, \href
  {http://adsabs.harvard.edu/abs/2007A%26A...476..893S} {476, 893}

\bibitem[\protect\citeauthoryear{{Siess}}{{Siess}}{2010}]{siess:10}
{Siess} L.,  2010, \mn@doi [\aap] {10.1051/0004-6361/200913556}, \href
  {http://adsabs.harvard.edu/abs/2010A%26A...512A..10S} {512, A10}

\bibitem[\protect\citeauthoryear{Straniero, Gallino, Busso, Chieffi, Raiteri,
  Salaris  \& Limongi}{Straniero et~al.}{1995}]{straniero:95}
Straniero O.,  Gallino R.,  Busso M.,  Chieffi A.,  Raiteri C.~M.,  Salaris M.,
    Limongi M.,  1995, ApJ, 440, L85

\bibitem[\protect\citeauthoryear{{Straniero}, {Cristallo}  \&
  {Piersanti}}{{Straniero} et~al.}{2014}]{straniero:14}
{Straniero} O.,  {Cristallo} S.,   {Piersanti} L.,  2014, \mn@doi [\apj]
  {10.1088/0004-637X/785/1/77}, \href
  {http://adsabs.harvard.edu/abs/2014ApJ...785...77S} {785, 77}

\bibitem[\protect\citeauthoryear{{Sukhbold} \& {Woosley}}{{Sukhbold} \&
  {Woosley}}{2014}]{sukhbold:14}
{Sukhbold} T.,  {Woosley} S.~E.,  2014, \mn@doi [\apj]
  {10.1088/0004-637X/783/1/10}, \href
  {http://adsabs.harvard.edu/abs/2014ApJ...783...10S} {783, 10}

\bibitem[\protect\citeauthoryear{{Sukhbold}, {Ertl}, {Woosley}, {Brown}  \&
  {Janka}}{{Sukhbold} et~al.}{2016}]{sukhbold:16}
{Sukhbold} T.,  {Ertl} T.,  {Woosley} S.~E.,  {Brown} J.~M.,   {Janka} H.-T.,
  2016, \mn@doi [\apj] {10.3847/0004-637X/821/1/38}, \href
  {http://ukads.nottingham.ac.uk/abs/2016ApJ...821...38S} {821, 38}

\bibitem[\protect\citeauthoryear{{The}, {El Eid}  \& {Meyer}}{{The}
  et~al.}{2007}]{the:07}
{The} L.,  {El Eid} M.~F.,   {Meyer} B.~S.,  2007, \mn@doi [ApJ]
  {10.1086/509753}, \href {http://adsabs.harvard.edu/abs/2007ApJ...655.1058T}
  {655, 1058}

\bibitem[\protect\citeauthoryear{{Thielemann}, {Arnould}  \&
  {Hillebrandt}}{{Thielemann} et~al.}{1979}]{thielemann:79}
{Thielemann} F.-K.,  {Arnould} M.,   {Hillebrandt} W.,  1979, \aap, \href
  {http://adsabs.harvard.edu/abs/1979A%26A....74..175T} {74, 175}

\bibitem[\protect\citeauthoryear{{Trippella}, {Busso}, {Palmerini}, {Maiorca}
  \& {Nucci}}{{Trippella} et~al.}{2016}]{trippella:16}
{Trippella} O.,  {Busso} M.,  {Palmerini} S.,  {Maiorca} E.,   {Nucci} M.~C.,
  2016, \mn@doi [\apj] {10.3847/0004-637X/818/2/125}, \href
  {http://adsabs.harvard.edu/abs/2016ApJ...818..125T} {818, 125}

\bibitem[\protect\citeauthoryear{{Ugliano}, {Janka}, {Marek}  \&
  {Arcones}}{{Ugliano} et~al.}{2012}]{ugliano:12}
{Ugliano} M.,  {Janka} H.-T.,  {Marek} A.,   {Arcones} A.,  2012, \mn@doi
  [\apj] {10.1088/0004-637X/757/1/69}, \href
  {http://adsabs.harvard.edu/abs/2012ApJ...757...69U} {757, 69}

\bibitem[\protect\citeauthoryear{Vassiliadis \& Wood}{Vassiliadis \&
  Wood}{1993}]{vassiliadis:93}
Vassiliadis E.,  Wood P.,  1993, ApJ, 413, 641

\bibitem[\protect\citeauthoryear{{Ventura} \& {D'Antona}}{{Ventura} \&
  {D'Antona}}{2011}]{ventura:11}
{Ventura} P.,  {D'Antona} F.,  2011, \mn@doi [\mnras]
  {10.1111/j.1365-2966.2010.17651.x}, \href
  {http://adsabs.harvard.edu/abs/2011MNRAS.410.2760V} {410, 2760}

\bibitem[\protect\citeauthoryear{{Ventura}, {Di Criscienzo}, {Carini}  \&
  {D'Antona}}{{Ventura} et~al.}{2013}]{ventura:13}
{Ventura} P.,  {Di Criscienzo} M.,  {Carini} R.,   {D'Antona} F.,  2013,
  \mn@doi [\mnras] {10.1093/mnras/stt444}, \href
  {http://adsabs.harvard.edu/abs/2013MNRAS.431.3642V} {431, 3642}

\bibitem[\protect\citeauthoryear{{Ventura}, {Karakas}, {Dell'Agli}, {Boyer},
  {Garc{\'{\i}}a-Hern{\'a}ndez}, {Di Criscienzo}  \& {Schneider}}{{Ventura}
  et~al.}{2015}]{ventura:15}
{Ventura} P.,  {Karakas} A.~I.,  {Dell'Agli} F.,  {Boyer} M.~L.,
  {Garc{\'{\i}}a-Hern{\'a}ndez} D.~A.,  {Di Criscienzo} M.,   {Schneider} R.,
  2015, \mn@doi [\mnras] {10.1093/mnras/stv918}, \href
  {http://adsabs.harvard.edu/abs/2015MNRAS.450.3181V} {450, 3181}

\bibitem[\protect\citeauthoryear{{Vink}, {de Koter}  \& {Lamers}}{{Vink}
  et~al.}{2001}]{vink:01}
{Vink} J.~S.,  {de Koter} A.,   {Lamers} H.~J.~G.~L.~M.,  2001, \mn@doi [\aap]
  {10.1051/0004-6361:20010127}, \href
  {http://adsabs.harvard.edu/abs/2001A%26A...369..574V} {369, 574}

\bibitem[\protect\citeauthoryear{{Weiss} \& {Ferguson}}{{Weiss} \&
  {Ferguson}}{2009}]{weiss:09}
{Weiss} A.,  {Ferguson} J.~W.,  2009, \mn@doi [\aap]
  {10.1051/0004-6361/200912043}, \href
  {http://adsabs.harvard.edu/abs/2009A%26A...508.1343W} {508, 1343}

\bibitem[\protect\citeauthoryear{{Willson}}{{Willson}}{2000}]{willson:00}
{Willson} L.~A.,  2000, \mn@doi [\araa] {10.1146/annurev.astro.38.1.573}, \href
  {http://adsabs.harvard.edu/abs/2000ARA%26A..38..573W} {38, 573}

\bibitem[\protect\citeauthoryear{{Woosley} \& {Hoffman}}{{Woosley} \&
  {Hoffman}}{1992}]{woosley:92}
{Woosley} S.~E.,  {Hoffman} R.~D.,  1992, \mn@doi [\apj] {10.1086/171644},
  \href {http://adsabs.harvard.edu/abs/1992ApJ...395..202W} {395, 202}

\bibitem[\protect\citeauthoryear{{Woosley} \& {Howard}}{{Woosley} \&
  {Howard}}{1978}]{woosley:78}
{Woosley} S.~E.,  {Howard} W.~M.,  1978, \mn@doi [\apjs] {10.1086/190501},
  \href {http://adsabs.harvard.edu/abs/1978ApJS...36..285W} {36, 285}

\bibitem[\protect\citeauthoryear{{Woosley} \& {Weaver}}{{Woosley} \&
  {Weaver}}{1995}]{woosley:95}
{Woosley} S.~E.,  {Weaver} T.~A.,  1995, APJS, 101, 181+

\bibitem[\protect\citeauthoryear{{Woosley}, {Arnett}  \& {Clayton}}{{Woosley}
  et~al.}{1973}]{woosley:73}
{Woosley} S.~E.,  {Arnett} W.~D.,   {Clayton} D.~D.,  1973, \mn@doi [\apjs]
  {10.1086/190282}, \href {http://adsabs.harvard.edu/abs/1973ApJS...26..231W}
  {26, 231}

\bibitem[\protect\citeauthoryear{Woosley, Heger  \& Weaver}{Woosley
  et~al.}{2002}]{woosley:02}
Woosley S.~E.,  Heger A.,   Weaver T.~A.,  2002, Rev.\ Mod.\ Phys., 74, 1015

\bibitem[\protect\citeauthoryear{{Young}, {Ellinger}, {Arnett}, {Fryer}  \&
  {Rockefeller}}{{Young} et~al.}{2009}]{young:09}
{Young} P.~A.,  {Ellinger} C.~I.,  {Arnett} D.,  {Fryer} C.~L.,   {Rockefeller}
  G.,  2009, \mn@doi [\apj] {10.1088/0004-637X/699/2/938}, \href
  {http://adsabs.harvard.edu/abs/2009ApJ...699..938Y} {699, 938}

\bibitem[\protect\citeauthoryear{{Zinner}}{{Zinner}}{2014}]{zinner:14}
{Zinner} E.,  2014, \mn@doi [Treatise on Geochemistry, 2nd edition]
  {10.1016/B978-0-08-095975-7.00101-7}, 1, 181

\bibitem[\protect\citeauthoryear{{de Jager}, {Nieuwenhuijzen}  \& {van der
  Hucht}}{{de Jager} et~al.}{1988}]{dejager:88}
{de Jager} C.,  {Nieuwenhuijzen} H.,   {van der Hucht} K.~A.,  1988, \aaps,
  \href {http://adsabs.harvard.edu/abs/1988A%26AS...72..259D} {72, 259}

\bibitem[\protect\citeauthoryear{{van Loon}}{{van Loon}}{2000}]{vanloon:00}
{van Loon} J.~T.,  2000, A\&A, \href
  {http://adsabs.harvard.edu/cgi-bin/nph-bib_query?bibcode=2000A%26A...354..125V&db_key=AST}
  {354, 125}

\bibitem[\protect\citeauthoryear{{van Loon}, {Cioni}, {Zijlstra}  \&
  {Loup}}{{van Loon} et~al.}{2005}]{vanloon:05b}
{van Loon} J.~T.,  {Cioni} M.-R.~L.,  {Zijlstra} A.~A.,   {Loup} C.,  2005,
  \mn@doi [A\&A] {10.1051/0004-6361:20042555}, \href
  {http://adsabs.harvard.edu/cgi-bin/nph-bib_query?bibcode=2005A%26A...438..273V&db_key=AST}
  {438, 273}

\makeatother
\end{thebibliography}

%\begin{thebibliography}{99}
%\end{thebibliography}

%%%%%%%%%%%%%%%%%%%%%%%%%%%%%%%%%%%%%%%%%%%%%%%%%%

%%%%%%%%%%%%%%%%% APPENDICES %%%%%%%%%%%%%%%%%%%%%

\appendix
\section{Data access}
\label{sec:appendix}

\begin{figure}
\centering
\includegraphics[width=0.46\textwidth]{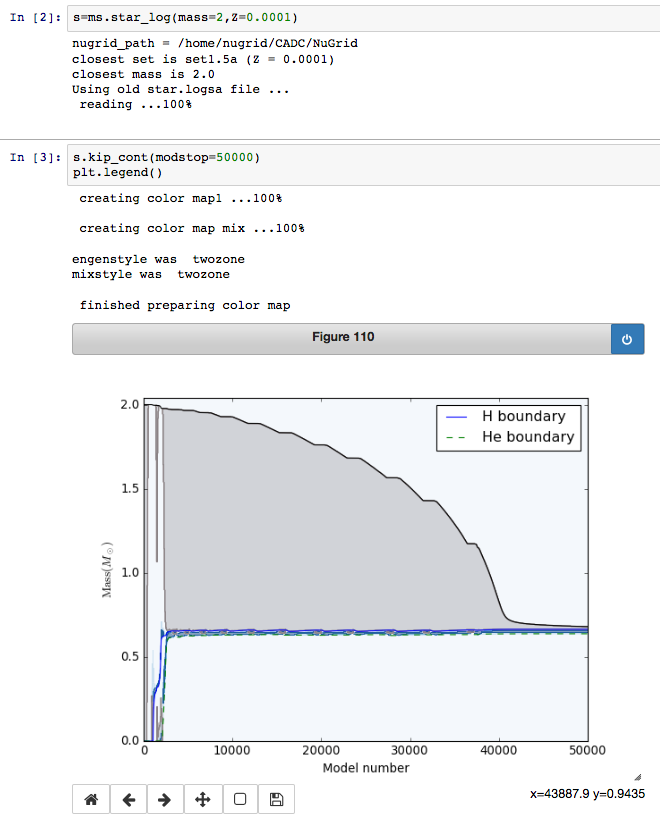}
\caption{Example of plotting a Kippenhahn diagram in the \textsc{WENDI} web exploration interface that provides Jupyter notebook-based analytic access to the NuGrid stellar evolution and nucleosynthesis data.}
\label{fig:wendi_interface}
\end{figure}

\noindent The NuGrid extended NuGrid Set 1 data (also refered to as set1 extension, or set1ext data) has been deposited at the Canadian Astronomical Data Center, DOI:10.11570/18.0002 (\url{http://www.canfar.phys.uvic.ca/vospace/nodes/AstroDataCitationDOI/CISTI.CANFAR/18.0002/18.0002.html?view=data}). All stellar evolution and post-processing data is accessible online
through NuGrid's \textsc{WENDI} interface at \url{http://wendi.nugridstars.org}.
%We provide widgets to interact with the data as well as plotting functions
%as shown in \fig{fig:wendi_interface}.
\textsc{WENDI} is a Cyberhubs application \citep{Herwig:2018gx}. Ipython notebooks allow analyzing the data via the command line and with plotting and data analytics
functions of NuGrid's python package NuGridPy (\fig{fig:wendi_interface}, \url{https://nugrid.github.io/NuGridPy}). 
NuGridPy provides various functions to read and analyze \MESA\ stellar evolution data as well as NuGrid post-processing data.
NuGridPy is available via the package manager pip and the source code and documentation is available on GitHub 
\url{https://github.com/NuGrid/NuGridPy}. 

For all stellar evolution tracks all profiles of $\rho$, $T$, $D$, $r$ and $m_\mathrm{r}$ are available for all time steps, as well as multiple scalar quantities as a function of model number, such as $T_\mathrm{eff}$, values of the stellar center, mass coordinates of H-, He-free cores and many others. For nucleosynthesis post-processing data complete isotopic profiles are available every 20 time steps.

In addition to plotting and analyzing the stellar evolution, nucleosynthesis and yield data online in the \textsc{WENDI} platform the raw data is also accessible online at \url{http://nugridstars.org/data-and-software/yields/set-1}. It can be read and analysed with the NuGridPy tools mentioned above. 
Stellar yield tables as shown in \tab{tab:yields_all_delay_set1_5} are provided for
all metallicities. Separate tables are available for contribution from stellar winds only, from winds plus pre-SN ejecta
and winds plus SN ejecta. Figures of overproduction factors for all elements and many isotopes such as in \fig{fig:cno_prodfac_zdep}
and in \fig{fig:wspr_prodfacs} are available for all metallicities. Stellar yields can be further applied and used in galactic chemical evolution models via the \emph{Stellar Yields for Galactic Modeling Applications} (\textsc{SYGMA}) Python code \citep{Ritter:2017uy} that is part of the \emph{NuGrid Python Chemical Evolution Environment} \textsc{NuPyCEE} package 
\citep[\url{http://nugrid.github.io/NuPyCEE}][]{Cote:2016iia}.

%%%%%%%%%%%%%%%%%%%%%%%%%%%%%%%%%%%%%%%%%%%%%%%%%%

% Don't change these lines
\bsp	% typesetting comment
\label{lastpage}
\end{document}